\documentclass[aps,prb,reprint,superscriptaddress,amsmath,amssymb,nofootinbib,floatfix]{revtex4-2}

\usepackage[T1]{fontenc}
\usepackage[utf8]{inputenc}
\usepackage{microtype}
\usepackage{amsmath,amssymb}
\usepackage{mathptmx}
\usepackage{amsthm}
\usepackage{graphicx}
\usepackage{xcolor}
\usepackage{tikz}
\usetikzlibrary{arrows.meta,calc,backgrounds}
\usepackage[hidelinks]{hyperref}

\newcommand{\ket}[1]{\lvert #1\rangle}
\newcommand{\bra}[1]{\langle #1\rvert}
\newcommand{\id}{\mathbf{1}}
\DeclareMathOperator{\End}{End}
\newtheorem{theorem}{Theorem}[section]
\newtheorem{corollary}[theorem]{Corollary}
\newtheorem{proposition}[theorem]{Proposition}

\hypersetup{
  pdftitle={A Universal Algebraic Construction of Hilbert-Space Fragmentation},
  pdfauthor={Pramod Padmanabhan and Vladimir Korepin}
}

\makeatletter
\let\auto@bib@innerbib\@empty
\makeatother

\begin{document}

\title{A Universal Algebraic Construction of Hilbert-Space Fragmentation}
\author{Pramod Padmanabhan}
\email{pramod23phys@gmail.com}
\affiliation{Department of Physics, School of Basic Sciences,
Indian Institute of Technology, Bhubaneswar, 752050, India}
\author{Vladimir Korepin}
\email{vladimir.korepin@stonybrook.edu}
\affiliation{C. N. Yang Institute for Theoretical Physics,
Stony Brook University, New York 11794, USA}

\begin{abstract}
We present a mechanism for obtaining Hilbert-space fragmentation from arbitrary quantum lattice Hamiltonians acting on finite-dimensional local spaces. We first express the Hamiltonian in an algebraic form using supersymmetry (SUSY) and parasupersymmetry (paraSUSY) generators. Enlarging their representations defines a lifting procedure, which we study in two forms, site-wise and bond-wise lifting. Site-wise lifting introduces immobile zero modes that generate exponentially many invariant sectors while retaining the original dynamics in the fully active sector. We determine the sector structure and establish strong fragmentation for uniform lifts at fixed intermediate zero-mode density. For nearest-neighbor chains, correlation functions reveal restricted propagation and, in the mixed-field example, persistent local memory. Bond-wise lifting produces mobile conserved strings, pair-flip dynamics and fluctuating dimers, and extends to fragmented unitary circuits. Different representations and embeddings realize different fragmentation mechanisms. Some of the fragmented models obtained by this method include lifted permutation and Ising spin chains, a lifted toric code, the \(XXC\) families of spin chains, and fragmented extensions derived from the integrable \(H_{1,11}\) and \(H_{1,3}\) seed densities. The last two models are derived from Hietarinta's classification of constant $4\times 4$ Yang-Baxter solutions.
\end{abstract}

\maketitle
\tableofcontents

\section{Introduction}
\label{sec:introduction}
How an isolated quantum system approaches thermal equilibrium depends on which parts of its Hilbert space the dynamics can explore. For generic interacting systems, the eigenstate thermalization hypothesis relates this relaxation to the properties of individual energy eigenstates~\cite{Deutsch1991,Srednicki1994}. Hilbert-space fragmentation (HSF) provides a different possibility, where the allowed interactions divide the Hilbert space into exponentially many dynamically disconnected sectors, often even after conventional global charges have been fixed~\cite{Sala2020,Khemani2020}. A state initially supported in one sector remains there, although it may still thermalize within that sector~\cite{Moudgalya2021}. Fragmentation therefore offers a setting in which persistent memory, restricted transport and sector-dependent thermalization can be studied without requiring disorder or integrability. Its relation to other departures from thermal behaviour is reviewed in~\cite{MoudgalyaReview2022}.

Several mechanisms now associated with HSF appeared before the modern terminology. An early example is the diffusing-dimer model, whose configurations split into exponentially many sectors labeled by conserved \emph{irreducible strings}. The latter are sequences obtained by deleting locally removable pairs~\cite{Menon1997}. The $XXC$ spin chains provide a related mechanism involving mobile internal labels~\cite{Maassarani1998}. Their exchange processes move states belonging to two sets through one another while preserving the ordering of labels within each set, up to cyclic identification on a ring. A different route to such behavior was shown explicitly in the spin-$1$ Ising model with single-ion anisotropy. Here immobile $S^z=0$ states, called holes, separate the chain into independent spin-$1/2$ segments~\cite{Yang2008}. Conserved strings also organize the invariant subspaces of the pair-flip chain, where local moves replace adjacent equal labels~\cite{Caha2018}.

The modern formulation of HSF emerged prominently in work on dipole-conserving quantum circuits and Hamiltonians~\cite{Pai2019,Sala2020,Khemani2020}. Here, locality together with conservation of charge and dipole moment forbids enough transitions to disconnect configurations that share the same global charges. Related work further described fragmentation through statistically localized integrals of motion, which label sectors even though the corresponding operators need not have fixed local support~\cite{Rakovszky2020}. Other restrictions on motion lead to fragmentation in the strong-interaction limit of fermionic chains~\cite{DeTomasi2019} and through strict confinement of Ising domain walls~\cite{Yang2020}. The folded $XXZ$ chain provides an integrable example with exponentially many jammed configurations and an exactly accessible sector structure~\cite{Zadnik2021,Pozsgay2021}.

Frustrated spin models supply examples in which locally conserved spin quantum numbers produce exponentially many invariant sectors~\cite{McClarty2020,Hahn2021}. Geometric frustration can also generate fragmentation within constrained low-energy descriptions, as in kagome spin models~\cite{Lee2021}. Beyond one dimension, ring-exchange models with conserved magnetization along rows and columns exhibit fragmentation and inactive structures that shield dynamically active regions~\cite{Khudorozhkov2022}. Further examples include prethermal fragmentation in the two-dimensional quantum Ising model~\cite{Hart2022} and the effective dynamics of strongly tilted Bose--Hubbard systems~\cite{Will2024}. These settings show how the geometry of the lattice and the allowed interaction supports shape the disconnected sectors.

An algebraic formulation characterized fragmentation through the exponential growth of the \emph{commutant}, the algebra of operators commuting with every specified interaction term, or with every gate in a circuit family~\cite{MoudgalyaMotrunich2022}. This formulation treats Hamiltonian and Floquet systems on the same footing. It also identified quantum fragmentation in Temperley--Lieb chains, whose invariant structure involves entangled states and non-Abelian commutants. A complementary framework uses local rewriting rules for words built from group or semigroup generators, connecting disconnected sectors with computational complexity and slow relaxation~\cite{Balasubramanian2024}.

Other constructions used flat bands and local symmetries to produce conserved charges or parities associated with compact localized states~\cite{Danieli2022,Nicolau2023}. These are single-particle eigenstates confined to finitely many sites by destructive interference. Their construction and relation to flat bands were developed by Flach and collaborators~\cite{Flach2014,Maimaiti2017}. More recent examples include hard-core constrained Hubbard models, where frozen particle arrangements block motion~\cite{Kwan2025}. Robustness has also been studied through prethermal conservation laws that protect frozen configurations to all finite orders in perturbation theory~\cite{Stephen2024}, and through group-valued loop models, where local moves, group relations and lattice topology obstruct transitions between sectors~\cite{Khudorozhkov2025}. These examples illustrate distinct routes to fragmentation rather than an exhaustive classification.

The experimental literature now spans several platforms. Tilted Fermi--Hubbard chains exhibit long-lived memory associated with emergent kinetic constraints~\cite{Scherg2021}. Quantum gas microscopy has revealed fragmentation and restricted defect motion in a two-dimensional tilted Bose--Hubbard system~\cite{Adler2024}, while superconducting processors have probed initial-state-dependent dynamics in Stark systems~\cite{Wang2025}. Rydberg arrays have demonstrated thermalization restricted to individual fragments~\cite{Zhao2025} and signatures of statistical localization in a constrained lattice gauge theory~\cite{Datla2026}. These experiments probe the dynamical consequences of fragmentation in regimes where the relevant constraints are realized to a controlled approximation.

It is useful to classify the different mechanisms of HSF according to their consequences. The first distinction is between \emph{strong} and \emph{weak} fragmentation. This distinction comes from the fraction occupied by the largest dynamical block~\cite{Sala2020}. We discuss the precise criterion in Sec.~\ref{sec:examples}. The next distinction is between \emph{classical} and \emph{quantum} fragmentation. This concerns the structure of the invariant sectors and their commutant. In the classical case the sector labels can be diagonal in a local product basis, whereas quantum fragmentation can require an entangled basis~\cite{MoudgalyaMotrunich2022}. Finally, {\it exact fragmentation} follows from exact conservation laws, while {\it emergent fragmentation} in an effective description can persist only over a finite, possibly long, time window~\cite{Khemani2020,Scherg2021}.

The mechanisms closest to the present work are those in which local conserved subspaces retain a record of the initial configuration and, in one dimension, can interrupt the active dynamics. The hole decomposition of Ref.~\cite{Yang2008} makes this picture especially transparent. Mixed diamond chains provide a related example, where conserved singlet dimers separate independent active spin clusters~\cite{Takano2009,Hida2009}. Another useful comparison is the analysis of $XXC$ fragmentation in Ref.~\cite{Katsura2025}. Here the relative ordering of mobile labels defines conserved irreducible strings, but the label at an individual site can change. That work also constructs perturbations that remove fragmentation while retaining isolated integrable sectors. In the site-wise construction developed here, the active or inactive character of each enlarged site is conserved. Our objective is to supply these conserved subspaces systematically for a prescribed seed Hamiltonian, including Hamiltonians whose interaction terms have only a scalar common commutant.

Our approach uses SUSY and paraSUSY representation lifting to construct fragmented Hamiltonians from prescribed seed interactions. We first give a complete sector analysis for a specified class of site-wise representations, where common local zero modes enforce conserved spatial patterns. We then realize an entire bond operator algebra through parasupercharges and enlarge its representation. In this bond-wise construction, the embedding into the physical tensor-product space affects how neighboring interactions overlap and which quantities they conserve. Section~\ref{sec:bondwise} develops explicit examples with mobile ordered words, pair-reduction strings and frozen unpaired symbols. The representation choice therefore supplies a common construction for distinct fragmentation mechanisms, while the scope of each conservation law must be established for the chosen embedding.

We will now outline this site-wise construction. We begin by expressing every local operator (operator acting on a site carrying the local Hilbert space) through supercharges or parasupercharges and their adjoints. This is always possible as supercharges or parasupercharges generate all possible operators acting on a finite dimensional space. Thus our starting point is this completeness property. We then enlarge the representations of these same generators by retaining active copies and adjoining a common zero-mode space. Substituting the enlarged representations into the operator expressions produces the lifted interactions. 

First we show the completeness of the local operator algebra realized using supercharges or parasupercharges according to the dimension of the finite-dimensional space. On $\mathbb{C}^{2}$, take the supercharge $q=\ket{0}\bra{1}$ and its adjoint. They realize the $\mathcal N=2$ supersymmetry (SUSY) algebra
\begin{equation}
 q^2=(q^\dagger)^2=0,
 \qquad \{q,q^\dagger\}=h_1,
 \qquad [h_1,q]=[h_1,q^\dagger]=0.
 \label{eq:intro-susy}
\end{equation}
Here $\mathcal N=2$ counts the two Hermitian supercharges, denoted by $\mathcal{Q}_1=q+q^\dagger$ and $\mathcal{Q}_2=i(q^\dagger-q)$, which obey $\{\mathcal Q_\alpha,\mathcal Q_\beta\}=2\delta_{\alpha\beta}h_1$. This is the algebra of supersymmetric quantum mechanics, developed in Witten's seminal work~\cite{Witten1981,Witten1982}; detailed accounts are available in Refs.~\cite{Cooper1995,Cooper2001}. In the present representation, $b=qq^\dagger$ and $f=q^\dagger q$ project onto $\ket{0}$ and $\ket{1}$, respectively, and the SUSY Hamiltonian is $h_1=b+f=\id_2$. The four operators $q$, $q^\dagger$, $b$ and $f$ form a basis of $\End(\mathbb{C}^{2})$. Consequently, every spin-$1/2$ operator can be expressed through these generators.

For a local space $\mathbb{C}^{n}$, the corresponding extension uses parasupersymmetry (paraSUSY) of order $p=n-1$. We use the Rubakov--Spiridonov--Khare form~\cite{Rubakov1988,Khare1992}, with a nilpotent parasupercharge $Q$ and a Hermitian Hamiltonian $h_p$ satisfying
\begin{equation}
 \begin{gathered}
 Q^{p+1}=0,\qquad [h_p,Q]=[h_p,Q^\dagger]=0,\\
 \sum_{k=0}^{p}Q^{p-k}Q^\dagger Q^k
       =p\,Q^{p-1}h_p,
 \end{gathered}
 \label{eq:intro-parasusy}
\end{equation}
and the adjoint relation. Our normalization reduces to Eq.~\eqref{eq:intro-susy} at $p=1$.\footnote{When the right-hand side of the paraSUSY relation is written as $2p\,Q^{p-1}H_{\mathrm{PSUSY}}$, our convention is $h_p=2H_{\mathrm{PSUSY}}$.} Thus the local operators of a spin-$1$ seed Hamiltonian use an order-two paraSUSY realization, with $Q^3=0$.

The completeness statement can be seen without assuming any property of the dynamics of the seed Hamiltonian. The normalized supercharge for $n=2$, or parasupercharge for $n>2$, is
\begin{equation}
 Q_n=\sum_{a=0}^{n-2}\ket{a}\bra{a+1},
 \qquad
 \Pi_0=Q_n^{n-1}(Q_n^\dagger)^{n-1}=\ket{0}\bra{0}.
 \label{eq:intro-ladder}
\end{equation}
$Q_n$ and its adjoint satisfy Eq.~\eqref{eq:intro-parasusy} with $p=n-1$ and $h_p=\id_n$. Thus the identity operator on the seed local Hilbert space can itself be viewed as the normalized SUSY or paraSUSY Hamiltonian. This local algebraic Hamiltonian is distinct from the many-body seed Hamiltonian $H$, whose dynamics we wish to lift. Every matrix unit is generated as
\begin{equation}
 E_{ab}=(Q_n^\dagger)^a\Pi_0 Q_n^b=\ket{a}\bra{b},
 \qquad 0\leq a,b<n.
 \label{eq:intro-matrix-units}
\end{equation}
Their linear combinations exhaust $\End(\mathbb{C}^{n})$, and their tensor products exhaust the operators on any finite set of sites. This completeness makes arbitrary Hamiltonian densities accessible to the construction.

\begin{figure*}[tp]
\centering
\begingroup
\definecolor{hsfFlowBlue}{HTML}{216B8B}
\definecolor{hsfFlowBlueWash}{HTML}{EDF5F8}
\definecolor{hsfFlowOrange}{HTML}{B36D18}
\definecolor{hsfFlowOrangeWash}{HTML}{FFF5E7}
\definecolor{hsfFlowInk}{HTML}{202A32}
\begin{tikzpicture}[
  x=1cm,y=1cm,font=\fontsize{10}{12}\selectfont,text=hsfFlowInk,
  >={Triangle[length=2.8mm,width=2.6mm]},
  hsfstep/.style={draw=hsfFlowBlue!60,fill=hsfFlowBlueWash,
    rounded corners=3pt,line width=.65pt,align=center,
    text width=12.8cm,minimum height=1cm,inner sep=4pt},
  outcome/.style={rounded corners=3pt,line width=.8pt,align=center,
    text width=6.7cm,minimum height=1.1cm,inner sep=4pt},
  flow/.style={->,draw=hsfFlowInk,line width=1.1pt,
    shorten <=1pt,shorten >=1.5pt}
]
\node[hsfstep] (input) at (0,0)
  {\textbf{1. Specify the seed Hamiltonian and interaction supports}\\[2pt]
   $\{h_\alpha,X_\alpha\}$ on local spaces $\mathcal H_i=\mathbb C^{n_i}$};
\node[hsfstep] (algebra) at (0,-1.75)
  {\textbf{2. Express local operators through SUSY or paraSUSY}\\[2pt]
   Supercharges for $n_i=2$, parasupercharges for $n_i>2$};
\node[hsfstep] (representation) at (0,-3.5)
  {\textbf{3. Enlarge the local representations}\\[2pt]
   $\phi_i(O)=(O\otimes\id_{r_i})\oplus0_{s_i},\qquad r_i,s_i\geq1$};
\node[hsfstep] (support) at (0,-5.25)
  {\textbf{4. Lift each density on its declared support}\\[2pt]
   Apply $\bigotimes_{i\in X_\alpha}\phi_i$ and retain full spectator identities};
\node[hsfstep] (conservation) at (0,-7.0)
  {\textbf{5. Obtain conserved local zero-mode projectors}\\[2pt]
   $[Z_i,\widetilde h_\alpha]=0$ for every site $i$ and interaction $\alpha$};
\foreach \a/\b in {input/algebra,algebra/representation,representation/support,support/conservation}
  \draw[flow] (\a.south) -- (\b.north);
\node[outcome,draw=hsfFlowOrange!75,fill=hsfFlowOrangeWash] (patterns) at (-3.75,-9.05)
  {\textbf{Exponential pattern fragmentation}\\[3pt]
   $\widetilde{\mathcal H}=\displaystyle\bigoplus_{A\subseteq\Lambda}\widetilde{\mathcal H}_A$
   with $2^L$ invariant patterns};
\node[outcome,draw=hsfFlowBlue!75,fill=hsfFlowBlueWash] (dynamics) at (3.75,-9.05)
  {\textbf{Inherited seed dynamics}\\[3pt]
   The fully active sector contains\\copies of the original Hamiltonian};
\coordinate (fork) at (0,-7.85);
\draw[draw=hsfFlowInk,line width=1.1pt,shorten <=1pt] (conservation.south) -- (fork);
\draw[flow,shorten <=0pt] (fork) -| (patterns.north);
\draw[flow,shorten <=0pt] (fork) -| (dynamics.north);
\end{tikzpicture}
\endgroup
\caption{The site-wise SUSY/paraSUSY lifting algorithm. At site $i$, paraSUSY has order $n_i-1$ when $n_i>2$. The enlarged space is $(\mathbb C^{n_i}\otimes\mathbb C^{r_i})\oplus\mathbb C^{s_i}$, and $Z_i=\id_{n_ir_i+s_i}-\phi_i(\id_{n_i})$ projects onto its zero-mode subspace. Taking $r_i,s_i\geq1$ at all $L$ sites gives the $2^L$ nonzero active-site patterns shown. The fully active sector carries $\prod_i r_i$ copies of the seed Hamiltonian. The construction allows arbitrary interaction supports and lattice geometry.}
\label{fig:lifting-algorithm}
\end{figure*}
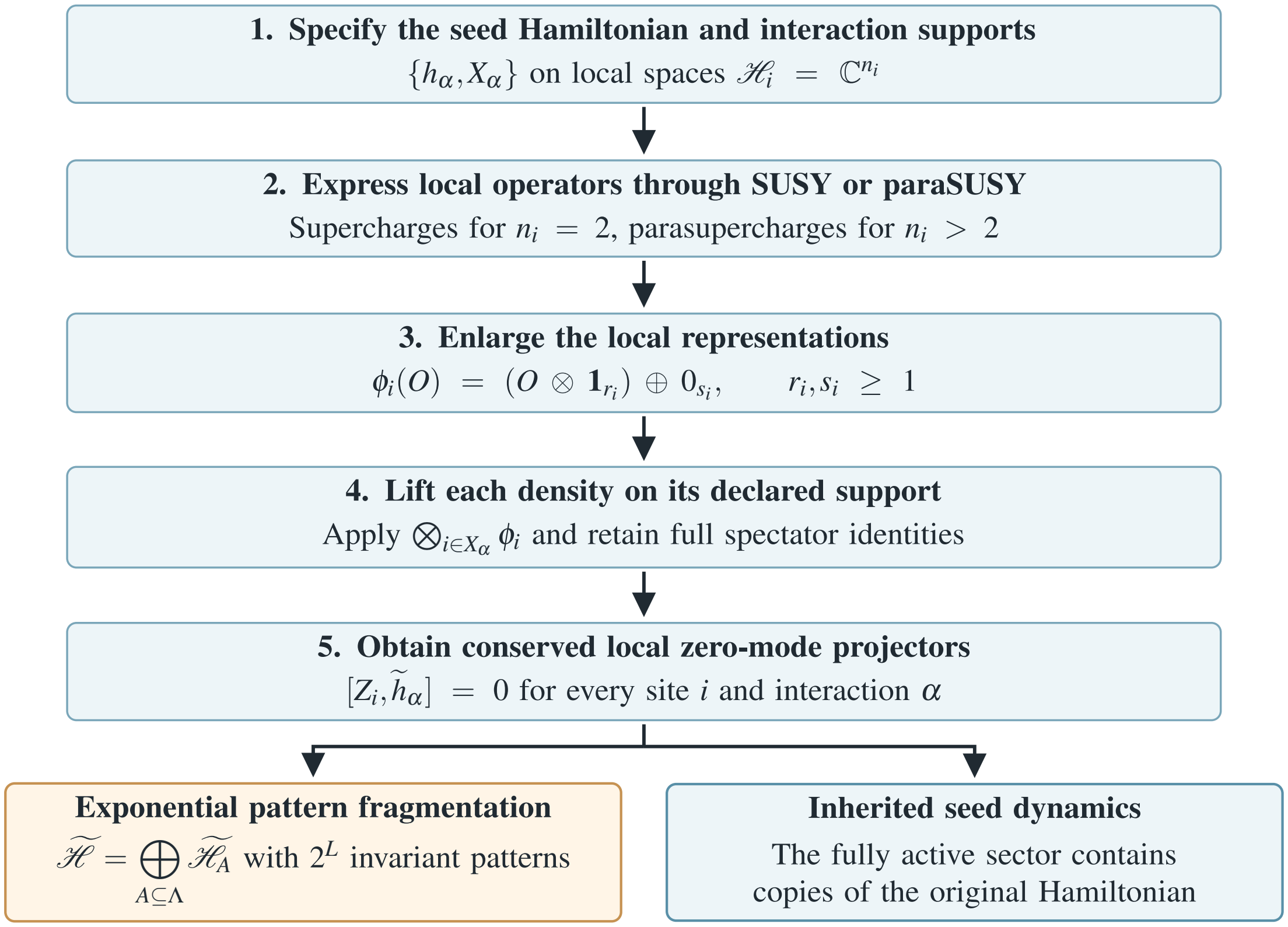

Once each interaction has been written in this algebraic form, we replace $Q_n$ and its adjoint by enlarged generators. Evaluating the same matrix-unit polynomials in these generators induces a map $\phi$ on every seed local operator $O\in\End(\mathbb C^n)$,
\begin{equation}
 \begin{gathered}
 \widetilde{\mathcal H}_{\mathrm{loc}}
   =(\mathbb C^n\otimes\mathbb C^r)\oplus\mathbb C^s,
 \qquad
 \widetilde Q_n=(Q_n\otimes\id_r)\oplus0_s,\\
 \phi(O)=(O\otimes\id_r)\oplus 0_s,
 \qquad r,s\geq1.
 \end{gathered}
 \label{eq:intro-lift}
\end{equation}
The enlarged local Hilbert space has dimension $d=\dim\widetilde{\mathcal H}_{\mathrm{loc}}=nr+s$, where $n$ is the dimension of the seed local Hilbert space, $r$ counts the active copies of the seed matrix-algebra representation, and $s$ is the dimension of the zero-mode subspace. The latter consists of states annihilated by the lifted supercharge or parasupercharge and its adjoint. The identity operator on the seed local Hilbert space, or equivalently $h_p$, is represented by the active projector $P=\phi(\id_n)=\phi(h_p)$. The enlarged local SUSY or paraSUSY Hamiltonian therefore has eigenvalue one on the active subspace and zero on the inactive subspace. Since $P$ is not the identity on the enlarged space, we specify the support of each seed interaction and apply the replacement on that support, leaving identity operators on all spectator sites unchanged. This convention makes the lifting prescription unambiguous for a given interaction decomposition and preserves its spatial support. Figure~\ref{fig:lifting-algorithm} summarizes the procedure and its consequences.

The projector $Z_i=\id_i-P_i$ onto the zero-mode subspace at site $i$ commutes with every lifted interaction. When all $L$ sites are enlarged, their active or inactive labels give $2^L$ nonzero invariant pattern sectors. The fully active sector contains copies of the original seed Hamiltonian. In any other pattern, only interactions whose declared supports lie entirely on active sites survive. Inactive states and active-copy labels supply multiplicities. The supercharges or parasupercharges thus provide the complete operator realization, and their chosen enlarged representations supply the common local kernels that produce the conserved patterns. The seed interactions determine the dynamics within each pattern. These local algebraic roles do not require the many-body Hamiltonian itself to have supersymmetry.

Having described the role of the local generators in our construction, we place it in the broader literature. Supersymmetry has a substantial history in quantum lattice systems, beginning with early constructions for spin systems~\cite{Nicolai1976}. In integrable models such as the supersymmetric $t$--$J$ chain, it appears as a graded Lie-algebra symmetry~\cite{EsslerKorepin1992}. Another family constructs the many-body Hamiltonian as the anticommutator of nilpotent supercharges, leading to constrained fermion models with exact $\mathcal N=2$ supersymmetry~\cite{FendleySchoutensDeBoer2003,FendleyNienhuisSchoutens2003}. Supersymmetry also permits exact results for strongly interacting lattice fermions with extensive ground-state entropy~\cite{FendleySchoutens2005}. Interacting Majorana chains provide examples with $\mathcal N=1$ supersymmetry, whose Hamiltonian is the square of a Hermitian supercharge and which can exhibit spontaneous SUSY breaking~\cite{SannomiyaKatsura2019}.

Spin chains admit a further realization through supercharges that relate Hilbert spaces with different numbers of sites. This \emph{dynamical lattice supersymmetry} occurs in the $XXZ$ chain at special anisotropy~\cite{YangFendley2004}, in the $XYZ$ chain along a special coupling manifold~\cite{HagendorfFendley2012}, and in higher-spin constructions that include the spin-$1$ Fateev--Zamolodchikov chain~\cite{Hagendorf2013}. Our construction instead uses local SUSY or paraSUSY generators to represent operators on a fixed set of sites, without requiring supersymmetry of the many-body Hamiltonian.

This algebraic viewpoint also connects with supersymmetric many-body systems constructed from partial symmetries~\cite{Padmanabhan2017}, where symmetric inverse semigroups realize local supercharges, and with representation-independent constructions of integrable models using SUSY algebras~\cite{Maity2026,Maity2025NonHermitian}. Here we use it to generate fragmentation from a prescribed seed Hamiltonian. The integrable densities of these constructions also provide non-Hermitian inputs for the bond-wise examples, without assuming that their integrability survives an arbitrary embedding.

ParaSUSY quantum mechanics was introduced in Ref.~\cite{Rubakov1988}. Early developments include an alternative formulation~\cite{Beckers1990}, higher-order constructions~\cite{Durand1991}, and an arbitrary-order extension~\cite{Khare1992}. Later studies explored alternative arbitrary-order formulations and the infinite-order limit~\cite{KhareMishraRajasekaran1993}, bosonic realizations using deformed oscillator algebras~\cite{QuesneVansteenkiste1998}, and connections with higher-order nonlinear supersymmetry in quantum many-body systems~\cite{Tanaka2007}.

We develop the site-wise mechanism through worked examples before stating its general theorem. Figure~\ref{fig:sitewise-lifting}, at the beginning of Sec.~\ref{sec:examples}, illustrates the algorithm for a spin-$1/2$ chain. Permutation and Ising models compare integrable and nonintegrable seed Hamiltonians, including interaction families with only a scalar common commutant. The examples also cover longer-range interactions and mixed spins, including selective lifting of one sublattice, before moving to higher-dimensional lattices. Their geometry shows when holes separate independent active regions and when surviving interactions reconnect them.

With fixed local representations, the same projectors remain conserved under arbitrary lifted time-dependent driving. A site-wise unitary gate lift likewise preserves the patterns through arbitrary circuit composition and reproduces the original circuit in the fully active sector. These constructions extend the mechanism to driven many-body dynamics without a high-frequency or adiabatic approximation. Section~\ref{subsec:applications-correlations} turns the sector decomposition into exact formulas for time-dependent correlation functions. For open nearest-neighbor chains, independently populated holes give an exponentially decaying spatial bound that is uniform in time. An explicit mixed-field Ising calculation also gives a nonzero lower bound on the infinite-time spin autocorrelation from isolated active sites.

The theorem identifies the restricted seed Hamiltonian and multiplicities in every pattern, and bounds the common commutant without counting additional symmetries of the seed interaction family. The guaranteed fragmentation is exact and classical. For uniform lifts, the largest connected block occupies an exponentially small fraction of the full Hilbert space. This suppression persists at fixed strictly intermediate hole density. The examples also examine additional global charges, including both sublattice hole numbers in a mixed-spin chain. These statements concern the specified comparison spaces, rather than further fragmentation within every individual pattern.

The lift preserves interaction norms and their declared supports, so quasi-local interaction-norm bounds are inherited. Every perturbation $V$ of the seed Hamiltonian also admits a lift using the same local representations and support convention. For any coupling $\lambda$,
\begin{equation}
 [Z_i,\widetilde H+\lambda\widetilde V]=0
 \qquad\text{for every site }i.
 \label{eq:intro-stability}
\end{equation}
Thus consistently lifted perturbations preserve the fragmentation even when they destroy integrability or global symmetries of the seed Hamiltonian. This stability holds within the constructed family. A general perturbation acting directly on the enlarged Hilbert space can couple active and zero-mode states and need not preserve the sectors.

Bond-wise lifting extends the construction beyond these conserved one-site patterns. Different lifts of the $XXX$ permutation density already give different conservation laws. A suitably shifted $XXZ$ density yields the $XXC$ families, including their diagonal interactions, and the $XYZ$ examples show how embeddings can preserve or destroy mobile-word conservation. The $H_{1,11}$ and $H_{1,3}$ examples extend the discussion to non-Hermitian input densities and to fluctuating colored dimers with frozen local defects. For selected embeddings, the conserved projectors commute with every lifted matrix unit and therefore survive for any choice of seed coefficients. Other conserved strings require special seed matrix elements. These criteria also govern bond-wise gate lifting, with the equal-pair embedding producing fragmented circuits from arbitrary two-qubit gates. We distinguish these sufficient results from a classification of all bond representations, and connectivity in one product basis does not exclude other sectors.

Here is a brief summary of the organization. Section~\ref{sec:examples} introduces the commutants and fragmentation criteria through examples of site-wise lifting and derives their consequences for correlation dynamics. Section~\ref{sec:theorem} proves the general site-wise theorem and establishes the sector-size, locality and stability bounds. Section~\ref{sec:bondwise} develops the bond-wise procedure, studies the $XXX$, $XXZ$, $XYZ$ and two Hietarinta examples, formulates sufficient conservation criteria and the remaining classification problem, and extends the construction to unitary circuits. Section~\ref{sec:conclusion} summarizes the results and discusses future directions, including alternative lifts of the Fredkin chain. Appendices~\ref{app:permutation-commutant} and~\ref{app:matrix-representations} give the permutation-commutant calculation and the representation-theoretic justification of the matrix-algebra lifts. Appendix~\ref{app:toric-sector-ground-states} complements the toric-code example with the geometry-dependent counting and construction of sector ground states.

\section{Examples of site-wise lifting}
\label{sec:examples}
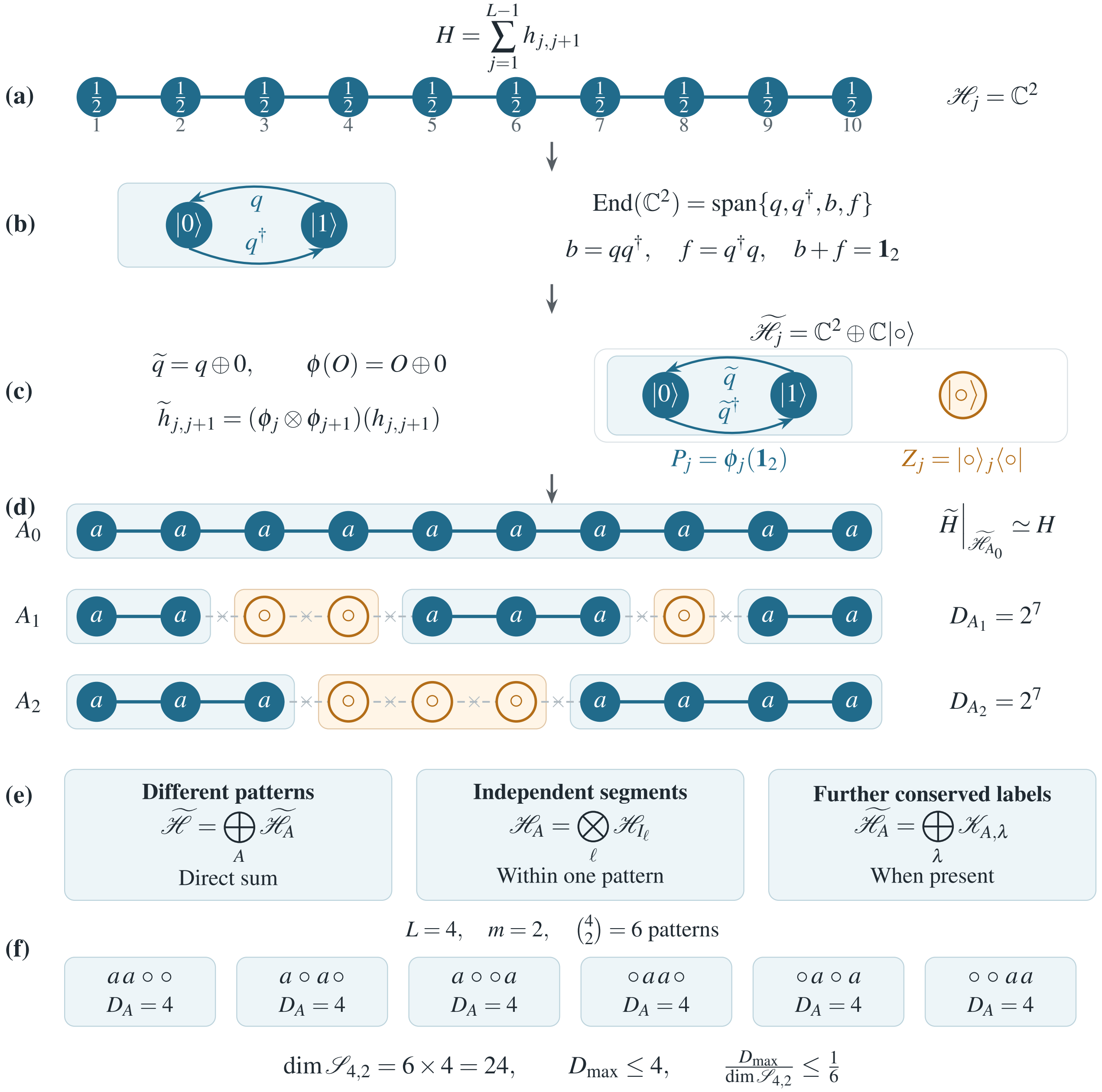
\begin{figure*}[tp]
\centering
\begingroup
\resizebox{\linewidth}{!}{%
\definecolor{hsfActive}{HTML}{216B8B}
\definecolor{hsfActiveWash}{HTML}{EDF5F8}
\definecolor{hsfHole}{HTML}{B36D18}
\definecolor{hsfHoleWash}{HTML}{FFF5E7}
\definecolor{hsfInk}{HTML}{202A32}
\definecolor{hsfMuted}{HTML}{59656D}
\definecolor{hsfGhost}{HTML}{B1BBC2}
\definecolor{hsfRule}{HTML}{DAE1E5}
\begin{tikzpicture}[
  x=1cm,y=1cm,font=\fontsize{10}{12}\selectfont,text=hsfInk,
  >={Stealth[length=2.1mm,width=1.6mm]},
  panel/.style={font=\fontsize{10}{12}\selectfont\bfseries,anchor=west},
  state/.style={circle,draw=hsfActive,fill=hsfActive,text=white,
    minimum size=6.4mm,inner sep=0pt},
  site/.style={circle,draw=hsfActive,fill=hsfActive,text=white,
    minimum size=5.5mm,inner sep=0pt},
  vacancy/.style={circle,draw=hsfHole,fill=hsfHoleWash,text=hsfHole,
    line width=1pt,minimum size=5.5mm,inner sep=0pt,
    font=\fontsize{12}{12}\selectfont},
  bond/.style={draw=hsfActive,line width=1.3pt},
  absent/.style={draw=hsfGhost,line width=.7pt,densely dashed},
  charge/.style={->,draw=hsfActive,line width=.9pt},
  flow/.style={->,draw=hsfInk!75,line width=1pt},
  activegroup/.style={rounded corners=4pt,fill=hsfActiveWash,draw=hsfActive!30},
  holegroup/.style={rounded corners=4pt,fill=hsfHoleWash,draw=hsfHole!40}
]
\useasboundingbox (0,.60) rectangle (15.8,-14.65);

\node[panel] at (.1,-.65) {(a)};
\node at (7.3,.18) {$H=\displaystyle\sum_{j=1}^{L-1}h_{j,j+1}$};
\foreach \j in {1,...,9}{
  \draw[bond] ({1.5+1.18*(\j-1)},-.65) -- ({1.5+1.18*\j},-.65);
}
\foreach \j in {1,...,10}{
  \node[site] at ({1.5+1.18*(\j-1)},-.65) {$\tfrac12$};
  \node[text=hsfMuted,font=\fontsize{8}{10}\selectfont]
    at ({1.5+1.18*(\j-1)},-1.04) {$\j$};
}
\node at (14.1,-.65) {$\mathcal H_j=\mathbb C^2$};
\draw[flow] (7.9,-1.28) -- (7.9,-1.70);

\node[panel] at (.1,-2.45) {(b)};
\filldraw[activegroup] (1.8,-1.86) rectangle (5.7,-3.04);
\node[state] (q0) at (2.8,-2.45) {$\ket{0}$};
\node[state] (q1) at (4.7,-2.45) {$\ket{1}$};
\draw[charge] (q1.north) to[bend right=22]
  node[below=0pt,text=hsfActive] {$q$} (q0.north);
\draw[charge] (q0.south) to[bend right=22]
  node[above=0pt,text=hsfActive] {$q^\dagger$} (q1.south);
\node at (10.45,-2.14)
  {$\End(\mathbb C^2)=\operatorname{span}\{q,q^\dagger,b,f\}$};
\node at (10.45,-2.78)
  {$b=qq^\dagger,\quad f=q^\dagger q,\quad b+f=\id_2$};
\draw[flow] (7.9,-3.28) -- (7.9,-3.70);

\node[panel] at (.1,-4.80) {(c)};
\node at (4.35,-4.42) {$\widetilde q=q\oplus0,\qquad\phi(O)=O\oplus0$};
\node at (4.35,-5.15)
  {$\widetilde h_{j,j+1}=(\phi_j\otimes\phi_{j+1})(h_{j,j+1})$};
\node at (11.85,-3.94)
  {$\widetilde{\mathcal H}_j=\mathbb C^2\oplus\mathbb C\ket{\circ}$};
\filldraw[rounded corners=5pt,fill=white,draw=hsfRule]
  (8.5,-4.19) rectangle (15.15,-5.50);
\filldraw[activegroup] (8.68,-4.29) rectangle (12.12,-5.40);
\node[state] (tq0) at (9.5,-4.84) {$\ket{0}$};
\node[state] (tq1) at (11.3,-4.84) {$\ket{1}$};
\draw[charge] (tq1.north) to[bend right=22]
  node[below=0pt,text=hsfActive] {$\widetilde q$} (tq0.north);
\draw[charge] (tq0.south) to[bend right=22]
  node[above=0pt,text=hsfActive] {$\widetilde q^\dagger$} (tq1.south);
\node[vacancy,minimum size=6.4mm] at (13.68,-4.84) {$\ket{\circ}$};
\node[text=hsfActive] at (10.4,-5.76) {$P_j=\phi_j(\id_2)$};
\node[text=hsfHole] at (13.68,-5.76) {$Z_j=\ket{\circ}_j\bra{\circ}$};
\draw[flow] (7.9,-5.95) -- (7.9,-6.37);

\node[panel] at (.1,-6.43) {(d)};

\filldraw[activegroup] (1.08,-6.37) rectangle (12.54,-7.13);
\foreach \j in {1,...,9}{
  \draw[bond] ({1.5+1.18*(\j-1)},-6.75) -- ({1.5+1.18*\j},-6.75);
}
\foreach \j in {1,...,10}{
  \node[site] at ({1.5+1.18*(\j-1)},-6.75) {$a$};
}
\node at (.55,-6.75) {$A_0$};
\node at (14.15,-6.75)
  {$\left.\widetilde H\right|_{\widetilde{\mathcal H}_{A_0}}\simeq H$};

\foreach \a/\b in {1/2,5/7,9/10}{
  \filldraw[activegroup] ({1.08+1.18*(\a-1)},-7.57)
    rectangle ({1.92+1.18*(\b-1)},-8.33);
}
\foreach \a/\b in {3/4,8/8}{
  \filldraw[holegroup] ({1.08+1.18*(\a-1)},-7.57)
    rectangle ({1.92+1.18*(\b-1)},-8.33);
}
\foreach \j in {2,3,4,7,8}{
  \draw[absent] ({1.5+1.18*(\j-1)},-7.95) -- ({1.5+1.18*\j},-7.95);
  \node[text=hsfGhost,font=\fontsize{9}{9}\selectfont]
    at ({2.09+1.18*(\j-1)},-7.95) {$\times$};
}
\foreach \j in {1,5,6,9}{
  \draw[bond] ({1.5+1.18*(\j-1)},-7.95) -- ({1.5+1.18*\j},-7.95);
}
\foreach \j in {1,2,5,6,7,9,10}{
  \node[site] at ({1.5+1.18*(\j-1)},-7.95) {$a$};
}
\foreach \j in {3,4,8}{
  \node[vacancy] at ({1.5+1.18*(\j-1)},-7.95) {$\circ$};
}
\node at (.55,-7.95) {$A_1$};
\node at (14.15,-7.95) {$D_{A_1}=2^7$};

\foreach \a/\b in {1/3,7/10}{
  \filldraw[activegroup] ({1.08+1.18*(\a-1)},-8.77)
    rectangle ({1.92+1.18*(\b-1)},-9.53);
}
\filldraw[holegroup] (4.62,-8.77) rectangle (7.82,-9.53);
\foreach \j in {3,4,5,6}{
  \draw[absent] ({1.5+1.18*(\j-1)},-9.15) -- ({1.5+1.18*\j},-9.15);
  \node[text=hsfGhost,font=\fontsize{9}{9}\selectfont]
    at ({2.09+1.18*(\j-1)},-9.15) {$\times$};
}
\foreach \j in {1,2,7,8,9}{
  \draw[bond] ({1.5+1.18*(\j-1)},-9.15) -- ({1.5+1.18*\j},-9.15);
}
\foreach \j in {1,2,3,7,8,9,10}{
  \node[site] at ({1.5+1.18*(\j-1)},-9.15) {$a$};
}
\foreach \j in {4,5,6}{
  \node[vacancy] at ({1.5+1.18*(\j-1)},-9.15) {$\circ$};
}
\node at (.55,-9.15) {$A_2$};
\node at (14.15,-9.15) {$D_{A_2}=2^7$};

\node[panel] at (.1,-10.48) {(e)};
\foreach \a/\b in {1.05/5.65,6.00/10.60,10.95/15.55}{
  \filldraw[activegroup] (\a,-10.10) rectangle (\b,-11.94);
}
\node[font=\fontsize{9}{11}\selectfont\bfseries] at (3.35,-10.44) {Different patterns};
\node at (3.35,-11.04) {$\widetilde{\mathcal H}=\displaystyle\bigoplus_A\widetilde{\mathcal H}_A$};
\node[font=\fontsize{9}{11}\selectfont] at (3.35,-11.62) {Direct sum};
\node[font=\fontsize{9}{11}\selectfont\bfseries] at (8.30,-10.44) {Independent segments};
\node at (8.30,-11.04) {$\mathcal H_A=\displaystyle\bigotimes_\ell\mathcal H_{I_\ell}$};
\node[font=\fontsize{9}{11}\selectfont] at (8.30,-11.62) {Within one pattern};
\node[font=\fontsize{9}{11}\selectfont\bfseries] at (13.25,-10.44) {Further conserved labels};
\node at (13.25,-11.04) {$\widetilde{\mathcal H}_A=\displaystyle\bigoplus_\lambda\mathcal K_{A,\lambda}$};
\node[font=\fontsize{9}{11}\selectfont] at (13.25,-11.62) {When present};

\node[panel] at (.1,-12.64) {(f)};
\node[font=\fontsize{9}{11}\selectfont] at (8.05,-12.35) {$L=4,\quad m=2,\quad\binom42=6\ \text{patterns}$};
\foreach \k/\word in {0/{a\,a\,\circ\,\circ},1/{a\,\circ\,a\,\circ},2/{a\,\circ\,\circ\,a},3/{\circ\,a\,a\,\circ},4/{\circ\,a\,\circ\,a},5/{\circ\,\circ\,a\,a}}{
  \filldraw[activegroup] ({1.05+2.42*\k},-12.74) rectangle ({3.17+2.42*\k},-13.71);
  \node at ({2.11+2.42*\k},-13.03) {$\word$};
  \node[font=\fontsize{9}{11}\selectfont] at ({2.11+2.42*\k},-13.43) {$D_A=4$};
}
\node at (8.05,-14.32) {$\dim\mathcal S_{4,2}=6\times4=24,\qquad D_{\max}\leq4,\qquad \frac{D_{\max}}{\dim\mathcal S_{4,2}}\leq\frac16$};
\end{tikzpicture}
}
\endgroup
\caption{A spin-$1/2$ chain illustrates the algorithm of Fig.~\ref{fig:lifting-algorithm}.
(a) An arbitrary nearest-neighbor Hamiltonian on an open chain with $L=10$.
(b) Its local operators are expressed through the SUSY generators.
(c) The minimal lift adds one zero mode $\ket{\circ}$ to the active two-state space. Blue denotes the active subspace and orange the zero mode, annihilated by the supercharge and its adjoint.
(d) Three invariant patterns are shown, with $a$ denoting the active subspace. Crossed dashed bonds vanish, so the surviving interactions act independently on the active segments. The fully active pattern reproduces the seed Hamiltonian. The other patterns have holes at $\{3,4,8\}$ and $\{4,5,6\}$. Both have dimension $2^7$, but their active segments have different lengths.
(e) Different patterns split the enlarged Hilbert space into a direct sum. Independent segments give tensor factors within a pattern, with $I_\ell$ denoting the segments and $\mathcal H_A$ the original-spin space on the active sites. Additional conserved labels $\lambda$ may split a pattern into smaller invariant subspaces $\mathcal K_{A,\lambda}$.
(f) Fixing only the total hole number still allows different spatial patterns. The six boxes for $L=4$, $m=2$ have equal dimension four. Each connected block lies inside one box, giving a largest-block fraction at most $1/6$. The same argument gives the general bound $\binom{L}{m}^{-1}$.}
\label{fig:sitewise-lifting}
\end{figure*}

In this long section we introduce the construction through examples of increasing generality. To make things easier we summarize the entire process in figure~\ref{fig:sitewise-lifting}. It illustrates the procedure on the simplest geometry, a nearest-neighbor spin-$1/2$ chain. As a consequence of the lift, hole (SUSY zero modes) patterns are generated in the spin chain that result in the Hilbert space fragmentation. These patterns separate the chain into independent active segments. The later examples explain how the conserved patterns persist beyond this setting. Panels (e) and (f) distinguish the sector structures and illustrate the largest-block bound used below.

\paragraph*{Three commutants.}
To compare the seed and lifted Hamiltonians, we first distinguish three commutants that will be used to describe these examples.
\begin{enumerate}
    \item Let $\mathcal H=\bigotimes_{i\in\Lambda}\mathcal H_i$ be the full many-body Hilbert space on a finite set $\Lambda$ of $L$ sites. For a chain we take $\Lambda=\{1,\ldots,L\}$. The \emph{local commutant of a single bond density} $h_{ij}$ is
\[
 \mathcal C_{\mathrm{local}}(h_{ij})
 =\{X\in\End(\mathcal H_i\otimes\mathcal H_j):
                         [X,h_{ij}]=0\}.
\]
Its elements act on the two-site space. A single Hermitian density on a space of dimension greater than one never has only a scalar commutant: a nonscalar $h_{ij}$ already commutes with both itself and the identity, while a scalar density commutes with every operator. For instance, in case of the spin-$1/2$ swap operator, the symmetric and antisymmetric subspaces have dimensions three and one, so this local commutant has dimension $3^2+1^2=10$.
\item The next type of commutant is the \emph{common commutant of the specified interaction family} $\mathcal B=\{h_\alpha\}$. It is defined by
\begin{equation}
 \mathcal C(\mathcal B)
 =\{O\in\End(\mathcal H):
                  [O,h_\alpha]=0\ \text{for every }\alpha\}.
 \label{eq:examples-commutant}
\end{equation}
Here each $h_\alpha$ is understood to act as the identity outside its support, and $O$ can act on the entire chain. For a family consisting of nearest-neighbor bond densities, we write
\[
 \mathcal C_{\mathrm{bonds}}
 =\{O\in\End(\mathcal H):
                    [O,h_{j,j+1}]=0\ \text{for every bond }j\}.
\]
Thus the word \emph{bond} specifies the operators with which $O$ must commute; it does not restrict the support of $O$. Equivalently, $\mathcal C(\mathcal B)$ is the commutant of the unital algebra generated by all the $h_\alpha$, often called the bond algebra~\cite{MoudgalyaMotrunich2022}. We use $\mathcal C(\mathcal B)$ when the family also contains one-site or longer-range terms.
\item Finally, the \emph{commutant of a fixed Hamiltonian} $H=\sum_\alpha J_\alpha h_\alpha$ is
\[
 \mathcal C(H)=\{O\in\End(\mathcal H):[O,H]=0\},
 \qquad
 \mathcal C(\mathcal B)\subseteq\mathcal C(H).
\]
Every operator counted by the common commutant is therefore conserved by $H$. The converse generally fails. For example, consider an open three-site permutation chain. Take $\mathcal B=\{\mathsf P_{12},\mathsf P_{23}\}$ and $H=\mathsf P_{12}+\mathsf P_{23}$. Choose the candidate operator $O=H$. It belongs to $\mathcal C(H)$ because $[O,H]=[H,H]=0$. To belong to $\mathcal C(\mathcal B)$, the same $O$ would have to commute with each bond separately. But
\begingroup
\small
\setlength{\jot}{2pt}
\[
[O,\mathsf P_{12}]=[H,\mathsf P_{12}]
   =[\mathsf P_{23},\mathsf P_{12}]\ne0,
 \qquad
 H\in\mathcal C(H)\setminus\mathcal C(\mathcal B).
\]
\endgroup
Thus $[H,\mathsf P_{12}]$ tests the membership of the chosen operator $O=H$ in the common bond commutant. More generally, $O\in\mathcal C(\mathcal B)$ precisely when it commutes with $\sum_\alpha J_\alpha h_\alpha$ for \emph{every independent choice} of the coefficients $J_\alpha$.
\end{enumerate}
We can now see why we use the termwise common commutant to characterize fragmentation~\cite{MoudgalyaMotrunich2022}. The commutant of one Hamiltonian is already large for a purely spectral reason: if its energy eigenspaces have dimensions $g_E$, then
\begingroup
\small
\setlength{\jot}{2pt}
\[
\dim\mathcal C(H)=\sum_E g_E^2\geq\dim\mathcal H.
\]
\endgroup
For a spin-$1/2$ chain this lower bound is $2^L$, even without fragmentation. By contrast, $\mathcal C(\mathcal B)$ detects constraints shared by every interaction on the full many-body space. A symmetry of one density need not commute with all overlapping densities, and conserved charges of an integrable Hamiltonian need not commute with its terms individually. These facts rule out the first and third commutants listed above.

\paragraph*{Strong and weak fragmentation.}
Before turning to the examples, we also specify how their fragmentation will be measured. In particular we will see that the number of dynamical blocks and the size of the largest block answer different questions. Exponentially many blocks can coexist with one block containing almost the entire comparison space. Their number alone therefore does not distinguish strong from weak fragmentation.

First we settle on a method to choose a comparison space $\mathcal S_L$ for a system of $L$ sites. Conventional global charges should be fixed when testing for fragmentation beyond those symmetries. In the lifted models, we will also fix the total number of holes while retaining all their compatible spatial arrangements. Fixing every local hole projector would instead select a single pattern and remove the splitting we wish to measure. If $\mathcal S_L$ decomposes into dynamical blocks $\mathcal K_\alpha$, define
\begingroup
\small
\setlength{\jot}{2pt}
\[
\begin{gathered}
 \mathcal S_L=\bigoplus_\alpha\mathcal K_\alpha,\qquad D_L=\dim\mathcal S_L,\\
 D_{\max}(L)=\max_\alpha\dim\mathcal K_\alpha,\qquad
 f_L=\frac{D_{\max}(L)}{D_L}.
\end{gathered}
\]
\endgroup
For a fragmented family, the standard distinction is~\cite{Sala2020,MoudgalyaMotrunich2022}
\begingroup
\small
\setlength{\jot}{2pt}
\[
\begin{aligned}
  \text{strong fragmentation}\qquad & f_L\longrightarrow0,\\
  \text{weak fragmentation}\qquad & f_L\longrightarrow1,
 \end{aligned}
 \qquad L\longrightarrow\infty.
\]
\endgroup
Thus, in the strong case even the largest block contains a vanishing fraction of the available states. Its dimension may nevertheless grow exponentially with $L$. In the weak case one block occupies an asymptotically unit fraction, although exponentially many smaller blocks may remain. A space consisting of only one block is unfragmented, rather than weakly fragmented merely because $f_L=1$.

The practical calculation is therefore to determine $D_L$, find or bound $D_{\max}(L)$, and examine their ratio along a specified sequence of system sizes and charge densities. For the classical fragmentation studied below, we use connected blocks under the allowed transitions in the stated local product basis. Their projectors commute with the specified interaction family. This is different from decomposing one fixed Hamiltonian into energy eigenspaces. A connected block need not be irreducible under the bond algebra. Any further splitting within these blocks can only reduce their largest dimension, so an upper bound that vanishes relative to $D_L$ is already sufficient to establish strong fragmentation in that comparison space.

A simple count illustrates the test for our construction. Suppose each lifted site has $n$ active states and one hole. Fix only the total hole number $m$ for this initial comparison, and denote the resulting space by $\mathcal S_{L,m}$. There are $\binom{L}{m}$ hole patterns, each of dimension $n^{L-m}$. Since the dynamics cannot connect different patterns,
\begingroup
\small
\setlength{\jot}{2pt}
\[
\begin{gathered}
 \dim\mathcal S_{L,m}=\binom{L}{m}n^{L-m},\quad
 D_{\max}(L,m)\leq n^{L-m},\\
 \frac{D_{\max}(L,m)}{\dim\mathcal S_{L,m}}\leq\binom{L}{m}^{-1}.
\end{gathered}
\]
\endgroup
At a strictly intermediate limiting hole density $m/L\to\rho\in(0,1)$, the binomial coefficient grows exponentially, so the largest-block fraction vanishes exponentially. Connectivity within each pattern need not be established for this conclusion. Figure~\ref{fig:sitewise-lifting}(f) illustrates the counting with four sites and two holes. When additional global charges are fixed, both dimensions must be evaluated in the smaller comparison space. The permutation and transverse-field Ising examples perform this check explicitly. The all-active sector has no hole-pattern splitting and retains the dynamics of the seed Hamiltonian. Section~\ref{subsec:general-counting} gives the corresponding bounds for more general local representations.

With these definitions, the examples are organized as follows. The commutants quoted below refer to the interaction families specified in the corresponding subsections.
\begin{enumerate}
 \item \textbf{Permutation chains: integrable nearest-neighbor seed Hamiltonians with polynomial common commutants.}
Section~\ref{subsec:permutation} treats the uniform spin-$1/2$ and spin-$1$ permutation chains~\cite{Faddeev1996,KorepinBogoliubovIzergin1993,Baxter1982,Sutherland1975}. For local dimension $n=2$ or $3$, their common bond commutant has dimension $\binom{L+n^2-1}{n^2-1}$. The explicit swap action illustrates SUSY and order-two paraSUSY lifting while retaining integrable dynamics on active segments.

 \item \textbf{Ising-type and mixed-spin chains with contrasting common commutants and interaction geometries.}
Section~\ref{subsec:ising} compares four seed Hamiltonians before applying their common lifting prescription. The transverse-field Ising chain has a two-dimensional common commutant and remains solvable with site-dependent couplings and fields~\cite{Pfeuty1970,McCoyWu1973}. Separate longitudinal fields make the common commutant scalar, with a nonintegrable uniform realization~\cite{Chiba2024}. Longer-range Ising couplings show how active segments can reconnect while holes remain fixed. An alternating spin-$1/2$ and spin-$1$ exchange chain with both field directions supplies unequal local dimensions and another scalar common commutant. Lifting only its spin-$1$ sublattice also illustrates selective enlargement. The comparison distinguishes inherited symmetries from the additional hole patterns.

 \item \textbf{Higher-dimensional lattices and many-site interactions.}
Section~\ref{subsec:applications-dimensions} treats the toric code~\cite{Kitaev2003}, a commuting family of star and plaquette interactions on edge qubits. Hole patterns illustrate active regions in two dimensions and determine which stabilizer interactions survive. The resulting sector ground spaces are examined separately in Appendix~\ref{app:toric-sector-ground-states}.

 \item \textbf{Time-dependent Hamiltonians and quantum circuits.}
Section~\ref{subsec:applications-time} derives conservation under arbitrary driving with fixed local representations and constructs unitary gate lifts. A controlled-NOT gate illustrates how the pattern decomposition survives circuit composition.

 \item \textbf{Correlation dynamics and observable consequences.}
Section~\ref{subsec:applications-correlations} expresses time-dependent correlations through the surviving seed dynamics. Open nearest-neighbor chains give an exact active-segment formula and an all-time spatial bound. An isolated-spin calculation in the mixed-field Ising example demonstrates persistent local memory.
\end{enumerate}
Furthermore, in the Ising examples, bonds and the two field directions are taken to be separate generators whenever present. Combining them into larger densities can change the common commutant. One-site fields retain their one-site supports under the lift.

We set the reduced Planck constant, $\hbar$ to one. In the nearest-neighbor chain examples, $\mathcal E$ denotes the bonds of a connected open chain or a ring. Operators written at different sites act on an ordinary tensor product of spin spaces. In particular, using local SUSY generators does not replace the ordinary permutation by a graded permutation or require a global fermionic supercharge.

\subsection{Permutation spin chains}
\label{subsec:permutation}

We treat the spin-$1/2$ and spin-$1$ seed Hamiltonians together, with local dimension $n=2$ and $n=3$, respectively. The matrix units $E_{ab}(Q_n)$ are the supercharge or parasupercharge polynomials of Eq.~\eqref{eq:intro-matrix-units}. Writing $q=\ket{\uparrow}\bra{\downarrow}$, $b=qq^\dagger$ and $f=q^\dagger q$, the two permutation operators are
\begin{equation}
\begin{aligned}
 \mathsf P^{(2)}_{ij}
   &=b_i b_j+f_i f_j+q_iq_j^\dagger+q_i^\dagger q_j\\
   &=2\boldsymbol s_i\cdot\boldsymbol s_j+\tfrac12\id_2\otimes\id_2,
 \qquad \boldsymbol s=\boldsymbol\sigma/2.
\end{aligned}
 \label{eq:perm-susy}
\end{equation}
The SUSY expression in Eq.~\eqref{eq:perm-susy} was first given in Ref.~\cite{Maity2026}. For $Q_3=\ket0\bra1+\ket1\bra2$, the corresponding order-two paraSUSY expression is
\begin{equation}
\begin{gathered}
 \mathsf P^{(3)}_{ij}
   =\sum_{a,b=0}^{2}E_{ab}(Q_3)_i E_{ba}(Q_3)_j\\
   =x_{ij}^2+x_{ij}-\id_3\otimes\id_3,
 \qquad x_{ij}=\boldsymbol S_i\cdot\boldsymbol S_j.
\end{gathered}
 \label{eq:perm-three-spin}
\end{equation}
where $\boldsymbol S$ are spin-$1$ matrices. Both operators exchange the two local states. 

The uniform seed Hamiltonians
\begin{equation}
 H_{\mathrm{perm}}^{(n)}
   =J\sum_{\{i,j\}\in\mathcal E}\mathsf P^{(n)}_{ij},
 \qquad n=2,3,\quad J\ne0,
 \label{eq:perm-seed}
\end{equation}
are the $XXX$ chain and the spin-$1$ bilinear--biquadratic chain at the Uimin--Lai--Sutherland point, up to normalization and additive constants~\cite{Faddeev1996,KorepinBogoliubovIzergin1993,Uimin1970,Lai1974,Sutherland1975}. Their rational $R$-matrix is $R^{(n)}(u)=u\id+\eta\mathsf P^{(n)}$, with $\eta\ne0$~\cite{Baxter1982,KorepinBogoliubovIzergin1993}. Periodic chains and chains with free open ends are integrable~\cite{Sklyanin1988}. 

Adjacent exchanges connect all configurations with the same global color occupations. These occupations are specified by $n-1$ independent diagonal charges of the global $SU(n)$ symmetry, together with the fixed total number of sites. There are only $\binom{L+n-1}{n-1}$ such occupation sectors. The common bond commutants also grow polynomially,
\begin{equation}
 \dim\mathcal C\bigl(\{\mathsf P^{(n)}_{ij}\}_{\mathcal E}\bigr)
   =\binom{L+n^2-1}{n^2-1}
   =\begin{cases}\binom{L+3}{3},&n=2,\\[2pt]\binom{L+8}{8},&n=3.
     \end{cases}
 \label{eq:perm-seed-commutant}
\end{equation}
In Appendix~\ref{app:permutation-commutant} we prove this operator-space count. Thus neither seed permutation chain has exponentially many disconnected product-state components after the global color occupations have been fixed.

\paragraph*{Enlarged representations and invariant sectors.}
For either seed local dimension, the minimal lift adds one hole,
\begingroup
\small
\setlength{\jot}{2pt}
\begin{equation}
\begin{gathered}
 \widetilde{\mathcal H}_{\mathrm{loc}}=\mathbb C^n\oplus\mathbb C\ket\circ,
 \quad \widetilde Q_n=Q_n\oplus0,\\
 \widetilde E_{ab}=E_{ab}\oplus0,\quad
 P=\id_n\oplus0,\quad Z=\id_{n+1}-P.
\end{gathered}
 \label{eq:perm-minimal-projectors}
\end{equation}
\endgroup
The hole is annihilated by the supercharge or parasupercharge and its adjoint. It is a zero-energy state of the normalized local SUSY or paraSUSY Hamiltonian $P$\footnote{We use $\mathsf P$ for the permutation operator and reserve $P$ for the SUSY/paraSUSY Hamiltonians or projectors to {\it active-subspaces} in the enlarged representations.}. A site is \emph{active} when its state lies in $\mathbb C^n$, equivalently when $P_i=1$, and \emph{inactive} when it contains the hole. Substitution of the enlarged generators gives
\begin{equation}
 \widetilde{\mathsf P}^{(n)}_{ij}
   =\sum_{a,b=0}^{n-1}\widetilde E_{ab}^{(i)}\widetilde E_{ba}^{(j)},
 \qquad
 \widetilde H_{\mathrm{perm}}^{(n)}
   =J\sum_{\{i,j\}\in\mathcal E}\widetilde{\mathsf P}^{(n)}_{ij}.
 \label{eq:perm-lift}
\end{equation}
Each bond exchanges two active labels and annihilates any state with a hole at either endpoint. In particular,
\begin{equation}
 (\widetilde{\mathsf P}^{(n)}_{ij})^2=P_iP_j,
 \qquad [Z_k,\widetilde{\mathsf P}^{(n)}_{ij}]=0
 \quad\text{for all }i,j,k.
 \label{eq:perm-lift-conservation}
\end{equation}
The incident interactions annihilate a hole, whereas the ordinary $(n+1)$-state permutation would exchange it with a neighboring active state.

Choosing the location of the zero-modes creates a {\it zero-mode pattern}. For every such pattern we can identify the set of active-sites collected in the set $A\subseteq\{1,\ldots,L\}$. On each such $A$, define
\begin{equation}
 R_A=\prod_{i\in A}P_i\prod_{j\notin A}Z_j,
 \qquad \widetilde{\mathcal H}_A=R_A\widetilde{\mathcal H},
 \qquad \dim\widetilde{\mathcal H}_A=n^{|A|}.
 \label{eq:perm-pattern-projectors}
\end{equation}
These $2^L$ mutually orthogonal projectors commute with every bond and sum to the identity. Their dimensions satisfy $\sum_A n^{|A|}=\sum_{N=0}^L\binom LN n^N=(n+1)^L$. Omitting the fixed hole factors identifies $\widetilde{\mathcal H}_A$ with the original-spin space $\mathcal H_A=\bigotimes_{i\in A}\mathbb C^{n_i}$, where $n_i=n$ here. We retain this notation when the local dimensions later vary between sites.

The enlarged dimensions are three and four, which may be viewed as spin $1$ and spin $3/2$. For the first lift, identify the active states with $m=\pm1$ and the hole with $m=0$. Then
\begingroup
\small
\setlength{\jot}{2pt}
\begin{equation}
\begin{gathered}
 \widetilde b=\frac{(S^z)^2+S^z}{2},\qquad
 \widetilde f=\frac{(S^z)^2-S^z}{2},\\
 \widetilde q=\frac{(S^+)^2}{2},\quad
 P=(S^z)^2,\quad Z=\id_3-(S^z)^2.
\end{gathered}
 \label{eq:perm-spinone-operators}
\end{equation}
\endgroup
and
\begin{equation}
 \widetilde{\mathsf P}^{(2)}_{ij}
 =\frac12\bigl[(S_i^z)^2(S_j^z)^2+S_i^zS_j^z\bigr]
 +\frac14\bigl[(S_i^+)^2(S_j^-)^2+(S_i^-)^2(S_j^+)^2\bigr].
 \label{eq:perm-spinone-bond}
\end{equation}
For the second lift, $\widetilde S^\alpha=S^\alpha\oplus0$ and $\widetilde x_{ij}=\sum_\alpha\widetilde S_i^\alpha\widetilde S_j^\alpha$ give $\widetilde{\mathsf P}^{(3)}_{ij}=\widetilde x_{ij}^2+\widetilde x_{ij}-P_iP_j$. These embedded operators retain the original active spin representation and a singlet, rather than replacing it with the irreducible spin representation of the enlarged dimension.

The same formulas apply on $(\mathbb C^n\otimes\mathbb C^r)\oplus\mathbb C^s$, with $d=nr+s$, representing $r$ copies of the active-subspace and $s$ zero-modes. Here $\widetilde Q_n=(Q_n\otimes\id_r)\oplus0_s$. On active states the exchange is $\ket{a,\mu}_i\ket{b,\nu}_j\mapsto\ket{b,\mu}_i\ket{a,\nu}_j$. The spin-state labels exchange while the copy labels remain at their sites. A pattern with $N$ active sites has dimension $(nr)^Ns^{L-N}$ and contains $r^Ns^{L-N}$ copies of the restricted seed dynamics. The minimal lifts used below have $r=s=1$.

\paragraph*{Geometry of the active segments and inherited dynamics.}
For nearest-neighbor bonds, a given zero-mode pattern splits $A$ into maximal consecutive active segments $I_1,\ldots,I_{k(A)}$. Removing the incident bonds at every hole gives, for both $n=2$ and $n=3$, the restricted Hamiltonian acting only on the active-subspace,
\begin{equation}
 \left.\widetilde H_{\mathrm{perm}}^{(n)}\right|_{\widetilde{\mathcal H}_A}
 \simeq\sum_{\ell=1}^{k(A)}
 \id\otimes\cdots\otimes H_{\mathrm{perm}}^{(n)}[I_\ell]
       \otimes\cdots\otimes\id.
 \label{eq:perm-segment-hamiltonian}
\end{equation}
Thus distinct patterns give a direct-sum decomposition, while disconnected segments within one pattern give tensor factors with independent dynamics. Segment eigenstates form a tensor-product eigenbasis, energies add and time evolution factorizes. Each segment inherits free ends, and an isolated active site has zero Hamiltonian because there are no one-site terms.

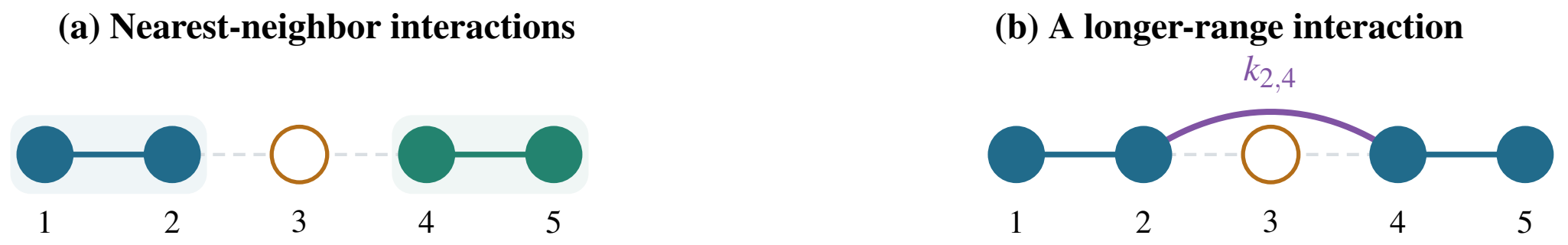
\begin{figure*}[tp]
\centering
\begingroup
\definecolor{geomBlue}{HTML}{216B8B}
\definecolor{geomTeal}{HTML}{23836F}
\definecolor{geomHole}{HTML}{B36D18}
\definecolor{geomPurple}{HTML}{8053A3}
\definecolor{geomGhost}{HTML}{D9DEE2}
\begin{tikzpicture}[x=1cm,y=1cm,
  font=\small,
  active/.style={circle,draw=geomBlue,fill=geomBlue,minimum size=4.8mm,inner sep=0pt},
  hole/.style={circle,draw=geomHole,fill=white,line width=1pt,minimum size=4.8mm,inner sep=0pt},
  bond/.style={draw=geomBlue,line width=1.4pt},
  absent/.style={draw=geomGhost,line width=.8pt,densely dashed}]
\useasboundingbox (-.35,-.85) rectangle (14.4,1.35);
\begin{scope}
 \node[anchor=west,font=\small\bfseries] at (0,1.08) {(a) Nearest-neighbor interactions};
 \fill[geomBlue!7,rounded corners=5pt] (-.30,-.34) rectangle (1.40,.34);
 \fill[geomTeal!7,rounded corners=5pt] (3.00,-.34) rectangle (4.70,.34);
 \draw[bond] (0,0)--(1.1,0);
 \draw[bond,geomTeal] (3.3,0)--(4.4,0);
 \draw[absent] (1.1,0)--(2.2,0)--(3.3,0);
 \foreach \i in {0,1}{\node[active] at (1.1*\i,0) {};}
 \foreach \i in {3,4}{\node[active,draw=geomTeal,fill=geomTeal] at (1.1*\i,0) {};}
 \node[hole] at (2.2,0) {};
 \foreach \i in {0,...,4}{\pgfmathtruncatemacro{\j}{\i+1}\node[font=\footnotesize] at (1.1*\i,-.58) {$\j$};}
\end{scope}
\begin{scope}[xshift=8.4cm]
 \node[anchor=west,font=\small\bfseries] at (-.3,1.08) {(b) A longer-range interaction};
 \draw[bond] (0,0)--(1.1,0);
 \draw[bond] (3.3,0)--(4.4,0);
 \draw[absent] (1.1,0)--(2.2,0)--(3.3,0);
 \draw[geomPurple,line width=1.5pt] (1.1,0) to[bend left=35]
       node[above=1pt,text=geomPurple] {$k_{2,4}$} (3.3,0);
 \foreach \i in {0,1,3,4}{\node[active] at (1.1*\i,0) {};}
 \node[hole] at (2.2,0) {};
 \foreach \i in {0,...,4}{\pgfmathtruncatemacro{\j}{\i+1}\node[font=\footnotesize] at (1.1*\i,-.58) {$\j$};}
\end{scope}
\end{tikzpicture}
\endgroup
\caption{Geometry of active segments in the permutation and Ising examples, for the open-chain pattern $A=\{1,2,4,5\}$. Filled circles are active sites, and the open orange circle is an inactive local zero mode. Solid lines denote surviving pair interactions, while dashed lines mark bonds removed by the hole. (a) Nearest-neighbor interactions leave two independent active segments. (b) An additional density $k_{2,4}$ with declared support $\{2,4\}$ joins them into one component. Its curved bond couples the endpoints without acting on the intervening site. The hole remains fixed in both panels. One-site fields, when present, are omitted.}
\label{fig:geometry-chain}
\end{figure*}

For example, a hole at site $3$ of an open five-site chain leaves $A=\{1,2,4,5\}$ and $H_A=J\mathsf P^{(n)}_{12}\otimes\id_{\{4,5\}}+\id_{\{1,2\}}\otimes J\mathsf P^{(n)}_{45}$. The sector has dimension $n^4$, as illustrated in Fig.~\ref{fig:geometry-chain}(a). On a ring, the all-active pattern retains periodic dynamics, one consecutive group of holes leaves one open active segment, and separated groups can leave several segments in cyclic order. Missing couplings may cause further disconnections. An additional exchange on the declared support $\{2,4\}$ reconnects the two segments of the five-site example without moving the hole, as in Fig.~\ref{fig:geometry-chain}(b). Section~\ref{subsec:ising} develops this range dependence for Ising couplings, and Sec.~\ref{subsec:general-theorem} gives the rule for arbitrary supports.

The support convention also fixes the treatment of additive constants. A seed bond $\mathsf P^{(n)}_{ij}-\id_n\otimes\id_n$ lifts to
\begin{equation}
 \widetilde{\mathsf P}^{(n)}_{ij}-P_iP_j.
 \label{eq:perm-constant-shift}
\end{equation}
The shift is constant within a pattern and counts its surviving bonds, so it can differ between patterns. This shows that a constant energy shift in the seed Hamiltonian need not lift to a constant shift for the enlarged Hamiltonian. Thus the declared support and the representation of its identity are part of the lifting prescription.

\paragraph*{Block sizes and strong fragmentation.}
Label the $n$ active states by colors $a=1,\ldots,n$. Let $j$ represent the index for an active segment. Then, on a given segment of length $\ell_j$, exchanges preserve the occupations $N_{j,a}$ and connect all their rearrangements. The connected block at fixed pattern and segment occupations consequently has dimension
\[
 \dim\mathcal K_{A,\{N_{j,a}\}}
   =\prod_{j=1}^{k(A)}\frac{\ell_j!}{\prod_{a=1}^{n}N_{j,a}!},
 \qquad \sum_aN_{j,a}=\ell_j.
\]
The empty product is one for the all-hole pattern. Longer-range interactions can connect active segments across holes while preserving the hole pattern. The sector counting changes accordingly.

Fix the total hole or zero-mode number to be $m$ and the global color occupations $\boldsymbol N=(N_1,\ldots,N_n)$, with $N=\sum_aN_a=L-m$. The comparison space and its largest connected block satisfy
\begingroup
\small
\setlength{\jot}{2pt}
\begin{equation}
\begin{gathered}
 \dim\mathcal S_{m,\boldsymbol N}
    =\binom Lm\frac{N!}{\prod_aN_a!},\quad
 D_{\max}^{\mathrm{conn}}=\frac{N!}{\prod_aN_a!},\\
 \frac{D_{\max}^{\mathrm{conn}}}{\dim\mathcal S_{m,\boldsymbol N}}
    =\binom Lm^{-1}.
\end{gathered}
 \label{eq:perm-strong}
\end{equation}
\endgroup
The binomial factor chooses the hole positions. The maximum is attained when all active sites form one connected segment, which is possible on an open chain or ring. Any further decomposition from the non-Abelian commutant only reduces the largest block. For $m/L\to\rho\in(0,1)$,
\[
\begin{gathered}
\binom Lm^{-1}=\exp[-Lh(\rho)+o(L)],\\
 h(\rho)=-\rho\log\rho-(1-\rho)\log(1-\rho)>0.
\end{gathered}
\]
Here $h$ is the \emph{binary entropy function}~\cite{Shannon1948}, with natural logarithms and $h(0)=h(1)=0$ by continuity. The same binary entropy function enters the Yang--Yang entropy through the counting of occupied and unoccupied Bethe quantum \mbox{numbers}~\cite{YangYang1969}. Both lifts are therefore strongly fragmented even with their global occupations fixed. The all-active pattern retains the seed dynamics, and the all-hole sector is one dimensional.

For comparison with the entire enlarged space, the largest connected block is fully active with its $n$ color occupations as equal as possible. Stirling's formula combines the two cases into
\[
 \frac{D_{\max}^{\mathrm{conn}}}{(n+1)^L}
 \sim\frac{n^{n/2}}{(2\pi L)^{(n-1)/2}}
       \left(\frac{n}{n+1}\right)^L,
 \qquad n=2,3.
\]
Section~\ref{subsec:general-counting} treats the sector-size bounds for larger multiplicities and more zero modes.

\subsection{Ising-type and mixed-spin chains}
\label{subsec:ising}
The next four seed Hamiltonians separate the roles of conventional symmetry, interaction range and unequal local dimensions. Their couplings are real and need not be translation invariant. In the first three models $\sigma_i^\alpha$ are Pauli matrices. In the fourth, $L=2M$ and $\mathbf T_i$ are spin-$j_i$ matrices with $j_i=1/2$ on odd sites and $j_i=1$ on even sites. The mixed-spin example uses a Heisenberg-type exchange. With $\mathcal E$ the nearest-neighbor bonds of an open chain or ring, the four Hamiltonians are
\begin{widetext}
\begin{align}
 H_{\mathrm I}
   &=-\sum_{\{i,j\}\in\mathcal E}J_{ij}\sigma_i^z\sigma_j^z
     -\sum_i g_i\sigma_i^x,
 &&\text{transverse-field Ising},
 \label{eq:ising-seed}\\
 H_{\mathrm M}
   &=H_{\mathrm I}-\sum_i h_i\sigma_i^z,
 &&\text{mixed-field Ising},
 \label{eq:mixed-seed}\\
 H_{\mathrm{lr}}
   &=-\sum_{i<j}J_{ij}\sigma_i^z\sigma_j^z
     -\sum_i(g_i\sigma_i^x+h_i\sigma_i^z),
 &&\text{longer-range Ising},
 \label{eq:longrange-seed}\\
 H_{\mathrm{alt}}
   &=\sum_{\{i,j\}\in\mathcal E}J_{ij}\mathbf T_i\cdot\mathbf T_j
     -\sum_i(g_iT_i^x+h_iT_i^z),
 &&\text{alternating spins}.
 \label{eq:applications-mixed-seed}
\end{align}
\end{widetext}
We assume $g_i\ne0$ throughout, $h_i\ne0$ in $H_{\mathrm M}$ and $H_{\mathrm{alt}}$, and either all $h_i=0$ or all $h_i\ne0$ in $H_{\mathrm{lr}}$. The nearest-neighbor exchanges are nonzero. For longer-range couplings, let $G$ be the graph whose edges are the pairs with $J_{ij}\ne0$.

The transverse-field Ising chain is converted to a free-fermion system by a standard Jordan-Wigner transformation, and thus is exactly solvable. This includes the case when there are spatially varying couplings and fields, and with parity fixed on a ring~\cite{Pfeuty1970,McCoyWu1973}. Detailed background and exact correlation results are available in Refs.~\cite{McCoy2010,McCoyPerkShrock1983,PerkCapelQuispelNijhoff1984}. The uniform mixed-field chain supplies a nonintegrable example when both fields and the exchange are nonzero~\cite{Chiba2024}. Longer-range transverse-field chains, including power-law interactions, have been studied in connection with phase transitions and entanglement~\cite{DuttaBhattacharjee2001,Koffel2012,FeySchmidt2016}. 

\paragraph*{Common commutants for the seed Hamiltonians.}
We take each exchange and each one-site field direction as a separate member of the specified interaction family. For the qubit models, an operator commuting with all transverse fields is a linear combination of $X_S=\prod_{i\in S}\sigma_i^x$. Such a product commutes with an Ising bond precisely when both endpoints or neither endpoint belong to $S$. On a connected interaction graph this leaves only the identity and the global spin flip $U=\prod_i\sigma_i^x$. Adding the individual longitudinal fields removes this symmetry. On the alternating chain, $T_i^z$ resolves the distinct magnetic levels and $T_i^x$ connects successive levels, so the two fields generate the full local matrix algebra. Hence
\begin{equation}
 \begin{aligned}
 \mathcal C(\mathcal B_{\mathrm I})
   &=\operatorname{span}\{\id,U\},\\
 \mathcal C(\mathcal B_{\mathrm M})
   &=\mathcal C(\mathcal B_{\mathrm{alt}})=\mathbb C\id,\\
 \mathcal C(\mathcal B_{\mathrm{lr}})
   &=\begin{cases}
       \operatorname{span}\{\id,U\},&h_i=0\text{ and }G\text{ connected},\\
       \mathbb C\id,&h_i\ne0\text{ at every site}.
     \end{cases}
 \end{aligned}
 \label{eq:ising-seed-commutants}
\end{equation}
If $G$ is disconnected in the transverse-field case, each component has its own spin-flip parity. These statements concern the common commutant of the interaction family, not the spectral commutant of its sum. In the magnetic product basis, the nonzero transverse fields connect all seed configurations. Ising interactions are diagonal, and the off-diagonal alternating-spin exchanges change two sites, so they cannot cancel the single-site field transitions.

\paragraph*{Lifting and invariant sectors.}
Next we lift all four models with the same minimal prescription, $\phi_i(O)=O\oplus0$. On qubit sites this is the SUSY representation $\widetilde q=q\oplus0$, while spin-$1$ sites use the parasupercharge, $\widetilde Q_3=Q_3\oplus0$. Thus uniform qubit spaces enlarge from dimension two to three, and the alternating chain has
\begin{equation}
 \mathcal H_{\mathrm{alt}}=(\mathbb C^2\otimes\mathbb C^3)^{\otimes M},
 \qquad
 \widetilde{\mathcal H}_{\mathrm{alt}}
   =\bigl[(\mathbb C^2\oplus\mathbb C)\otimes
          (\mathbb C^3\oplus\mathbb C)\bigr]^{\otimes M},
 \label{eq:applications-mixed-spaces}
\end{equation}
with dimensions $6^M$ and $12^M$. The maps act on every local matrix unit, so they also lift arbitrary $6\times6$ densities on odd--even bonds, with reversed tensor factors on even--odd bonds.

For the qubit models, $\sigma^x=q+q^\dagger$ and $\sigma^z=b-f$. Equation~\eqref{eq:perm-spinone-operators} gives $\widetilde\sigma^z=S^z$ and $\widetilde\sigma^x=(S^x)^2-(S^y)^2$. Their three lifts can therefore be written together as
\begin{equation}
 \widetilde H_G
   =-\sum_{\{i,j\}\in E(G)}J_{ij}S_i^zS_j^z
     -\sum_i\bigl\{g_i[(S_i^x)^2-(S_i^y)^2]+h_iS_i^z\bigr\},
 \label{eq:mixed-lift}
\end{equation}
with $E(G)$ denoting the set of edges of $G$.
Nearest-neighbor $G$ with $h_i=0$ gives $\widetilde H_{\mathrm I}$, nearest-neighbor $G$ with both fields gives $\widetilde H_{\mathrm M}$, and general $G$ gives $\widetilde H_{\mathrm{lr}}$. On the alternating chain, put $\widetilde T_i^\alpha=T_i^\alpha\oplus0$ to obtain
\begin{equation}
 \widetilde H_{\mathrm{alt}}
   =\sum_{\{i,j\}\in\mathcal E}J_{ij}
                \widetilde{\mathbf T}_i\cdot\widetilde{\mathbf T}_j
      -\sum_i(g_i\widetilde T_i^x+h_i\widetilde T_i^z).
 \label{eq:applications-mixed-lift}
\end{equation}
Here the odd and even operators act as spin $1/2$ plus a singlet and spin $1$ plus a singlet, respectively. One-site fields are lifted on their own supports, with no additional projector on neighboring sites.

The uniform nearest-neighbor transverse-field lift is a known spin-$1$ Ising model with single-ion anisotropy~\cite{Yang2008}. Its precise identification follows from
\begin{equation}
 -g[(S^x)^2-(S^y)^2]
   =-2g(S^x)^2-g(S^z)^2+2g\id_3.
 \label{eq:ising-single-ion}
\end{equation}
Thus it equals the Hamiltonian of~\cite{Yang2008} at $D_x=D_z=g$, plus $2gL\id$. The decomposition into hole-separated transverse-field Ising segments is also established in~\cite{Yang2008}. Thus our mechanism reproduces it and accommodates other couplings and local dimensions above.

Each added hole is annihilated by the lifted local operators. Consequently every $Z_i$ and every pattern projector $R_A$ of Eq.~\eqref{eq:perm-pattern-projectors} commutes with all lifted interactions. Write the seed Hamiltonian as a sum of pair terms $k_{ij}$ and one-site terms $u_i$. After removing the fixed hole factors, all four restricted Hamiltonians have the common form
\begin{equation}
 H_A=\sum_{\substack{\{i,j\}\in E(G)\\i,j\in A}}k_{ij}
          +\sum_{i\in A}u_i,
 \qquad D_A=\prod_{i\in A}n_i.
 \label{eq:ising-common-restriction}
\end{equation}
Here $n_i=2$ for the qubit models and alternates between two and three for $H_{\mathrm{alt}}$, whose graph is the nearest-neighbor chain. The all-active pattern carries the complete seed Hamiltonian. There are $2^L$ patterns, with $D_A=2^{|A|}$ for the qubit models and $D_A=2^a3^b$ when the alternating pattern has $a$ active odd sites and $b$ active even sites. In the latter case $\sum_{a,b}\binom Ma\binom Mb2^a3^b=12^M$.

\begin{figure*}[tp]
\centering
\begingroup
\resizebox{\linewidth}{!}{%
\definecolor{mixBlue}{HTML}{216B8B}
\definecolor{mixWash}{HTML}{EDF5F8}
\definecolor{mixHole}{HTML}{B36D18}
\definecolor{mixHoleWash}{HTML}{FFF5E7}
\definecolor{mixGrey}{HTML}{B1BBC2}
\definecolor{mixInk}{HTML}{202A32}
\begin{tikzpicture}[x=1cm,y=1cm,font=\fontsize{10}{12}\selectfont,text=mixInk,
 odd/.style={circle,draw=mixBlue,fill=mixBlue,text=white,minimum size=6.5mm,inner sep=0pt},
 even/.style={rectangle,rounded corners=1pt,draw=mixBlue,fill=mixBlue,text=white,minimum size=6.5mm,inner sep=0pt},
 ohole/.style={odd,draw=mixHole,fill=mixHoleWash,text=mixHole,line width=1pt},
 ehole/.style={even,draw=mixHole,fill=mixHoleWash,text=mixHole,line width=1pt},
 bond/.style={draw=mixBlue,line width=1.2pt},
 absent/.style={draw=mixGrey,densely dashed,line width=.8pt},
 panel/.style={font=\fontsize{10}{12}\selectfont\bfseries,anchor=west},
 note/.style={font=\fontsize{9}{11}\selectfont},
 card/.style={draw=mixBlue!30,fill=mixWash,rounded corners=3pt}]
\useasboundingbox (0,.25) rectangle (15.8,-8.2);

\node[panel] at (.05,0) {(a) Alternating spins and local representation enlargements};
\foreach \j in {1,...,7}{\draw[bond] ({2.2+1.6*(\j-1)},-.75)--({2.2+1.6*\j},-.75);}
\foreach \j in {1,3,5,7}{\node[odd] at ({2.2+1.6*(\j-1)},-.75) {$\frac12$};}
\foreach \j in {2,4,6,8}{\node[even] at ({2.2+1.6*(\j-1)},-.75) {$1$};}
\foreach \j in {1,...,8}{\node[note,text=mixInk!65] at ({2.2+1.6*(\j-1)},-1.23) {$\j$};}
\filldraw[card] (.65,-1.60) rectangle (7.55,-2.80);
\filldraw[card] (8.05,-1.60) rectangle (14.95,-2.80);
\node[note] at (4.1,-1.86) {Odd sites, SUSY};
\node at (4.1,-2.25) {$\widetilde q=q\oplus0,\quad \mathbb C^2\longrightarrow\mathbb C^2\oplus\mathbb C$};
\node[note,text=mixBlue] at (4.1,-2.59) {$2\longrightarrow3$};
\node[note] at (11.5,-1.86) {Even sites, order-two paraSUSY};
\node at (11.5,-2.25) {$\widetilde Q_3=Q_3\oplus0,\quad \mathbb C^3\longrightarrow\mathbb C^3\oplus\mathbb C$};
\node[note,text=mixBlue] at (11.5,-2.59) {$3\longrightarrow4$};

\node[panel] at (.05,-3.30) {(b) Three active sites, different segment dimensions};
\draw[mixGrey!50] (7.9,-3.65)--(7.9,-5.45);
\foreach \j in {1,2}{\draw[bond] ({1.55+.95*(\j-1)},-4.18)--({1.55+.95*\j},-4.18);}
\foreach \j in {3,4,5}{\draw[absent] ({1.55+.95*(\j-1)},-4.18)--({1.55+.95*\j},-4.18);}
\foreach \j in {1,3}{\node[odd] at ({1.55+.95*(\j-1)},-4.18) {$\frac12$};}
\node[even] at (2.50,-4.18) {$1$};
\foreach \j in {4,6}{\node[ehole] at ({1.55+.95*(\j-1)},-4.18) {$\circ$};}
\node[ohole] at (5.35,-4.18) {$\circ$};
\foreach \j in {1,...,6}{\node[note,text=mixInk!65] at ({1.55+.95*(\j-1)},-4.64) {$\j$};}
\node at (3.92,-5.14) {$D_A=2\cdot3\cdot2=12$};

\foreach \j in {2,3}{\draw[bond] ({9.50+.95*(\j-1)},-4.18)--({9.50+.95*\j},-4.18);}
\foreach \j in {1,4,5}{\draw[absent] ({9.50+.95*(\j-1)},-4.18)--({9.50+.95*\j},-4.18);}
\foreach \j in {1,5}{\node[ohole] at ({9.50+.95*(\j-1)},-4.18) {$\circ$};}
\foreach \j in {2,4}{\node[even] at ({9.50+.95*(\j-1)},-4.18) {$1$};}
\node[odd] at (11.40,-4.18) {$\frac12$};
\node[ehole] at (14.25,-4.18) {$\circ$};
\foreach \j in {1,...,6}{\node[note,text=mixInk!65] at ({9.50+.95*(\j-1)},-4.64) {$\j$};}
\node at (11.87,-5.14) {$D_A=3\cdot2\cdot3=18$};

\node[panel] at (.05,-5.82) {(c) Selective lifting of the even sites};
\foreach \j in {1,2,7}{\draw[bond] ({2.2+1.6*(\j-1)},-6.62)--({2.2+1.6*\j},-6.62);}
\foreach \j in {3,4,5,6}{\draw[absent] ({2.2+1.6*(\j-1)},-6.62)--({2.2+1.6*\j},-6.62);}
\foreach \j in {1,3,5,7}{\node[odd] at ({2.2+1.6*(\j-1)},-6.62) {$\frac12$};}
\foreach \j in {2,8}{\node[even] at ({2.2+1.6*(\j-1)},-6.62) {$1$};}
\foreach \j in {4,6}{\node[ehole] at ({2.2+1.6*(\j-1)},-6.62) {$\circ$};}
\foreach \j in {1,...,8}{\node[note,text=mixInk!65] at ({2.2+1.6*(\j-1)},-7.10) {$\j$};}
\node at (7.9,-7.78) {$i\ \text{odd}\quad\mathbb C^2,\qquad i\ \text{even}\quad\mathbb C^3\oplus\mathbb C\ket{\circ}$};
\end{tikzpicture}
}
\endgroup
\caption{Mixed-spin chains under minimal site-wise lifting. Circles denote original spin-$1/2$ sites and squares original spin-$1$ sites. Blue symbols are active, orange symbols are holes, and the small numbers label sites. Solid bonds survive and dashed bonds vanish in the displayed pattern. One-site fields are omitted.
(a) Odd and even sites use SUSY and order-two paraSUSY, respectively, giving local dimensions three and four after enlargement.
(b) Two patterns on an open six-site chain have one three-site active segment each. Starting on an odd or even site gives dimensions $12$ and $18$. Their sublattice hole counts differ, illustrating why the composition of an active segment matters.
(c) When only even sites are enlarged, holes can occur only on that sublattice. The displayed eight-site pattern has holes at sites $4$ and $6$. All odd sites remain active, and the isolated active site $5$ retains its one-site fields. The shapes and spin labels indicate the original active representations, rather than irreducible spins of the enlarged dimensions.}
\label{fig:mixed-spin-lifting}
\end{figure*}
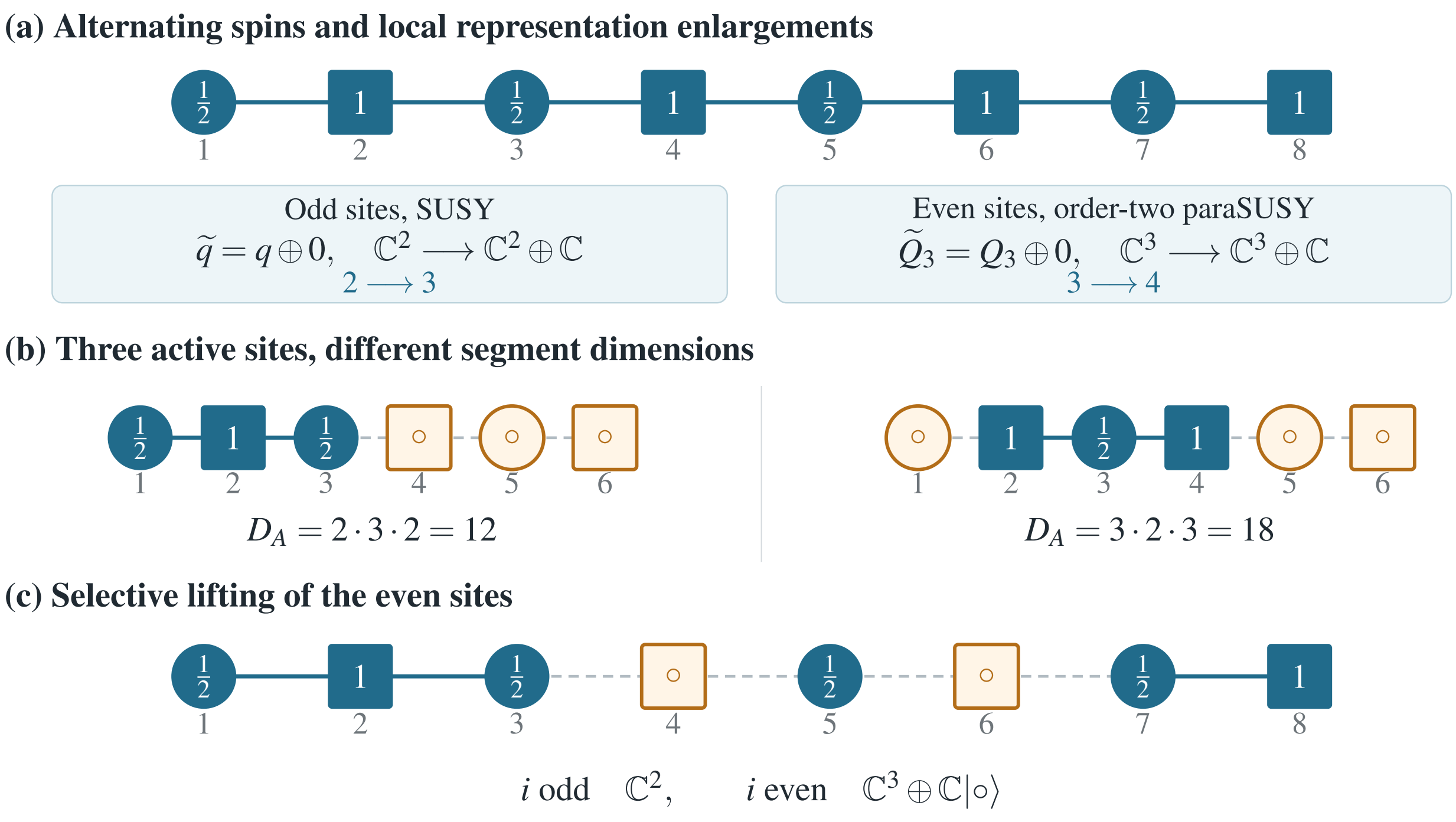

\paragraph*{Geometry and inherited dynamics.}
Nearest-neighbor interactions give the independent active segments described in Sec.~\ref{subsec:permutation} and Fig.~\ref{fig:geometry-chain}(a). The additional feature for the Ising-type systems is that every active site retains its one-site fields. In particular, an isolated qubit evolves under $-g_i\sigma_i^x-h_i\sigma_i^z$, with $h_i=0$ in the transverse-field case. For alternating spins, segment composition also matters. Three active sites starting on an odd or even site have dimensions $2\cdot3\cdot2=12$ or $3\cdot2\cdot3=18$, as illustrated in Fig.~\ref{fig:mixed-spin-lifting}. Related decompositions occur in mixed diamond chains, where conserved singlet dimers separate active spin clusters~\cite{Takano2009,Hida2009}. Here the inactive states are supplied by the representation lift.

For longer-range interactions, a bond survives whenever its two endpoints are active, regardless of the intervening sites. The coupling $J_{24}\ne0$ therefore bridges the hole at site $3$ in Fig.~\ref{fig:geometry-chain}(b). To describe several holes, let $R$ be an upper bound on the pair separation, measured in lattice spacings, so $J_{ij}=0$ for $|i-j|>R$. On an open chain, two consecutive active segments separated by a gap of $g$ holes have their closest endpoints $g+1$ sites apart. Thus $g\geq R$ guarantees separation across that gap. When $g<R$, a bridge exists only if an appropriate nonzero coupling is actually present. The gap width is not the total number of holes, and endpoint holes merely shorten the active part.

For example, holes at $3,4,7$ of a nine-site chain leave segments $\{1,2\}$, $\{5,6\}$ and $\{8,9\}$ with gaps of widths two and one. At range $R=2$, $J_{68}\ne0$ can join the last two segments, while at $R=3$ an additional $J_{25}\ne0$ can join all three. Alternatively, adding only $J_{28}\ne0$ to the bonds within the segments joins the outer two and leaves the middle one independent. An interacting component therefore need not be a contiguous interval.

If every pair at separation at most $R$ is coupled, the components are obtained by cutting the ordered active sites at gaps with $g\geq R$. Sparser coupling graphs may have additional disconnections, as when distance-two bonds separate odd and even sites. On a ring, surviving paths around the other side must also be checked. With unbounded-range interactions no finite gap alone guarantees separation, and a nonzero coupling between every pair leaves one interacting component for every nonempty active set. Small couplings still count for exact connectivity, while the holes remain immobile in all cases.

Quasi-local examples include exponentially decaying couplings $|J_{ij}|\leq J_0e^{-|i-j|/\xi}$ and bounded fields. More generally, bounded pair interaction strength per site follows from $\sup_i\sum_{j\ne i}|J_{ij}|<\infty$. The lift preserves operator norms and hence these interaction bounds. No decay condition is needed at finite $L$. Sections~\ref{subsec:general-theorem} and \ref{subsec:general-scope} extend these support and norm statements to arbitrary many-site interactions.

\paragraph*{Common commutants after lifting.}
All four lifted models possess the conserved hole-pattern projectors
$R_A$, where $A\subseteq\Lambda$ is the set of active sites. The
transverse-field models have additional conserved operators associated
with the parity of each active component. We now explain why these
operators exhaust the respective common commutants and determine
their dimensions.

First, every operator in the common commutant preserves each hole
pattern. For the qubit models, commuting with the individual lifted
transverse fields implies commuting with
\(
 (\widetilde\sigma_i^x)^2=P_i.
\)
For the alternating-spin chain, the two field directions generate
the local spin algebra, including the Casimir
\(
 \sum_{\alpha=x,y,z}(\widetilde T_i^\alpha)^2
       =j_i(j_i+1)P_i.
\)
Thus, in all four models, any common-commutant operator $O$ commutes
with every $P_i$ and consequently with every pattern projector
$R_A$ defined in Eq.~\eqref{eq:perm-pattern-projectors}. Since
different pattern projectors are orthogonal,
\[
 R_AOR_B=OR_AR_B=0,\qquad A\ne B.
\]
There are therefore no common-commutant operators connecting distinct patterns. We are then left with computing the commutant separately on each pattern subspace $\widetilde{\mathcal H}_A=R_A\widetilde{\mathcal H}$ and adding the resulting dimensions.

Consider first the transverse-field model, with no longitudinal fields.
For a fixed pattern $A$, let $I_1,\ldots,I_{k(A)}$ be the nonempty
connected components of the surviving interaction graph $G[A]$.
For nearest-neighbor interactions these are the active segments.
Each component has a spin-flip parity
\[
 U_\ell=\prod_{i\in I_\ell}\sigma_i^x,
 \qquad \ell=1,\ldots,k(A),
\]
where the Pauli matrices act on the active qubit spaces.
This operator commutes with every transverse field and every
surviving Ising bond. Indeed, a bond within $I_\ell$ has both
endpoints flipped by $U_\ell$, leaving its product of two
$\sigma^z$ operators unchanged.

To see that these parities generate the full common commutant
within $\widetilde{\mathcal H}_A$, an operator commuting with all individual
transverse fields must be a linear combination of products
$\prod_{i\in S}\sigma_i^x$, with $S\subseteq A$.
Commutation with every surviving Ising bond requires that either
both endpoints or neither endpoint belong to $S$. Hence $S$ must
be a union of complete connected components. The allowed products
are precisely
\[
 \prod_{\ell=1}^{k(A)}U_\ell^{\epsilon_\ell},
 \qquad \epsilon_\ell\in\{0,1\}.
\]
Each component provides two independent choices, its identity or
its parity. The resulting $2^{k(A)}$ products are linearly
independent because they are distinct Pauli strings on the active
sites. For example, two components give the four operators
$\id_A$, $U_1$, $U_2$ and $U_1U_2$, where $\id_A$ is the identity
on $\widetilde{\mathcal H}_A$. Extending these operators by zero outside the
pattern turns $\id_A$ into $R_A$. Thus the pattern projector is
already included in this count.

When both field directions are present at every active site,
they generate the full local matrix algebra. This holds for
both the qubit models and the alternating-spin chain.
Their common commutant within a fixed pattern is therefore
scalar, irrespective of the connectivity of $G[A]$. Each pattern
contributes only its identity, represented on the enlarged
Hilbert space by $R_A$.

Adding the contributions from the mutually orthogonal pattern
subspaces gives
\begin{equation}
 \begin{aligned}
 \dim\mathcal C(\widetilde{\mathcal B}_{\mathrm{transverse}})
   &=\sum_{A\subseteq\Lambda}2^{k(A)},\\
 \mathcal C(\widetilde{\mathcal B}_{\mathrm{both\ fields}})
   &=\operatorname{span}\{R_A:A\subseteq\Lambda\},\\
 \dim\mathcal C(\widetilde{\mathcal B}_{\mathrm{both\ fields}})
   &=\sum_{A\subseteq\Lambda}1=2^L.
 \end{aligned}
 \label{eq:mixed-exact-commutant}
\end{equation}
Here $k(\varnothing)=0$, so the one-dimensional all-hole pattern
contributes one operator. The $2^L$ pattern projectors are nonzero
and mutually orthogonal, and hence linearly independent.
The second and third lines apply to the mixed-field,
alternating-spin and longer-range models with both field
directions present at every site. Their scalar seed common
commutants thus become exponentially large after lifting.
The transverse-field models have still larger common commutants,
because each hole pattern is further resolved by the independent
parities of its active components.

\paragraph*{Block sizes and strong fragmentation.}
Nonzero transverse fields connect every magnetic product-state configuration within a fixed pattern. For all three qubit families, fixing only $m$ holes therefore gives $\binom Lm$ connected blocks of dimension $2^{L-m}$ and
\begin{equation}
 \dim\mathcal S_m=\binom Lm2^{L-m},
 \qquad
 \frac{D_{\max}^{\mathrm{conn}}(\mathcal S_m)}{\dim\mathcal S_m}
     =\binom Lm^{-1}.
 \label{eq:mixed-strong}
\end{equation}
For the models with both field directions these blocks are also irreducible under the specified interaction algebra. For the transverse-field models, the component parities refine them. This finer decomposition can be resolved while retaining the same strong-fragmentation conclusion.

For a connected seed graph, assign parity $+1$ to a hole and define $U_{\mathrm{glob}}=\prod_i(\widetilde\sigma_i^x+Z_i)$. Fix $m$ and $U_{\mathrm{glob}}=u=\pm1$, with $N=L-m>0$. A component of length $\ell$ has $2^{\ell-1}$ states of each parity, since spin reversal pairs distinct magnetic configurations. Fixing all $k(A)$ component parities therefore gives a block of dimension $2^{N-k(A)}$. There are $2^{k(A)-1}$ assignments consistent with their fixed product $u$, so every pattern contributes $2^{N-1}$ states. Consequently,
\begin{equation}
 \dim\mathcal S_{m,u}=\binom Lm2^{N-1},
 \qquad
 \frac{D_{\max}^{\mathrm{parity}}(\mathcal S_{m,u})}
      {\dim\mathcal S_{m,u}}\leq\binom Lm^{-1}.
 \label{eq:ising-parity-strong}
\end{equation}
Equality holds when a connected active pattern of size $N$ exists, in particular on the open chain, ring, or any connected pair-interaction graph. 

For example, consider an open chain with $L=6$ and $m=2$,
and fix the global parity to $u=\pm1$. Placing the holes at
sites $5$ and $6$ leaves the single active segment
$A=\{1,2,3,4\}$. Its $2^4=16$ states split into two
eight-dimensional parity blocks, of which one belongs to
the chosen global parity.

Instead, placing the holes at sites $3$ and $6$ leaves
two active segments, $\{1,2\}$ and $\{4,5\}$. Each segment
has two states of each parity. The constraint $u_1u_2=u$
allows two assignments of the segment parities, and each
assignment gives a block of dimension $2\cdot2=4$.
Thus this pattern contributes two four-dimensional blocks
at the chosen global parity.

Other hole arrangements can produce different segment
lengths or three active segments. Nevertheless, every
pattern contains four active qubits and contributes
$2^{4-1}=8$ states at fixed global parity, distributed
among its segment-parity blocks. There are
$\binom{6}{2}=15$ patterns, so the comparison space has
dimension $15\cdot8=120$. The largest block has dimension
eight, attained by the single-segment arrangement above,
giving the fraction $8/120=1/15$.

For $m=0$, a connected interaction graph has only the
seed parity blocks. With our convention for global
parity, the all-hole state has parity $+1$.

For the alternating chain, fix $m_o$ odd-site holes and $m_e$ even-site holes. Their positions can still be chosen in $\binom M{m_o}\binom M{m_e}$ ways, and every such pattern has dimension $2^{M-m_o}3^{M-m_e}$. Hence
\begin{equation}
 \begin{gathered}
 \dim\mathcal K_{m_o,m_e}
   =\binom M{m_o}\binom M{m_e}\,2^{M-m_o}3^{M-m_e},\\
 \frac{D_{\max;m_o,m_e}^{\mathrm{conn}}}{\dim\mathcal K_{m_o,m_e}}
   \leq\left[\binom M{m_o}\binom M{m_e}\right]^{-1}.
 \end{gathered}
 \label{eq:applications-mixed-strong}
\end{equation}
The inequality is saturated for the displayed Hamiltonian because its transverse fields connect each pattern. It also applies to arbitrary interactions on the alternating spaces, which may split the patterns further. This explains why fixing both total hole numbers does not remove fragmentation, as their spatial arrangements remain distinct conserved labels.

Using the binary entropy defined in Sec.~\ref{subsec:permutation}, the qubit ratios decay as $\exp[-Lh(\rho)+o(L)]$ when $m/L\to\rho\in(0,1)$. For the alternating chain with $m_o/M\to\rho_o$ and $m_e/M\to\rho_e$, Eq.~\eqref{eq:applications-mixed-strong} gives the bound
\[
 \exp\!\left[-\frac L2\{h(\rho_o)+h(\rho_e)\}+o(L)\right].
\]
An intermediate hole density on either sublattice is sufficient for exponential decay. For example, $m_o=M/2$, $m_e=0$ gives $\binom M{M/2}^{-1}\sim\sqrt{\pi M/2}\,2^{-M}$, while one odd-site hole and no even-site holes gives $1/M$. These are exact classical fragmentation patterns with strong fragmentation in the stated comparison spaces and limits. In the entire enlarged spaces, the largest magnetic-basis connected-block fractions are $(2/3)^L$ for the qubit models and $6^M/12^M=2^{-M}$ for the alternating chain.

\paragraph*{Selective lifting.}
We may also enlarge only the spin-$1$ sites of the alternating chain, as in Fig.~\ref{fig:mixed-spin-lifting}(c). The local dimensions then alternate between two and four, giving total dimension $8^M$ and $2^M$ hole patterns. Every spin-$1/2$ site remains active, and the fully active sector still has dimension $6^M$ and reproduces the original seed Hamiltonian. More generally, lifting only a chosen subset $\Omega\subseteq\Lambda$ gives $2^{|\Omega|}$ patterns. Interactions supported entirely outside $\Omega$ retain their original form in every pattern. Enlarging a positive fraction of the sites therefore generates exponentially many patterns in the total system size, whereas enlarging a fixed number generates only finitely many additional patterns. The general theorem includes this possibility by taking $r_i=1$ and $s_i=0$ at unlifted sites.

\subsection{Higher-dimensional lattices: Kitaev's toric code}
\label{subsec:applications-dimensions}
\label{subsubsec:applications-toric}
The site-wise construction also applies to Hamiltonians defined on higher-dimensional lattices, with many-site interaction terms. We illustrate both through the toric code, whose commuting star and plaquette terms show how the lifting prescription acts on four-edge interaction supports~\cite{Kitaev2003}. Place a qubit on each edge of a periodic square lattice\footnote{The toric code is well defined on any triangulation of the two dimensional space. This is one of the reasons why it is associated with topological physics.} and take positive couplings. Denote the edges incident on a vertex $v$ by $\operatorname{star}(v)$ and those surrounding a plaquette $p$ by $\partial p$. The seed Hamiltonian is
\begingroup
\small
\setlength{\jot}{2pt}
\begin{equation}
\begin{gathered}
 H_{\mathrm{TC}}=-\sum_v J_v\mathsf S_v-\sum_p K_p B_p,\\
 \mathsf S_v=\prod_{e\in\operatorname{star}(v)}\sigma_e^x,
 \qquad B_p=\prod_{e\in\partial p}\sigma_e^z.
\end{gathered}
 \label{eq:applications-toric-seed}
\end{equation}
\endgroup
Let $N_V,N_E,N_P$ count the vertices, edges and plaquettes. On the torus, $N_V-N_E+N_P=0$. The seed stabilizers obey the two independent relations
\[
 \prod_v\mathsf S_v=\id,
 \qquad
 \prod_p B_p=\id.
\]
There are consequently $N_V-1$ independent star constraints and $N_P-1$ independent plaquette constraints. Their common $+1$ eigenspace has dimension $2^{N_E-(N_V-1)-(N_P-1)}=4$.

\paragraph*{SUSY lift and conserved patterns.}
Enlarge each edge space to $\mathbb C^2\oplus\mathbb C\ket{\circ}$ and take $\widetilde q_e=\ket{0}_e\bra{1}$, with both the supercharge and its adjoint annihilating the hole. To distinguish the lifted Pauli matrices from the hole projector $Z_e$, write
\begingroup
\small
\setlength{\jot}{2pt}
\begin{equation}
\begin{gathered}
 \widehat\sigma_e^x=\widetilde q_e+\widetilde q_e^\dagger,\quad
 \widehat\sigma_e^z=\widetilde q_e\widetilde q_e^\dagger
                  -\widetilde q_e^\dagger\widetilde q_e,\\
 P_e=\{\widetilde q_e,\widetilde q_e^\dagger\},
 \qquad Z_e=\id_3-P_e.
\end{gathered}
 \label{eq:applications-toric-operators}
\end{equation}
\endgroup
Replacing every Pauli matrix by its hatted counterpart gives
\begingroup
\small
\setlength{\jot}{2pt}
\begin{equation}
\begin{gathered}
 \widetilde H_{\mathrm{TC}}
 =-\sum_v J_v\widetilde{\mathsf S}_v-\sum_p K_p\widetilde B_p,\\
 \widetilde{\mathsf S}_v=\prod_{e\in\operatorname{star}(v)}\widehat\sigma_e^x,
 \qquad \widetilde B_p=\prod_{e\in\partial p}\widehat\sigma_e^z.
\end{gathered}
 \label{eq:applications-toric-lift}
\end{equation}
\endgroup
The lifted stars and plaquettes still commute. At a shared edge the lifted Pauli matrices anticommute, and each star and plaquette share an even number of edges. Their squares are now projectors
\begin{equation}
 \widetilde{\mathsf S}_v^{\,2}
       =\prod_{e\in\operatorname{star}(v)}P_e,
 \qquad
 \widetilde B_p^{\,2}
       =\prod_{e\in\partial p}P_e,
 \label{eq:applications-toric-squares}
\end{equation}
supported on the edges sharing the vertex and plaquette respectively.
Every $Z_e$ commutes with every lifted interaction. For an active-edge set $A$, the pattern projector $R_A$ fixes all edges outside $A$ in the hole state. Omitting those fixed factors leaves
\begin{equation}
 H_{\mathrm{TC},A}
 =-\sum_{\operatorname{star}(v)\subseteq A}J_v\mathsf S_v
  -\sum_{\partial p\subseteq A}K_pB_p.
 \label{eq:applications-toric-restricted}
\end{equation}
Only complete active supports survive. A hole removes each star or plaquette containing that edge, without turning it into a three-edge interaction. The fully active sector reproduces the original toric code.

\paragraph*{Geometry of active regions.}
Two active edges belong to the same interacting region when a sequence of surviving star or plaquette supports connects them, with successive supports sharing an active edge. Ordinary adjacency in the lattice drawing is therefore insufficient. An active edge contained in no surviving support is a free qubit and forms a component by itself.

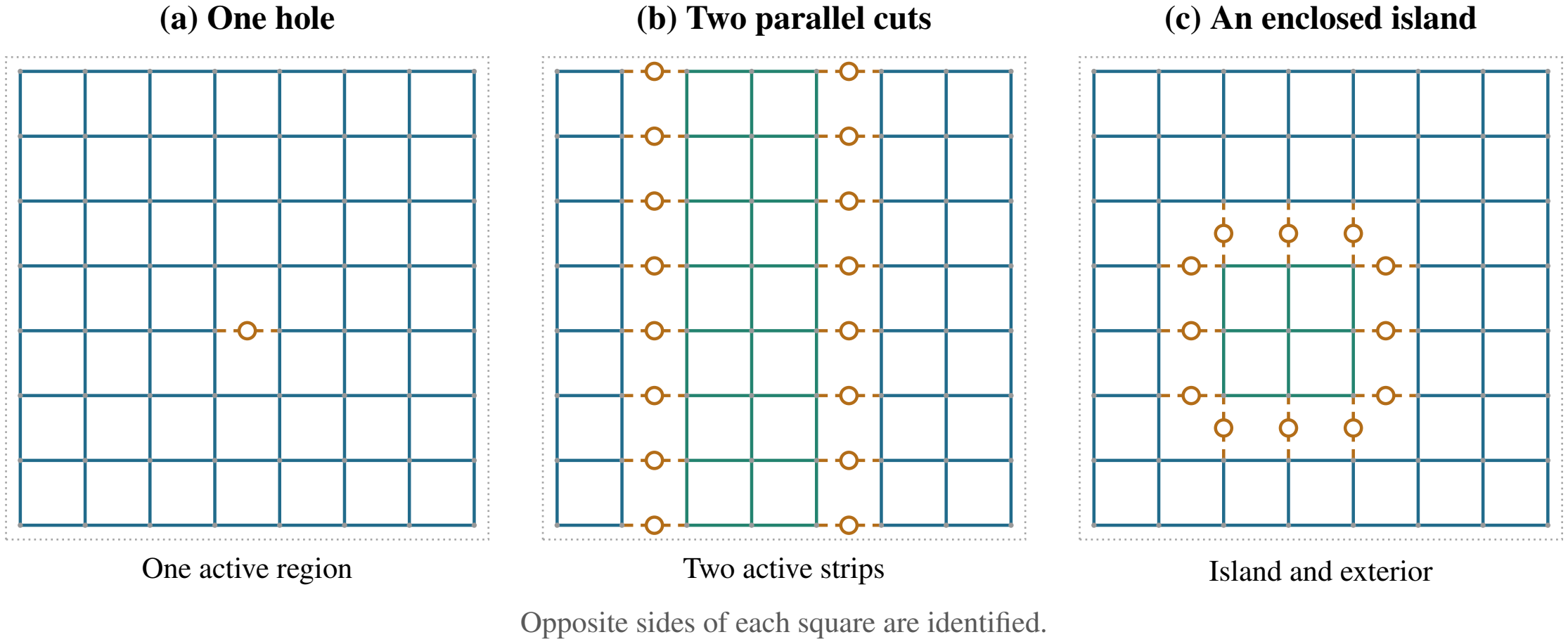
\begin{figure*}[tp]
\centering
\begingroup
\definecolor{geomBlue}{HTML}{216B8B}
\definecolor{geomTeal}{HTML}{23836F}
\definecolor{geomHole}{HTML}{B36D18}
\begin{tikzpicture}[x=1cm,y=1cm,font=\small]
\useasboundingbox (-.2,-1.08) rectangle (14.7,4.95);
\begin{scope}[shift={(0.0,0)}]
\node[font=\small\bfseries] at (2.135,4.75) {(a) One hole};
\begin{scope}[scale=.61]
\draw[draw=geomBlue,line width=.85pt] (0,0) -- (1,0);
\draw[draw=geomBlue,line width=.85pt] (0,1) -- (1,1);
\draw[draw=geomBlue,line width=.85pt] (0,2) -- (1,2);
\draw[draw=geomBlue,line width=.85pt] (0,3) -- (1,3);
\draw[draw=geomBlue,line width=.85pt] (0,4) -- (1,4);
\draw[draw=geomBlue,line width=.85pt] (0,5) -- (1,5);
\draw[draw=geomBlue,line width=.85pt] (0,6) -- (1,6);
\draw[draw=geomBlue,line width=.85pt] (0,7) -- (1,7);
\draw[draw=geomBlue,line width=.85pt] (1,0) -- (2,0);
\draw[draw=geomBlue,line width=.85pt] (1,1) -- (2,1);
\draw[draw=geomBlue,line width=.85pt] (1,2) -- (2,2);
\draw[draw=geomBlue,line width=.85pt] (1,3) -- (2,3);
\draw[draw=geomBlue,line width=.85pt] (1,4) -- (2,4);
\draw[draw=geomBlue,line width=.85pt] (1,5) -- (2,5);
\draw[draw=geomBlue,line width=.85pt] (1,6) -- (2,6);
\draw[draw=geomBlue,line width=.85pt] (1,7) -- (2,7);
\draw[draw=geomBlue,line width=.85pt] (2,0) -- (3,0);
\draw[draw=geomBlue,line width=.85pt] (2,1) -- (3,1);
\draw[draw=geomBlue,line width=.85pt] (2,2) -- (3,2);
\draw[draw=geomBlue,line width=.85pt] (2,3) -- (3,3);
\draw[draw=geomBlue,line width=.85pt] (2,4) -- (3,4);
\draw[draw=geomBlue,line width=.85pt] (2,5) -- (3,5);
\draw[draw=geomBlue,line width=.85pt] (2,6) -- (3,6);
\draw[draw=geomBlue,line width=.85pt] (2,7) -- (3,7);
\draw[draw=geomBlue,line width=.85pt] (3,0) -- (4,0);
\draw[draw=geomBlue,line width=.85pt] (3,1) -- (4,1);
\draw[draw=geomBlue,line width=.85pt] (3,2) -- (4,2);
\draw[draw=geomHole,line width=.85pt,densely dashed] (3,3) -- (4,3);
\draw[draw=geomHole,fill=white,line width=.85pt] (3.5,3.0) circle[radius=1.25mm];
\draw[draw=geomBlue,line width=.85pt] (3,4) -- (4,4);
\draw[draw=geomBlue,line width=.85pt] (3,5) -- (4,5);
\draw[draw=geomBlue,line width=.85pt] (3,6) -- (4,6);
\draw[draw=geomBlue,line width=.85pt] (3,7) -- (4,7);
\draw[draw=geomBlue,line width=.85pt] (4,0) -- (5,0);
\draw[draw=geomBlue,line width=.85pt] (4,1) -- (5,1);
\draw[draw=geomBlue,line width=.85pt] (4,2) -- (5,2);
\draw[draw=geomBlue,line width=.85pt] (4,3) -- (5,3);
\draw[draw=geomBlue,line width=.85pt] (4,4) -- (5,4);
\draw[draw=geomBlue,line width=.85pt] (4,5) -- (5,5);
\draw[draw=geomBlue,line width=.85pt] (4,6) -- (5,6);
\draw[draw=geomBlue,line width=.85pt] (4,7) -- (5,7);
\draw[draw=geomBlue,line width=.85pt] (5,0) -- (6,0);
\draw[draw=geomBlue,line width=.85pt] (5,1) -- (6,1);
\draw[draw=geomBlue,line width=.85pt] (5,2) -- (6,2);
\draw[draw=geomBlue,line width=.85pt] (5,3) -- (6,3);
\draw[draw=geomBlue,line width=.85pt] (5,4) -- (6,4);
\draw[draw=geomBlue,line width=.85pt] (5,5) -- (6,5);
\draw[draw=geomBlue,line width=.85pt] (5,6) -- (6,6);
\draw[draw=geomBlue,line width=.85pt] (5,7) -- (6,7);
\draw[draw=geomBlue,line width=.85pt] (6,0) -- (7,0);
\draw[draw=geomBlue,line width=.85pt] (6,1) -- (7,1);
\draw[draw=geomBlue,line width=.85pt] (6,2) -- (7,2);
\draw[draw=geomBlue,line width=.85pt] (6,3) -- (7,3);
\draw[draw=geomBlue,line width=.85pt] (6,4) -- (7,4);
\draw[draw=geomBlue,line width=.85pt] (6,5) -- (7,5);
\draw[draw=geomBlue,line width=.85pt] (6,6) -- (7,6);
\draw[draw=geomBlue,line width=.85pt] (6,7) -- (7,7);
\draw[draw=geomBlue,line width=.85pt] (0,0) -- (0,1);
\draw[draw=geomBlue,line width=.85pt] (0,1) -- (0,2);
\draw[draw=geomBlue,line width=.85pt] (0,2) -- (0,3);
\draw[draw=geomBlue,line width=.85pt] (0,3) -- (0,4);
\draw[draw=geomBlue,line width=.85pt] (0,4) -- (0,5);
\draw[draw=geomBlue,line width=.85pt] (0,5) -- (0,6);
\draw[draw=geomBlue,line width=.85pt] (0,6) -- (0,7);
\draw[draw=geomBlue,line width=.85pt] (1,0) -- (1,1);
\draw[draw=geomBlue,line width=.85pt] (1,1) -- (1,2);
\draw[draw=geomBlue,line width=.85pt] (1,2) -- (1,3);
\draw[draw=geomBlue,line width=.85pt] (1,3) -- (1,4);
\draw[draw=geomBlue,line width=.85pt] (1,4) -- (1,5);
\draw[draw=geomBlue,line width=.85pt] (1,5) -- (1,6);
\draw[draw=geomBlue,line width=.85pt] (1,6) -- (1,7);
\draw[draw=geomBlue,line width=.85pt] (2,0) -- (2,1);
\draw[draw=geomBlue,line width=.85pt] (2,1) -- (2,2);
\draw[draw=geomBlue,line width=.85pt] (2,2) -- (2,3);
\draw[draw=geomBlue,line width=.85pt] (2,3) -- (2,4);
\draw[draw=geomBlue,line width=.85pt] (2,4) -- (2,5);
\draw[draw=geomBlue,line width=.85pt] (2,5) -- (2,6);
\draw[draw=geomBlue,line width=.85pt] (2,6) -- (2,7);
\draw[draw=geomBlue,line width=.85pt] (3,0) -- (3,1);
\draw[draw=geomBlue,line width=.85pt] (3,1) -- (3,2);
\draw[draw=geomBlue,line width=.85pt] (3,2) -- (3,3);
\draw[draw=geomBlue,line width=.85pt] (3,3) -- (3,4);
\draw[draw=geomBlue,line width=.85pt] (3,4) -- (3,5);
\draw[draw=geomBlue,line width=.85pt] (3,5) -- (3,6);
\draw[draw=geomBlue,line width=.85pt] (3,6) -- (3,7);
\draw[draw=geomBlue,line width=.85pt] (4,0) -- (4,1);
\draw[draw=geomBlue,line width=.85pt] (4,1) -- (4,2);
\draw[draw=geomBlue,line width=.85pt] (4,2) -- (4,3);
\draw[draw=geomBlue,line width=.85pt] (4,3) -- (4,4);
\draw[draw=geomBlue,line width=.85pt] (4,4) -- (4,5);
\draw[draw=geomBlue,line width=.85pt] (4,5) -- (4,6);
\draw[draw=geomBlue,line width=.85pt] (4,6) -- (4,7);
\draw[draw=geomBlue,line width=.85pt] (5,0) -- (5,1);
\draw[draw=geomBlue,line width=.85pt] (5,1) -- (5,2);
\draw[draw=geomBlue,line width=.85pt] (5,2) -- (5,3);
\draw[draw=geomBlue,line width=.85pt] (5,3) -- (5,4);
\draw[draw=geomBlue,line width=.85pt] (5,4) -- (5,5);
\draw[draw=geomBlue,line width=.85pt] (5,5) -- (5,6);
\draw[draw=geomBlue,line width=.85pt] (5,6) -- (5,7);
\draw[draw=geomBlue,line width=.85pt] (6,0) -- (6,1);
\draw[draw=geomBlue,line width=.85pt] (6,1) -- (6,2);
\draw[draw=geomBlue,line width=.85pt] (6,2) -- (6,3);
\draw[draw=geomBlue,line width=.85pt] (6,3) -- (6,4);
\draw[draw=geomBlue,line width=.85pt] (6,4) -- (6,5);
\draw[draw=geomBlue,line width=.85pt] (6,5) -- (6,6);
\draw[draw=geomBlue,line width=.85pt] (6,6) -- (6,7);
\draw[draw=geomBlue,line width=.85pt] (7,0) -- (7,1);
\draw[draw=geomBlue,line width=.85pt] (7,1) -- (7,2);
\draw[draw=geomBlue,line width=.85pt] (7,2) -- (7,3);
\draw[draw=geomBlue,line width=.85pt] (7,3) -- (7,4);
\draw[draw=geomBlue,line width=.85pt] (7,4) -- (7,5);
\draw[draw=geomBlue,line width=.85pt] (7,5) -- (7,6);
\draw[draw=geomBlue,line width=.85pt] (7,6) -- (7,7);
\fill[black!40] (0,0) circle[radius=.4mm];
\fill[black!40] (0,1) circle[radius=.4mm];
\fill[black!40] (0,2) circle[radius=.4mm];
\fill[black!40] (0,3) circle[radius=.4mm];
\fill[black!40] (0,4) circle[radius=.4mm];
\fill[black!40] (0,5) circle[radius=.4mm];
\fill[black!40] (0,6) circle[radius=.4mm];
\fill[black!40] (0,7) circle[radius=.4mm];
\fill[black!40] (1,0) circle[radius=.4mm];
\fill[black!40] (1,1) circle[radius=.4mm];
\fill[black!40] (1,2) circle[radius=.4mm];
\fill[black!40] (1,3) circle[radius=.4mm];
\fill[black!40] (1,4) circle[radius=.4mm];
\fill[black!40] (1,5) circle[radius=.4mm];
\fill[black!40] (1,6) circle[radius=.4mm];
\fill[black!40] (1,7) circle[radius=.4mm];
\fill[black!40] (2,0) circle[radius=.4mm];
\fill[black!40] (2,1) circle[radius=.4mm];
\fill[black!40] (2,2) circle[radius=.4mm];
\fill[black!40] (2,3) circle[radius=.4mm];
\fill[black!40] (2,4) circle[radius=.4mm];
\fill[black!40] (2,5) circle[radius=.4mm];
\fill[black!40] (2,6) circle[radius=.4mm];
\fill[black!40] (2,7) circle[radius=.4mm];
\fill[black!40] (3,0) circle[radius=.4mm];
\fill[black!40] (3,1) circle[radius=.4mm];
\fill[black!40] (3,2) circle[radius=.4mm];
\fill[black!40] (3,3) circle[radius=.4mm];
\fill[black!40] (3,4) circle[radius=.4mm];
\fill[black!40] (3,5) circle[radius=.4mm];
\fill[black!40] (3,6) circle[radius=.4mm];
\fill[black!40] (3,7) circle[radius=.4mm];
\fill[black!40] (4,0) circle[radius=.4mm];
\fill[black!40] (4,1) circle[radius=.4mm];
\fill[black!40] (4,2) circle[radius=.4mm];
\fill[black!40] (4,3) circle[radius=.4mm];
\fill[black!40] (4,4) circle[radius=.4mm];
\fill[black!40] (4,5) circle[radius=.4mm];
\fill[black!40] (4,6) circle[radius=.4mm];
\fill[black!40] (4,7) circle[radius=.4mm];
\fill[black!40] (5,0) circle[radius=.4mm];
\fill[black!40] (5,1) circle[radius=.4mm];
\fill[black!40] (5,2) circle[radius=.4mm];
\fill[black!40] (5,3) circle[radius=.4mm];
\fill[black!40] (5,4) circle[radius=.4mm];
\fill[black!40] (5,5) circle[radius=.4mm];
\fill[black!40] (5,6) circle[radius=.4mm];
\fill[black!40] (5,7) circle[radius=.4mm];
\fill[black!40] (6,0) circle[radius=.4mm];
\fill[black!40] (6,1) circle[radius=.4mm];
\fill[black!40] (6,2) circle[radius=.4mm];
\fill[black!40] (6,3) circle[radius=.4mm];
\fill[black!40] (6,4) circle[radius=.4mm];
\fill[black!40] (6,5) circle[radius=.4mm];
\fill[black!40] (6,6) circle[radius=.4mm];
\fill[black!40] (6,7) circle[radius=.4mm];
\fill[black!40] (7,0) circle[radius=.4mm];
\fill[black!40] (7,1) circle[radius=.4mm];
\fill[black!40] (7,2) circle[radius=.4mm];
\fill[black!40] (7,3) circle[radius=.4mm];
\fill[black!40] (7,4) circle[radius=.4mm];
\fill[black!40] (7,5) circle[radius=.4mm];
\fill[black!40] (7,6) circle[radius=.4mm];
\fill[black!40] (7,7) circle[radius=.4mm];
\draw[black!35,densely dotted,line width=.5pt] (-.22,-.22) rectangle (7.22,7.22);
\end{scope}
\node[font=\footnotesize] at (2.135,-.43) {One active region};
\end{scope}
\begin{scope}[shift={(5.05,0)}]
\node[font=\small\bfseries] at (2.135,4.75) {(b) Two parallel cuts};
\begin{scope}[scale=.61]
\draw[draw=geomBlue,line width=.85pt] (0,0) -- (1,0);
\draw[draw=geomBlue,line width=.85pt] (0,1) -- (1,1);
\draw[draw=geomBlue,line width=.85pt] (0,2) -- (1,2);
\draw[draw=geomBlue,line width=.85pt] (0,3) -- (1,3);
\draw[draw=geomBlue,line width=.85pt] (0,4) -- (1,4);
\draw[draw=geomBlue,line width=.85pt] (0,5) -- (1,5);
\draw[draw=geomBlue,line width=.85pt] (0,6) -- (1,6);
\draw[draw=geomBlue,line width=.85pt] (0,7) -- (1,7);
\draw[draw=geomHole,line width=.85pt,densely dashed] (1,0) -- (2,0);
\draw[draw=geomHole,fill=white,line width=.85pt] (1.5,0.0) circle[radius=1.25mm];
\draw[draw=geomHole,line width=.85pt,densely dashed] (1,1) -- (2,1);
\draw[draw=geomHole,fill=white,line width=.85pt] (1.5,1.0) circle[radius=1.25mm];
\draw[draw=geomHole,line width=.85pt,densely dashed] (1,2) -- (2,2);
\draw[draw=geomHole,fill=white,line width=.85pt] (1.5,2.0) circle[radius=1.25mm];
\draw[draw=geomHole,line width=.85pt,densely dashed] (1,3) -- (2,3);
\draw[draw=geomHole,fill=white,line width=.85pt] (1.5,3.0) circle[radius=1.25mm];
\draw[draw=geomHole,line width=.85pt,densely dashed] (1,4) -- (2,4);
\draw[draw=geomHole,fill=white,line width=.85pt] (1.5,4.0) circle[radius=1.25mm];
\draw[draw=geomHole,line width=.85pt,densely dashed] (1,5) -- (2,5);
\draw[draw=geomHole,fill=white,line width=.85pt] (1.5,5.0) circle[radius=1.25mm];
\draw[draw=geomHole,line width=.85pt,densely dashed] (1,6) -- (2,6);
\draw[draw=geomHole,fill=white,line width=.85pt] (1.5,6.0) circle[radius=1.25mm];
\draw[draw=geomHole,line width=.85pt,densely dashed] (1,7) -- (2,7);
\draw[draw=geomHole,fill=white,line width=.85pt] (1.5,7.0) circle[radius=1.25mm];
\draw[draw=geomTeal,line width=.85pt] (2,0) -- (3,0);
\draw[draw=geomTeal,line width=.85pt] (2,1) -- (3,1);
\draw[draw=geomTeal,line width=.85pt] (2,2) -- (3,2);
\draw[draw=geomTeal,line width=.85pt] (2,3) -- (3,3);
\draw[draw=geomTeal,line width=.85pt] (2,4) -- (3,4);
\draw[draw=geomTeal,line width=.85pt] (2,5) -- (3,5);
\draw[draw=geomTeal,line width=.85pt] (2,6) -- (3,6);
\draw[draw=geomTeal,line width=.85pt] (2,7) -- (3,7);
\draw[draw=geomTeal,line width=.85pt] (3,0) -- (4,0);
\draw[draw=geomTeal,line width=.85pt] (3,1) -- (4,1);
\draw[draw=geomTeal,line width=.85pt] (3,2) -- (4,2);
\draw[draw=geomTeal,line width=.85pt] (3,3) -- (4,3);
\draw[draw=geomTeal,line width=.85pt] (3,4) -- (4,4);
\draw[draw=geomTeal,line width=.85pt] (3,5) -- (4,5);
\draw[draw=geomTeal,line width=.85pt] (3,6) -- (4,6);
\draw[draw=geomTeal,line width=.85pt] (3,7) -- (4,7);
\draw[draw=geomHole,line width=.85pt,densely dashed] (4,0) -- (5,0);
\draw[draw=geomHole,fill=white,line width=.85pt] (4.5,0.0) circle[radius=1.25mm];
\draw[draw=geomHole,line width=.85pt,densely dashed] (4,1) -- (5,1);
\draw[draw=geomHole,fill=white,line width=.85pt] (4.5,1.0) circle[radius=1.25mm];
\draw[draw=geomHole,line width=.85pt,densely dashed] (4,2) -- (5,2);
\draw[draw=geomHole,fill=white,line width=.85pt] (4.5,2.0) circle[radius=1.25mm];
\draw[draw=geomHole,line width=.85pt,densely dashed] (4,3) -- (5,3);
\draw[draw=geomHole,fill=white,line width=.85pt] (4.5,3.0) circle[radius=1.25mm];
\draw[draw=geomHole,line width=.85pt,densely dashed] (4,4) -- (5,4);
\draw[draw=geomHole,fill=white,line width=.85pt] (4.5,4.0) circle[radius=1.25mm];
\draw[draw=geomHole,line width=.85pt,densely dashed] (4,5) -- (5,5);
\draw[draw=geomHole,fill=white,line width=.85pt] (4.5,5.0) circle[radius=1.25mm];
\draw[draw=geomHole,line width=.85pt,densely dashed] (4,6) -- (5,6);
\draw[draw=geomHole,fill=white,line width=.85pt] (4.5,6.0) circle[radius=1.25mm];
\draw[draw=geomHole,line width=.85pt,densely dashed] (4,7) -- (5,7);
\draw[draw=geomHole,fill=white,line width=.85pt] (4.5,7.0) circle[radius=1.25mm];
\draw[draw=geomBlue,line width=.85pt] (5,0) -- (6,0);
\draw[draw=geomBlue,line width=.85pt] (5,1) -- (6,1);
\draw[draw=geomBlue,line width=.85pt] (5,2) -- (6,2);
\draw[draw=geomBlue,line width=.85pt] (5,3) -- (6,3);
\draw[draw=geomBlue,line width=.85pt] (5,4) -- (6,4);
\draw[draw=geomBlue,line width=.85pt] (5,5) -- (6,5);
\draw[draw=geomBlue,line width=.85pt] (5,6) -- (6,6);
\draw[draw=geomBlue,line width=.85pt] (5,7) -- (6,7);
\draw[draw=geomBlue,line width=.85pt] (6,0) -- (7,0);
\draw[draw=geomBlue,line width=.85pt] (6,1) -- (7,1);
\draw[draw=geomBlue,line width=.85pt] (6,2) -- (7,2);
\draw[draw=geomBlue,line width=.85pt] (6,3) -- (7,3);
\draw[draw=geomBlue,line width=.85pt] (6,4) -- (7,4);
\draw[draw=geomBlue,line width=.85pt] (6,5) -- (7,5);
\draw[draw=geomBlue,line width=.85pt] (6,6) -- (7,6);
\draw[draw=geomBlue,line width=.85pt] (6,7) -- (7,7);
\draw[draw=geomBlue,line width=.85pt] (0,0) -- (0,1);
\draw[draw=geomBlue,line width=.85pt] (0,1) -- (0,2);
\draw[draw=geomBlue,line width=.85pt] (0,2) -- (0,3);
\draw[draw=geomBlue,line width=.85pt] (0,3) -- (0,4);
\draw[draw=geomBlue,line width=.85pt] (0,4) -- (0,5);
\draw[draw=geomBlue,line width=.85pt] (0,5) -- (0,6);
\draw[draw=geomBlue,line width=.85pt] (0,6) -- (0,7);
\draw[draw=geomBlue,line width=.85pt] (1,0) -- (1,1);
\draw[draw=geomBlue,line width=.85pt] (1,1) -- (1,2);
\draw[draw=geomBlue,line width=.85pt] (1,2) -- (1,3);
\draw[draw=geomBlue,line width=.85pt] (1,3) -- (1,4);
\draw[draw=geomBlue,line width=.85pt] (1,4) -- (1,5);
\draw[draw=geomBlue,line width=.85pt] (1,5) -- (1,6);
\draw[draw=geomBlue,line width=.85pt] (1,6) -- (1,7);
\draw[draw=geomTeal,line width=.85pt] (2,0) -- (2,1);
\draw[draw=geomTeal,line width=.85pt] (2,1) -- (2,2);
\draw[draw=geomTeal,line width=.85pt] (2,2) -- (2,3);
\draw[draw=geomTeal,line width=.85pt] (2,3) -- (2,4);
\draw[draw=geomTeal,line width=.85pt] (2,4) -- (2,5);
\draw[draw=geomTeal,line width=.85pt] (2,5) -- (2,6);
\draw[draw=geomTeal,line width=.85pt] (2,6) -- (2,7);
\draw[draw=geomTeal,line width=.85pt] (3,0) -- (3,1);
\draw[draw=geomTeal,line width=.85pt] (3,1) -- (3,2);
\draw[draw=geomTeal,line width=.85pt] (3,2) -- (3,3);
\draw[draw=geomTeal,line width=.85pt] (3,3) -- (3,4);
\draw[draw=geomTeal,line width=.85pt] (3,4) -- (3,5);
\draw[draw=geomTeal,line width=.85pt] (3,5) -- (3,6);
\draw[draw=geomTeal,line width=.85pt] (3,6) -- (3,7);
\draw[draw=geomTeal,line width=.85pt] (4,0) -- (4,1);
\draw[draw=geomTeal,line width=.85pt] (4,1) -- (4,2);
\draw[draw=geomTeal,line width=.85pt] (4,2) -- (4,3);
\draw[draw=geomTeal,line width=.85pt] (4,3) -- (4,4);
\draw[draw=geomTeal,line width=.85pt] (4,4) -- (4,5);
\draw[draw=geomTeal,line width=.85pt] (4,5) -- (4,6);
\draw[draw=geomTeal,line width=.85pt] (4,6) -- (4,7);
\draw[draw=geomBlue,line width=.85pt] (5,0) -- (5,1);
\draw[draw=geomBlue,line width=.85pt] (5,1) -- (5,2);
\draw[draw=geomBlue,line width=.85pt] (5,2) -- (5,3);
\draw[draw=geomBlue,line width=.85pt] (5,3) -- (5,4);
\draw[draw=geomBlue,line width=.85pt] (5,4) -- (5,5);
\draw[draw=geomBlue,line width=.85pt] (5,5) -- (5,6);
\draw[draw=geomBlue,line width=.85pt] (5,6) -- (5,7);
\draw[draw=geomBlue,line width=.85pt] (6,0) -- (6,1);
\draw[draw=geomBlue,line width=.85pt] (6,1) -- (6,2);
\draw[draw=geomBlue,line width=.85pt] (6,2) -- (6,3);
\draw[draw=geomBlue,line width=.85pt] (6,3) -- (6,4);
\draw[draw=geomBlue,line width=.85pt] (6,4) -- (6,5);
\draw[draw=geomBlue,line width=.85pt] (6,5) -- (6,6);
\draw[draw=geomBlue,line width=.85pt] (6,6) -- (6,7);
\draw[draw=geomBlue,line width=.85pt] (7,0) -- (7,1);
\draw[draw=geomBlue,line width=.85pt] (7,1) -- (7,2);
\draw[draw=geomBlue,line width=.85pt] (7,2) -- (7,3);
\draw[draw=geomBlue,line width=.85pt] (7,3) -- (7,4);
\draw[draw=geomBlue,line width=.85pt] (7,4) -- (7,5);
\draw[draw=geomBlue,line width=.85pt] (7,5) -- (7,6);
\draw[draw=geomBlue,line width=.85pt] (7,6) -- (7,7);
\fill[black!40] (0,0) circle[radius=.4mm];
\fill[black!40] (0,1) circle[radius=.4mm];
\fill[black!40] (0,2) circle[radius=.4mm];
\fill[black!40] (0,3) circle[radius=.4mm];
\fill[black!40] (0,4) circle[radius=.4mm];
\fill[black!40] (0,5) circle[radius=.4mm];
\fill[black!40] (0,6) circle[radius=.4mm];
\fill[black!40] (0,7) circle[radius=.4mm];
\fill[black!40] (1,0) circle[radius=.4mm];
\fill[black!40] (1,1) circle[radius=.4mm];
\fill[black!40] (1,2) circle[radius=.4mm];
\fill[black!40] (1,3) circle[radius=.4mm];
\fill[black!40] (1,4) circle[radius=.4mm];
\fill[black!40] (1,5) circle[radius=.4mm];
\fill[black!40] (1,6) circle[radius=.4mm];
\fill[black!40] (1,7) circle[radius=.4mm];
\fill[black!40] (2,0) circle[radius=.4mm];
\fill[black!40] (2,1) circle[radius=.4mm];
\fill[black!40] (2,2) circle[radius=.4mm];
\fill[black!40] (2,3) circle[radius=.4mm];
\fill[black!40] (2,4) circle[radius=.4mm];
\fill[black!40] (2,5) circle[radius=.4mm];
\fill[black!40] (2,6) circle[radius=.4mm];
\fill[black!40] (2,7) circle[radius=.4mm];
\fill[black!40] (3,0) circle[radius=.4mm];
\fill[black!40] (3,1) circle[radius=.4mm];
\fill[black!40] (3,2) circle[radius=.4mm];
\fill[black!40] (3,3) circle[radius=.4mm];
\fill[black!40] (3,4) circle[radius=.4mm];
\fill[black!40] (3,5) circle[radius=.4mm];
\fill[black!40] (3,6) circle[radius=.4mm];
\fill[black!40] (3,7) circle[radius=.4mm];
\fill[black!40] (4,0) circle[radius=.4mm];
\fill[black!40] (4,1) circle[radius=.4mm];
\fill[black!40] (4,2) circle[radius=.4mm];
\fill[black!40] (4,3) circle[radius=.4mm];
\fill[black!40] (4,4) circle[radius=.4mm];
\fill[black!40] (4,5) circle[radius=.4mm];
\fill[black!40] (4,6) circle[radius=.4mm];
\fill[black!40] (4,7) circle[radius=.4mm];
\fill[black!40] (5,0) circle[radius=.4mm];
\fill[black!40] (5,1) circle[radius=.4mm];
\fill[black!40] (5,2) circle[radius=.4mm];
\fill[black!40] (5,3) circle[radius=.4mm];
\fill[black!40] (5,4) circle[radius=.4mm];
\fill[black!40] (5,5) circle[radius=.4mm];
\fill[black!40] (5,6) circle[radius=.4mm];
\fill[black!40] (5,7) circle[radius=.4mm];
\fill[black!40] (6,0) circle[radius=.4mm];
\fill[black!40] (6,1) circle[radius=.4mm];
\fill[black!40] (6,2) circle[radius=.4mm];
\fill[black!40] (6,3) circle[radius=.4mm];
\fill[black!40] (6,4) circle[radius=.4mm];
\fill[black!40] (6,5) circle[radius=.4mm];
\fill[black!40] (6,6) circle[radius=.4mm];
\fill[black!40] (6,7) circle[radius=.4mm];
\fill[black!40] (7,0) circle[radius=.4mm];
\fill[black!40] (7,1) circle[radius=.4mm];
\fill[black!40] (7,2) circle[radius=.4mm];
\fill[black!40] (7,3) circle[radius=.4mm];
\fill[black!40] (7,4) circle[radius=.4mm];
\fill[black!40] (7,5) circle[radius=.4mm];
\fill[black!40] (7,6) circle[radius=.4mm];
\fill[black!40] (7,7) circle[radius=.4mm];
\draw[black!35,densely dotted,line width=.5pt] (-.22,-.22) rectangle (7.22,7.22);
\end{scope}
\node[font=\footnotesize] at (2.135,-.43) {Two active strips};
\end{scope}
\begin{scope}[shift={(10.1,0)}]
\node[font=\small\bfseries] at (2.135,4.75) {(c) An enclosed island};
\begin{scope}[scale=.61]
\draw[draw=geomBlue,line width=.85pt] (0,0) -- (1,0);
\draw[draw=geomBlue,line width=.85pt] (0,1) -- (1,1);
\draw[draw=geomBlue,line width=.85pt] (0,2) -- (1,2);
\draw[draw=geomBlue,line width=.85pt] (0,3) -- (1,3);
\draw[draw=geomBlue,line width=.85pt] (0,4) -- (1,4);
\draw[draw=geomBlue,line width=.85pt] (0,5) -- (1,5);
\draw[draw=geomBlue,line width=.85pt] (0,6) -- (1,6);
\draw[draw=geomBlue,line width=.85pt] (0,7) -- (1,7);
\draw[draw=geomBlue,line width=.85pt] (1,0) -- (2,0);
\draw[draw=geomBlue,line width=.85pt] (1,1) -- (2,1);
\draw[draw=geomHole,line width=.85pt,densely dashed] (1,2) -- (2,2);
\draw[draw=geomHole,fill=white,line width=.85pt] (1.5,2.0) circle[radius=1.25mm];
\draw[draw=geomHole,line width=.85pt,densely dashed] (1,3) -- (2,3);
\draw[draw=geomHole,fill=white,line width=.85pt] (1.5,3.0) circle[radius=1.25mm];
\draw[draw=geomHole,line width=.85pt,densely dashed] (1,4) -- (2,4);
\draw[draw=geomHole,fill=white,line width=.85pt] (1.5,4.0) circle[radius=1.25mm];
\draw[draw=geomBlue,line width=.85pt] (1,5) -- (2,5);
\draw[draw=geomBlue,line width=.85pt] (1,6) -- (2,6);
\draw[draw=geomBlue,line width=.85pt] (1,7) -- (2,7);
\draw[draw=geomBlue,line width=.85pt] (2,0) -- (3,0);
\draw[draw=geomBlue,line width=.85pt] (2,1) -- (3,1);
\draw[draw=geomTeal,line width=.85pt] (2,2) -- (3,2);
\draw[draw=geomTeal,line width=.85pt] (2,3) -- (3,3);
\draw[draw=geomTeal,line width=.85pt] (2,4) -- (3,4);
\draw[draw=geomBlue,line width=.85pt] (2,5) -- (3,5);
\draw[draw=geomBlue,line width=.85pt] (2,6) -- (3,6);
\draw[draw=geomBlue,line width=.85pt] (2,7) -- (3,7);
\draw[draw=geomBlue,line width=.85pt] (3,0) -- (4,0);
\draw[draw=geomBlue,line width=.85pt] (3,1) -- (4,1);
\draw[draw=geomTeal,line width=.85pt] (3,2) -- (4,2);
\draw[draw=geomTeal,line width=.85pt] (3,3) -- (4,3);
\draw[draw=geomTeal,line width=.85pt] (3,4) -- (4,4);
\draw[draw=geomBlue,line width=.85pt] (3,5) -- (4,5);
\draw[draw=geomBlue,line width=.85pt] (3,6) -- (4,6);
\draw[draw=geomBlue,line width=.85pt] (3,7) -- (4,7);
\draw[draw=geomBlue,line width=.85pt] (4,0) -- (5,0);
\draw[draw=geomBlue,line width=.85pt] (4,1) -- (5,1);
\draw[draw=geomHole,line width=.85pt,densely dashed] (4,2) -- (5,2);
\draw[draw=geomHole,fill=white,line width=.85pt] (4.5,2.0) circle[radius=1.25mm];
\draw[draw=geomHole,line width=.85pt,densely dashed] (4,3) -- (5,3);
\draw[draw=geomHole,fill=white,line width=.85pt] (4.5,3.0) circle[radius=1.25mm];
\draw[draw=geomHole,line width=.85pt,densely dashed] (4,4) -- (5,4);
\draw[draw=geomHole,fill=white,line width=.85pt] (4.5,4.0) circle[radius=1.25mm];
\draw[draw=geomBlue,line width=.85pt] (4,5) -- (5,5);
\draw[draw=geomBlue,line width=.85pt] (4,6) -- (5,6);
\draw[draw=geomBlue,line width=.85pt] (4,7) -- (5,7);
\draw[draw=geomBlue,line width=.85pt] (5,0) -- (6,0);
\draw[draw=geomBlue,line width=.85pt] (5,1) -- (6,1);
\draw[draw=geomBlue,line width=.85pt] (5,2) -- (6,2);
\draw[draw=geomBlue,line width=.85pt] (5,3) -- (6,3);
\draw[draw=geomBlue,line width=.85pt] (5,4) -- (6,4);
\draw[draw=geomBlue,line width=.85pt] (5,5) -- (6,5);
\draw[draw=geomBlue,line width=.85pt] (5,6) -- (6,6);
\draw[draw=geomBlue,line width=.85pt] (5,7) -- (6,7);
\draw[draw=geomBlue,line width=.85pt] (6,0) -- (7,0);
\draw[draw=geomBlue,line width=.85pt] (6,1) -- (7,1);
\draw[draw=geomBlue,line width=.85pt] (6,2) -- (7,2);
\draw[draw=geomBlue,line width=.85pt] (6,3) -- (7,3);
\draw[draw=geomBlue,line width=.85pt] (6,4) -- (7,4);
\draw[draw=geomBlue,line width=.85pt] (6,5) -- (7,5);
\draw[draw=geomBlue,line width=.85pt] (6,6) -- (7,6);
\draw[draw=geomBlue,line width=.85pt] (6,7) -- (7,7);
\draw[draw=geomBlue,line width=.85pt] (0,0) -- (0,1);
\draw[draw=geomBlue,line width=.85pt] (0,1) -- (0,2);
\draw[draw=geomBlue,line width=.85pt] (0,2) -- (0,3);
\draw[draw=geomBlue,line width=.85pt] (0,3) -- (0,4);
\draw[draw=geomBlue,line width=.85pt] (0,4) -- (0,5);
\draw[draw=geomBlue,line width=.85pt] (0,5) -- (0,6);
\draw[draw=geomBlue,line width=.85pt] (0,6) -- (0,7);
\draw[draw=geomBlue,line width=.85pt] (1,0) -- (1,1);
\draw[draw=geomBlue,line width=.85pt] (1,1) -- (1,2);
\draw[draw=geomBlue,line width=.85pt] (1,2) -- (1,3);
\draw[draw=geomBlue,line width=.85pt] (1,3) -- (1,4);
\draw[draw=geomBlue,line width=.85pt] (1,4) -- (1,5);
\draw[draw=geomBlue,line width=.85pt] (1,5) -- (1,6);
\draw[draw=geomBlue,line width=.85pt] (1,6) -- (1,7);
\draw[draw=geomBlue,line width=.85pt] (2,0) -- (2,1);
\draw[draw=geomHole,line width=.85pt,densely dashed] (2,1) -- (2,2);
\draw[draw=geomHole,fill=white,line width=.85pt] (2.0,1.5) circle[radius=1.25mm];
\draw[draw=geomTeal,line width=.85pt] (2,2) -- (2,3);
\draw[draw=geomTeal,line width=.85pt] (2,3) -- (2,4);
\draw[draw=geomHole,line width=.85pt,densely dashed] (2,4) -- (2,5);
\draw[draw=geomHole,fill=white,line width=.85pt] (2.0,4.5) circle[radius=1.25mm];
\draw[draw=geomBlue,line width=.85pt] (2,5) -- (2,6);
\draw[draw=geomBlue,line width=.85pt] (2,6) -- (2,7);
\draw[draw=geomBlue,line width=.85pt] (3,0) -- (3,1);
\draw[draw=geomHole,line width=.85pt,densely dashed] (3,1) -- (3,2);
\draw[draw=geomHole,fill=white,line width=.85pt] (3.0,1.5) circle[radius=1.25mm];
\draw[draw=geomTeal,line width=.85pt] (3,2) -- (3,3);
\draw[draw=geomTeal,line width=.85pt] (3,3) -- (3,4);
\draw[draw=geomHole,line width=.85pt,densely dashed] (3,4) -- (3,5);
\draw[draw=geomHole,fill=white,line width=.85pt] (3.0,4.5) circle[radius=1.25mm];
\draw[draw=geomBlue,line width=.85pt] (3,5) -- (3,6);
\draw[draw=geomBlue,line width=.85pt] (3,6) -- (3,7);
\draw[draw=geomBlue,line width=.85pt] (4,0) -- (4,1);
\draw[draw=geomHole,line width=.85pt,densely dashed] (4,1) -- (4,2);
\draw[draw=geomHole,fill=white,line width=.85pt] (4.0,1.5) circle[radius=1.25mm];
\draw[draw=geomTeal,line width=.85pt] (4,2) -- (4,3);
\draw[draw=geomTeal,line width=.85pt] (4,3) -- (4,4);
\draw[draw=geomHole,line width=.85pt,densely dashed] (4,4) -- (4,5);
\draw[draw=geomHole,fill=white,line width=.85pt] (4.0,4.5) circle[radius=1.25mm];
\draw[draw=geomBlue,line width=.85pt] (4,5) -- (4,6);
\draw[draw=geomBlue,line width=.85pt] (4,6) -- (4,7);
\draw[draw=geomBlue,line width=.85pt] (5,0) -- (5,1);
\draw[draw=geomBlue,line width=.85pt] (5,1) -- (5,2);
\draw[draw=geomBlue,line width=.85pt] (5,2) -- (5,3);
\draw[draw=geomBlue,line width=.85pt] (5,3) -- (5,4);
\draw[draw=geomBlue,line width=.85pt] (5,4) -- (5,5);
\draw[draw=geomBlue,line width=.85pt] (5,5) -- (5,6);
\draw[draw=geomBlue,line width=.85pt] (5,6) -- (5,7);
\draw[draw=geomBlue,line width=.85pt] (6,0) -- (6,1);
\draw[draw=geomBlue,line width=.85pt] (6,1) -- (6,2);
\draw[draw=geomBlue,line width=.85pt] (6,2) -- (6,3);
\draw[draw=geomBlue,line width=.85pt] (6,3) -- (6,4);
\draw[draw=geomBlue,line width=.85pt] (6,4) -- (6,5);
\draw[draw=geomBlue,line width=.85pt] (6,5) -- (6,6);
\draw[draw=geomBlue,line width=.85pt] (6,6) -- (6,7);
\draw[draw=geomBlue,line width=.85pt] (7,0) -- (7,1);
\draw[draw=geomBlue,line width=.85pt] (7,1) -- (7,2);
\draw[draw=geomBlue,line width=.85pt] (7,2) -- (7,3);
\draw[draw=geomBlue,line width=.85pt] (7,3) -- (7,4);
\draw[draw=geomBlue,line width=.85pt] (7,4) -- (7,5);
\draw[draw=geomBlue,line width=.85pt] (7,5) -- (7,6);
\draw[draw=geomBlue,line width=.85pt] (7,6) -- (7,7);
\fill[black!40] (0,0) circle[radius=.4mm];
\fill[black!40] (0,1) circle[radius=.4mm];
\fill[black!40] (0,2) circle[radius=.4mm];
\fill[black!40] (0,3) circle[radius=.4mm];
\fill[black!40] (0,4) circle[radius=.4mm];
\fill[black!40] (0,5) circle[radius=.4mm];
\fill[black!40] (0,6) circle[radius=.4mm];
\fill[black!40] (0,7) circle[radius=.4mm];
\fill[black!40] (1,0) circle[radius=.4mm];
\fill[black!40] (1,1) circle[radius=.4mm];
\fill[black!40] (1,2) circle[radius=.4mm];
\fill[black!40] (1,3) circle[radius=.4mm];
\fill[black!40] (1,4) circle[radius=.4mm];
\fill[black!40] (1,5) circle[radius=.4mm];
\fill[black!40] (1,6) circle[radius=.4mm];
\fill[black!40] (1,7) circle[radius=.4mm];
\fill[black!40] (2,0) circle[radius=.4mm];
\fill[black!40] (2,1) circle[radius=.4mm];
\fill[black!40] (2,2) circle[radius=.4mm];
\fill[black!40] (2,3) circle[radius=.4mm];
\fill[black!40] (2,4) circle[radius=.4mm];
\fill[black!40] (2,5) circle[radius=.4mm];
\fill[black!40] (2,6) circle[radius=.4mm];
\fill[black!40] (2,7) circle[radius=.4mm];
\fill[black!40] (3,0) circle[radius=.4mm];
\fill[black!40] (3,1) circle[radius=.4mm];
\fill[black!40] (3,2) circle[radius=.4mm];
\fill[black!40] (3,3) circle[radius=.4mm];
\fill[black!40] (3,4) circle[radius=.4mm];
\fill[black!40] (3,5) circle[radius=.4mm];
\fill[black!40] (3,6) circle[radius=.4mm];
\fill[black!40] (3,7) circle[radius=.4mm];
\fill[black!40] (4,0) circle[radius=.4mm];
\fill[black!40] (4,1) circle[radius=.4mm];
\fill[black!40] (4,2) circle[radius=.4mm];
\fill[black!40] (4,3) circle[radius=.4mm];
\fill[black!40] (4,4) circle[radius=.4mm];
\fill[black!40] (4,5) circle[radius=.4mm];
\fill[black!40] (4,6) circle[radius=.4mm];
\fill[black!40] (4,7) circle[radius=.4mm];
\fill[black!40] (5,0) circle[radius=.4mm];
\fill[black!40] (5,1) circle[radius=.4mm];
\fill[black!40] (5,2) circle[radius=.4mm];
\fill[black!40] (5,3) circle[radius=.4mm];
\fill[black!40] (5,4) circle[radius=.4mm];
\fill[black!40] (5,5) circle[radius=.4mm];
\fill[black!40] (5,6) circle[radius=.4mm];
\fill[black!40] (5,7) circle[radius=.4mm];
\fill[black!40] (6,0) circle[radius=.4mm];
\fill[black!40] (6,1) circle[radius=.4mm];
\fill[black!40] (6,2) circle[radius=.4mm];
\fill[black!40] (6,3) circle[radius=.4mm];
\fill[black!40] (6,4) circle[radius=.4mm];
\fill[black!40] (6,5) circle[radius=.4mm];
\fill[black!40] (6,6) circle[radius=.4mm];
\fill[black!40] (6,7) circle[radius=.4mm];
\fill[black!40] (7,0) circle[radius=.4mm];
\fill[black!40] (7,1) circle[radius=.4mm];
\fill[black!40] (7,2) circle[radius=.4mm];
\fill[black!40] (7,3) circle[radius=.4mm];
\fill[black!40] (7,4) circle[radius=.4mm];
\fill[black!40] (7,5) circle[radius=.4mm];
\fill[black!40] (7,6) circle[radius=.4mm];
\fill[black!40] (7,7) circle[radius=.4mm];
\draw[black!35,densely dotted,line width=.5pt] (-.22,-.22) rectangle (7.22,7.22);
\end{scope}
\node[font=\footnotesize] at (2.135,-.43) {Island and exterior};
\end{scope}
\node[font=\footnotesize,text=black!65] at (7.185,-.94) {Opposite sides of each square are identified.};
\end{tikzpicture}
\endgroup
\caption{Active regions of the minimally lifted toric code on a $7\times7$ periodic square lattice. Colored edge segments represent active qubits, and open orange circles mark hole edges. Blue and green distinguish components connected by surviving star or plaquette supports. Boundary edges and vertices are repeated to show the periodic identification. (a) One hole leaves the remaining active edges connected. (b) Two parallel noncontractible cuts separate two active strips. A single such cut would leave one connected cylinder. (c) Hole edges crossing a contractible boundary isolate an active island from its exterior. Components are determined by the surviving interaction supports, not by visual adjacency alone.}
\label{fig:geometry-toric}
\end{figure*}

Figure~\ref{fig:geometry-toric} illustrates three possible patterns on a torus. An isolated hole leaves the other active edges in one component, because surviving supports connect them around the defect. A single noncontractible cut of hole edges likewise leaves a connected cylinder. Two separated parallel cuts can instead divide the active edges into independent strips. A contractible closed cut can isolate an active island from its exterior. These examples show why both the hole pattern and the boundary conditions matter.

Every surviving term in Eq.~\eqref{eq:applications-toric-restricted} acts inside one such region. The restricted Hamiltonian is consequently a sum of region Hamiltonians, and its time-evolution operator factorizes over their Hilbert spaces, as formalized in Eq.~\eqref{eq:general-geometric-factorization}. Additional interactions would have to be checked on their own supports and could reconnect these regions without moving the holes. The general range-dependent criterion is given in Sec.~\ref{subsec:general-geometry}.

\paragraph*{Pattern sizes and strong fragmentation.}
The $2^{N_E}$ invariant patterns have dimensions $2^{|A|}$. At fixed hole number $m$, the largest connected block lies within a single pattern, so
\begin{equation}
 \dim\mathcal S_m=\binom{N_E}{m}2^{N_E-m},
 \qquad
 \frac{D_{\max}^{\mathrm{conn}}(\mathcal S_m)}{\dim\mathcal S_m}
 \leq\binom{N_E}{m}^{-1}.
 \label{eq:applications-toric-count}
\end{equation}
This gives the same exponential suppression at intermediate hole density as in the preceding examples. The commuting stabilizers provide further invariant subspaces within each pattern, so the pattern projectors give only a lower bound on the common commutant.

The surviving constraints also determine the lowest-energy states within each fixed hole pattern. Their degeneracy can depend on the relative positions of the holes, even at fixed hole number. This \emph{sector ground-state degeneracy} is distinct from both the number of invariant patterns and the ground-state degeneracy of the full lifted Hamiltonian. For the positive couplings assumed here, the full lifted Hamiltonian retains its four global ground states in the fully active sector. Appendix~\ref{app:toric-sector-ground-states} gives the explicit one-hole ground states, compares two-hole configurations and derives the counting formula for arbitrary hole patterns.

\subsection{Time-dependent Hamiltonians and quantum circuits}
\label{subsec:applications-time}

The preceding examples have static couplings, but the conservation
of hole patterns does not require a time-independent Hamiltonian.
The essential condition is that every instantaneous interaction
preserves the same active and inactive local subspaces. The dynamics
inside each pattern can then change with time, while transitions
between different patterns remain forbidden.

To make this precise, keep the local representations fixed and use
the same maps
\[
 \phi_i(O)=(O\otimes\id_{r_i})\oplus0_{s_i},
 \qquad
 P_i=\phi_i(\id_{n_i}),
 \qquad
 Z_i=\id-P_i
\]
throughout the evolution. Here $r_i,s_i\geq1$, as in the preceding
examples. The projectors $P_i$ and $Z_i$ are therefore independent
of time. Keep the declared interaction supports and their spectator
identities fixed as well, but allow the supported seed densities to
vary with time,
\begin{equation}
 H_\Lambda(t)=c_\Lambda(t)\id
       +\sum_\alpha h_\alpha(t)
                    \otimes\id_{\Lambda\setminus X_\alpha}.
 \label{eq:applications-time-seed}
\end{equation}
The strengths and operator content of these interactions may change, and individual terms may be switched on or off\footnote{At finite volume, piecewise-continuous Hermitian densities suffice to define the unitary evolution.}.
For example, the exchange couplings and magnetic fields in the
Ising models of Sec.~\ref{subsec:ising} may depend arbitrarily on
time. Applying the same maps $\phi_i$ at every instant still makes
each lifted local operator preserve the active subspace and
annihilate the inactive subspace. Consequently,
\[
 [\widetilde H_\Lambda(t),Z_i]=0
 \qquad\text{for every }i\text{ and every }t.
\]
This instantaneous commutation relation is the input needed to
establish conservation under the complete time evolution.

The propagator satisfies
\[
 i\frac{d}{dt}\widetilde U(t,t_0)
       =\widetilde H_\Lambda(t)\widetilde U(t,t_0),
 \qquad
 \widetilde U(t_0,t_0)=\id.
\]
Using this equation and its adjoint gives
\begin{equation}
 \begin{gathered}
 \widetilde U(t,t_0)
   =\mathcal T\exp\!\left[
       -i\int_{t_0}^{t}\widetilde H_\Lambda(s)\,ds\right],
 \\[2pt]
 \begin{aligned}
 \frac{d}{dt}
   \left(\widetilde U^\dagger Z_i\widetilde U\right)
   &=
   i\widetilde U^\dagger\widetilde H_\Lambda(t)Z_i\widetilde U
   -i\widetilde U^\dagger Z_i\widetilde H_\Lambda(t)\widetilde U
   \\
   &=i\widetilde U^\dagger
       [\widetilde H_\Lambda(t),Z_i]\widetilde U
   =0.
 \end{aligned}
 \end{gathered}
 \label{eq:applications-time-conservation}
\end{equation}
The time arguments of the propagators have been suppressed in the
derivative. Since its initial value is $Z_i$, we obtain
\[
 \widetilde U(t,t_0)^\dagger Z_i\widetilde U(t,t_0)=Z_i,
 \qquad
 [\widetilde U(t,t_0),Z_i]=0.
\]
No commutativity between Hamiltonians at different times is required. Even when
$[\widetilde H_\Lambda(t),\widetilde H_\Lambda(t')]\ne0$,
each instantaneous Hamiltonian preserves the same hole subspaces,
and so does their time-ordered evolution. Furthermore, no slow-driving or
high-frequency approximation enters this argument.

The same conclusion holds for every pattern projector
\[
 R_A=\prod_{i\in A}P_i\prod_{j\notin A}Z_j,
\]
where $A$ is the set of active sites, as defined in
Eq.~\eqref{eq:perm-pattern-projectors}. In particular,
\[
 [\widetilde U(t,t_0),R_A]=0,
 \qquad
 R_B\widetilde U(t,t_0)R_A=0
 \quad\text{for }A\ne B.
\]
A state initially supported in pattern $A$ therefore remains in
that pattern. For a superposition of different patterns, the
probability of each pattern remains constant, although its
internal state and phase may evolve.

We can also identify the evolution within each pattern explicitly.
Fix $A$ and omit the inactive local factors. Every interaction
touching a hole vanishes, leaving the effective Hamiltonian
\[
 H_A(t)=c_\Lambda(t)\id_A
       +\sum_{\alpha:\,X_\alpha\subseteq A}
          h_\alpha(t)\otimes\id_{A\setminus X_\alpha}.
\]
This Hamiltonian acts on the original seed degrees of freedom at
the active sites. Copies of the active-space and hole-flavour labels belong
to the multiplicity space
\[
 \mathcal M_A=
 \left(\bigotimes_{i\in A}\mathbb C^{r_i}\right)
 \otimes
 \left(\bigotimes_{j\notin A}\mathbb C^{s_j}\right).
\]
Every lifted interaction acts as the identity on these labels.
Restricting the Schr\"odinger equation to the conserved patterns
therefore gives
\begin{equation}
\begin{gathered}
\widetilde U(t,t_0)
   \simeq
   \bigoplus_{A\subseteq\Lambda}
      \left[U_A(t,t_0)\otimes\id_{\mathcal M_A}\right],
 \\
 U_A(t,t_0)=\mathcal T\exp\!\left[
                  -i\int_{t_0}^{t}H_A(s)\,ds\right].
\end{gathered}
 \label{eq:applications-time-blocks}
\end{equation}
For the minimal lift, $r_i=s_i=1$, the multiplicity space is one
dimensional and its identity factor can be omitted. The fully
active pattern retains every seed interaction and hence reproduces
the complete evolution of the original time-dependent Hamiltonian.

For illustration, consider an open nearest-neighbor chain of five
sites and fix a hole at site $3$. The surviving active segments
are $\{1,2\}$ and $\{4,5\}$. Their restricted Hamiltonian is
\[
 H_A(t)=c_\Lambda(t)\id_A
          +H_{\{1,2\}}(t)+H_{\{4,5\}}(t),
\]
where each segment retains its original internal bond and any
one-site fields. Its propagator factorizes as
\[
 U_A(t,t_0)
   =e^{-i\int_{t_0}^{t}c_\Lambda(s)\,ds}
      U_{\{1,2\}}(t,t_0)\otimes U_{\{4,5\}}(t,t_0).
\]
Each segment can undergo nontrivial driven dynamics, including
evolution generated by Hamiltonians that fail to commute at
different times. The hole remains at site $3$ and prevents
dynamical coupling between the segments.

Persistence of a local zero mode does not mean that the entire
many-body state is stationary. The active degrees of freedom can
evolve while the inactive local factor remains unchanged. An
entirely inactive state acquires only the scalar phase
$\exp[-i\int_{t_0}^{t}c_\Lambda(s)\,ds]$. When $c_\Lambda=0$,
such a state is unchanged by the propagator and has eigenvalue
one. A unitary propagator itself has no zero eigenvalues.

For periodic driving of period $T$, the Floquet operator
$\widetilde U_F=\widetilde U(t_0+T,t_0)$ consequently preserves
every pattern, at every driving frequency and after arbitrarily
many periods. This conclusion continues to rely on fixed local
representations. If the active and inactive subspaces themselves
rotate with time, then
\[
 \frac{d}{dt}
 \left(\widetilde U^\dagger Z_i(t)\widetilde U\right)
 =
 \widetilde U^\dagger
 \left(\dot Z_i(t)
       +i[\widetilde H_\Lambda(t),Z_i(t)]\right)
 \widetilde U.
\]
Instantaneous commutation alone would then leave the additional
$\dot Z_i(t)$ term. The construction considered here allows
arbitrary driving of the seed interactions within fixed enlarged
representations.

\subsubsection{Unitary gate lifting}

The preceding time-evolution argument also suggests how to lift
individual unitary gates and construct fragmented quantum circuits.
The essential distinction is that, while a lifted Hamiltonian density
vanishes on an inactive subspace, we want the lifted unitary evolution to act
as the identity there. A direct replacement of a unitary gate by
its image under $\phi$ must therefore be supplemented by an
identity action on the inactive subspace.

Consider first the minimal enlargement of one qubit,
\[
\begin{gathered}
\widetilde{\mathcal H}_i
   =\operatorname{span}\{\ket{0},\ket{1},\ket{\circ}\},
 \\
 P_i=\ket{0}\bra{0}+\ket{1}\bra{1},
 \\
 Z_i=\ket{\circ}\bra{\circ}.
\end{gathered}
\]
The operator map is $\phi_i(O)=O\oplus0$. For the spin-flip gate
$u=\sigma^x=q+q^\dagger$, this gives
\[
 \phi_i(u)=\widetilde q_i+\widetilde q_i^\dagger
   =
 \begin{pmatrix}
  0&1&0\\
  1&0&0\\
  0&0&0
 \end{pmatrix}
\]
in the ordered basis $(\ket{0},\ket{1},\ket{\circ})$.
Although it performs the required spin flip on the active states,
it sends $\ket{\circ}$ to the zero vector and therefore fails to
preserve its norm. It is not a unitary operator on the enlarged
space. Adding the identity action on the hole gives
\[
 \widehat u_i=Z_i+\phi_i(u)
   =
 \begin{pmatrix}
  0&1&0\\
  1&0&0\\
  0&0&1
 \end{pmatrix}.
\]
The active states now undergo the original gate, while the hole
remains unchanged. This can be viewed as a norm-preserving lift. 

The same construction applies to a gate acting on several sites.
Let $u_X$ be a seed unitary gate on a declared support $X$, and
define
\begin{equation}
 \phi_X=\bigotimes_{i\in X}\phi_i,
 \qquad
 P_X=\phi_X(\id_X)=\prod_{i\in X}P_i.
 \label{eq:applications-gate-projector}
\end{equation}
The projector $P_X$ selects the subspace in which every site of
$X$ is active. Its complement selects all configurations containing
at least one hole in $X$. Since the local maps do not preserve the
identity, the image of the seed gate satisfies
\[
 \phi_X(u_X)^\dagger\phi_X(u_X)
   =\phi_X(u_X^\dagger u_X)
   =\phi_X(\id_X)
   =P_X.
\]
Thus $\phi_X(u_X)$ is unitary on the fully active subspace, but
annihilates its complement. We choose the identity action on that
complement and define
\begin{equation}
 \widehat u_X
   =\id_{\widetilde{\mathcal H}_X}-P_X+\phi_X(u_X).
 \label{eq:applications-unitary-lift}
\end{equation}

When every site in $X$ is active, this operator performs the
original gate, with identity action on the active-copy labels.
When any site in $X$ is inactive, the entire gate acts as the
identity. In particular, a two-site gate with one inactive site
does not become a one-site gate on its active partner. The whole
supported gate becomes the identity, just as a Hamiltonian density
vanishes when its declared support contains a hole. This rule
defines a single linear operator and applies coherently to
superpositions of different hole patterns. No measurement of the
hole positions is required.

To verify unitarity, observe that the two terms in
Eq.~\eqref{eq:applications-unitary-lift} act on orthogonal
subspaces,
\[
 (\id-P_X)\phi_X(u_X)
   =\phi_X(u_X)(\id-P_X)=0.
\]
Their cross terms therefore vanish. Using multiplicativity and
adjoint preservation gives
\begin{equation}
 \begin{aligned}
 \widehat u_X^\dagger\widehat u_X
   &=(\id-P_X)^2+\phi_X(u_X)^\dagger\phi_X(u_X)\\
   &=(\id-P_X)+P_X=\id,
 \end{aligned}
 \qquad
 [\widehat u_X,Z_i]=0.
 \label{eq:applications-gate-unitarity}
\end{equation}
The reverse product gives
$\widehat u_X\widehat u_X^\dagger=\id$ as well.
The commutation relation follows because both pieces preserve
every local active and inactive subspace. Sites outside $X$
carry full spectator identities, so the gate commutes with
$Z_i$ at every site of the lattice.

This unitary completion agrees exactly with the Hamiltonian
lifting prescription. Suppose the seed gate is generated by a
Hermitian pulse,
$u_X=e^{-i\tau h_X}$. Expanding the exponential of the lifted
density gives
\begin{equation}
 \begin{aligned}
 e^{-i\tau\phi_X(h_X)}
   &=\id+\sum_{k=1}^{\infty}
          \frac{(-i\tau)^k}{k!}\phi_X(h_X^k)\\
   &=\id+\phi_X\!\left(e^{-i\tau h_X}-\id_X\right)\\
   &=\id-P_X+\phi_X(u_X)
    =\widehat u_X.
 \end{aligned}
 \label{eq:applications-gate-exponential}
\end{equation}
Every positive power is mapped multiplicatively, whereas the
constant term of the exponential is the full identity on the
enlarged space. This explains the additional $\id-P_X$ in the
gate lift. The direct formula in
Eq.~\eqref{eq:applications-unitary-lift} also allows us to lift
a given unitary without first choosing a Hamiltonian logarithm.

A controlled-NOT gate gives a two-site example in the SUSY
notation. Its seed expression is
\[
 u_{ij}=b_i\id_j+f_i(q_j+q_j^\dagger),
\]
where $b_i=\ket{0}_i\bra{0}$ and $f_i=\ket{1}_i\bra{1}$ resolve
the control state. Both sites belong to the declared gate support,
so the identity on site $j$ is represented by $P_j$ under the
operator map. The completed minimal qutrit lift is consequently
\begin{equation}
\begin{gathered}
\widehat u_{ij}
    =\widetilde b_iP_j
       +\widetilde f_i(\widetilde q_j+\widetilde q_j^\dagger)
       +\id_{ij}-P_iP_j,
 \\
 \widetilde b_i=\widetilde q_i\widetilde q_i^\dagger,
 \quad
 \widetilde f_i=\widetilde q_i^\dagger\widetilde q_i.
\end{gathered}
 \label{eq:applications-cnot}
\end{equation}
On the four active states it acts as
\[
 \ket{00}\longmapsto\ket{00},
 \qquad
 \ket{01}\longmapsto\ket{01},
 \qquad
 \ket{10}\longleftrightarrow\ket{11}.
\]
Each of the remaining five two-qutrit basis states,
\[
 \ket{0\circ},\quad
 \ket{1\circ},\quad
 \ket{\circ0},\quad
 \ket{\circ1},\quad
 \ket{\circ\circ},
\]
is unchanged. The hole therefore remains at its site without
spoiling the unitarity of the gate.

We can now compose these gates into a circuit. Take an arbitrary
ordered sequence of seed gates,
$\mathcal U=u_{X_K}\cdots u_{X_1}$, and lift each gate separately
on its declared support. The supports may overlap and the gates
need not commute. Since every lifted gate commutes with every
pattern projector, their product satisfies
\begin{equation}
 \widehat{\mathcal U}
     =\widehat u_{X_K}\cdots\widehat u_{X_1},
 \qquad
 [\widehat{\mathcal U},R_A]=0,
 \qquad
 \widehat{\mathcal U}\big|_{\widetilde{\mathcal H}_A}
     \simeq\mathcal U_A\otimes\id_{\mathcal M_A}.
 \label{eq:applications-circuit}
\end{equation}
Here $\mathcal U_A$ retains the original gates whose supports
lie entirely in $A$, in their original order. Every other gate
becomes an identity. The multiplicity labels remain spectators,
as in the Hamiltonian evolution.

\begin{figure*}[tp]
\centering
\begingroup
\resizebox{\linewidth}{!}{%
\definecolor{circBlue}{HTML}{216B8B}
\definecolor{circWash}{HTML}{EDF5F8}
\definecolor{circHole}{HTML}{B36D18}
\definecolor{circHoleWash}{HTML}{FFF5E7}
\definecolor{circGrey}{HTML}{89959C}
\definecolor{circInk}{HTML}{202A32}
\begin{tikzpicture}[
 x=1cm,y=1cm,
 font=\fontsize{10}{12}\selectfont,
 text=circInk,
 >={Stealth[length=2mm,width=1.5mm]},
 wire/.style={
   draw=circBlue,line width=.85pt},
 gate/.style={
   draw=circBlue,fill=circWash,
   rounded corners=2pt,line width=.85pt},
 idle/.style={
   draw=circGrey,fill=white,densely dashed,
   rounded corners=2pt,line width=.85pt},
 panel/.style={
   font=\fontsize{10}{12}\selectfont\bfseries,
   anchor=west}
]
\useasboundingbox (0,.3) rectangle (15.8,-5.75);

\node[panel] at (.05,0) {(a) Original circuit};
\node[panel] at (8.05,0)
  {(b) Lifted circuit in the pattern $A=\{1,2,4\}$};

\foreach \j in {1,...,4}{
 \draw[wire]
   (.85,{-1.0-.85*(\j-1)})
   --(7.35,{-1.0-.85*(\j-1)});
 \node[anchor=east]
   at (.64,{-1.0-.85*(\j-1)}) {$\j$};
}
\foreach \j in {1,2,4}{
 \draw[wire]
   (8.8,{-1.0-.85*(\j-1)})
   --(15.30,{-1.0-.85*(\j-1)});
 \node[anchor=east,text=circBlue]
   at (8.60,{-1.0-.85*(\j-1)}) {$a$};
}
\draw[circHole,line width=1.1pt,densely dashed]
  (8.8,-2.70)--(15.30,-2.70);
\node[anchor=east,text=circHole]
  at (8.60,-2.70) {$\circ$};

\filldraw[gate] (1.25,-.67) rectangle (2.25,-2.18);
\node at (1.75,-1.425) {$u_{12}$};

\filldraw[gate] (2.75,-1.52) rectangle (3.75,-3.03);
\node at (3.25,-2.275) {$u_{23}$};

\filldraw[gate] (4.25,-2.37) rectangle (5.25,-3.88);
\node at (4.75,-3.125) {$u_{34}$};

\filldraw[gate] (5.75,-3.22) rectangle (6.75,-3.88);
\node at (6.25,-3.55) {$u_4$};

\filldraw[gate] (9.20,-.67) rectangle (10.20,-2.18);
\node at (9.70,-1.425) {$u_{12}$};

\filldraw[idle] (10.70,-1.52) rectangle (11.70,-3.03);
\node[text=circGrey] at (11.20,-2.12) {$\id$};
\draw[circHole,line width=1.1pt,densely dashed]
  (10.70,-2.70)--(11.70,-2.70);

\filldraw[idle] (12.20,-2.37) rectangle (13.20,-3.88);
\node[text=circGrey] at (12.70,-3.23) {$\id$};
\draw[circHole,line width=1.1pt,densely dashed]
  (12.20,-2.70)--(13.20,-2.70);

\filldraw[gate] (13.70,-3.22) rectangle (14.70,-3.88);
\node at (14.20,-3.55) {$u_4$};

\node[
 circle,draw=circHole,fill=circHoleWash,text=circHole,
 inner sep=1pt,minimum size=4.5mm
] at (15.30,-2.70) {$\circ$};

\draw[->,circInk!65]
  (.85,-4.28)--(7.35,-4.28)
  node[midway,below=2pt] {time};
\draw[->,circInk!65]
  (8.80,-4.28)--(15.30,-4.28)
  node[midway,below=2pt] {time};

\node at (4.1,-5.05)
  {$\mathcal U=u_4u_{34}u_{23}u_{12}$};
\node at (12.05,-5.05)
  {$\mathcal U_A=u_4u_{12}$};

\node[
 text=circInk!75,
 font=\fontsize{9}{11}\selectfont
] at (7.9,-5.60)
  {$\widehat u_X=\id-P_X+\phi_X(u_X),
    \qquad [\widehat u_X,Z_i]=0$};

\end{tikzpicture}
}
\endgroup
\caption{\small
Conserved hole occupations in a lifted quantum circuit.
Time runs from left to right and operator products act from
right to left.
(a) A four-site circuit with arbitrary unitary gates on supports
$\{1,2\}$, $\{2,3\}$, $\{3,4\}$ and $\{4\}$, with spectator
identities understood.
(b) In the minimal lift, fix a hole at site $3$ and take sites
$1,2,4$ active. The symbol $a$ indicates an active local subspace,
not a particular basis state. Gates touching site $3$ act as the
identity, shown by dashed boxes. The gates on $\{1,2\}$ and
$\{4\}$ retain their original action, so the restricted circuit
is $\mathcal U_A=u_4u_{12}$. The orange wire records the fixed
hole throughout the sequence.
Equation~\eqref{eq:applications-unitary-lift} supplies the
identity action on inactive supports and ensures unitarity.
In the fully active pattern all four gates act as in the
original circuit.}
\label{fig:fragmented-circuit}
\end{figure*}
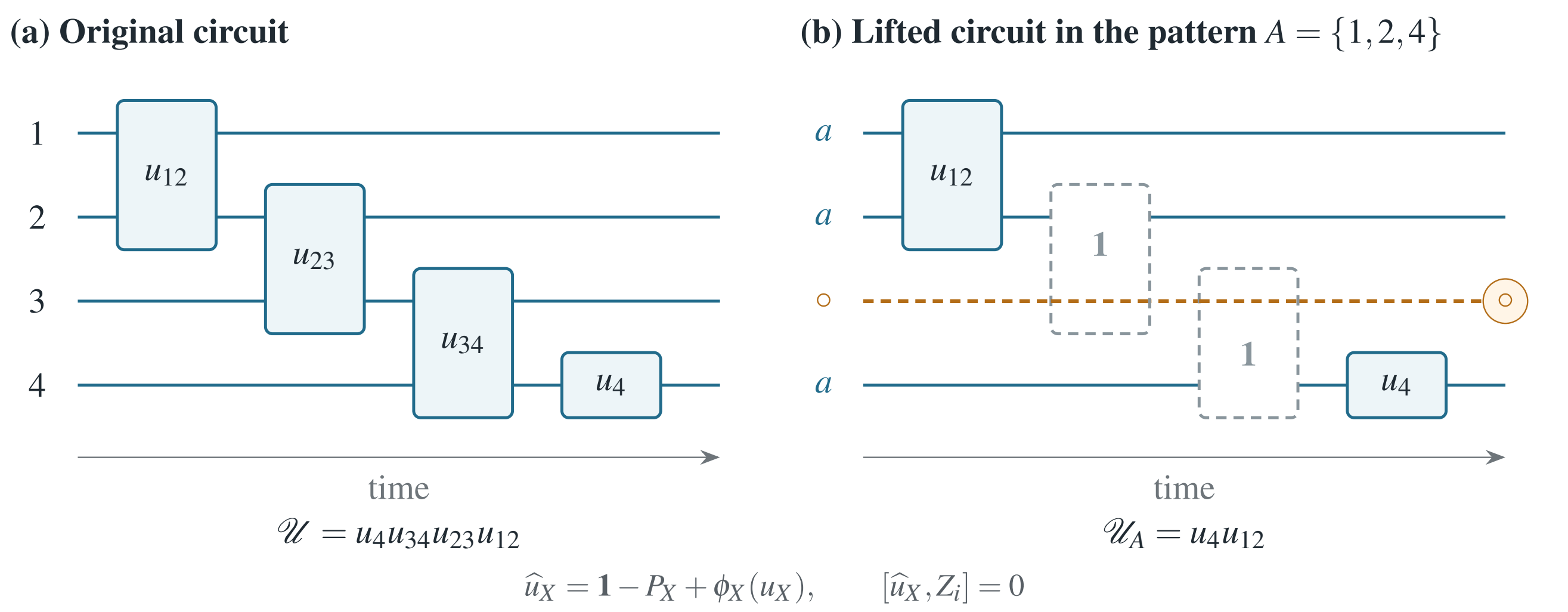

Figure~\ref{fig:fragmented-circuit} illustrates this rule for
the four-site circuit
\[
 \mathcal U=u_4u_{34}u_{23}u_{12}.
\]
Fixing a hole at site $3$ gives $A=\{1,2,4\}$.
The gates on $\{2,3\}$ and $\{3,4\}$ become identities, while
the gates on $\{1,2\}$ and $\{4\}$ retain their original action.
The restricted circuit is therefore
\[
 \mathcal U_A=u_4u_{12}.
\]
In the fully active pattern, all four gates survive and the
original circuit is reproduced exactly.

Every lifted gate commutes with every $R_A$, supplying at least
$2^L$ nonzero invariant pattern sectors when all sites are
enlarged. This is the circuit version of the common-commutant
formulation~\cite{MoudgalyaMotrunich2022}. For uniform lifts,
the same fixed-hole-number bounds establish strong fragmentation
at intermediate hole density, while particular gate sets may
split the patterns further. Replacing the seed gates by other
unitary gates lifted through the same representations preserves
the pattern decomposition, with no restriction on circuit depth
or gate ordering. Bond-wise gate lifting can instead preserve mobile
strings or dimer sectors, as discussed in Sec.~\ref{subsec:bond-circuits}.

\subsection{Time-dependent correlations and constrained propagation}
\label{subsec:applications-correlations}

The conserved patterns also determine observable consequences of site-wise lifting. We now compute time-dependent correlations in terms of the surviving seed Hamiltonians and derive bounds on propagation and local memory. Throughout this subsection the Hamiltonian is Hermitian and the local representations are fixed. The first reduction allows arbitrary interaction supports and local dimensions. The explicit propagation and memory results then specialize to the nearest-neighbor chains already studied above.

\paragraph*{Correlations inherited from the active sectors.}
Consider the stationary ensemble
\begin{equation}
 \varrho_p=\bigotimes_{i\in\Lambda}
    \left(\frac{1-p}{n_i r_i}P_i+\frac{p}{s_i}Z_i\right),
 \qquad 0<p<1.
 \label{eq:correlations-ensemble}
\end{equation}
Each site is independently inactive with probability $p$, and its active and inactive subspaces are each maximally mixed. This assumes $r_i,s_i\geq1$. The state commutes with every lifted interaction because it is scalar within each pattern subspace. For uniform local dimensions $d=nr+s$, the choice $p=s/d$ gives the full infinite-temperature state $\id/d^L$. Other values prescribe different populations of the conserved hole sectors.

Let $O_i$ and $O_j$ be Hermitian, traceless one-site seed observables of operator norm at most one. Their lifts $\widetilde O_i=\phi_i(O_i)$ vanish on the local holes. Write $\widetilde O_i(t)=\widetilde U(t,0)^\dagger\widetilde O_i\widetilde U(t,0)$ and define
\begin{equation}
\begin{aligned}
 F_{ij}(t;p)
   &=\operatorname{Tr}\!\left[\varrho_p\widetilde O_i(t)
                                         \widetilde O_j\right]\\
   &=\sum_{A\supseteq\{i,j\}}
       \frac{p^{L-|A|}(1-p)^{|A|}}{d_A}\\
   &\quad\times\operatorname{Tr}_{A}\!\left[
          U_A(t,0)^\dagger O_iU_A(t,0)O_j\right],\\
 d_A&=\prod_{a\in A}n_a.
\end{aligned}
 \label{eq:correlations-pattern-sum}
\end{equation}
Here $\operatorname{Tr}_A$ is over the seed degrees of freedom on the active sites, and $U_A$ is their restricted propagator from Eq.~\eqref{eq:applications-time-blocks}. The active-copy and hole-flavour traces cancel their normalization factors. A pattern contributes only when both observed sites are active. Since the one-point functions vanish, $F_{ij}$ is also the connected two-point function. This identity reduces a correlation function of the enlarged model to correlations of the original interactions on their surviving supports. It does not assume that the seed Hamiltonian is integrable.

\paragraph*{Exact reduction to active segments.}
For an open nearest-neighbor chain, different active segments evolve independently. If a hole separates $i$ and $j$, the trace in Eq.~\eqref{eq:correlations-pattern-sum} factorizes and vanishes by tracelessness. For $i\leq j$, group the remaining patterns by their unique maximal active segment $I=[a,b]$ containing both sites. Its probability is
\[
 w_{a,b}(p)=(1-p)^{b-a+1}p^{\chi(a>1)+\chi(b<L)},
\]
where $\chi$ is one when its condition holds and zero otherwise. All sites in $I$ must be active, and each endpoint not at the physical boundary must be followed by a hole. The states outside these specified sites are unrestricted and their probabilities sum to one. Consequently,
\begin{equation}
 \begin{aligned}
 F_{ij}(t;p)&=\sum_{a=1}^{i}\sum_{b=j}^{L}
                   w_{a,b}(p)\,f_{ij}^{[a,b]}(t),\\
 f_{ij}^{I}(t)&=\frac{1}{d_I}\operatorname{Tr}_{I}
       \left[U_I(t,0)^\dagger O_iU_I(t,0)O_j\right].
 \end{aligned}
 \label{eq:correlations-segment-sum}
\end{equation}
The segment Hamiltonian retains all original fields on $I$ and all bonds inside it. Thus the correlation can be computed from finite seed segments, without diagonalizing the enlarged chain or enumerating every hole pattern. The formula includes spatially varying couplings. For a static segment with eigenstates $\ket{\mu}$ and energies $E_\mu$, its time dependence is explicitly
\[
 f_{ij}^{I}(t)=\frac{1}{d_I}\sum_{\mu,\nu}
   e^{i(E_\mu-E_\nu)t}
   \langle\mu|O_i|\nu\rangle\langle\nu|O_j|\mu\rangle.
\]
When the active-segment Hamiltonians retain integrability,
the quantum inverse scattering method provides tools for
evaluating the spectral data and correlation functions
entering this reduction~\cite{KorepinBogoliubovIzergin1993}.

\paragraph*{An all-time bound on propagation.}
Let $d_{ij}=|i-j|$, measured in lattice spacings. The total probability of the contributing segments is $(1-p)^{d_{ij}+1}$, since every site from $i$ through $j$, including both endpoints, must be active. Each normalized segment correlation has magnitude at most one. The same geometric restriction controls the squared commutator, a measure of operator spreading,
\begin{equation}
 C_{ij}(t;p)=\operatorname{Tr}\!\left[
   \varrho_p[\widetilde O_i(t),\widetilde O_j]^\dagger
             [\widetilde O_i(t),\widetilde O_j]\right].
 \label{eq:correlations-squared-commutator}
\end{equation}
Within any pattern where a hole separates the two sites, the commutator vanishes as an operator. In each remaining pattern its squared norm is at most four. We therefore obtain
\begin{equation}
\begin{gathered}
|F_{ij}(t;p)|\leq(1-p)^{d_{ij}+1},
 \\
 0\leq C_{ij}(t;p)\leq4(1-p)^{d_{ij}+1}
 \\\text{for all }t.
\end{gathered}
 \label{eq:correlations-propagation-bound}
\end{equation}
The spatial envelope has length $\xi(p)=[-\log(1-p)]^{-1}$ and is uniform in time. These are ensemble bounds and coexist with the complete seed dynamics in the fully active pattern. Vanishing two-point correlations across distinct components uses the ensemble in Eq.~\eqref{eq:correlations-ensemble}, whereas vanishing commutators across them is independent of the state.

At fixed hole number $m$, a uniform mixture of patterns gives the same bounds with $(1-p)^{d_{ij}+1}$ replaced by $\binom{L-d_{ij}-1}{m}/\binom{L}{m}$, taken to be zero when the numerator has too few sites for the holes. This counts the patterns with every hole outside the interval joining the observables and connects the propagation calculation directly to the fixed-hole-number fragmentation analysis.

\paragraph*{Explicit local memory in the mixed-field chain.}
Return to the static mixed-field Ising model of Eq.~\eqref{eq:mixed-seed}, with uniform $g,h\ne0$, and take $O_i=\sigma_i^z$ at an interior site. The event that $i$ is active and both neighbors are holes has probability $p^2(1-p)$. On this event the spin evolves under $H_{\{i\}}=-g\sigma_i^x-h\sigma_i^z$. Putting $\Omega=\sqrt{g^2+h^2}$ gives the exact single-site autocorrelation
\begin{equation}
 f_{ii}^{\{i\}}(t)
    =\frac12\operatorname{Tr}\!\left[
       e^{iH_{\{i\}}t}\sigma_i^z
       e^{-iH_{\{i\}}t}\sigma_i^z\right]
    =\frac{h^2+g^2\cos(2\Omega t)}{g^2+h^2}.
 \label{eq:correlations-isolated-spin}
\end{equation}
It contributes $p^2(1-p)f_{ii}^{\{i\}}(t)$ to the full autocorrelation. Longer segments supply the other terms in Eq.~\eqref{eq:correlations-segment-sum}, and can have different relaxation behavior. For each static segment, the infinite-time average of a Hermitian autocorrelation is nonnegative. The spectral sum retains only equal-energy pairs and becomes $d_I^{-1}\sum_{E_\mu=E_\nu}|\langle\mu|O_i|\nu\rangle|^2$. Keeping just the isolated-spin contribution therefore proves
\begin{equation}
 \overline F_{ii}(p)
   =\lim_{T\to\infty}\frac1T\int_0^T F_{ii}(t;p)\,dt
   \geq p^2(1-p)\frac{h^2}{g^2+h^2}>0.
 \label{eq:correlations-memory-bound}
\end{equation}
The bound is independent of chain length and holds for an interior site on every finite open chain. For the minimal qubit lift at full infinite temperature, $p=1/3$. If $g=h\ne0$, it gives $\overline F_{ii}\geq1/27$, or $\overline F_{ii}/F_{ii}(0)\geq1/18$ because $F_{ii}(0)=2/3$. Thus a physical spin observable retains local memory even when the connected seed chain is nonintegrable. The mechanism is the finite probability of an isolated active spin, together with the component of its magnetization along its static local field. This is consistent with the broader use of autocorrelation bounds to diagnose memory in fragmented systems~\cite{MoudgalyaMotrunich2022}.

The correlation reduction and propagation bound also apply to arbitrary time-dependent nearest-neighbor driving and to lifted circuits made from one-site and nearest-neighbor gates. Only the last memory bound assumes a static Hamiltonian. For longer-range interactions, a ring or higher-dimensional lattices, the corresponding geometric factor is the probability that the two observed sites lie in the same surviving interaction component. It need not equal the open-chain interval probability. Mobile-string and pair-flip bond lifts require their own correlation analysis, since their conserved labels do not generally provide immobile barriers.

\section{General site-wise construction and fragmentation theorem}
\label{sec:theorem}

The examples and extensions in Sec.~\ref{sec:examples} have exhibited a common mechanism. Enlarging the supercharge or parasupercharge representation retains the original operators on active copies and makes every lifted local operator vanish on an added zero-mode space. We now collect the assumptions on these representations and the declared interaction supports, and prove the static pattern decomposition in full generality. The theorem identifies the restricted dynamics and multiplicities and gives a lower bound on the common commutant. We then derive sector-size, locality and stability bounds. The time-dependent and circuit constructions in Sec.~\ref{subsec:applications-time} use the same termwise conservation and sector structure, with time ordering or the unitary completion appropriate to gates. The correlation formulas in Sec.~\ref{subsec:applications-correlations} follow from this same decomposition after specifying the state and observables.

\subsection{Local operator algebras and their enlarged site representations}
\label{subsec:general-local}

Let $\Lambda$ be a finite set of $L$ sites, each labelling a local Hilbert-space factor. It may describe an open or periodic chain, a lattice in any dimension, or an arbitrary finite graph. No metric is required. The interaction supports, specified in Sec.~\ref{subsec:general-support}, determine which sites are coupled. The seed many-body Hilbert space is
\begin{equation}
 \mathcal H_\Lambda=\bigotimes_{i\in\Lambda}\mathbb C^{n_i},
 \qquad n_i\geq2.
 \label{eq:general-seed-space}
\end{equation}
The dimensions $n_i$ need not be equal. At site $i$, choose the supercharge or parasupercharge $Q_{n_i}$ of Eq.~\eqref{eq:intro-ladder}. The matrix units
$E_{ab}^{(i)}=\ket{a}_i\bra{b}$ obey
\begin{equation}
 E_{ab}^{(i)}E_{cd}^{(i)}=\delta_{bc}E_{ad}^{(i)},
 \qquad
 (E_{ab}^{(i)})^\dagger=E_{ba}^{(i)},
 \qquad
 \sum_{a=0}^{n_i-1}E_{aa}^{(i)}=\id_{n_i}.
 \label{eq:general-matrix-relations}
\end{equation}
Equation~\eqref{eq:intro-matrix-units} expresses every one of them as a polynomial in $Q_{n_i}$ and $Q_{n_i}^\dagger$. Thus SUSY, for $n_i=2$, and paraSUSY of order $n_i-1$, for $n_i>2$, generate the complete local operator algebra in the chosen supercharge or parasupercharge realization.

The representation enlargement is performed on these supercharges or parasupercharges. We retain $r_i$ copies of $Q_{n_i}$ and adjoin an $s_i$-dimensional space annihilated by both enlarged generators, with $r_i,s_i\geq1$. Evaluating the matrix-unit polynomials in this representation induces the local operator map $\phi_i$. The resulting space and map are
\begin{equation}
 \begin{gathered}
 \widetilde{\mathcal H}_i
   =(\mathbb C^{n_i}\otimes\mathbb C^{r_i})\oplus\mathbb C^{s_i},
 \qquad d_i=n_ir_i+s_i,\\
 \phi_i(O)=(O\otimes\id_{r_i})\oplus0_{s_i},
 \qquad
 P_i=\phi_i(\id_{n_i}),\qquad Z_i=\id_{d_i}-P_i.
 \end{gathered}
 \label{eq:general-local-map}
\end{equation}
The local lifting map $\phi_i$ sends $\End(\mathbb C^{n_i})$ into $\End(\widetilde{\mathcal H}_i)$. It applies $O$ identically on each of the $r_i$ active copies, leaves the copy label unchanged, and vanishes on the $s_i$-dimensional inactive summand. The projectors $P_i$ and $Z_i$ select these two summands. The enlarged supercharge or parasupercharge and local algebraic Hamiltonian are
\begin{equation}
 \widetilde Q_i=(Q_{n_i}\otimes\id_{r_i})\oplus0_{s_i},
 \qquad
 \widetilde h_{n_i-1}=P_i.
 \label{eq:general-ladder-map}
\end{equation}
They satisfy Eqs.~\eqref{eq:intro-susy} or \eqref{eq:intro-parasusy} on every active copy, while both sides of those relations vanish on the inactive summand. Each inactive vector is annihilated by $\widetilde Q_i$, $\widetilde Q_i^\dagger$ and $\widetilde h_{n_i-1}$. Evaluating the matrix-unit polynomials gives
\begin{equation}
 \widetilde E_{ab}^{(i)}
 =(\widetilde Q_i^\dagger)^a
   \widetilde Q_i^{\,n_i-1}
   (\widetilde Q_i^\dagger)^{n_i-1}
   \widetilde Q_i^{\,b}
 =\phi_i(E_{ab}^{(i)}).
 \label{eq:general-lifted-units}
\end{equation}
For a general seed local operator $O=\sum_{a,b}O_{ab}E_{ab}^{(i)}$, the substitution therefore gives $\sum_{a,b}O_{ab}\widetilde E_{ab}^{(i)}=\phi_i(O)$. The map $\phi_i$ records the replacement of the supercharges or parasupercharges in every operator expression. The matrix-unit relations allow the ensuing many-body proof to be written uniformly for all local dimensions. The common kernel, and hence the guaranteed pattern conservation, comes from the chosen enlarged representation.

The map $\phi_i$ is linear, injective and preserves adjoints. For any seed local operators $O$ and $O'$, it satisfies
\begin{equation}
\begin{gathered}
\phi_i(OO')=\phi_i(O)\phi_i(O'),
 \\
 \|\phi_i(O)\|=\|O\|,
 \\
 Z_i\phi_i(O)=\phi_i(O)Z_i=0.
\end{gathered}
 \label{eq:general-map-properties}
\end{equation}
It is a \emph{nonunital} representation on the enlarged space. This means that $\phi_i(\id_{n_i})=P_i\ne\id_{d_i}$. The inactive states were introduced by adjoining the summand $\mathbb C^{s_i}$ in Eq.~\eqref{eq:general-local-map}. Nonunitality expresses their decoupling algebraically. Since $O=\id_{n_i}O\id_{n_i}$ and $\phi_i$ preserves multiplication,
\[
 \phi_i(O)=P_i\phi_i(O)P_i.
\]
Every lifted seed local operator therefore acts entirely within the active subspace and has no matrix elements connecting it to the inactive subspace. A supported interaction built from these operators vanishes whenever any site in its support is inactive.

This also makes interaction supports essential. A one-site field $O_i$ acts independently of a spectator site $j$, so its lift is $\phi_i(O_i)\otimes\id_{d_j}$. Lifting the spectator identity would instead give $\phi_i(O_i)\otimes P_j$, causing a hole at $j$ to switch off the field at $i$. We therefore apply the maps only on declared supports, retaining full identities elsewhere.

In this paper we use the particular representation family in Eq.~\eqref{eq:general-local-map}, consisting of $r_i$ copies of the normalized $n_i$-state supercharge or parasupercharge realization and $s_i$ one-dimensional zero-mode summands. Other admissible partitions of the enlarged local dimension $d_i$ into SUSY or paraSUSY representation blocks can give different lifted models. We will address those alternatives elsewhere. The matrix-algebra structure of the present choice is established in Appendix~\ref{app:matrix-representations}.

\subsection{Supported interactions and the site-wise lifting prescription}
\label{subsec:general-support}

Specify the seed Hamiltonian by an interaction family with declared supports,
\begin{equation}
 H_\Lambda=c_\Lambda\id_{\mathcal H_\Lambda}
       +\sum_{\alpha\in\mathcal I_\Lambda}
          h_\alpha\otimes\id_{\Lambda\setminus X_\alpha}.
 \label{eq:general-seed-interactions}
\end{equation}
Here $c_\Lambda$ is real and $\mathcal I_\Lambda$ is a finite set of interaction labels, not sites. Each label $\alpha$ specifies a nonempty support $X_\alpha\subseteq\Lambda$ and a Hermitian operator $h_\alpha$ on $\bigotimes_{i\in X_\alpha}\mathbb C^{n_i}$. The sum runs over terms, with each support determined by its label.

Different labels may have the same support. For example, the transverse and longitudinal fields $g_i\sigma_i^x$ and $h_i\sigma_i^z$ can be labelled by $\alpha=(i,x)$ and $\alpha=(i,z)$, respectively, while both have $X_\alpha=\{i\}$. Keeping these terms separate also specifies the interaction family whose common commutant is used in the theorem. This notation accommodates fields, bond terms and arbitrary many-site interactions. Their coefficients and operator forms may vary from place to place. In particular, translational invariance is not assumed for the Hamiltonian in Eq.~\eqref{eq:general-seed-interactions}.

For a nonempty support $X$, define $\phi_X=\bigotimes_{i\in X}\phi_i$. The lifted Hamiltonian is
\begingroup
\small
\setlength{\jot}{2pt}
\begin{equation}
\begin{gathered}
 \widetilde H_\Lambda
   =c_\Lambda\id_{\widetilde{\mathcal H}_\Lambda}
      +\sum_{\alpha\in\mathcal I_\Lambda}\widetilde h_\alpha,\\
 \widetilde h_\alpha=\phi_{X_\alpha}(h_\alpha)
       \otimes\id_{\Lambda\setminus X_\alpha},\quad
 \widetilde{\mathcal H}_\Lambda
   =\bigotimes_{i\in\Lambda}\widetilde{\mathcal H}_i.
\end{gathered}
 \label{eq:general-lifted-interactions}
\end{equation}
\endgroup
The spectator identity outside $X_\alpha$ acts on the full enlarged complement $\bigotimes_{j\notin X_\alpha}\widetilde{\mathcal H}_j$. An identity factor inside the declared support is instead mapped to $P_i$. The separately specified global scalar remains a global scalar.

To see the effect of lifting the spectator identities as well, apply $\phi_\Lambda=\bigotimes_{i\in\Lambda}\phi_i$ to an entire interaction term. This gives
\[
 \phi_\Lambda\!\left(h_\alpha\otimes\id_{\Lambda\setminus X_\alpha}\right)
 =\phi_{X_\alpha}(h_\alpha)
       \otimes\bigotimes_{j\notin X_\alpha}P_j.
\]
Every spectator identity has become the nontrivial projector $P_j=\widetilde h_{n_j-1}$, making a nonzero density act on the whole lattice. A hole anywhere outside its original support now switches it off. Such factors occur in \emph{Almost local integrable models from supersymmetry algebras}~\cite{Maity2026}, where nearest-neighbor operators are multiplied by local SUSY Hamiltonians on the remaining sites, as in Eq.~(2.23) of that work. The present prescription preserves the declared support by retaining the spectator identities.

Operationally, expand each $h_\alpha$ in tensor products of matrix units, express them through supercharges or parasupercharges, and substitute the enlarged generators. Multiplicativity makes this independent of algebraic simplifications \emph{within a fixed support}. It is not independent of the support assignment, as the one-site-field example above and the bond constant in Eq.~\eqref{eq:perm-constant-shift} demonstrate. The input is therefore a supported interaction family, rather than only the total matrix $H_\Lambda$.

\subsection{The fragmentation theorem and its proof}
\label{subsec:general-theorem}

Let $A\subseteq\Lambda$ denote the set of active sites in a fixed pattern. The sites in $\Lambda\setminus A$ are inactive. Define the corresponding seed Hilbert space by
$\mathcal H_A=\bigotimes_{i\in A}\mathbb C^{n_i}$, with
$\mathcal H_\varnothing=\mathbb C$, and define the restricted seed Hamiltonian
\begin{equation}
 H_A=c_\Lambda\id_{\mathcal H_A}
       +\sum_{\alpha:\,X_\alpha\subseteq A}
           h_\alpha\otimes\id_{A\setminus X_\alpha}.
 \label{eq:general-restricted-seed}
\end{equation}
Thus $H_A$ retains precisely the seed interactions whose declared supports lie entirely in the active-site set $A$.

\begin{theorem}[Exact fragmentation by local zero modes]
\label{thm:fragmentation}
Let the seed interaction family be as in Eq.~\eqref{eq:general-seed-interactions}, and let every local representation have the form in Eq.~\eqref{eq:general-local-map}, with $r_i,s_i\geq1$. Then the lift in Eq.~\eqref{eq:general-lifted-interactions} has the following properties.

\begin{enumerate}
 \item Every local zero-mode projector is conserved term by term:
 \begin{equation}
  [Z_i,\widetilde h_\alpha]=0,
  \qquad [Z_i,\widetilde H_\Lambda]=0.
  \label{eq:general-conserved-projectors}
 \end{equation}

 \item The $2^L$ active-site patterns define nonzero, mutually orthogonal invariant sectors,
 \begin{equation}
  R_A=\prod_{i\in A}P_i\prod_{j\notin A}Z_j,\qquad
  \widetilde{\mathcal H}_\Lambda
       =\bigoplus_{A\subseteq\Lambda}\widetilde{\mathcal H}_A,
  \qquad
  \widetilde{\mathcal H}_A=R_A\widetilde{\mathcal H}_\Lambda.
  \label{eq:general-pattern-decomposition}
 \end{equation}

 \item After a reordering of tensor factors, the Hamiltonian in each sector is exactly a restricted seed Hamiltonian with a multiplicity space:
 \begingroup
\small
\setlength{\jot}{2pt}
\begin{equation}
\begin{gathered}
 \widetilde{\mathcal H}_A\simeq\mathcal H_A\otimes\mathcal M_A,
 \quad \widetilde H_\Lambda\big|_{\widetilde{\mathcal H}_A}
       \simeq H_A\otimes\id_{\mathcal M_A},\\
 \mathcal M_A=\left(\bigotimes_{i\in A}\mathbb C^{r_i}\right)
       \otimes\left(\bigotimes_{j\notin A}\mathbb C^{s_j}\right),\\
 m_A=\dim\mathcal M_A=\prod_{i\in A}r_i\prod_{j\notin A}s_j.
\end{gathered}
 \label{eq:general-sector-hamiltonian}
\end{equation}
\endgroup
 In particular, the fully active sector contains $\prod_i r_i$ copies of the original $H_\Lambda$.

 \item The common commutant of the lifted interaction family contains
 \begin{equation}
\begin{gathered}
\bigoplus_{A\subseteq\Lambda}
      \left(\id_{\mathcal H_A}\otimes\End(\mathcal M_A)\right)
       \ \subseteq\ \mathcal C(\widetilde{\mathcal B}),
  \\
  \dim\mathcal C(\widetilde{\mathcal B})
       \geq\prod_{i\in\Lambda}(r_i^2+s_i^2)\geq2^L,
\end{gathered}
 \label{eq:general-commutant-bound}
\end{equation}
 where $\widetilde{\mathcal B}=\{\widetilde h_\alpha\}_{\alpha\in\mathcal I_\Lambda}$. The guaranteed pattern labels are diagonal in an on-site product basis.
\end{enumerate}
No assumption is made about integrability, translation invariance, interaction range, spatial dimension or the commutant of the seed interaction family.
\end{theorem}

\begin{proof}
\emph{Local conservation.}
If $i\notin X_\alpha$, the operator $\widetilde h_\alpha$ acts as the identity at site $i$ and hence commutes with $Z_i$. If $i\in X_\alpha$, expand $h_\alpha$ in tensor products of matrix units. Every resulting factor at site $i$ is annihilated on both sides by $Z_i$, by Eq.~\eqref{eq:general-map-properties}. Consequently,
$Z_i\widetilde h_\alpha=\widetilde h_\alpha Z_i=0$ in this case. These two cases prove Eq.~\eqref{eq:general-conserved-projectors}.

\emph{Resolution into patterns.}
At each site $P_i$ and $Z_i$ are orthogonal projections satisfying $P_i+Z_i=\id_{d_i}$. Projectors on different sites commute. Their products therefore obey
\begin{equation}
 R_A R_B=\delta_{A,B}R_A,
 \qquad
 \sum_{A\subseteq\Lambda}R_A
       =\prod_{i\in\Lambda}(P_i+Z_i)
       =\id_{\widetilde{\mathcal H}_\Lambda}.
 \label{eq:general-resolution}
\end{equation}
Both local ranges have positive dimension, so every $R_A$ is nonzero. Local conservation implies that each of these ranges is preserved by every interaction.

\emph{Dynamics in a fixed pattern.}
If $X_\alpha\not\subseteq A$, some site in the support is inactive, and the corresponding factor in $\phi_{X_\alpha}(h_\alpha)$ annihilates that sector. If $X_\alpha\subseteq A$, all its supported factors act on active spaces as the original matrix units tensored with identities on the copy labels. Reordering the seed local Hilbert-space factors, active-copy factors and inactive factors gives
\begin{equation}
 \widetilde h_\alpha\big|_{\widetilde{\mathcal H}_A}
 \simeq
 \begin{cases}
  (h_\alpha\otimes\id_{A\setminus X_\alpha})
        \otimes\id_{\mathcal M_A},&X_\alpha\subseteq A,\\
  0,&X_\alpha\not\subseteq A.
 \end{cases}
 \label{eq:general-term-restriction}
\end{equation}
Summing this identity and adding the scalar term proves Eq.~\eqref{eq:general-sector-hamiltonian}.

\emph{The common commutant.}
Within a fixed sector, every lifted term is the identity on $\mathcal M_A$. Hence every operator $\id_{\mathcal H_A}\otimes T_A$, with $T_A\in\End(\mathcal M_A)$, commutes with every restricted interaction. Extending it by zero on the other sectors proves the algebra inclusion in Eq.~\eqref{eq:general-commutant-bound}. The $m_A$-dimensional multiplicity space admits $m_A^2$ independent operators, while their action on $\mathcal H_A$ is fixed to be the identity. Operators extended from distinct patterns are linearly independent because those sectors are orthogonal. Their dimensions therefore add, giving
\begin{equation}
 \sum_{A\subseteq\Lambda}m_A^2
   =\sum_{A\subseteq\Lambda}
        \prod_{i\in A}r_i^2\prod_{j\notin A}s_j^2
   =\prod_{i\in\Lambda}(r_i^2+s_i^2).
 \label{eq:general-commutant-count}
\end{equation}
The last equality follows by expanding the product and choosing $r_i^2$ at active sites and $s_i^2$ at inactive sites. Each choice corresponds to exactly one subset $A\subseteq\Lambda$. Equation~\eqref{eq:general-commutant-count} therefore gives a lower bound on $\dim\mathcal C(\widetilde{\mathcal B})$. It counts only the operators supplied by the multiplicity spaces and does not include additional symmetries of the restricted seed Hamiltonians $H_A$ that commute with every surviving interaction term.

Finally, choose at each site a basis of the active space followed by a basis of the zero-mode space. All $P_i$, $Z_i$ and $R_A$ are diagonal in the resulting product basis. This proves the last statement.
\end{proof}

The $2^L$ patterns are guaranteed invariant sectors, not necessarily minimal blocks. Seed symmetries and multiplicity spaces can refine them, and their classical labels do not require the full commutant to be Abelian.

\paragraph*{When the commutant bound is saturated.}
An additional simple hypothesis makes the commutant bound exact. Suppose the specified interaction family contains, at every site, enough one-site terms to generate $\End(\mathbb C^{n_i})$ as an algebra. Their lifts generate the active matrix units and $P_i$, so any common commuting operator must preserve the active and inactive local spaces. On the active space it can act only on the copy factor; on the inactive space it is arbitrary. Thus
\begin{equation}
\begin{aligned}
 \mathcal C(\widetilde{\mathcal B})
 &=\bigotimes_{i\in\Lambda}
       \left[
        (\id_{n_i}\otimes\End(\mathbb C^{r_i}))
          \oplus\End(\mathbb C^{s_i})
       \right]\\
 &\simeq\bigoplus_A
       \left(\id_{\mathcal H_A}\otimes\End(\mathcal M_A)\right).
\end{aligned}
 \label{eq:general-exact-commutant}
\end{equation}
Every other lifted interaction commutes with these multiplicity operators, so adding such interactions cannot reduce this algebra. In the minimal case $r_i=s_i=1$, the common commutant is exactly $\operatorname{span}\{R_A\}$ and is Abelian of dimension $2^L$. This recovers the algebraic conclusion of the mixed-field example. Without this extra hypothesis, Eq.~\eqref{eq:general-commutant-bound} remains the general statement.

\paragraph*{Dynamics and selective lifting.}
The sector decomposition also gives the evolution and spectrum,
\begin{equation}
\begin{gathered}
e^{-it\widetilde H_\Lambda}
  \simeq\bigoplus_{A\subseteq\Lambda}
      \left(e^{-itH_A}\otimes\id_{\mathcal M_A}\right),
 \\
 \operatorname{spec}(\widetilde H_\Lambda)
  =\biguplus_{A\subseteq\Lambda}
      \operatorname{spec}(H_A)^{[m_A]}.
\end{gathered}
 \label{eq:general-dynamics-spectrum}
\end{equation}
where $[m_A]$ denotes repetition with multiplicity $m_A$, including any degeneracies of $H_A$. For $A=\varnothing$, the energy is $c_\Lambda$. The term \emph{local zero mode} refers to the local algebraic Hamiltonian $P_i$, independently of this global energy shift.

Setting $r_i=1$, $s_i=0$ at unlifted sites includes selective lifting as discussed in Sec.~\ref{subsec:ising}, since only patterns with those sites active remain nonzero. The geometry of the surviving interactions determines any further spatial factorization within a pattern.

\paragraph*{Geometry of the surviving interactions.}
\label{subsec:general-geometry}
\label{subsec:example-geometry}

The one-dimensional examples in Sec.~\ref{subsec:ising} and the toric-code example in Sec.~\ref{subsec:applications-dimensions} are consequences of the same support rule. For a fixed active-site set $A$, the theorem retains precisely the nonzero seed interaction terms whose declared supports satisfy $X_\alpha\subseteq A$. A term with any inactive site in its support vanishes as a whole. It does not become an interaction on the remaining sites. Inactive sites outside its support do not affect it, even when they lie spatially between sites on which it acts.

Regard the active sites as vertices and each retained support as a set joining its sites. Pair interactions give an ordinary graph. A many-site interaction gives a \emph{hyperedge}, which can join more than two vertices, and the resulting collection is a hypergraph. Two active sites lie in the same component if a sequence of retained supports connects them, with successive supports sharing at least one site. An active site with no surviving interaction involving another site is a component by itself, whether or not a one-site field acts there.

Let $A_1,\ldots,A_k$ be these components. Every retained term is supported entirely within one component, so the active seed Hilbert space and restricted Hamiltonian satisfy
\begin{equation}
 \mathcal H_A=\bigotimes_{\ell=1}^k\mathcal H_{A_\ell},
 \qquad
 H_A=c_\Lambda\id_{\mathcal H_A}
        +\sum_{\ell=1}^k
          H^{\mathrm{int}}_{A_\ell}
              \otimes\id_{A\setminus A_\ell}.
 \label{eq:general-geometric-factorization}
\end{equation}
Here $H^{\mathrm{int}}_{A_\ell}$ includes the surviving interactions and fields within $A_\ell$, but excludes the global scalar. Terms on distinct components act on different tensor factors, giving
\[
 e^{-itH_A}=e^{-itc_\Lambda}
          \bigotimes_{\ell=1}^k e^{-itH^{\mathrm{int}}_{A_\ell}}.
\]
The multiplicity space $\mathcal M_A$ remains a spectator in the lifted sector. Independent evolution of the components does not preclude correlations in a state initially entangled between them. For $A=\varnothing$, only the global scalar acts. This spatial factorization need not be the finest possible decomposition, since additional operator structure or symmetries can split a component further.

Interaction range gives useful sufficient conditions for spatial separation. For a specified metric $\operatorname{dist}$ on the lattice, define the \emph{diameter} of a nonempty declared support by
\[
 \operatorname{diam}(X_\alpha)
   =\max_{i,j\in X_\alpha}\operatorname{dist}(i,j).
\]
It measures the greatest separation between sites in the support, rather than their number. A one-site support has diameter zero. A finite-range family has a common bound $\operatorname{diam}(X_\alpha)\leq R$, with $R$ independent of system size. Two candidate active regions whose minimum mutual distance exceeds $R$ cannot be joined by one retained interaction. To conclude that they belong to different components, one must also exclude a sequence of interactions connecting them through other active sites.

On an open chain $\Lambda=\{1,\ldots,L\}$ with distance measured in lattice spacings, this definition reduces to
\[
 \operatorname{diam}(X_\alpha)=\max X_\alpha-\min X_\alpha,
\]
where $\max X_\alpha$ and $\min X_\alpha$ are the rightmost and leftmost site labels in the support. A term spanning a gap of $g$ consecutive holes must have diameter at least $g+1$. The condition $g\geq R$ therefore separates the two sides even for many-site interactions, extending the pair-coupling argument of Sec.~\ref{subsec:ising}. If $g<R$, the actual declared supports must be examined. For instance, $\{2,4\}$ and $\{2,3,4\}$ both have diameter two, but a hole at site $3$ removes only the term declared on the latter support. Equal interaction ranges need not produce the same active geometry.

In higher dimensions or with periodic boundaries, active paths may go around an inactive region, as in Sec.~\ref{subsec:applications-dimensions}. The components of the retained interaction graph or hypergraph supply the general criterion in every geometry. For interactions of unbounded range, a gap or inactive boundary alone gives no universal separation criterion. The direct-sum decomposition into conserved patterns nevertheless persists, independently of whether the active dynamics factorizes spatially.

\subsection{Sector sizes and strong fragmentation}
\label{subsec:general-counting}

The dimensions in the theorem are explicit:
\begin{equation}
 D_A=\dim\widetilde{\mathcal H}_A
       =\prod_{i\in A}(n_ir_i)\prod_{j\notin A}s_j,
 \qquad
 \sum_A D_A=\prod_{i\in\Lambda}(n_ir_i+s_i).
 \label{eq:general-sector-dimensions}
\end{equation}
Since no matrix element connects two different patterns, every connected component of the Hamiltonian graph in the adapted product basis lies within one of these sectors. The following bound therefore requires no assumption about connectivity inside an active sector.

\begin{corollary}[Size of the largest dynamical block]
\label{cor:general-strong}
For uniform local dimensions $n_i=n$, $r_i=r$ and $s_i=s$, with fixed $n,r,s$ and $r,s\geq1$, let $D_{\max}^{\mathrm{conn}}$ be the size of the largest connected component in the adapted product basis. Then
\begin{equation}
 \frac{D_{\max}^{\mathrm{conn}}}{(nr+s)^L}
 \leq
 \frac{\max_A D_A}{(nr+s)^L}
 =
 \left(\frac{\max\{nr,s\}}{nr+s}\right)^L
 \xrightarrow[L\to\infty]{}0.
 \label{eq:general-strong-bound}
\end{equation}
Within the sector of fixed total number of inactive sites,
$N_0=\sum_i Z_i=m$, one has
\begin{equation}
 \dim\mathcal K_m=\binom{L}{m}(nr)^{L-m}s^m,
 \qquad
 \frac{D_{\max,m}^{\mathrm{conn}}}{\dim\mathcal K_m}
       \leq\binom{L}{m}^{-1}.
 \label{eq:general-fixed-number}
\end{equation}
For $m/L\to\rho$ with $0<\rho<1$, the latter ratio vanishes exponentially.
\end{corollary}

\begin{proof}
For $m$ inactive sites, Eq.~\eqref{eq:general-sector-dimensions} gives $D_A=(nr)^{L-m}s^m$. Maximizing over $m$ gives $\max\{nr,s\}^L$. At fixed $m$, the $\binom{L}{m}$ patterns have equal dimensions, and a connected component lies in only one of them. These observations prove the two bounds. The entropy estimate $\log\binom{L}{m}=Lh(\rho)+o(L)$ then gives exponential suppression for $0<\rho<1$.
\end{proof}

By the criterion in the preamble to Sec.~\ref{sec:examples}, these bounds establish strong fragmentation in the full Hilbert space and at fixed strictly intermediate inactive-site density. They do not assert strong fragmentation within every further symmetry sector. In particular, the fully active sector retains the seed Hamiltonian without additional hole-pattern splitting.

For site-dependent dimensions the corresponding estimate is
\begin{equation}
 \frac{D_{\max}^{\mathrm{conn}}}{\dim\widetilde{\mathcal H}_\Lambda}
 \leq\prod_{i\in\Lambda}
    \frac{\max\{n_ir_i,s_i\}}{n_ir_i+s_i}.
 \label{eq:general-inhomogeneous-bound}
\end{equation}
It gives exponential decay whenever these factors are uniformly bounded above by a constant smaller than one. Fixed local dimensions satisfy this condition. Allowing dimensions to grow without a uniform bound requires examining this product separately.

\subsection{Interaction norms and stability under perturbations}
\label{subsec:general-scope}

The algebraic statement has no range restriction at finite volume. Moreover, for every supported operator,
\begin{equation}
\begin{gathered}
 \|\phi_X(h_X)\|=\|h_X\|,
 \qquad
 \operatorname{supp}\bigl(\widetilde h_X\bigr)\subseteq X\\
 \text{as a declared interaction support}.
\end{gathered}
 \label{eq:general-locality-norm}
\end{equation}
Indeed, on the all-active part of $X$ the image is $h_X$ tensored with copy identities, and it vanishes on the remaining patterns in $X$. The support is not enlarged and the largest singular value is unchanged. Therefore any prescribed weighted summability bound on the local interaction norms is inherited term by term. For example, for nonnegative support weights $w(X)$,
\begin{equation}
 \sup_i\sum_{\alpha:\,i\in X_\alpha}
       w(X_\alpha)\|\phi_{X_\alpha}(h_\alpha)\|
 =
 \sup_i\sum_{\alpha:\,i\in X_\alpha}
       w(X_\alpha)\|h_\alpha\|.
 \label{eq:general-summability}
\end{equation}
Quasi-local interactions can consequently be treated in the same finite-volume construction while retaining their interaction-norm bounds. A thermodynamic limit additionally requires the convergence assumptions appropriate to the chosen interaction class; the finite-volume theorem by itself does not assert that an arbitrary non-summable interaction has such a limit. The finite-volume pattern count is likewise not an assertion of a finite-rank direct-sum decomposition of an infinite-system Hilbert space.

Finally, let
$V_\Lambda=v_\Lambda\id+\sum_\beta v_\beta\otimes\id_{\Lambda\setminus Y_\beta}$
be any Hermitian perturbation of the seed Hamiltonian with declared supports, and lift it using the same $\phi_i$. The theorem applies to the combined interaction family. For every real $\lambda$,
\begin{equation}
 \begin{gathered}
 {}[Z_i,\widetilde H_\Lambda+\lambda\widetilde V_\Lambda]=0,\\
 (\widetilde H_\Lambda+\lambda\widetilde V_\Lambda)
       \big|_{\widetilde{\mathcal H}_A}
       \simeq (H_A+\lambda V_A)\otimes\id_{\mathcal M_A},
 \end{gathered}
 \label{eq:general-perturbation}
\end{equation}
where $V_A$ keeps the perturbation terms with $Y_\beta\subseteq A$ and its global scalar. No smallness assumption on $\lambda$ is needed. Perturbations of the seed Hamiltonian may break its integrability, translation invariance or symmetries, but their lifts cannot change the local active/inactive labels. Extra fragmentation inherited from the seed interaction family can disappear under these perturbations; the pattern decomposition survives. A lifted perturbation can nevertheless join previously independent active regions if it introduces a surviving interaction between them.

This is stability within a specified algebraic family, rather than stability against every operator on the enlarged space. For example, if $\ket{a,\mu}$ is an active state and $\ket{z}$ an inactive state at site $i$, the Hermitian perturbation
$\ket{z}\bra{a,\mu}+\ket{a,\mu}\bra{z}$ does not commute with $Z_i$ and can connect different patterns.

For a uniform spin-$j_0$ seed local space, $n=2j_0+1$, every target dimension $d>n$ admits $r=1$, $s=d-n$. It can be interpreted as a spin-$(d-1)/2$ space. The resulting fragmented model need not retain the polynomial form or rotational symmetry of a conventional higher-spin Hamiltonian, since its interactions are specified by the SUSY or paraSUSY lift.

\section{Examples of bond-wise lifting}
\label{sec:bondwise}
\begingroup
\postdisplaypenalty=100

The site-wise construction enlarges the representation of the operator algebra attached to each physical site. Its common local zero modes supply conserved pattern projectors for any choice of the seed interactions. We now apply the same procedure to an entire two-site operator algebra. This introduces an additional choice, namely how the enlarged representation is embedded into the tensor product of the two physical sites. The resulting conserved structures can be local patterns, ordered strings of mobile labels, or irreducible strings associated with pair flips. The examples below establish these structures directly. They also show why the site-wise theorem does not extend to arbitrary bond embeddings without an additional argument. We conclude this section by applying the same conservation criteria to unitary circuits.

Throughout this section we use open chains and fix the qubit bond basis as $e_0=\ket{00}$, $e_1=\ket{01}$, $e_2=\ket{10}$ and $e_3=\ket{11}$. Periodic boundaries require cyclic versions of the string labels and are not used in the counting below. The order-three parasupercharge $Q_4$ generates the seed bond matrix units through
\begingroup
\small
\setlength{\jot}{2pt}
\begin{equation}
\begin{gathered}
 E_{ab}(Q)=(Q^\dagger)^a Q^3(Q^\dagger)^3Q^b,\\
 E_{ab}(Q_4)=\ket{e_a}\bra{e_b},\qquad 0\leq a,b\leq3.
\end{gathered}
 \label{eq:bond-matrix-units}
\end{equation}
\endgroup
Thus a density $h=\sum_{a,b}h_{ab}E_{ab}(Q_4)$ is first written as a polynomial in the parasupercharge and its adjoint. To enlarge the representation on physical sites of dimension $d$, choose
\begingroup
\small
\setlength{\jot}{2pt}
\begin{equation}
\begin{gathered}
 d^2=4r+s,\qquad r\geq1,\quad s\geq0,\\
 \widetilde Q=W\bigl[(Q_4\otimes\id_r)\oplus0_s\bigr]W^\dagger,\\
 \Phi(O)=W\bigl[(O\otimes\id_r)\oplus0_s\bigr]W^\dagger,\\
 \widetilde h=\sum_{a,b}h_{ab}E_{ab}(\widetilde Q)=\Phi(h),
 \quad \widetilde H=\sum_{j=1}^{L-1}\widetilde h_{j,j+1}.
\end{gathered}
 \label{eq:bond-prescription}
\end{equation}
\endgroup
Here $W$ is a unitary identification with $\mathbb C^d\otimes\mathbb C^d$. The lifted normalized paraSUSY Hamiltonian is $\Phi(\id_4)$, the projector onto the $4r$-dimensional active representation space. We allow $s=0$, since the bond examples need not obtain fragmentation from common zero modes of the parasupercharges. The integer $d$ labels a local spin-$(d-1)/2$ space, without implying that the lifted density is a bilinear higher-spin interaction.

For an explicit embedding of the original four-dimensional bond
space into $\mathbb C^d\otimes\mathbb C^d$, choose an ordered
orthonormal quartet $\boldsymbol v=(v_0,v_1,v_2,v_3)$ and map
$\ket{e_a}\mapsto\ket{v_a}$ for $a=0,1,2,3$.
The choice of these four vectors and their ordering specifies
the physical embedding, with the corresponding parasupercharge
and operator map given by
\begin{equation}
 Q[\boldsymbol v]=\sum_{a=0}^{2}\ket{v_a}\bra{v_{a+1}},
 \qquad
 \Phi_{\boldsymbol v}(O)
 =\sum_{a,b=0}^3O_{ab}\ket{v_a}\bra{v_b}.
 \label{eq:bond-quartet}
\end{equation}
Within a quartet, we abbreviate the two-site product state
$\ket{ab}=\ket a\otimes\ket b$ by $ab$. Several mutually orthogonal quartets supply the $r$ copies in Eq.~\eqref{eq:bond-prescription}, and their orthogonal complement supplies the $s$ zero modes.

Two compatibility issues distinguish general bond-wise lifting from the site-wise construction. First, a bond zero-mode projector commutes with its own lifted density, but need not commute with a density on an overlapping bond. Bond zero modes therefore do not by themselves guarantee conserved many-body sectors. Second, an invariant copy of the original chain requires compatible embeddings on overlapping bonds. Such compatibility can hold for a whole seed operator family or only for particular densities, and does not follow from an isolated bond embedding.

We therefore establish fragmentation by identifying projectors or strings conserved by every overlapping lifted interaction. Integrability must also be checked at the many-body level, since an arbitrary bond embedding need not preserve the overlapping operator relations that ensure it. The exact $XXC$ identification below provides a case in which integrability is independently known.

As in the site-wise lifting case, replacing $h$ by $h+c\id_4$ changes its lift by $c\Phi(\id_4)$ and this need not be a constant on the enlarged bond space. Similarly, a difference of one-site terms that telescopes in the seed chain need not telescope after a general bond embedding. We explicitly specify the density before carrying out each lift.

\subsection{Different lifts of the \texorpdfstring{$XXX$}{XXX} permutation chain}
\label{subsec:bond-xxx}
The permutation density of Sec.~\ref{subsec:permutation} has the bond polynomial
\begin{equation}
\begin{gathered}
\mathsf P^{(2)}=F(Q_4),\\
 F(Q)=E_{00}(Q)+E_{33}(Q)+E_{12}(Q)+E_{21}(Q).
\end{gathered}
 \label{eq:bond-permutation-polynomial}
\end{equation}
For one quartet this becomes
\begin{equation}
 F(Q[\boldsymbol v])=
 \ket{v_0}\bra{v_0}+\ket{v_3}\bra{v_3}
 +\ket{v_1}\bra{v_2}+\ket{v_2}\bra{v_1}.
 \label{eq:bond-permutation-quartet}
\end{equation}
We now change the representation and embedding while retaining this same polynomial. Figure~\ref{fig:bond-lifting-scope} illustrates the three choices.

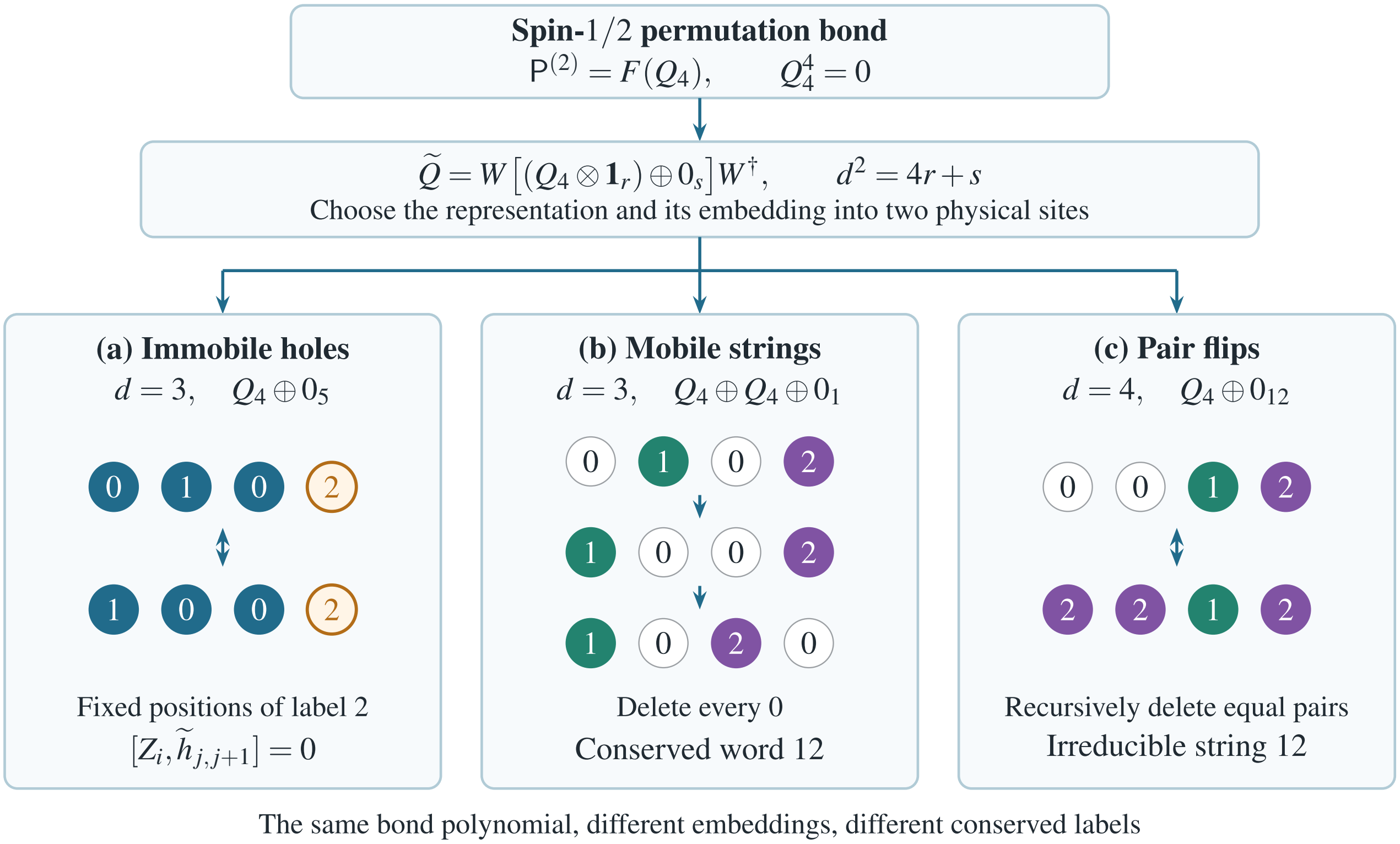
\begin{figure*}[tp]
\centering
\begingroup
\resizebox{\linewidth}{!}{%
\definecolor{bondBlue}{HTML}{216B8B}
\definecolor{bondWash}{HTML}{EDF5F8}
\definecolor{bondHole}{HTML}{B36D18}
\definecolor{bondHoleWash}{HTML}{FFF5E7}
\definecolor{bondPurple}{HTML}{8053A3}
\definecolor{bondTeal}{HTML}{23836F}
\definecolor{bondInk}{HTML}{202A32}
\begin{tikzpicture}[x=1cm,y=1cm,font=\fontsize{10}{12}\selectfont,text=bondInk,
 >={Stealth[length=2mm,width=1.5mm]},
 card/.style={draw=bondBlue!35,fill=bondWash!45,rounded corners=4pt,line width=.8pt},
 flow/.style={->,draw=bondBlue,line width=.95pt},
 state/.style={circle,draw=bondBlue,fill=bondBlue,text=white,minimum size=5.4mm,inner sep=0pt},
 zero/.style={state,fill=white,text=bondInk,draw=bondInk!45},
 one/.style={state,fill=bondTeal,draw=bondTeal},
 two/.style={state,fill=bondPurple,draw=bondPurple},
 hole/.style={state,fill=bondHoleWash,draw=bondHole,text=bondHole,line width=1pt},
 title/.style={font=\fontsize{10}{12}\selectfont\bfseries},
 note/.style={font=\fontsize{9}{11}\selectfont}]
\useasboundingbox (0,.2) rectangle (15.8,-9.25);
\filldraw[card] (3.40,0) rectangle (12.40,-1.02);
\node[title] at (7.9,-.30) {Spin-$1/2$ permutation bond};
\node at (7.9,-.73) {$\mathsf P^{(2)}=F(Q_4),\qquad Q_4^4=0$};
\draw[flow] (7.9,-1.02)--(7.9,-1.50);
\filldraw[card] (1.75,-1.50) rectangle (14.05,-2.55);
\node at (7.9,-1.86) {$\widetilde Q=W\bigl[(Q_4\otimes\id_r)\oplus0_s\bigr]W^\dagger,\qquad d^2=4r+s$};
\node[note] at (7.9,-2.28) {Choose the representation and its embedding into two physical sites};
\draw[bondBlue,line width=.95pt] (7.9,-2.55)--(7.9,-2.92);
\draw[bondBlue,line width=.95pt] (2.65,-2.92)--(13.15,-2.92);
\foreach \x in {2.65,7.9,13.15}{\draw[flow] (\x,-2.92)--(\x,-3.40);}
\filldraw[card] (.25,-3.40) rectangle (5.05,-8.62);
\filldraw[card] (5.50,-3.40) rectangle (10.30,-8.62);
\filldraw[card] (10.75,-3.40) rectangle (15.55,-8.62);
\node[title] at (2.65,-3.80) {(a) Immobile holes};
\node[title] at (7.9,-3.80) {(b) Mobile strings};
\node[title] at (13.15,-3.80) {(c) Pair flips};
\node at (2.65,-4.25) {$d=3,\quad Q_4\oplus0_5$};
\node at (7.9,-4.25) {$d=3,\quad Q_4\oplus Q_4\oplus0_1$};
\node at (13.15,-4.25) {$d=4,\quad Q_4\oplus0_{12}$};

\foreach \x/\v in {1.45/0,2.25/1,3.05/0}{\node[state] at (\x,-5.30) {$\v$};}
\node[hole] at (3.85,-5.30) {$2$};
\draw[<->,bondBlue,line width=.95pt] (2.65,-5.75)--(2.65,-6.18);
\foreach \x/\v in {1.45/1,2.25/0,3.05/0}{\node[state] at (\x,-6.65) {$\v$};}
\node[hole] at (3.85,-6.65) {$2$};
\node[note,align=center] at (2.65,-7.75) {Fixed positions of label $2$};
\node at (2.65,-8.18) {$[Z_i,\widetilde h_{j,j+1}]=0$};

\node[zero] at (6.70,-5.02) {$0$}; \node[one] at (7.50,-5.02) {$1$};
\node[zero] at (8.30,-5.02) {$0$}; \node[two] at (9.10,-5.02) {$2$};
\draw[flow] (7.90,-5.39)--(7.90,-5.65);
\node[one] at (6.70,-6.02) {$1$}; \node[zero] at (7.50,-6.02) {$0$};
\node[zero] at (8.30,-6.02) {$0$}; \node[two] at (9.10,-6.02) {$2$};
\draw[flow] (7.90,-6.39)--(7.90,-6.65);
\node[one] at (6.70,-7.02) {$1$}; \node[zero] at (7.50,-7.02) {$0$};
\node[two] at (8.30,-7.02) {$2$}; \node[zero] at (9.10,-7.02) {$0$};
\node[note] at (7.90,-7.75) {Delete every $0$};
\node at (7.90,-8.18) {Conserved word $12$};

\node[zero] at (11.95,-5.30) {$0$}; \node[zero] at (12.75,-5.30) {$0$};
\node[one] at (13.55,-5.30) {$1$}; \node[two] at (14.35,-5.30) {$2$};
\draw[<->,bondBlue,line width=.95pt] (13.15,-5.75)--(13.15,-6.18);
\node[two] at (11.95,-6.65) {$2$}; \node[two] at (12.75,-6.65) {$2$};
\node[one] at (13.55,-6.65) {$1$}; \node[two] at (14.35,-6.65) {$2$};
\node[note] at (13.15,-7.75) {Recursively delete equal pairs};
\node at (13.15,-8.18) {Irreducible string $12$};
\node[note] at (7.9,-9.04) {The same bond polynomial, different embeddings, different conserved labels};
\end{tikzpicture}
}
\endgroup
\caption{Three bond-wise paraSUSY lifts of the same spin-$1/2$ permutation density. The order-three parasupercharge $Q_4$ generates the operator algebra on the four-dimensional original bond space. Changing its representation and its embedding $W$ into the physical two-site space gives the three densities discussed below. The charge blocks in the panels are displayed up to the corresponding unitary embedding, and $0_s$ denotes an $s$-dimensional zero block.
(a) The $d=3$ embedding with one active quartet reproduces the site-wise lift. Label $2$ is an immobile hole, while $0$ and $1$ exchange.
(b) A different $d=3$ embedding has two active quartets. Labels $1$ and $2$ move through the $0$ background while retaining the word $12$ obtained by deleting all zeros.
(c) The $d=4$ embedding produces equal-pair flips. Both configurations shown reduce to the irreducible string $12$ after deleting equal adjacent pairs. The arrows show allowed off-diagonal processes. The choice of embedding determines which pattern or string conservation law applies. The pair-flip construction also extends to arbitrary bond densities through Eq.~\eqref{eq:bond-general-pair}.}
\label{fig:bond-lifting-scope}
\end{figure*}

\subsubsection*{Immobile holes}
For $d=3$, choose the quartet $(00,01,10,11)$ and let the parasupercharge vanish on the five remaining bond states. The lifted density exchanges labels $0$ and $1$, while every bond containing $2$ has zero action. Therefore $Z_j=\ket{2}_j\bra{2}$ commutes with every density. This bond embedding reproduces the minimal site-wise permutation lift, including its independent active segments and the pattern-sector analysis of Sec.~\ref{subsec:permutation}.

\subsubsection*{Mobile labels and conserved words}
On the same three-state local space, use two orthogonal quartets,
\begin{equation}
 \widetilde Q_{\mathrm{str}}
 =Q[00,01,10,11]+Q[12,02,20,21],
 \qquad \widetilde Q_{\mathrm{str}}\ket{22}=0.
 \label{eq:bond-string-charge}
\end{equation}
The decomposition of the bond representation is now $9=4+4+1$. Writing $\widetilde h_{\mathrm{stat}}$ for the preceding lifted density, we obtain
\begin{equation}
 \widetilde h_{\mathrm{str}}
 =\widetilde h_{\mathrm{stat}}
 +\ket{12}\bra{12}+\ket{21}\bra{21}
 +\ket{02}\bra{20}+\ket{20}\bra{02}.
 \label{eq:bond-string-density}
\end{equation}
The only off-diagonal processes are $01\leftrightarrow10$ and $02\leftrightarrow20$. Both labels $1$ and $2$ move through the $0$ background, but they cannot pass each other. Deleting every zero from a configuration therefore leaves a conserved word $w$. The projector
\begin{equation}
 \Pi_w=\sum_{\boldsymbol\sigma\,:
      \operatorname{del}_0(\boldsymbol\sigma)=w}
 \ket{\boldsymbol\sigma}\bra{\boldsymbol\sigma}
 \label{eq:bond-word-projector}
\end{equation}
commutes with every bond density. At fixed populations $N_0,N_1,N_2$, there are $\binom{N_1+N_2}{N_1}$ words, each spanning $\binom{L}{N_0}$ configurations. Adjacent exchanges connect all placements of a fixed word. The ratio of a word-sector dimension to the fixed-population space is consequently
\begin{equation}
 \frac{\binom{L}{N_0}}{L!/(N_0!N_1!N_2!)}
 =\binom{N_1+N_2}{N_1}^{-1}.
 \label{eq:bond-xxx-word-fraction}
\end{equation}
It vanishes exponentially when both $N_1$ and $N_2$ have positive limiting densities. This is the mobile-word conservation familiar from $XXC$ chains~\cite{Maassarani1998,Katsura2025}. The diagonal terms in Eq.~\eqref{eq:bond-string-density} preserve these words. Note that this is not exactly the integrable $XXC$ density, but rather just shares the fragmentation mechanism with the latter. For an exact identification with the $XXC$ density we also require equivalent dynamics. This construction is shown in the next subsection.

This example also demonstrates why bond zero modes cannot simply be treated as static holes. The neighboring exchange $\ket{220}\leftrightarrow\ket{202}$ changes whether the first bond is in $\ket{22}$. Its zero-mode projector is not conserved, although the ordered word $22$ is conserved.

\subsubsection*{Pair flips}
The pair-flip construction uses an embedding defined in the eigenbasis of the seed permutation.
We first express the seed permutation through its action on the
symmetric and antisymmetric states of two qubits. The permutation
acts as
\[
 \mathsf P^{(2)}\ket{ab}=\ket{ba},
 \qquad a,b\in\{0,1\}.
\]
Define
\[
 \ket{\psi_\pm}
 =\frac{\ket{01}\pm\ket{10}}{\sqrt2}.
\]
The three symmetric states $\ket{00}$, $\ket{\psi_+}$ and
$\ket{11}$ have eigenvalue $+1$, while the antisymmetric state
$\ket{\psi_-}$ has eigenvalue $-1$. Thus
\[
 \begin{aligned}
 \mathsf P^{(2)}
 &=\ket{00}\bra{00}
   +\ket{\psi_+}\bra{\psi_+}
   +\ket{11}\bra{11}
   -\ket{\psi_-}\bra{\psi_-}\\
 &=\id_4-2\ket{\psi_-}\bra{\psi_-}.
 \end{aligned}
\]
The factor of two changes the eigenvalue on the antisymmetric
state from $+1$ to $-1$, while leaving the symmetric subspace
unchanged.

We now choose four-state physical sites, so that the enlarged
bond space has dimension $16$. Within this space, take the
equal-pair subspace
\[
 \mathcal D_4
 =\operatorname{span}
   \{\ket{00},\ket{11},\ket{22},\ket{33}\},
 \qquad
 \ket{\Omega}
 =\frac12\sum_{a=0}^3\ket{aa}.
\]
The kets in this expression refer to the two enlarged physical
sites. Our aim is to map the antisymmetric seed state to
$\ket{\Omega}$. Its projector then connects every equal-pair
configuration to every other one.

Choose orthonormal vectors $u_0,u_1,u_2$ spanning
$\mathcal D_4\cap\ket{\Omega}^{\perp}$. We specify the embedding
by the mappings
\[
 \begin{aligned}
 \ket{00}&\longmapsto\ket{u_0},
 &\qquad \ket{\psi_+}&\longmapsto\ket{u_1},\\
 \ket{11}&\longmapsto\ket{u_2},
 &\qquad \ket{\psi_-}&\longmapsto\ket{\Omega}.
 \end{aligned}
\]
To specify the ordered quartet, we now determine the images of $e_0=\ket{00}$, $e_1=\ket{01}$, $e_2=\ket{10}$ and $e_3=\ket{11}$.
Since
\[
 \ket{01}
 =\frac{\ket{\psi_+}+\ket{\psi_-}}{\sqrt2},
 \qquad
 \ket{10}
 =\frac{\ket{\psi_+}-\ket{\psi_-}}{\sqrt2},
\]
their images are
\[
 \ket{v_1}
 =\frac{\ket{u_1}+\ket{\Omega}}{\sqrt2},
 \qquad
 \ket{v_2}
 =\frac{\ket{u_1}-\ket{\Omega}}{\sqrt2}.
\]
Together with $v_0=u_0$ and $v_3=u_2$, these form an
orthonormal quartet. The corresponding order-three
parasupercharge is therefore
\begin{equation}
 \widetilde Q_{\mathrm{PF}}
 =Q\!\left[
 u_0,\frac{u_1+\Omega}{\sqrt2},
     \frac{u_1-\Omega}{\sqrt2},u_2
 \right].
 \label{eq:bond-pair-charge}
\end{equation}
It acts within $\mathcal D_4$ and vanishes on its
twelve-dimensional orthogonal complement. This realizes one
active quartet and twelve zero-mode singlets.

Under this embedding, the seed identity becomes the projector
onto $\mathcal D_4$, while the antisymmetric projector becomes
the projector onto $\Omega$,
\[
 \Phi(\id_4)=K_{\mathcal D_4}
 =\sum_{a=0}^3\ket{aa}\bra{aa},
 \qquad
 \Phi\!\left(\ket{\psi_-}\bra{\psi_-}\right)
 =\ket{\Omega}\bra{\Omega}.
\]
Applying the lift to the spectral decomposition of the seed
permutation gives
\begin{equation}
 \begin{aligned}
 \widetilde h_{\mathrm{PF}}
 =K_{\mathcal D_4}-2\ket{\Omega}\bra{\Omega}
 =\sum_{a=0}^3\ket{aa}\bra{aa}
       -\frac12\sum_{a,b=0}^3\ket{aa}\bra{bb}.
 \end{aligned}
 \label{eq:bond-permutation-pair}
\end{equation}
The second equality follows from
$\ket{\Omega}\bra{\Omega}
=\frac14\sum_{a,b}\ket{aa}\bra{bb}$.
The resulting density is independent of the particular
orthonormal basis $u_0,u_1,u_2$, because the seed permutation
has eigenvalue $+1$ throughout the symmetric subspace.

The pair-flip mechanism is now explicit. Acting on an equal
pair gives
\begingroup
\small
\setlength{\jot}{2pt}
\[
\begin{aligned}
 \widetilde h_{\mathrm{PF}}\ket{bb}
 =\ket{bb}-\frac12\sum_{a=0}^3\ket{aa}
 =\frac12\ket{bb}
   -\frac12\sum_{\substack{a=0\\a\ne b}}^3\ket{aa}.
 \end{aligned}
\]
\endgroup
Hence every equal pair can change into every other equal pair,
with matrix element
\(
 \bra{aa}\widetilde h_{\mathrm{PF}}\ket{bb}
 =-\frac12,
 a\ne b.
\)
An unequal pair lies outside $\mathcal D_4$ and is annihilated,
\(
 \widetilde h_{\mathrm{PF}}\ket{ab}=0,
  a\ne b.
\)
The only off-diagonal processes are therefore
$bb\longleftrightarrow aa$, as illustrated in
Fig.~\ref{fig:bond-lifting-scope}(c).

These moves produce the established pair-flip fragmentation
mechanism~\cite{Caha2018,MoudgalyaMotrunich2022}.
For a product configuration $\boldsymbol\sigma$, let
$\mathcal R(\boldsymbol\sigma)$ denote the unique irreducible
word obtained by repeatedly deleting equal adjacent pairs.
A pair flip preserves this word because
\begingroup
\small
\setlength{\jot}{2pt}
\[
\mathcal R(x\,aa\,y)
 =\mathcal R(xy)
 =\mathcal R(x\,bb\,y),
\]
\endgroup
where $x$ and $y$ are the words on either side of the pair.
The projectors
\begin{equation}
 \Pi_\omega^{\mathrm{PF}}
 =\sum_{\mathcal R(\boldsymbol\sigma)=\omega}
       \ket{\boldsymbol\sigma}\bra{\boldsymbol\sigma}
 \label{eq:bond-pair-projectors}
\end{equation}
therefore commute with every lifted bond density.
In particular, every open-chain configuration without equal
neighbors is annihilated by all bond densities and is frozen.
There are $4$ choices for its first symbol and $3$ choices for
each subsequent symbol, giving $4\cdot3^{L-1}$ such states.

The preceding example suggests a more general construction,
because the pair-flip constraint follows from the embedding
into the equal-pair subspace. For a seed bond space of
dimension $m$, the identification $\ket{e_a}\mapsto\ket{aa}$
sends every matrix unit $\ket{e_a}\bra{e_b}$ to
$\ket{aa}\bra{bb}$. An arbitrary seed bond density therefore
lifts to a linear combination of diagonal equal-pair terms
and pair-flip processes. Changing its matrix elements changes
the amplitudes of these processes, while every allowed
transition continues to preserve the irreducible word obtained
by deleting equal adjacent pairs. The corresponding conserved
projectors thus survive for any choice of the seed bond
densities, leading to the following proposition.
\begin{proposition}[Pair-flip lift of an arbitrary bond density]
\label{prop:bond-pair}
Let a seed bond Hilbert space have dimension $m\geq3$, and let $h_j$ be any family of matrices acting on it. On physical sites of dimension $m$, embed the seed basis by $V\ket{a}=\ket{aa}$. The order-$(m-1)$ parasupercharge on this embedded subspace, extended by zero to its complement, gives
\begin{equation}
 \widetilde h_j=Vh_jV^\dagger
       =\sum_{a,b=0}^{m-1}(h_j)_{ab}\ket{aa}\bra{bb}.
 \label{eq:bond-general-pair}
\end{equation}
Every irreducible-string projector commutes with every $\widetilde h_j$. In particular, the common commutant has dimension at least $m(m-1)^{L-1}$ for an open chain of length $L\geq2$.
\end{proposition}
\begin{proof}
The only possible off-diagonal moves replace $bb$ by $aa$. Either pair can be deleted in the reduction, so the resulting irreducible word is unchanged. Thus the projectors in Eq.~\eqref{eq:bond-pair-projectors}, with $m$ colors, commute with every lifted density. A word with no equal neighbors is annihilated by all these densities. Its rank-one projector is therefore in the common commutant. There are $m$ choices for its first symbol and $m-1$ choices for each subsequent symbol, proving the bound.
\end{proof}
For a qubit nearest-neighbor density, $m=4$ and the lift uses order-three paraSUSY. Hermiticity is preserved when the seed density is Hermitian. The invariant-projector statement also holds for non-Hermitian densities. Vanishing coefficients in the seed density can produce further splitting. The pair-flip family with arbitrary bond-dependent coefficients and its fragmentation were studied in Ref.~\cite{MoudgalyaMotrunich2022}, following the pair-flip model of Ref.~\cite{Caha2018}. The above proposition gives a paraSUSY realization of this family from any prescribed seed bond density.

\subsection{Lifting \texorpdfstring{$XXZ$}{XXZ} to the
\texorpdfstring{$XXC$}{XXC} families}
\label{subsec:bond-xxz}
The $XXC$ families can be obtained by appropriately lifting the $XXZ$ spin chains.
For this we choose a bond embedding that allows labels from two
different sets to exchange while preserving their ordering
within each set. The choice is guided by the two seed states
coupled by the spin-$1/2$ $XXZ$ density. In the ordered basis
$(e_0,e_1,e_2,e_3)
=(\ket{00},\ket{01},\ket{10},\ket{11})$, take
\begin{equation}
 \begin{split}
 h_\Delta
 &=\frac12\bigl(\sigma^x\otimes\sigma^x
                 +\sigma^y\otimes\sigma^y\bigr)
   +\frac{\Delta}{2}
      \bigl(\sigma^z\otimes\sigma^z-\id_4\bigr)\\
 &=E_{12}+E_{21}-\Delta(E_{11}+E_{22}).
 \end{split}
 \label{eq:bond-xxz-seed}
\end{equation}
Its action is
\begingroup
\small
\setlength{\jot}{2pt}
\[
\begin{gathered}
 h_\Delta\ket{e_1}=\ket{e_2}-\Delta\ket{e_1},\quad
 h_\Delta\ket{e_2}=\ket{e_1}-\Delta\ket{e_2},\\
 h_\Delta\ket{e_0}=h_\Delta\ket{e_3}=0.
\end{gathered}
\]
\endgroup
Thus only $e_1$ and $e_2$ participate in the non-trivial action of the density.
The identity shift makes the other two seed states have zero
energy, although they remain part of the nontrivial
four-dimensional paraSUSY representation.
This energy convention is fixed before lifting, since the
seed identity becomes an active-bond projector.
In particular, $h_1=\mathsf P^{(2)}-\id_4$.

Divide the enlarged local basis into two disjoint sets
$\mathcal A$ and $\mathcal B$, with
$|\mathcal A|=p$, $|\mathcal B|=q$ and $d=p+q$.
For each $a\in\mathcal A$ and $b\in\mathcal B$, we want the
exchange $\ket{ab}\leftrightarrow\ket{ba}$.
We therefore assign one copy of the four-dimensional seed
representation to each pair $(a,b)$.
Its embedding $V_{ab}$ is specified by
\begingroup
\small
\setlength{\jot}{2pt}
\[
\begin{aligned}
 V_{ab}\ket{e_0}&=\ket{u_{ab}},&\qquad V_{ab}\ket{e_1}&=\ket{ab},\\
 V_{ab}\ket{e_2}&=\ket{ba},& V_{ab}\ket{e_3}&=\ket{v_{ab}}.
\end{aligned}
\]
\endgroup
The pair $(a,b)$ labels the copy of the representation.
The images of $e_1,e_2$ implement the desired exchange.
The vectors $u_{ab},v_{ab}$ are additional chosen states
that complete this copy to an orthonormal quartet.

To choose these completion vectors, first observe that the
$2pq$ mixed-set states $\ket{ab},\ket{ba}$ are mutually
orthonormal. Their orthogonal complement is the same-set
subspace
\[
 \mathcal H_\parallel
 =
 (\mathbb C^{\mathcal A}\otimes\mathbb C^{\mathcal A})
 \oplus
 (\mathbb C^{\mathcal B}\otimes\mathbb C^{\mathcal B}),
\]
of dimension $p^2+q^2$.
Choose all $2pq$ vectors $u_{ab},v_{ab}$ mutually orthonormal
within this subspace. There is always enough room because
\[
 p^2+q^2-2pq=(p-q)^2\geq0.
\]
Each completed quartet is orthogonal to every other quartet.
The corresponding parasupercharge is
\begin{equation}
 \widetilde Q_{\mathcal A\mathcal B}
 =
 \sum_{a\in\mathcal A,\,b\in\mathcal B}
 Q[u_{ab},ab,ba,v_{ab}].
 \label{eq:bond-xxc-charge}
\end{equation}
There are $pq$ four-dimensional copies, and the remaining
states are common zero modes of the parasupercharge and its
adjoint. Hence
\begin{equation}
 r=pq,\qquad s=(p-q)^2,\qquad
 d^2=4pq+(p-q)^2.
 \label{eq:bond-xxc-dimensions}
\end{equation}

For a concrete example, take
$\mathcal A=\{0\}$ and $\mathcal B=\{1,2\}$.
The mixed-set states are $\ket{01},\ket{10},\ket{02},\ket{20}$.
One possible choice of completion vectors is
\[
 \begin{aligned}
 \ket{u_{01}}&=\ket{00},
 &\qquad \ket{v_{01}}&=\ket{11},\\
 \ket{u_{02}}&=\ket{12},
 &\qquad \ket{v_{02}}&=\ket{21}.
 \end{aligned}
\]
This gives the two quartets
\[
 (00,01,10,11),
 \qquad
 (12,02,20,21),
\]
with $\ket{22}$ as the remaining zero-mode singlet.
It is precisely the $9=4+4+1$ representation used in
Eq.~\eqref{eq:bond-string-charge}.

We can now evaluate the lifted density. The operator map
acts identically on the mutually orthogonal copies,
\[
 \Phi(O)
 =
 \sum_{a\in\mathcal A,\,b\in\mathcal B}
 V_{ab}OV_{ab}^{\dagger}.
\]
Within one copy, the seed density becomes
\[
 \begin{aligned}
 V_{ab}h_\Delta V_{ab}^{\dagger}
 =
 \ket{ab}\bra{ba}+\ket{ba}\bra{ab}
 -\Delta\bigl(
 \ket{ab}\bra{ab}+\ket{ba}\bra{ba}
 \bigr).
 \end{aligned}
\]
The completion vectors $u_{ab},v_{ab}$ do not appear because
the rows and columns associated with $e_0,e_3$ vanish in
$h_\Delta$. Consequently, every permitted choice of these
completion vectors gives the same lifted density, even
though it can give a different parasupercharge.
Summing over all copies yields
\begin{equation}
 \begin{gathered}
 \widetilde h_\Delta
 =T_{\mathcal A\mathcal B}-\Delta P_\times,\\
 T_{\mathcal A\mathcal B}
 =
 \sum_{a\in\mathcal A,\,b\in\mathcal B}
 \bigl(
 \ket{ab}\bra{ba}+\ket{ba}\bra{ab}
 \bigr),\\
 P_\times
 =
 \sum_{a\in\mathcal A,\,b\in\mathcal B}
 \bigl(
 \ket{ab}\bra{ab}+\ket{ba}\bra{ba}
 \bigr),
 \qquad
 P_\parallel=\id_{d^2}-P_\times.
 \end{gathered}
 \label{eq:bond-exact-xxc}
\end{equation}
This is precisely the untwisted $(p,q)$ $XXC$ density, up to the conventional additive constant and overall energy scale~\cite{Maassarani1998}. To make the identification explicit, take $\Delta=\cos\eta$. The known $XXC$ braid $R$-matrix can be written in our convention as
\begin{equation}
\begin{gathered}
\check R(u)=\frac{\sin(\eta+u)P_\parallel
                      +\sin\eta\,P_\times
                      +\sin u\,T_{\mathcal A\mathcal B}}{\sin\eta},
 \\
 \sin\eta\,\check R'(0)-\cos\eta\,\id_{d^2}
 =\widetilde h_\Delta.
\end{gathered}
 \label{eq:bond-xxc-rmatrix}
\end{equation}
The parameter in the convention of Ref.~\cite{Maassarani1998} is $\gamma=\pi-\eta$. Equation~\eqref{eq:bond-xxc-rmatrix} is regular for $\sin\eta\neq0$, with the isotropic cases obtained by a rational limit. Thus the identification concerns the full density, including its diagonal terms. Independent unit-modulus exchange phases can also be obtained by choosing relative phases for $\ket{ab}$ and $\ket{ba}$ in each quartet. This realizes the Hermitian phase-twisted subfamily. General non-unit-modulus twists are not asserted to follow from these unitary embeddings.

The fragmentation follows from the allowed exchanges $ab\leftrightarrow ba$, illustrated for $p=1,q=2$ in Fig.~\ref{fig:bond-lifting-scope}(b). Deleting all $\mathcal B$ symbols gives an invariant ordered word $w_{\mathcal A}$, and deleting all $\mathcal A$ symbols gives a second invariant word $w_{\mathcal B}$. Neither word is tied to fixed sites. Let $N_{\mathcal A}=\sum_{a\in\mathcal A}N_a$ and $N_{\mathcal B}=\sum_{b\in\mathcal B}N_b=L-N_{\mathcal A}$ denote the total populations in the two sets. Every configuration of the two words is connected by adjacent exchanges, so its configuration-sector dimension is $\binom{L}{N_{\mathcal A}}$. At fixed populations $\{N_c\}_{c\in\mathcal A\cup\mathcal B}$, the number of compatible word pairs and the corresponding fraction are
\begin{equation}
 \begin{gathered}
 \mathcal N_{\mathrm{words}}
 =\frac{N_{\mathcal A}!}{\prod_{a\in\mathcal A}N_a!}
  \frac{N_{\mathcal B}!}{\prod_{b\in\mathcal B}N_b!},\\
 \frac{\binom{L}{N_{\mathcal A}}}{L!/\prod_c N_c!}
 =\mathcal N_{\mathrm{words}}^{-1}.
 \end{gathered}
 \label{eq:bond-xxc-count}
\end{equation}
The ratio is exponentially small whenever an extensive set contains at least two colors with positive limiting fractions. This establishes strong fragmentation in that fixed-population comparison space. For $p=q=1$, both words are determined by the populations and one recovers the ordinary $XXZ$ chain, without this extra fragmentation. The conserved-word interpretation agrees with the $XXC$ analysis in Ref.~\cite{Katsura2025}.

For $p=1,q=2$, the charge in Eq.~\eqref{eq:bond-string-charge} supplies an explicit realization of Eq.~\eqref{eq:bond-xxc-charge}. For balanced sets with $p=q\geq2$, Eq.~\eqref{eq:bond-xxc-dimensions} instead has $s=0$. The chain still has exponentially many word sectors even though the bond parasupercharge and its adjoint have no common zero-mode singlets. Bond-wise lifting can therefore realize fragmentation through the embedding and ordered labels alone.

\subsection{Lifting the \texorpdfstring{$XYZ$}{XYZ} spin chain and the dependence on the embedding}
\label{subsec:bond-xyz}
An embedded spin-$1/2$ $XYZ$ chain on spin-$1$ sites, with conserved projectors onto the local active spaces, was discussed in Sec.~2.3.1 of Ref.~\cite{ClassenHowes2024}. It provides a precedent for fragmentation into immobile inactive-site patterns. The bond-wise embeddings below can preserve pair-reduction strings or mobile words, or produce only polynomially many connected components in the displayed product basis. Write
\begin{equation}
 \begin{gathered}
 h_{\mathrm{XYZ}}=J_x\sigma^x\otimes\sigma^x
                 +J_y\sigma^y\otimes\sigma^y
                 +J_z\sigma^z\otimes\sigma^z\\
 =J_z(E_{00}+E_{33}-E_{11}-E_{22})
       +t(E_{12}+E_{21})+\delta(E_{03}+E_{30}),\\
 t=J_x+J_y,\qquad\delta=J_x-J_y.
 \end{gathered}
 \label{eq:bond-xyz-seed}
\end{equation}
We keep the seed density in Eq.~\eqref{eq:bond-xyz-seed}
fixed and compare three embeddings. Its two off-diagonal
processes are $e_1\leftrightarrow e_2$, with amplitude $t$,
and $e_0\leftrightarrow e_3$, with amplitude $\delta$.
Their physical interpretation after lifting depends on the
images of these four seed states.

\begin{enumerate}

\item \emph{Equal-pair embedding.}

Choose four-state physical sites and embed the seed bond basis
by $\ket{e_a}\mapsto\ket{aa}$ for $a=0,1,2,3$.
The parasupercharge is $Q[00,11,22,33]$, with one active
quartet and twelve zero-mode singlets. The lifted density is
\[
 \begin{aligned}
 \widetilde h_{\mathrm{XYZ}}^{\mathrm{PF}}
 ={}&J_z\bigl(
       \ket{00}\bra{00}+\ket{33}\bra{33}
       -\ket{11}\bra{11}-\ket{22}\bra{22}
       \bigr)\\
 &+t\bigl(
       \ket{11}\bra{22}+\ket{22}\bra{11}
       \bigr)\\
 &+\delta\bigl(
       \ket{00}\bra{33}+\ket{33}\bra{00}
       \bigr).
 \end{aligned}
\]
The two seed processes become the pair flips
\[
 11\longleftrightarrow22,
 \qquad
 00\longleftrightarrow33.
\]
Every transition preserves the irreducible word obtained by
deleting equal adjacent pairs. Proposition~\ref{prop:bond-pair}
therefore applies for arbitrary $J_x,J_y,J_z$.
In particular, all $4\cdot3^{L-1}$ open-chain configurations
without equal neighbors are frozen.

Only a subset of the possible four-color pair flips occurs
in this density. Consequently, the irreducible-word sectors
can split further, but different irreducible words remain
dynamically disconnected.

\item \emph{Three-state embedding that loses mobile-word
conservation.}

On three-state physical sites, use the two quartets from
Eq.~\eqref{eq:bond-string-charge},
\[
 (00,01,10,11),
 \qquad
 (12,02,20,21),
\]
with $\ket{22}$ as the remaining zero-mode singlet.
The representation content is $9=4+4+1$.
The seed exchange $e_1\leftrightarrow e_2$ produces
\[
 01\longleftrightarrow10,
 \qquad
 02\longleftrightarrow20,
\]
while the additional $XYZ$ process
$e_0\leftrightarrow e_3$ produces
\[
 00\longleftrightarrow11,
 \qquad
 12\longleftrightarrow21.
\]
Explicitly,
\begin{equation}
 \begin{split}
 \widetilde h_{\mathrm{XYZ}}^{\mathrm{str}}
 ={}&J_z\bigl(
       \ket{00}\bra{00}+\ket{11}\bra{11}
       +\ket{12}\bra{12}+\ket{21}\bra{21}
       \bigr)\\
 &-J_z\bigl(
       \ket{01}\bra{01}+\ket{10}\bra{10}
       +\ket{02}\bra{02}+\ket{20}\bra{20}
       \bigr)\\
 &+t\bigl(
       \ket{01}\bra{10}+\ket{02}\bra{20}
       +\mathrm{h.c.}
       \bigr)\\
 &+\delta\bigl(
       \ket{00}\bra{11}+\ket{12}\bra{21}
       +\mathrm{h.c.}
       \bigr).
 \end{split}
 \label{eq:bond-xyz-string-lift}
\end{equation}
When $\delta=0$, deleting all zeros gives a conserved ordered
word of labels $1,2$. For nonzero $\delta$, the process
$12\leftrightarrow21$ changes the order of this word, while
$00\leftrightarrow11$ changes its length.

For $t\delta\neq0$, the connected components in the displayed
product basis can be determined completely.
All three exchanges
$01\leftrightarrow10$, $02\leftrightarrow20$ and
$12\leftrightarrow21$ are available, so all arrangements
with fixed populations are connected.
The conversion $00\leftrightarrow11$ additionally changes
$N_1$ by two while preserving $N_2$.
Thus the components are precisely labeled by
\begin{equation}
 m=N_2,\qquad \epsilon=N_1\bmod2.
 \label{eq:bond-xyz-labels}
\end{equation}
Here $m$ is the number of sites carrying label $2$.
For $m<L$, choose its $m$ positions in $\binom{L}{m}$ ways.
The remaining $L-m$ sites carry labels $0,1$, and exactly
half of their binary configurations have either specified
parity. Therefore
\begin{equation}
 D_{m,\epsilon}
 =\binom{L}{m}2^{L-m-1},
 \qquad m<L.
 \label{eq:bond-xyz-components}
\end{equation}
Each $m=0,\ldots,L-1$ gives two components, while $m=L$
gives only $\ket{22\cdots2}$. There are consequently
$2L+1$ product-configuration components.
After fixing the population and parity labels in
Eq.~\eqref{eq:bond-xyz-labels}, each such space is connected.

This establishes the absence of exponential configuration fragmentation in the displayed product basis when $t\delta\neq0$. It does not rule out additional invariant sectors that become apparent in another local product basis or require an entangled basis. Determining the growth of the full bond commutant requires a separate analysis.

\item \emph{Rotated three-state embedding preserving mobile
words.}

We now retain the same local dimension and the same
representation content $9=4+4+1$, but change the images of
$e_0$ and $e_3$ within each quartet. Define
\begingroup
\small
\setlength{\jot}{2pt}
\begin{equation}
\begin{gathered}
 u_\pm=\frac{\ket{00}\pm\ket{11}}{\sqrt2},\qquad
 v_\pm=\frac{\ket{12}\pm\ket{21}}{\sqrt2},\\
 \widetilde Q_{\mathrm{rot}}
 =Q[u_+,01,10,u_-]+Q[v_+,02,20,v_-],\\
 \widetilde Q_{\mathrm{rot}}\ket{22}=0.
\end{gathered}
 \label{eq:bond-rotated-string-charge}
\end{equation}
\endgroup
The images of $e_1,e_2$ are unchanged, so the processes
$01\leftrightarrow10$ and $02\leftrightarrow20$ retain
amplitude $t$.
The rotation instead converts the
$e_0\leftrightarrow e_3$ process into diagonal terms.
Indeed,
\[
 \begin{aligned}
 \ket{u_+}\bra{u_-}+\ket{u_-}\bra{u_+}
 &=\ket{00}\bra{00}-\ket{11}\bra{11},\\
 \ket{v_+}\bra{v_-}+\ket{v_-}\bra{v_+}
 &=\ket{12}\bra{12}-\ket{21}\bra{21}.
 \end{aligned}
\]
Thus the terms that previously changed the word now assign
diagonal energies to its configurations.

Using $T_{\mathcal A\mathcal B}$ and $P_\times$ from
Eq.~\eqref{eq:bond-exact-xxc}, with
$\mathcal A=\{0\}$ and $\mathcal B=\{1,2\}$, define
\[
 \begin{aligned}
 K&=\Phi_{\mathrm{rot}}(\id_4)
   =\id_9-\ket{22}\bra{22},\\
 D&=\ket{00}\bra{00}-\ket{11}\bra{11}
      +\ket{12}\bra{12}-\ket{21}\bra{21}.
 \end{aligned}
\]
The lift of the same unshifted seed density is
\begin{equation}
 \Phi_{\mathrm{rot}}(h_{\mathrm{XYZ}})
 =J_zK+tT_{\mathcal A\mathcal B}
       -2J_zP_\times+\delta D.
 \label{eq:bond-tmd-unshifted}
\end{equation}
The $K$ term is retained because the seed identity lifts to
the active-bond projector. It is diagonal, as are
$P_\times$ and $D$. Only $T_{\mathcal A\mathcal B}$ changes
product configurations, by exchanging a zero with a label
$1$ or $2$.

Deleting all zeros therefore gives a conserved ordered word
for arbitrary $J_x,J_y,J_z$.
When $t\neq0$, adjacent exchanges connect every placement
of a fixed word, so the corresponding sector dimension is
$\binom{L}{N_0}$. At fixed populations,
\[
 \frac{D_{\mathrm{word}}}{D_{\mathrm{fixed}}}
 =
 \frac{\binom{L}{N_0}}{L!/(N_0!N_1!N_2!)}
 =
 \binom{N_1+N_2}{N_1}^{-1}.
\]
This is the ratio in
Eq.~\eqref{eq:bond-xxx-word-fraction}.
It is exponentially small when both labels $1,2$ have
positive limiting densities, establishing strong
fragmentation in that fixed-population space.
If $t=0$, the density is entirely diagonal and the word
sectors split further.

The rotated lift thus has the same mobile-word connectivity
as the preceding $XXX$ and $XXZ$ examples, while its
additional diagonal terms modify the dynamics within each
sector. This argument establishes fragmentation without
asserting Yang--Baxter integrability of the lifted
$XYZ$ chain.

\end{enumerate}

These constructions distinguish the role of the seed density
from that of its embedding. The equal-pair embedding preserves
pair-reduction strings for arbitrary seed coefficients.
The other two constructions show that, even at fixed local
dimension and representation content, changing the embedding
can destroy or restore mobile-word conservation.
Although their bond densities are related by a unitary
transformation on the bond, this transformation mixes product
states into entangled states and is not a product of one-site
basis changes. The overlapping bond embeddings therefore do
not automatically define globally equivalent chains.

\subsection{The non-Hermitian \texorpdfstring{$H_{1,11}$}{H1,11} density}
\label{subsec:bond-h111}
The next two examples use seed densities derived from Hietarinta's classification of constant $4\times4$ solutions of the Yang--Baxter equation~\cite{Hietarinta1992,Hietarinta1993}. The construction also applies to non-Hermitian seed densities. Consider the $H_{1,11}$ Hamiltonian obtained by Baxterizing the non-invertible Hietarinta representative $R_{1,11}$ in Ref.~\cite{Maity2025NonHermitian}. With its overall coupling absorbed into the energy scale, the density is
\begin{equation}
 h_{1,11}=\alpha E_{01}+\beta E_{02}
 =\begin{pmatrix}
 0&\alpha&\beta&0\\
 0&0&0&0\\
 0&0&0&0\\
 0&0&0&0
 \end{pmatrix},\qquad h_{1,11}^2=0.
 \label{eq:bond-h111-seed}
\end{equation}
Its direct quartet embedding $(00,01,10,11)$ in $d=3$ reproduces immobile occurrences of the unused symbol $2$. A different quartet on the same local space produces pair flips. Choose
\begingroup
\small
\setlength{\jot}{2pt}
\begin{equation}
\begin{gathered}
 u_0=\frac{\ket{00}+\ket{11}+\ket{22}}{\sqrt3},\quad
 u_1=\frac{\ket{00}-\ket{11}}{\sqrt2},\\
 u_2=\frac{\ket{00}+\ket{11}-2\ket{22}}{\sqrt6},\quad
 u_3=\ket{01},\\
 \widetilde Q_{1,11}=Q[u_0,u_1,u_2,u_3].
\end{gathered}
 \label{eq:bond-h111-charge}
\end{equation}
\endgroup
The fourth state completes the parasupercharge representation but does not enter the density, since the fourth row and column of Eq.~\eqref{eq:bond-h111-seed} vanish. Direct substitution gives
\begin{equation}
 \begin{gathered}
 \widetilde h_{1,11}
 =\ket{u_0}\bigl(\alpha\bra{u_1}+\beta\bra{u_2}\bigr)
 =\frac1{\sqrt3}\sum_{a,b=0}^2c_b\ket{aa}\bra{bb},\\
 (c_0,c_1,c_2)
 =\left(\frac{\alpha}{\sqrt2}+\frac{\beta}{\sqrt6},
        -\frac{\alpha}{\sqrt2}+\frac{\beta}{\sqrt6},
        -\frac{2\beta}{\sqrt6}\right).
 \end{gathered}
 \label{eq:bond-h111-pair}
\end{equation}
For instance, $\alpha=\beta=1$ makes every $c_b$ nonzero. Every equal pair can then be converted to every other equal pair, although the forward and reverse amplitudes are unequal. These moves follow the pair-flip mechanism illustrated in Fig.~\ref{fig:bond-lifting-scope}(c), here with three colors. The projectors onto pair-reduction strings commute with every bond density. In particular, the $3\cdot2^{L-1}$ words with no equal neighbors are frozen. For nonzero $c_b$, the connected configuration sectors are the usual three-color pair-flip sectors~\cite{Caha2018,MoudgalyaMotrunich2022}. Their number is $2^{L+1}-1$, obtained by summing the irreducible words of lengths $L,L-2,\ldots$.

The same algebraic representation content, $9=4+5$, has therefore produced fixed one-site holes in one embedding and conserved pair-reduction strings in another. These are exact reducing subspaces of the non-Hermitian Hamiltonian and its exponential evolution. No unitary time evolution is assumed here. Also, the local nilpotency $\widetilde h_{1,11}^2=0$ does not determine the nilpotency or spectrum of the sum on overlapping bonds.

\subsection{The \texorpdfstring{$H_{1,3}$}{H1,3} density and fluctuating colored dimers}
\label{subsec:bond-h13}
For the invertible Hietarinta family, take the density obtained directly from the $H_{1,3}$ braid representative in Ref.~\cite{Maity2026},
\begin{equation}
 h_{1,3}=\begin{pmatrix}
 \alpha^2&-\alpha\beta&\alpha\beta&\beta\gamma\\
 0&0&\alpha^2&\alpha\gamma\\
 0&\alpha^2&0&-\alpha\gamma\\
 0&0&0&\alpha^2
 \end{pmatrix}.
 \label{eq:bond-h13-seed}
\end{equation}
We retain this density before removing terms that telescope in the
original chain. Proposition~\ref{prop:bond-pair} immediately supplies
an equal-pair lift on $d=4$ sites. We now choose a different embedding
on $d=5$ sites, which produces fluctuating colored dimers separated
by frozen unpaired symbols.

\paragraph*{Embedding and allowed transitions.}
Each enlarged site has the five states
$\ket{0},\ket{1},\ket{2},\ket{3},\ket{4}$, so the enlarged bond space
has dimension $25$. Within this space, choose
\[
 \mathcal W
 =\operatorname{span}
 \{\ket{00},\ket{12},\ket{13},\ket{14}\}.
\]
We embed the four seed basis vectors $\ket{e_a}$ into an orthonormal
basis $\ket{u_a}$ of $\mathcal W$. Specifically, define
\begingroup
\small
\setlength{\jot}{2pt}
\begin{equation}
\begin{gathered}
 \ket{w_0}=\ket{00},\quad \ket{w_1}=\ket{12},\quad
 \ket{w_2}=\ket{13},\quad \ket{w_3}=\ket{14},\\
 U=\frac12\begin{pmatrix}
 1&1&1&1\\
 1&1&-1&-1\\
 1&-1&1&-1\\
 1&-1&-1&1
 \end{pmatrix},\qquad
 \begin{gathered}
 \ket{u_a}=\sum_{b=0}^{3}U_{ba}\ket{w_b},\\
 \ket{e_a}\longmapsto\ket{u_a},
 \end{gathered}\\
 \widetilde Q_{1,3}=Q[u_0,u_1,u_2,u_3]
 =\sum_{a=0}^{2}\ket{u_a}\bra{u_{a+1}}.
\end{gathered}
 \label{eq:bond-h13-charge}
\end{equation}
\endgroup
The parasupercharge and its adjoint vanish on the orthogonal
complement of $\mathcal W$. This is the representation choice
$25=4+21$, with one four-dimensional active representation and
$21$ common zero modes on the bond.

The embedding maps each seed matrix unit
$\ket{e_a}\bra{e_b}$ to $\ket{u_a}\bra{u_b}$. Expressing the result
in the physical product basis $\{\ket{w_a}\}$ gives
\begin{equation}
 \widetilde h_{1,3}
 =\sum_{a,b=0}^{3}K_{ab}\ket{w_a}\bra{w_b},
 \qquad K=Uh_{1,3}U^\dagger.
 \label{eq:bond-h13-lift}
\end{equation}
Thus the rotation $U$ determines which physical configurations
are coupled by the lifted density. For generic values of $\alpha,\beta,\gamma$, all off-diagonal
entries of $K$ are nonzero. We first consider this regime,
in which the allowed local transitions are
\begin{equation}
 00\longleftrightarrow1a,
 \qquad
 1a\longleftrightarrow1b,
 \qquad a,b\in\{2,3,4\}.
 \label{eq:bond-dimer-moves}
\end{equation}
We interpret $1$ as a dimer head and $a\in\{2,3,4\}$ as its
tail color. A dimer is therefore the oriented product-state
configuration $1a$, rather than an entangled spin singlet. Dimers cannot
overlap. The first transition creates or annihilates a dimer,
while the second changes its color.

Although no individual transition translates a dimer, successive
moves can displace it through an intermediate empty configuration.
For example,
\[
 120\longrightarrow000\longrightarrow013.
\]
These processes are illustrated in
Fig.~\ref{fig:bond-dimer-fragmentation}(a).
The arrows indicate nonzero transition amplitudes, which need
not agree in opposite directions because the density is
non-Hermitian. The symbol $0$ is available for dimer creation
and does not represent an immobile one-site zero mode.

\paragraph*{Frozen symbols and invariant sectors.}
We consider an open chain in the following counting arguments.
A head $1$ can participate in a transition only when its
immediate right neighbor is a tail. Likewise, a tail can
participate only when its immediate left neighbor is $1$.
An unpaired head or tail is therefore frozen.

A neighboring transition cannot supply the missing partner.
For example, creating a dimer immediately to the right of an
unpaired head places another head next to it, rather than a
tail. Similarly, creating a dimer immediately to the left of
an unpaired tail places another tail next to it. The positions
and identities of all unpaired symbols consequently remain
unchanged.

This conservation can be expressed through local projectors.
Let $n_j^{(a)}=\ket{a}_j\bra{a}$ and
$n_j^R=\sum_{a=2}^{4}n_j^{(a)}$. Then
\begin{equation}
 F_j^{(1)}
 =n_j^{(1)}(\id-n_{j+1}^R),
 \qquad
 F_j^{(a)}
 =(\id-n_{j-1}^{(1)})n_j^{(a)},
 \quad a=2,3,4,
 \label{eq:bond-dimer-projectors}
\end{equation}
commute with every bond density. Missing neighboring occupations
are taken to be zero at the open boundaries. The above projectors
detect an unpaired head or an unpaired tail of a specified color.
Their conservation holds for every matrix unit
$\ket{w_a}\bra{w_b}$ and is therefore independent of the
coefficients in $K$. Unlike the site-wise construction, the
inactivity of a symbol is determined by its neighborhood.

To label the sectors, replace every dimer $12$, $13$ or $14$
by $00$ at the same positions. Denote the resulting canonical
configuration by $\mathcal C$. This replacement preserves all
site positions. It neither deletes sites nor closes gaps, and
it cannot create new dimers. Since dimers do not overlap, the
canonical configuration is unique.

Every allowed transition preserves $\mathcal C$. Conversely,
for nonzero amplitudes, every
configuration can reach its canonical configuration by
annihilating its dimers. Reversing these steps reaches any
other configuration with the same $\mathcal C$. Thus each
canonical configuration labels one connected configuration
sector, whose projector commutes with every bond density.

The nonzero symbols remaining in $\mathcal C$ are frozen.
They separate active segments in which dimers can fluctuate.
For example,
\[
 \mathcal C=2\,0000\,1\,0000\,3
\]
contains two active segments of length four, separated and
bounded by frozen symbols. Every bond touching a frozen
symbol vanishes within this sector. The restricted Hamiltonian
is therefore the sum of the independent segment Hamiltonians,
with identities on the other segments.

\paragraph*{Dimensions of the invariant sectors.}
Let $f_\ell$ be the number of configurations in an active
segment of length $\ell$. These configurations are tilings
by empty sites $0$ and nonoverlapping dimers $12$, $13$ and $14$.
The first site is either empty, leaving $f_{\ell-1}$ possibilities,
or begins a dimer, giving three choices of color followed by
$f_{\ell-2}$ possibilities. Hence
\begin{equation}
 \begin{gathered}
 f_0=f_1=1,\qquad
 f_\ell=f_{\ell-1}+3f_{\ell-2},\\
 f_\ell
 =\sum_{k=0}^{\lfloor\ell/2\rfloor}
   \binom{\ell-k}{k}3^k.
 \end{gathered}
 \label{eq:bond-dimer-segment-count}
\end{equation}
The sum counts configurations with exactly $k$ dimers.
These dimers occupy $2k$ sites, leaving $\ell-2k$ empty sites.
Treating each dimer as one object gives $\ell-k$ objects
altogether. Choosing which $k$ objects are dimers gives
$\binom{\ell-k}{k}$, while their independent colors give $3^k$.

As an example we find for a four-site segment,
\[
 f_4
 =\underbrace{1}_{\text{zero dimers}}
  +\underbrace{3\cdot3}_{\text{one dimer}}
  +\underbrace{3^2}_{\text{two dimers}}
 =19.
\]
If a canonical configuration has active-segment lengths
$\ell_1,\ldots,\ell_r$, the configurations in distinct segments
can be chosen independently. Its sector dimension is therefore
\begin{equation}
 D_{\mathcal C}=\prod_{\nu=1}^{r}f_{\ell_\nu}.
 \label{eq:bond-dimer-sector-size}
\end{equation}
In particular, the canonical configuration
$\mathcal C=2\,0000\,1\,0000\,3$ has dimension $19^2=361$,
as illustrated in Fig.~\ref{fig:bond-dimer-fragmentation}(b).

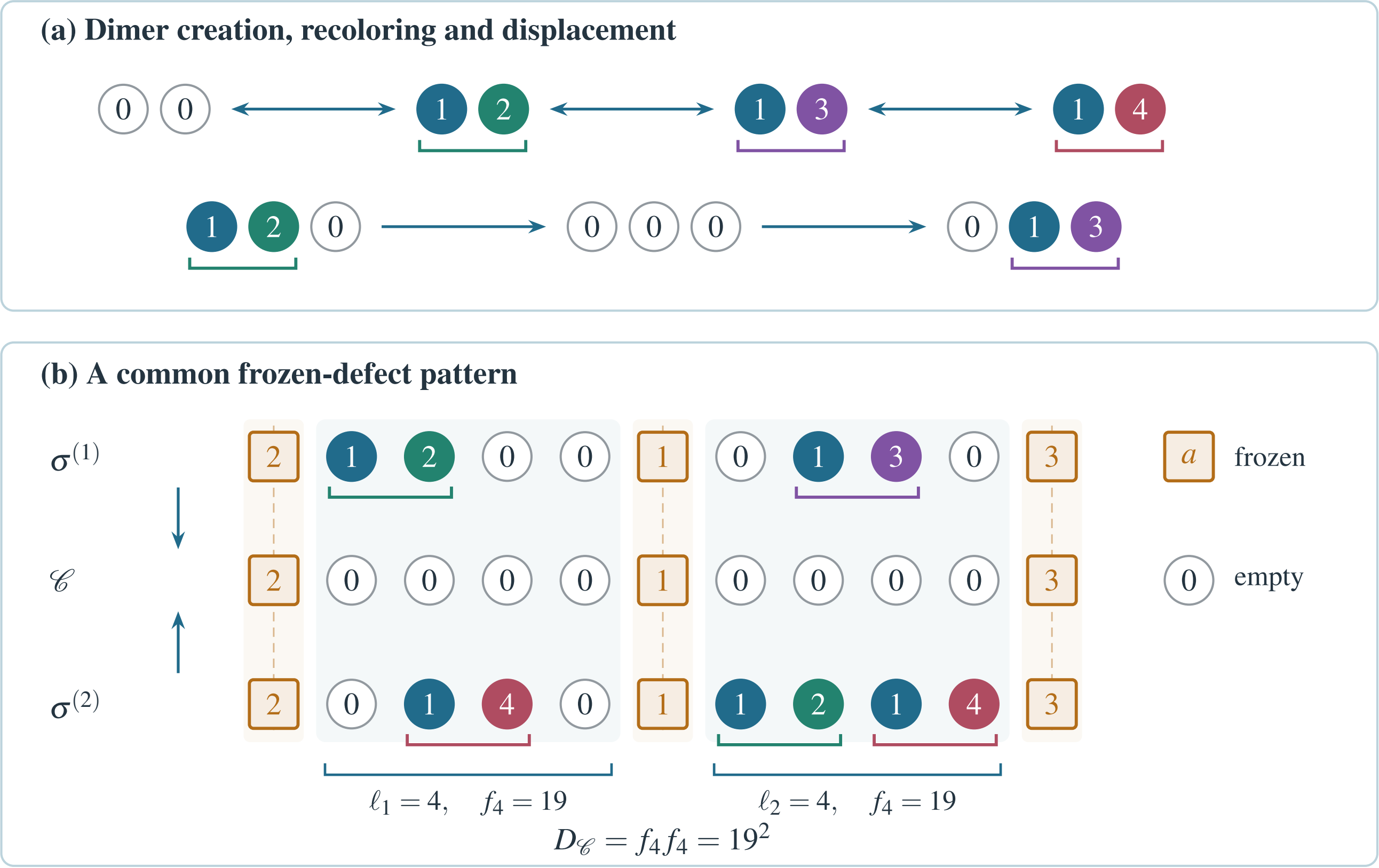
\begin{figure*}[tp]
\centering
\begingroup
\resizebox{\linewidth}{!}{%
\definecolor{dimerInk}{HTML}{243440}
\definecolor{dimerBlue}{HTML}{216B8B}
\definecolor{dimerTeal}{HTML}{23836F}
\definecolor{dimerPurple}{HTML}{8053A3}
\definecolor{dimerRed}{HTML}{AF4C61}
\definecolor{dimerGold}{HTML}{B36D18}
\begin{tikzpicture}[x=1cm,y=1cm,font=\fontsize{10}{12}\selectfont,text=dimerInk,
 >={Stealth[length=2mm,width=1.5mm]},
 panel/.style={draw=dimerBlue!30,fill=white,rounded corners=4pt,line width=.7pt},
 site/.style={circle,draw=dimerInk!50,fill=white,minimum size=5.5mm,inner sep=0pt,line width=.7pt},
 head/.style={site,draw=dimerBlue,fill=dimerBlue,text=white},
 tailtwo/.style={site,draw=dimerTeal,fill=dimerTeal,text=white},
 tailthree/.style={site,draw=dimerPurple,fill=dimerPurple,text=white},
 tailfour/.style={site,draw=dimerRed,fill=dimerRed,text=white},
 frozen/.style={site,rectangle,rounded corners=1.3pt,draw=dimerGold,fill=dimerGold!12,text=dimerGold,line width=1pt},
 move/.style={draw=dimerBlue,line width=.9pt},
 title/.style={anchor=west,font=\fontsize{10}{12}\selectfont\bfseries},
 note/.style={font=\fontsize{9}{11}\selectfont}]
\useasboundingbox (0,.1) rectangle (15.8,-9.9);
\path[panel] (.12,0) rectangle (15.68,-3.50);
\node[title] at (.45,-.35) {(a) Dimer creation, recoloring and displacement};

\foreach \x in {1.50,2.20}{\node[site] at (\x,-1.22) {$0$};}
\node[head] at (5.10,-1.22) {$1$};
\node[tailtwo] at (5.80,-1.22) {$2$};
\node[head] at (8.70,-1.22) {$1$};
\node[tailthree] at (9.40,-1.22) {$3$};
\node[head] at (12.30,-1.22) {$1$};
\node[tailfour] at (13.00,-1.22) {$4$};
\draw[move,<->] (2.72,-1.22)--(4.58,-1.22);
\draw[move,<->] (6.32,-1.22)--(8.18,-1.22);
\draw[move,<->] (9.92,-1.22)--(11.78,-1.22);
\draw[dimerTeal,line width=1pt] (4.85,-1.57)--(4.85,-1.69)--(6.05,-1.69)--(6.05,-1.57);
\draw[dimerPurple,line width=1pt] (8.45,-1.57)--(8.45,-1.69)--(9.65,-1.69)--(9.65,-1.57);
\draw[dimerRed,line width=1pt] (12.05,-1.57)--(12.05,-1.69)--(13.25,-1.69)--(13.25,-1.57);

\node[head] at (2.50,-2.55) {$1$};
\node[tailtwo] at (3.20,-2.55) {$2$};
\node[site] at (3.90,-2.55) {$0$};
\foreach \x in {6.80,7.50,8.20}{\node[site] at (\x,-2.55) {$0$};}
\node[site] at (11.10,-2.55) {$0$};
\node[head] at (11.80,-2.55) {$1$};
\node[tailthree] at (12.50,-2.55) {$3$};
\draw[move,->] (4.42,-2.55)--(6.28,-2.55);
\draw[move,->] (8.72,-2.55)--(10.58,-2.55);
\draw[dimerTeal,line width=1pt] (2.25,-2.90)--(2.25,-3.02)--(3.45,-3.02)--(3.45,-2.90);
\draw[dimerPurple,line width=1pt] (11.55,-2.90)--(11.55,-3.02)--(12.75,-3.02)--(12.75,-2.90);

\path[panel] (.12,-3.86) rectangle (15.68,-9.78);
\node[title] at (.45,-4.24) {(b) A common frozen-defect pattern};
\foreach \xa/\xb in {3.68/7.12,8.08/11.52}{
 \fill[dimerBlue!5,rounded corners=3pt] (\xa,-4.73) rectangle (\xb,-8.38);
}
\foreach \x in {3.20,7.60,12.00}{
 \fill[dimerGold!5,rounded corners=2pt] ({\x-.34},-4.73) rectangle ({\x+.34},-8.38);
 \draw[dimerGold!50,densely dashed,line width=.5pt] (\x,-4.80)--(\x,-8.30);
}
\node[anchor=west] at (.55,-5.15) {$\boldsymbol\sigma^{(1)}$};
\node[anchor=west] at (.55,-6.55) {$\mathcal C$};
\node[anchor=west] at (.55,-7.95) {$\boldsymbol\sigma^{(2)}$};
\draw[move,->] (2.12,-5.50)--(2.12,-6.20);
\draw[move,->] (2.12,-7.60)--(2.12,-6.90);

\foreach \j/\v/\sty in {0/2/frozen,1/1/head,2/2/tailtwo,3/0/site,4/0/site,5/1/frozen,6/0/site,7/1/head,8/3/tailthree,9/0/site,10/3/frozen}{
 \node[\sty] at ({3.2+.88*\j},-5.15) {$\v$};
}
\draw[dimerTeal,line width=1pt] (3.83,-5.49)--(3.83,-5.61)--(5.21,-5.61)--(5.21,-5.49);
\draw[dimerPurple,line width=1pt] (9.11,-5.49)--(9.11,-5.61)--(10.49,-5.61)--(10.49,-5.49);

\foreach \j/\v/\sty in {0/2/frozen,1/0/site,2/0/site,3/0/site,4/0/site,5/1/frozen,6/0/site,7/0/site,8/0/site,9/0/site,10/3/frozen}{
 \node[\sty] at ({3.2+.88*\j},-6.55) {$\v$};
}

\foreach \j/\v/\sty in {0/2/frozen,1/0/site,2/1/head,3/4/tailfour,4/0/site,5/1/frozen,6/1/head,7/2/tailtwo,8/1/head,9/4/tailfour,10/3/frozen}{
 \node[\sty] at ({3.2+.88*\j},-7.95) {$\v$};
}
\draw[dimerRed,line width=1pt] (4.71,-8.29)--(4.71,-8.41)--(6.09,-8.41)--(6.09,-8.29);
\draw[dimerTeal,line width=1pt] (8.23,-8.29)--(8.23,-8.41)--(9.61,-8.41)--(9.61,-8.29);
\draw[dimerRed,line width=1pt] (9.99,-8.29)--(9.99,-8.41)--(11.37,-8.41)--(11.37,-8.29);

\foreach \xa/\xb in {3.78/7.02,8.18/11.42}{
 \draw[dimerBlue,line width=.8pt] (\xa,-8.62)--(\xa,-8.75)--(\xb,-8.75)--(\xb,-8.62);
}
\node[note] at (5.40,-9.07) {$\ell_1=4,\quad f_4=19$};
\node[note] at (9.80,-9.07) {$\ell_2=4,\quad f_4=19$};
\node at (7.60,-9.48) {$D_{\mathcal C}=f_4f_4=19^2$};

\node[frozen] at (13.55,-5.15) {$a$};
\node[note,anchor=west] at (13.95,-5.15) {frozen};
\node[site] at (13.55,-6.55) {$0$};
\node[note,anchor=west] at (13.95,-6.55) {empty};
\end{tikzpicture}
}
\endgroup
\caption{Dimer fragmentation in the lifted $H_{1,3}$ example.
(a) Colored brackets mark the oriented product-state dimers $1a$, with $a=2,3,4$. The upper row shows representative creation, annihilation and recoloring moves. The lower row shows displacement through an intermediate empty configuration. The symbol $0$ is available for pair creation and is not an immobile zero mode. Arrows denote allowed transitions, without implying equal forward and reverse amplitudes.
(b) Both configurations reduce to the middle canonical configuration $\mathcal C=2\,0000\,1\,0000\,3$ by replacing each dimer with $00$ at the same positions. The arrows can represent several annihilation steps. Orange squares mark unpaired symbols whose positions and identities are fixed. They separate two independent four-site active segments, each with $f_4=19$ configurations, giving $D_{\mathcal C}=19^2$. A symbol is frozen because of its neighboring occupations, not its one-site label alone.}
\label{fig:bond-dimer-fragmentation}
\end{figure*}

\paragraph*{Number of sectors and largest sector dimension.}
Canonical configurations are precisely the length-$L$ words
over $\{0,1,2,3,4\}$ that avoid the adjacent patterns $12$, $13$
and $14$. Let $N_L$ denote their number. Appending any of the
five symbols to an allowed word of length $L-1$ gives
$5N_{L-1}$ candidates. A forbidden extension consists of an
allowed prefix of length $L-2$, followed by one of $12$, $13$
or $14$. There are exactly $3N_{L-2}$ such extensions. Thus
\begin{equation}
 \begin{gathered}
 N_0=1,\qquad N_1=5,\qquad
 N_L=5N_{L-1}-3N_{L-2},\\
 N_L\sim C
 \left(\frac{5+\sqrt{13}}{2}\right)^L,
 \qquad C>0.
 \end{gathered}
 \label{eq:bond-dimer-sector-number}
\end{equation}
This establishes exponentially many invariant sectors.
To determine whether fragmentation is strong, we must also
compare the largest sector dimension with the dimension of
the chosen comparison space.

The all-empty canonical configuration
$\mathcal C=00\cdots0$ has no frozen symbols. Its sector
contains all dimer tilings of the full chain and therefore
has dimension $f_L$. No other canonical sector can be larger.
Indeed, replacing the frozen symbols of a configuration by
zeros, while leaving its dimers unchanged, gives a valid
tiling of the full chain. Within a fixed canonical sector,
this map is injective because its frozen positions and labels
are fixed and can be restored. Consequently,
\[
 D_{\mathcal C}\leq f_L.
\]
The largest connected configuration sector has dimension
exactly $f_L$. Any further invariant decomposition can only
reduce the largest block dimension.

Solving the recurrence in
Eq.~\eqref{eq:bond-dimer-segment-count} gives
\[
 f_L
 =\frac{\lambda_+^{L+1}-\lambda_-^{L+1}}{\sqrt{13}},
 \qquad
 \lambda_\pm=\frac{1\pm\sqrt{13}}{2}.
\]
Thus $f_L$ grows as $\lambda_+^L$, with
$\lambda_+\simeq2.303$. Relative to the full Hilbert space,
whose dimension is $5^L$, we obtain
\[
 \frac{D_{\max}}{5^L}
 \leq\frac{f_L}{5^L}
 =O\!\left[
 \left(\frac{1+\sqrt{13}}{10}\right)^L
 \right]
 \longrightarrow0.
\]

\paragraph*{Strong fragmentation at fixed global charges.}
A stronger comparison resolves the two conserved global charges
\begin{equation}
 C_0=\sum_j\bigl(n_j^{(1)}-n_j^R\bigr),
 \qquad
 M_0=\sum_j(-1)^j
       \bigl(n_j^{(1)}+n_j^R\bigr).
 \label{eq:bond-dimer-global-charges}
\end{equation}
The first measures the difference between head and tail numbers.
Creating a dimer adds one head and one tail, leaving $C_0$
unchanged. The second measures the staggered occupation.
A dimer occupies two neighboring sites with opposite
staggering signs, leaving $M_0$ unchanged. Annihilation and
recoloring also preserve both charges.

For even $L$, consider the comparison space with $C_0=M_0=0$.
We do not need its exact dimension. A lower bound follows by
counting the fully occupied configurations containing $L/2$
heads and $L/2$ tails. Equal head and tail numbers give $C_0=0$,
while full occupancy gives
$M_0=\sum_{j=1}^{L}(-1)^j=0$.

The head positions can be chosen in $\binom{L}{L/2}$ ways,
and each remaining site independently carries one of three
tail colors. These configurations therefore give
\[
 D_{C_0=M_0=0}
 \geq\binom{L}{L/2}3^{L/2}.
\]
They need not consist entirely of paired dimers. Unpaired
symbols are allowed because fixing the two global charges
does not fix the canonical configuration. Configurations
containing empty sites can also belong to this comparison
space, which is why the expression is a lower bound.

The all-empty canonical sector lies entirely within
$C_0=M_0=0$, since each of its configurations consists only
of empty sites and dimers. Its dimension is $f_L$, and no
other connected configuration sector is larger. Combining
the largest-block bound with the lower bound on the
comparison-space dimension gives
\begin{equation}
 \frac{D_{\max}}{D_{C_0=M_0=0}}
 \leq
 \frac{f_L}{\binom{L}{L/2}3^{L/2}}
 =
 O\!\left[
 \sqrt L
 \left(\frac{1+\sqrt{13}}{4\sqrt3}\right)^L
 \right]
 \longrightarrow0.
 \label{eq:bond-dimer-strong}
\end{equation}
Here we used
$\binom{L}{L/2}\sim2^L/\sqrt{\pi L/2}$.
The exponential base is approximately $0.665$, so the
largest-block fraction vanishes exponentially even after
these two global charges have been resolved.

Constrained creation and annihilation within an active segment has a precedent in hard-boson models~\cite{Fendley2004}. Fragmentation into regions separated by frozen local motifs occurs in the full PXP Hilbert space, before restriction to the Rydberg-blockaded sector~\cite{Ljubotina2023}. Diffusing-dimer models preserve irreducible strings under a different set of local moves~\cite{Menon1997}. These are related mechanisms with different local moves and conserved labels from the colored-dimer embedding used here. The present calculation identifies the explicit paraSUSY embedding, conserved projectors and sector decomposition for the $H_{1,3}$-derived density. The frozen-defect argument applies to every density supported on $\operatorname{span}\{00,12,13,14\}$. Special coefficients can further split the sectors. Integrability of this lifted chain has not been established by the bond embedding.

\subsection{Fragmentation criteria and the role of bond embeddings}
\label{subsec:bond-classification}

The preceding examples raise a general question. Which
conservation laws, or equivalently, fragmentation mechanisms, follow from the bond embedding itself,
and which depend on the particular seed Hamiltonian density?
The equal-pair and colored-dimer embeddings preserve their
sector labels for arbitrary seed coefficients. In contrast,
the conserved words of the $XXC$ construction rely on the
transitions present in the chosen $XXZ$ density. Distinguishing
these possibilities helps identify sufficient conditions for
fragmentation and clarifies what remains to be classified.

In the site-wise construction, the local active and inactive
projectors commute with every lifted interaction. Their
spatial patterns therefore provide invariant sectors directly.
For bond-wise lifting, neighboring bonds overlap. The active
projector on one bond commutes with operators lifted on that
bond, but need not commute with operators lifted on an
adjoining bond. A decomposition of an individual enlarged
bond space therefore does not automatically define invariant
sectors of the entire chain. We must find operators that
commute with the lifted interactions on all bonds
simultaneously.

\paragraph*{Conservation independent of the seed density.}
Fix a family of bond embeddings $\Phi_j$. The chosen enlarged
parasupercharges and their physical embeddings determine
the lifted matrix units $\Phi_j(E_{ab})$, where
$E_{ab}=\ket{e_a}\bra{e_b}$ are matrix units on the seed bond
space. Every seed Hamiltonian density and its lift can be written as
\[
 h_j=\sum_{a,b}(h_j)_{ab}E_{ab},
 \qquad
 \widetilde h_j
 =\sum_{a,b}(h_j)_{ab}\Phi_j(E_{ab}).
\]
To identify conserved operators that survive every choice
of the coefficients $(h_j)_{ab}$, define
\begin{equation}
 \mathcal C_{\mathrm{all}}(\Phi)
 =
 \left\{
 O\,\middle|\,
 [O,\Phi_j(E_{ab})]=0
 \text{ for every }j,a,b
 \right\}.
 \label{eq:bond-universal-commutant}
\end{equation}
Here $O$ acts on the entire chain. Each lifted matrix unit
acts on its physical bond and as the identity outside that
bond. Thus this is a common commutant on the full chain,
rather than the commutant of operators on one isolated bond.

For any $O\in\mathcal C_{\mathrm{all}}(\Phi)$, linearity gives
\[
 [O,\widetilde h_j]
 =\sum_{a,b}(h_j)_{ab}
       [O,\Phi_j(E_{ab})]
 =0.
\]
This holds independently of the seed coefficients, which
may also vary from bond to bond. Consequently,
\begin{equation}
 \mathcal C_{\mathrm{all}}(\Phi)
 \subseteq
 \mathcal C_{\mathrm{bonds}}(\widetilde H)
 \label{eq:bond-universal-inclusion}
\end{equation}
for every Hamiltonian obtained through these embeddings.
The left-hand side contains conservation laws shared by
the entire lifted family. The right-hand side is the common
commutant after particular seed densities have been chosen,
and can contain additional conserved operators.

\begin{proposition}[A sufficient condition for fragmentation
for every seed density]
\label{prop:bond-criterion}
Suppose that, for chains of length $L$, the common commutant
$\mathcal C_{\mathrm{all}}(\Phi)$ contains $N_L$ mutually
orthogonal nonzero projectors, where $N_L$ grows exponentially
with $L$. Every Hamiltonian obtained through these embeddings
has at least $N_L$ mutually orthogonal invariant subspaces,
and
\[
 \dim\mathcal C_{\mathrm{bonds}}(\widetilde H)\geq N_L.
\]
If these projectors are diagonal in a local product basis,
they define classical invariant sector labels. Strong
fragmentation additionally requires a bound showing that
the largest block occupies a vanishing fraction of the
specified global-symmetry comparison space.
\end{proposition}

\begin{proof}
Let the projectors be $\Pi_\mu$, with
$\mu=1,\ldots,N_L$. By assumption,
\[
 \Pi_\mu\Pi_\nu=\delta_{\mu\nu}\Pi_\mu,
 \qquad
 [\Pi_\mu,\Phi_j(E_{ab})]=0
\]
for every $j,a,b$. Each projector therefore commutes with
every lifted density. In particular,
\[
 \Pi_\nu\widetilde h_j\Pi_\mu
 =\Pi_\nu\Pi_\mu\widetilde h_j
 =0,
 \qquad \mu\neq\nu.
\]
Thus no bond term connects the ranges of distinct projectors,
and each range is invariant under the full Hamiltonian.
If the projectors do not sum to the identity, their
orthogonal complement supplies an additional invariant
subspace.

The projectors are also linearly independent. Multiplying
$\sum_\mu c_\mu\Pi_\mu=0$ by $\Pi_\nu$ gives
$c_\nu\Pi_\nu=0$, and hence $c_\nu=0$ because $\Pi_\nu$ is
nonzero. Equation~\eqref{eq:bond-universal-inclusion} then
implies
$\dim\mathcal C_{\mathrm{bonds}}(\widetilde H)\geq N_L$.
\end{proof}

The proposition isolates conservation enforced by the
embedding, without requiring a choice of seed coefficients.


\paragraph*{Application to the preceding examples.}
For the equal-pair embedding, every lifted matrix unit has
the form
\[
 \Phi(E_{ab})=\ket{aa}\bra{bb}.
\]
Each such operator preserves the pair-reduction string.
The associated string projectors therefore commute with
every lifted matrix unit, independently of the seed
coefficients. Together with the exponential sector count,
this realizes the criterion in
Proposition~\ref{prop:bond-criterion}.

The colored-dimer embedding provides another realization.
Its image algebra consists of operators supported on
\[
 \mathcal W
 =\operatorname{span}
 \{\ket{00},\ket{12},\ket{13},\ket{14}\}.
\]
Every matrix unit within this subspace preserves the
canonical configuration obtained by replacing dimers with
$00$. The canonical-pattern projectors consequently belong
to $\mathcal C_{\mathrm{all}}(\Phi)$. Their conservation does
not depend on the parameters of the $H_{1,3}$ density.
Special coefficients may remove transitions and split the
canonical sectors further, but cannot connect different
canonical patterns.

The words in the $XXC$ families illustrate a different situation. Their conservation follows from the particular matrix elements
present in the lifted $XXZ$ density. Additional seed matrix
elements can introduce transitions that change these words,
as illustrated by the unrotated $XYZ$ embedding. The word
projectors can therefore belong to
$\mathcal C_{\mathrm{bonds}}(\widetilde H)$ without belonging
to $\mathcal C_{\mathrm{all}}(\Phi)$.

Accordingly, Proposition~\ref{prop:bond-criterion} is a
sufficient condition for fragmentation throughout a lifted
family. It is not a necessary condition for fragmentation
of a particular lifted Hamiltonian. A polynomial dimension
of $\mathcal C_{\mathrm{all}}(\Phi)$ would not exclude an
exponentially larger common commutant for special seed
densities. Moreover, connectivity in a specified product
basis alone cannot exclude quantum fragmentation associated
with invariant subspaces not spanned by subsets of that
basis~\cite{MoudgalyaMotrunich2022}.

\paragraph*{Representation data and physical embeddings.}
The classification requires two steps. We must first specify
the abstract representation of the seed matrix algebra.
We must then determine how that representation is embedded
into the physical two-site tensor product, since this
embedding controls the local transitions and their
compatibility on overlapping bonds. 

For the multiplicative, adjoint-preserving lifts of
the full seed matrix algebra used here, the multiplicities
$r,s$ in Eq.~\eqref{eq:bond-prescription} specify the abstract
representation, as discussed in
Appendix~\ref{app:matrix-representations}. For a four-dimensional
seed bond space, these multiplicities satisfy $d^2=4r+s$.
They specify the number of active copies and the dimension
of the inactive subspace, but do not determine how these
spaces sit inside the physical tensor product
$\mathbb C^d\otimes\mathbb C^d$.

For a single active copy, write
\[
 \Phi(O)=VOV^\dagger,
 \qquad V^\dagger V=\id_4.
\]
The isometry $V$ specifies both the active support and the
identification of seed basis vectors with vectors in that
support. Replacing $V$ by $VU$, where $U$ is unitary on the
seed bond space, changes a fixed lifted density to
\[
 \Phi_U(h)=V(UhU^\dagger)V^\dagger.
\]
This can change the conservation laws of the selected
Hamiltonian even though the active support is unchanged.

There is, however, an important distinction between the
selected density and the full image algebra. Since unitary
conjugation maps the full matrix algebra onto itself,
\[
 VU\,M_4(\mathbb C)\,U^\dagger V^\dagger
 =V\,M_4(\mathbb C)\,V^\dagger.
\]
Such a rotation therefore leaves
$\mathcal C_{\mathrm{all}}(\Phi)$ unchanged, while potentially
changing the additional conservation laws of a particular
seed density. The classification of conservation laws shared
by all seed densities must consequently be distinguished
from the classification of conservation laws for selected
seed densities.

Physical equivalences must also respect the overlap of
neighboring bonds. For example, embeddings related by
$V'=(g\otimes g)V$ produce homogeneous chains related by the
global conjugation $g^{\otimes L}$. Site-dependent changes
of basis are likewise consistent when the same transformation
at a shared site is used on both adjoining bonds. A general
unitary transformation of the entire bond space need not
arise from such compatible one-site transformations.
Equivalent abstract parasupercharge representations can
therefore lead to inequivalent many-body constructions.

The results above provide explicit embedding families and
sufficient conservation criteria. Determining all embeddings
that enforce fragmentation for arbitrary seed densities,
and those that produce fragmentation only for selected
seed densities, remains open. A classification of paraSUSY
realizations outside the full-matrix-algebra class used here
is a further problem. Extensions to larger interaction
supports are discussed in Sec.~\ref{sec:conclusion}.
\subsection{Unitary circuits from bond-wise lifting}
\label{subsec:bond-circuits}

The identity completion used for site-wise gates in
Eq.~\eqref{eq:applications-unitary-lift} also applies to a bond.
For a two-qubit seed gate $u\in U(4)$ and the fixed embedding
$\Phi_j$ on sites $(j,j+1)$, define
\begin{equation}
 \begin{gathered}
 K_j=\Phi_j(\id_4),\qquad
 \widehat u_j=\id_{d^2}-K_j+\Phi_j(u),\\
 \Phi_j(u)=\sum_{a,b=0}^3u_{ab}E_{ab}(\widetilde Q_j).
 \end{gathered}
 \label{eq:bond-unitary-lift}
\end{equation}
Here $K_j$ projects onto the active bond representation space.
The two summands act on orthogonal subspaces, giving
$\widehat u_j^\dagger\widehat u_j=(\id_{d^2}-K_j)+K_j=\id_{d^2}$.
For a Hermitian seed density $h$,
$e^{-i\tau\Phi_j(h)}=\id_{d^2}-K_j+\Phi_j(e^{-i\tau h})$.
The bond zero modes therefore acquire identity action under the
completed gate. They need not remain inactive under neighboring gates,
since $K_j$ need not commute with gates on overlapping bonds.

The common commutant in Eq.~\eqref{eq:bond-universal-commutant}
nevertheless controls arbitrary circuit composition.
Any projector $\Pi_\omega\in\mathcal C_{\mathrm{all}}(\Phi)$
commutes with both $K_j$ and $\Phi_j(u)$, and hence with every
$\widehat u_j$. It therefore commutes with every circuit formed
from these gates, with arbitrary gate choices, ordering and depth.
Conversely, commutation with all completed gates implies commutation
with all lifted Hermitian densities by differentiating at $\tau=0$,
and thus with every lifted matrix unit. The common commutant of
this full gate family is exactly $\mathcal C_{\mathrm{all}}(\Phi)$,
so Proposition~\ref{prop:bond-criterion} extends directly to circuits.

For example, the equal-pair embedding of
Proposition~\ref{prop:bond-pair} gives, on four-state physical sites,
\begin{equation}
 \widehat u_{\mathrm{PF}}
 =\sum_{a,b=0}^3u_{ab}\ket{aa}\bra{bb}
  +\sum_{\substack{a,b=0\\a\ne b}}^3\ket{ab}\bra{ab}.
 \label{eq:bond-pair-gate}
\end{equation}
Equal pairs undergo the arbitrary unitary rotation $u$, while unequal
pairs remain unchanged. Every gate preserves the pair-reduction strings
and their projectors in Eq.~\eqref{eq:bond-pair-projectors}.
In particular, the $4\cdot3^{L-1}$ open-chain words without equal
neighbors are fixed by every gate. Thus any nearest-neighbor
two-qubit circuit admits a gate-by-gate lift with exponentially many
invariant sectors. Pair-flip circuits have been studied previously~\cite{Logaric2025}.
Here they follow from the paraSUSY realization of prescribed seed gates.
The colored-dimer embedding likewise preserves its frozen-defect
sectors for arbitrary unitary rotations within
$\operatorname{span}\{00,12,13,14\}$, completed by the identity outside.

For the $XXC$ embedding, a restricted gate family preserves the mobile
words. With the notation of Eq.~\eqref{eq:bond-exact-xxc} and real
$\Delta,\tau$, exponentiation gives
\begin{equation}
 \widehat u_\Delta(\tau)
 =P_\parallel+e^{i\Delta\tau}
  \bigl(\cos\tau\,P_\times-i\sin\tau\,T_{\mathcal A\mathcal B}\bigr).
 \label{eq:bond-xxc-gate}
\end{equation}
This follows from $T_{\mathcal A\mathcal B}^2=P_\times$.
Only unlike-set pairs are exchanged, so both ordered words remain
conserved under arbitrary compositions of these gates.
A general $U(4)$ seed gate in the same embedding need not preserve
them. Likewise, the non-Hermitian densities discussed above do not
generally exponentiate to unitary gates. Finally, unlike site-wise
lifting, a bond-wise circuit lift need not contain a common invariant
copy of the original many-body circuit, because overlapping embeddings
require additional compatibility.

\endgroup

\section{Conclusion}
\label{sec:conclusion}
We have developed an algebraic construction of Hilbert-space fragmentation using SUSY and paraSUSY representation lifting. Supercharges on two-state spaces and parasupercharges on larger local spaces generate the full local operator algebras, allowing a prescribed quantum lattice Hamiltonian to be expressed algebraically before its representations are enlarged. Our site-wise theorem determines the invariant sectors and their dynamics. Bond-wise lifting also depends on the embeddings into physical sites. Although a universal fragmentation theorem for these embeddings remains open, the examples in Sec.~\ref{sec:bondwise} realize several distinct mechanisms.

In the site-wise construction, the enlarged local representations contain active copies of the original local Hilbert space and common zero modes. Lifting each interaction on its declared support gives conserved projectors onto the spatial patterns of these zero modes. Enlarging every site therefore produces exponentially many invariant pattern sectors, even when the original interaction family has only a scalar common commutant. Within each pattern, exactly those original interactions supported entirely on active sites survive. The fully active sector contains copies of the original Hamiltonian. The resulting fragmentation is exact and classical, and for uniform lifts we establish strong fragmentation even at fixed strictly intermediate hole density.

This construction requires neither integrability nor translation invariance. It accommodates mixed spins, arbitrary geometry and interaction supports, while preserving quasi-local interaction-norm bounds. The same pattern projectors survive consistently lifted perturbations, time-dependent driving and unitary gate composition. These results provide a controlled way to retain prescribed dynamics within fragmented Hilbert spaces, including in quantum circuits.

The sector structure also makes physical correlation functions accessible through the surviving seed dynamics. For open nearest-neighbor chains, a stationary ensemble with independently populated holes has an all-time exponential bound on spatial correlations and operator spreading. In the static mixed-field Ising example, isolated active sites give a size-independent positive lower bound on the infinite-time spin autocorrelation. These results distinguish dynamics within an active segment from the obstruction to propagation between segments.

Bond-wise lifting introduces additional possibilities because the enlarged bond representation must be embedded into overlapping physical sites. The lifted $XXX$ examples realize immobile holes, mobile conserved words and pair-flip dynamics. An appropriate lift of the $XXZ$ density gives the $XXC$ families exactly, while the lifted $XYZ$ examples show that changing the embedding can preserve or destroy mobile-word conservation. The non-Hermitian $H_{1,11}$ and $H_{1,3}$ examples give pair-flip sectors and fluctuating colored dimers separated by frozen defects. We also establish a pair-flip construction for arbitrary seed bond densities and a sufficient conservation criterion that is independent of their coefficients. These results distinguish conservation enforced by an embedding from conservation requiring particular seed matrix elements. The same distinction applies to unitary circuits. The equal-pair embedding preserves its strings for arbitrary two-qubit seed gates, while the $XXC$ words survive for a restricted gate family.

The construction studied here opens up some interesting possibilities. Consider Hamiltonian densities with interactions on three or more sites. Such densities admit additional choices of how to group and lift the interacting sites. The spin-$1/2$ Fredkin chain~\cite{SalbergerKorepin2017} is a natural setting in which to compare them. Its bulk dynamics involves conditional exchanges on three neighboring sites, and its ground state with the usual boundary terms has a description in terms of Dyck paths~\cite{SalbergerKorepin2017}. Applying the site-wise theorem already guarantees immobile zero-mode patterns in a lifted Fredkin chain. With the original three-site supports retained, a hole removes every bulk term containing that site and separates the remaining dynamics into active segments. Each segment inherits the surviving bulk interactions. Boundary fields survive only where the corresponding original boundary terms act, so an internal segment need not have the same boundary conditions or ground state as the original open chain.

The same three-site Fredkin density can instead be treated by grouping two sites and the remaining single site, using order-three paraSUSY and SUSY respectively, or by realizing the entire eight-dimensional operator algebra with an order-seven parasupercharge. These choices allow different embeddings into the enlarged three-site space. For example, embedding the eight basis vectors into eight equal-label triples on an eight-state local space would restrict the lifted bulk matrix elements to triple flips and diagonal terms. This provides a concrete extension of the pair-flip construction in which to investigate strings reduced by deleting equal adjacent triples. More general embeddings could combine the conditional Fredkin moves with mobile internal labels or frozen local motifs. Their conserved quantities, sector dimensions and fragmentation strength must be established from the overlapping interactions, with compatible boundary terms included. The present work supplies the algebraic prescription, while a systematic comparison of these alternatives remains open.

A broader objective is to classify bond and multi-site embeddings according to their common matrix-unit commutants and the conservation laws they impose. Such a classification should distinguish projectors conserved for every seed density from those requiring special coefficients. It should also identify embeddings with quantum fragmentation, for which product-state connectivity is insufficient, and determine when an invariant copy of the original many-body dynamics or Yang--Baxter integrability survives. The finite-dimensional matrix-algebra result in Appendix~\ref{app:matrix-representations} fixes the abstract representation content within the class considered here, but leaves these physical embedding questions unresolved.

Further dynamical questions include transport and entanglement within individual sectors, correlations in the mobile-string and pair-flip bond lifts, and the lifetimes of sector memory under perturbations that mix the conserved sectors. In higher dimensions, the toric-code construction in Sec.~\ref{subsec:applications-dimensions} and the sector ground-state analysis in Appendix~\ref{app:toric-sector-ground-states} motivate a closer examination of how hole geometry changes the surviving constraints and logical operators, and which features remain topologically protected. Together with the Fredkin extensions, these questions offer routes to understanding how the choice of representation and its physical embedding determine constrained many-body dynamics.

\section*{Acknowledgements}
We acknowledge the use of ChatGPT (OpenAI) for assistance with mathematical discussions, algebraic checks, and the organization and editing of this manuscript. The authors reviewed the resulting material and take full responsibility for the scientific content and conclusions.

\appendix
\numberwithin{equation}{section}
\section{Common commutant of permutation bonds}
\label{app:permutation-commutant}

We first prove the spin-$1/2$ case of Eq.~\eqref{eq:perm-seed-commutant} by constructing an explicit basis of the common commutant, then extend the count to general local dimension. Let
$\mathcal H_L=(\mathbb C^2)^{\otimes L}$ and let $\mathcal E$ contain the nearest-neighbor bonds of a connected open chain or a ring. The operator space of interest is
\begin{equation}
 \mathcal C_L
 =\{O\in\End(\mathcal H_L):
       [O,\mathsf P^{(2)}_{ij}]=0
           \ \text{for every }\{i,j\}\in\mathcal E\}.
 \label{eq:app-perm-definition}
\end{equation}
All operators $O$ here act on the full chain. We count them over $\mathbb C$, with the requirement of commutation imposed separately for every bond.

First, a swap satisfies $(\mathsf P^{(2)}_{ij})^2=\id$, so
\begin{equation}
 [O,\mathsf P^{(2)}_{ij}]=0
 \quad\Longleftrightarrow\quad
 \mathsf P^{(2)}_{ij}O\mathsf P^{(2)}_{ij}=O.
 \label{eq:app-perm-conjugation}
\end{equation}
Conjugation by the swap exchanges the two corresponding tensor factors of an operator. The adjacent swaps of an open chain generate every permutation of the $L$ sites: any ordering can be obtained by moving each entry to its desired position through successive adjacent exchanges. Therefore Eq.~\eqref{eq:app-perm-conjugation} for all bonds is equivalent to invariance under conjugation by every site permutation. A ring contains these same adjacent swaps; its additional closing bond imposes no further constraint. This is why the result is the same for both boundary conditions.

To express this invariance in components, choose the local operator basis
$\sigma^0=\id_2,\sigma^x,\sigma^y,\sigma^z$. Its tensor products form a basis of $\End(\mathcal H_L)$:
\begin{equation}
 W_{\boldsymbol\alpha}
   =\sigma^{\alpha_1}\otimes\cdots\otimes\sigma^{\alpha_L},
 \qquad
 O=\sum_{\boldsymbol\alpha\in\{0,x,y,z\}^{L}}
         c_{\boldsymbol\alpha}W_{\boldsymbol\alpha}.
 \label{eq:app-perm-expansion}
\end{equation}
The expansion is unique. If $s_j\boldsymbol\alpha$ denotes the sequence obtained by exchanging $\alpha_j$ and $\alpha_{j+1}$, then
\begin{equation}
\begin{gathered}
\mathsf P^{(2)}_{j,j+1}
       W_{\boldsymbol\alpha}\mathsf P^{(2)}_{j,j+1}
       =W_{s_j\boldsymbol\alpha},
 \\
 O\in\mathcal C_L
 \quad\Longleftrightarrow\quad
 c_{\boldsymbol\alpha}=c_{s_j\boldsymbol\alpha}
       \ \text{for every }j,\boldsymbol\alpha.
\end{gathered}
 \label{eq:app-perm-coefficients}
\end{equation}
The coefficients must consequently be equal whenever their label sequences are permutations of one another.

Such a permutation class is specified exactly by the numbers of occurrences of the four labels. Write these numbers as
\begin{equation}
 \boldsymbol n=(n_0,n_x,n_y,n_z),
 \qquad
 n_\mu\geq0,\qquad n_0+n_x+n_y+n_z=L.
 \label{eq:app-perm-counts}
\end{equation}
Two label sequences are permutations of one another exactly when these four counts agree.

For each admissible $\boldsymbol n$, let $\Omega_{\boldsymbol n}$ be the set of all distinct label sequences with these counts, and define
\begin{equation}
 T_{\boldsymbol n}
    =\sum_{\boldsymbol\alpha\in\Omega_{\boldsymbol n}}
         W_{\boldsymbol\alpha}.
 \label{eq:app-perm-orbit-basis}
\end{equation}
For example, at $L=3$,
\[
 T_{1,2,0,0}
 =\id_2\otimes\sigma^x\otimes\sigma^x
  +\sigma^x\otimes\id_2\otimes\sigma^x
  +\sigma^x\otimes\sigma^x\otimes\id_2.
\]
Every $T_{\boldsymbol n}$ belongs to $\mathcal C_L$, since a swap only rearranges its summands. Conversely, Eq.~\eqref{eq:app-perm-coefficients} implies that any $O\in\mathcal C_L$ has one coefficient for each $\Omega_{\boldsymbol n}$, and hence is a linear combination of the $T_{\boldsymbol n}$.

These operators are also linearly independent. Different $\Omega_{\boldsymbol n}$ contain disjoint sets of basis strings, so coefficients from different sums cannot cancel. Equivalently, Pauli-string orthogonality gives
\begin{equation}
 \operatorname{tr}\!\left(T_{\boldsymbol n}^{\dagger}
                              T_{\boldsymbol m}\right)
 =2^L\delta_{\boldsymbol n,\boldsymbol m}
       \frac{L!}{n_0!\,n_x!\,n_y!\,n_z!}.
 \label{eq:app-perm-orthogonality}
\end{equation}

It remains to count the quadruples in Eq.~\eqref{eq:app-perm-counts}. Distributing $L$ identical objects among four labelled boxes gives precisely these nonnegative integer solutions. Equivalently, place three separators among a sequence of $L$ objects; choosing their positions among the $L+3$ available positions gives
\begin{equation}
 \dim\mathcal C_L
   =\#\{\boldsymbol n\in\mathbb Z_{\geq0}^{4}:
                           n_0+n_x+n_y+n_z=L\}
   =\binom{L+3}{3}.
 \label{eq:app-perm-result}
\end{equation}
This proves the $n=2$ case of Eq.~\eqref{eq:perm-seed-commutant}.

As a check, at $L=2$ the basis contains one string with two identities, three symmetrized combinations with one identity, and six combinations with no identities. The last six are the three operators $\sigma^\alpha\otimes\sigma^\alpha$ and the three operators
$\sigma^\alpha\otimes\sigma^\beta+\sigma^\beta\otimes\sigma^\alpha$ with $\alpha<\beta$ in $\{x,y,z\}$. Their total is $1+3+6=10=\binom{5}{3}$, agreeing with the triplet--singlet calculation of the two-site swap commutant.

The argument extends immediately to a local space $\mathbb C^n$. Choose any basis of its $n^2$-dimensional operator space. Conjugation by a swap again permutes the tensor factors, and a symmetrized basis operator is now labelled by $n^2$ nonnegative occupation numbers summing to $L$. Consequently,
\begin{equation}
 \dim\mathcal C\bigl(\{\mathsf P^{(n)}_{ij}\}_{\mathcal E}\bigr)
       =\binom{L+n^2-1}{n^2-1}.
 \label{eq:app-perm-general}
\end{equation}
For $n=3$ this gives $\binom{L+8}{8}$, as used for the spin-$1$ permutation chain. At fixed $n$ the growth is polynomial in $L$. This is a count of independent full-chain operators commuting with every permutation bond; it is distinct from the count of product-state occupation sectors. The proof depends only on the exchange action of the bonds and does not invoke integrability.

\section{Sector ground states of the lifted toric code}
\label{app:toric-sector-ground-states}

This appendix complements the site-wise lifting example of Sec.~\ref{subsec:applications-dimensions}. We keep the same periodic square lattice, positive couplings and minimal SUSY lift. For a fixed hole set, the restricted Hamiltonian is Eq.~\eqref{eq:applications-toric-restricted}. Its \emph{sector ground-state degeneracy}, or sector GSD, is the dimension of its lowest-energy eigenspace. This count fixes the hole positions and is distinct from the ground-state degeneracy of the full lifted Hamiltonian and from the number of invariant hole patterns.

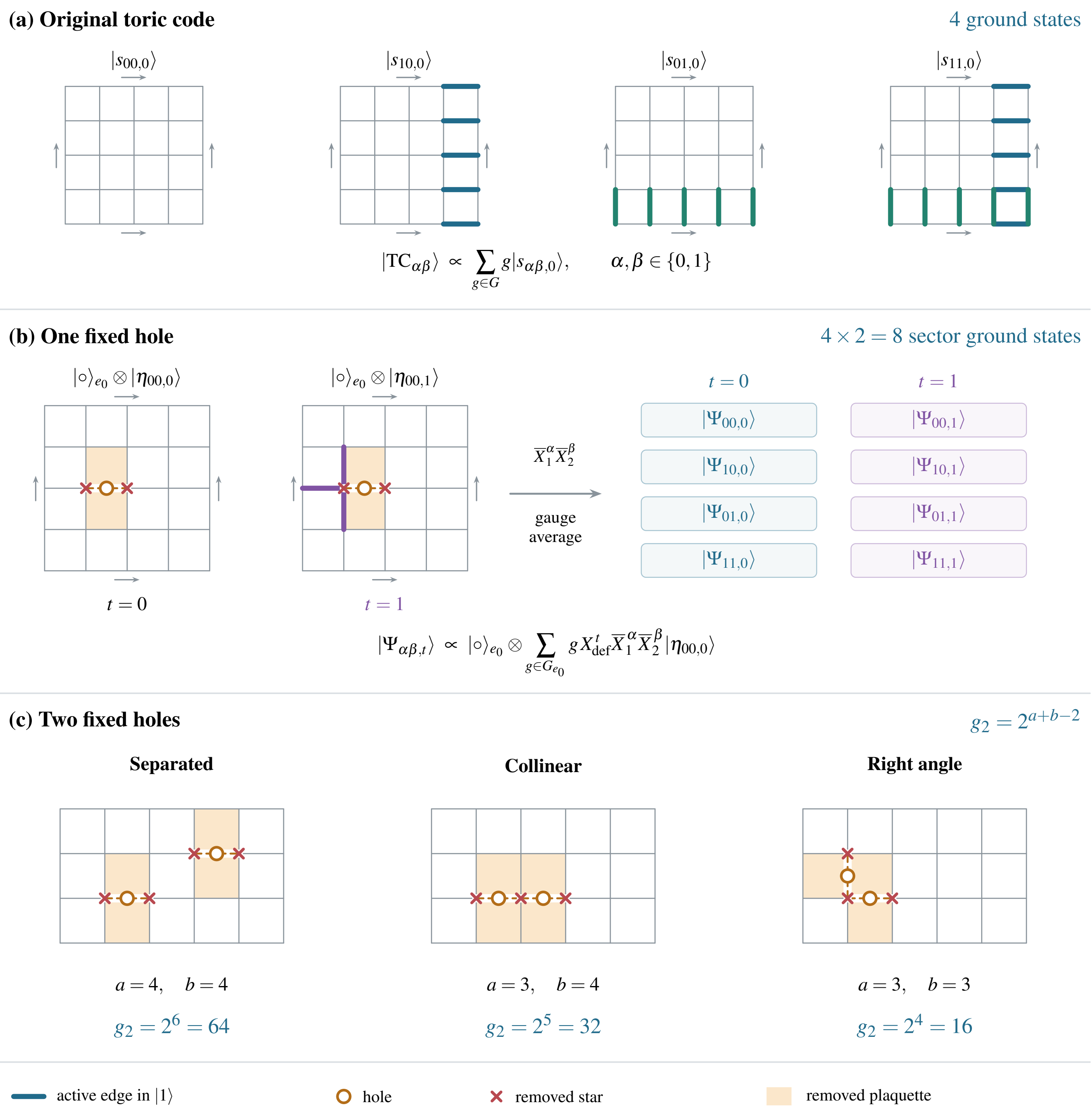
\begin{figure*}[tp]
\centering
\resizebox{\textwidth}{!}{%
\begingroup%
\definecolor{tcBlue}{HTML}{216B8B}%
\definecolor{tcTeal}{HTML}{23836F}%
\definecolor{tcHole}{HTML}{B36D18}%
\definecolor{tcPurple}{HTML}{8053A3}%
\definecolor{tcRed}{HTML}{B84950}%
\definecolor{tcFace}{HTML}{FBE7CB}%
\definecolor{tcGrid}{HTML}{89939B}%
\definecolor{tcRule}{HTML}{D9DEE2}%
\begin{tikzpicture}[x=1cm,y=1cm,line cap=round,line join=round,
  every node/.style={font=\footnotesize,inner sep=2pt}]

\node[anchor=west,font=\small] at (0.050,16.250) {\textbf{(a) Original toric code}};
\node[anchor=east,font=\small,text=tcBlue] at (15.800,16.250) {$4$ ground states};
\node[] at (1.950,15.670) {$\ket{s_{00,0}}$};
\draw[tcGrid,line width=.45pt] (0.950,13.300) -- (0.950,15.300);
\draw[tcGrid,line width=.45pt] (1.450,13.300) -- (1.450,15.300);
\draw[tcGrid,line width=.45pt] (1.950,13.300) -- (1.950,15.300);
\draw[tcGrid,line width=.45pt] (2.450,13.300) -- (2.450,15.300);
\draw[tcGrid,line width=.45pt] (2.950,13.300) -- (2.950,15.300);
\draw[tcGrid,line width=.45pt] (0.950,13.300) -- (2.950,13.300);
\draw[tcGrid,line width=.45pt] (0.950,13.800) -- (2.950,13.800);
\draw[tcGrid,line width=.45pt] (0.950,14.300) -- (2.950,14.300);
\draw[tcGrid,line width=.45pt] (0.950,14.800) -- (2.950,14.800);
\draw[tcGrid,line width=.45pt] (0.950,15.300) -- (2.950,15.300);
\draw[tcGrid,-{Stealth[length=3pt]},line width=.55pt] (1.770,13.170) -- (2.130,13.170);
\draw[tcGrid,-{Stealth[length=3pt]},line width=.55pt] (1.770,15.430) -- (2.130,15.430);
\draw[tcGrid,-{Stealth[length=3pt]},line width=.55pt] (0.820,14.120) -- (0.820,14.480);
\draw[tcGrid,-{Stealth[length=3pt]},line width=.55pt] (3.080,14.120) -- (3.080,14.480);
\node[] at (5.950,15.670) {$\ket{s_{10,0}}$};
\draw[tcGrid,line width=.45pt] (4.950,13.300) -- (4.950,15.300);
\draw[tcGrid,line width=.45pt] (5.450,13.300) -- (5.450,15.300);
\draw[tcGrid,line width=.45pt] (5.950,13.300) -- (5.950,15.300);
\draw[tcGrid,line width=.45pt] (6.450,13.300) -- (6.450,15.300);
\draw[tcGrid,line width=.45pt] (6.950,13.300) -- (6.950,15.300);
\draw[tcGrid,line width=.45pt] (4.950,13.300) -- (6.950,13.300);
\draw[tcGrid,line width=.45pt] (4.950,13.800) -- (6.950,13.800);
\draw[tcGrid,line width=.45pt] (4.950,14.300) -- (6.950,14.300);
\draw[tcGrid,line width=.45pt] (4.950,14.800) -- (6.950,14.800);
\draw[tcGrid,line width=.45pt] (4.950,15.300) -- (6.950,15.300);
\draw[tcBlue,line width=2.1pt] (6.450,13.300) -- (6.950,13.300);
\draw[tcBlue,line width=2.1pt] (6.450,13.800) -- (6.950,13.800);
\draw[tcBlue,line width=2.1pt] (6.450,14.300) -- (6.950,14.300);
\draw[tcBlue,line width=2.1pt] (6.450,14.800) -- (6.950,14.800);
\draw[tcBlue,line width=2.1pt] (6.450,15.300) -- (6.950,15.300);
\draw[tcGrid,-{Stealth[length=3pt]},line width=.55pt] (5.770,13.170) -- (6.130,13.170);
\draw[tcGrid,-{Stealth[length=3pt]},line width=.55pt] (5.770,15.430) -- (6.130,15.430);
\draw[tcGrid,-{Stealth[length=3pt]},line width=.55pt] (4.820,14.120) -- (4.820,14.480);
\draw[tcGrid,-{Stealth[length=3pt]},line width=.55pt] (7.080,14.120) -- (7.080,14.480);
\node[] at (9.950,15.670) {$\ket{s_{01,0}}$};
\draw[tcGrid,line width=.45pt] (8.950,13.300) -- (8.950,15.300);
\draw[tcGrid,line width=.45pt] (9.450,13.300) -- (9.450,15.300);
\draw[tcGrid,line width=.45pt] (9.950,13.300) -- (9.950,15.300);
\draw[tcGrid,line width=.45pt] (10.450,13.300) -- (10.450,15.300);
\draw[tcGrid,line width=.45pt] (10.950,13.300) -- (10.950,15.300);
\draw[tcGrid,line width=.45pt] (8.950,13.300) -- (10.950,13.300);
\draw[tcGrid,line width=.45pt] (8.950,13.800) -- (10.950,13.800);
\draw[tcGrid,line width=.45pt] (8.950,14.300) -- (10.950,14.300);
\draw[tcGrid,line width=.45pt] (8.950,14.800) -- (10.950,14.800);
\draw[tcGrid,line width=.45pt] (8.950,15.300) -- (10.950,15.300);
\draw[tcTeal,line width=2.1pt] (8.950,13.300) -- (8.950,13.800);
\draw[tcTeal,line width=2.1pt] (9.450,13.300) -- (9.450,13.800);
\draw[tcTeal,line width=2.1pt] (9.950,13.300) -- (9.950,13.800);
\draw[tcTeal,line width=2.1pt] (10.450,13.300) -- (10.450,13.800);
\draw[tcTeal,line width=2.1pt] (10.950,13.300) -- (10.950,13.800);
\draw[tcGrid,-{Stealth[length=3pt]},line width=.55pt] (9.770,13.170) -- (10.130,13.170);
\draw[tcGrid,-{Stealth[length=3pt]},line width=.55pt] (9.770,15.430) -- (10.130,15.430);
\draw[tcGrid,-{Stealth[length=3pt]},line width=.55pt] (8.820,14.120) -- (8.820,14.480);
\draw[tcGrid,-{Stealth[length=3pt]},line width=.55pt] (11.080,14.120) -- (11.080,14.480);
\node[] at (13.950,15.670) {$\ket{s_{11,0}}$};
\draw[tcGrid,line width=.45pt] (12.950,13.300) -- (12.950,15.300);
\draw[tcGrid,line width=.45pt] (13.450,13.300) -- (13.450,15.300);
\draw[tcGrid,line width=.45pt] (13.950,13.300) -- (13.950,15.300);
\draw[tcGrid,line width=.45pt] (14.450,13.300) -- (14.450,15.300);
\draw[tcGrid,line width=.45pt] (14.950,13.300) -- (14.950,15.300);
\draw[tcGrid,line width=.45pt] (12.950,13.300) -- (14.950,13.300);
\draw[tcGrid,line width=.45pt] (12.950,13.800) -- (14.950,13.800);
\draw[tcGrid,line width=.45pt] (12.950,14.300) -- (14.950,14.300);
\draw[tcGrid,line width=.45pt] (12.950,14.800) -- (14.950,14.800);
\draw[tcGrid,line width=.45pt] (12.950,15.300) -- (14.950,15.300);
\draw[tcBlue,line width=2.1pt] (14.450,13.300) -- (14.950,13.300);
\draw[tcBlue,line width=2.1pt] (14.450,13.800) -- (14.950,13.800);
\draw[tcBlue,line width=2.1pt] (14.450,14.300) -- (14.950,14.300);
\draw[tcBlue,line width=2.1pt] (14.450,14.800) -- (14.950,14.800);
\draw[tcBlue,line width=2.1pt] (14.450,15.300) -- (14.950,15.300);
\draw[tcTeal,line width=2.1pt] (12.950,13.300) -- (12.950,13.800);
\draw[tcTeal,line width=2.1pt] (13.450,13.300) -- (13.450,13.800);
\draw[tcTeal,line width=2.1pt] (13.950,13.300) -- (13.950,13.800);
\draw[tcTeal,line width=2.1pt] (14.450,13.300) -- (14.450,13.800);
\draw[tcTeal,line width=2.1pt] (14.950,13.300) -- (14.950,13.800);
\draw[tcGrid,-{Stealth[length=3pt]},line width=.55pt] (13.770,13.170) -- (14.130,13.170);
\draw[tcGrid,-{Stealth[length=3pt]},line width=.55pt] (13.770,15.430) -- (14.130,15.430);
\draw[tcGrid,-{Stealth[length=3pt]},line width=.55pt] (12.820,14.120) -- (12.820,14.480);
\draw[tcGrid,-{Stealth[length=3pt]},line width=.55pt] (15.080,14.120) -- (15.080,14.480);
\node[] at (7.950,12.620) {$\displaystyle\ket{\mathrm{TC}_{\alpha\beta}}\ \propto\ \sum_{g\in G}g\ket{s_{\alpha\beta,0}},\qquad \alpha,\beta\in\{0,1\}$};
\draw[tcRule,line width=.6pt] (0.000,12.060) -- (15.850,12.060);
\node[anchor=west,font=\small] at (0.050,11.650) {\textbf{(b) One fixed hole}};
\node[anchor=east,font=\small,text=tcBlue] at (15.800,11.650) {$4\times2=8$ sector ground states};
\node[] at (1.850,11.060) {$\ket{\circ}_{e_0}\otimes\ket{\eta_{00,0}}$};
\fill[tcFace] (1.250,8.860) rectangle (1.850,9.460);
\fill[tcFace] (1.250,9.460) rectangle (1.850,10.060);
\draw[tcGrid,line width=.45pt] (0.650,8.260) -- (0.650,10.660);
\draw[tcGrid,line width=.45pt] (1.250,8.260) -- (1.250,10.660);
\draw[tcGrid,line width=.45pt] (1.850,8.260) -- (1.850,10.660);
\draw[tcGrid,line width=.45pt] (2.450,8.260) -- (2.450,10.660);
\draw[tcGrid,line width=.45pt] (3.050,8.260) -- (3.050,10.660);
\draw[tcGrid,line width=.45pt] (0.650,8.260) -- (3.050,8.260);
\draw[tcGrid,line width=.45pt] (0.650,8.860) -- (3.050,8.860);
\draw[tcGrid,line width=.45pt] (0.650,9.460) -- (3.050,9.460);
\draw[tcGrid,line width=.45pt] (0.650,10.060) -- (3.050,10.060);
\draw[tcGrid,line width=.45pt] (0.650,10.660) -- (3.050,10.660);
\draw[white,line width=3.4pt] (1.250,9.460) -- (1.850,9.460);
\draw[tcHole,densely dashed,line width=.8pt] (1.250,9.460) -- (1.850,9.460);
\draw[tcHole,fill=white,line width=1.1pt] (1.550,9.460) circle (.092);
\draw[tcRed,line width=1.2pt] (1.185,9.395) -- (1.315,9.525);
\draw[tcRed,line width=1.2pt] (1.185,9.525) -- (1.315,9.395);
\draw[tcRed,line width=1.2pt] (1.785,9.395) -- (1.915,9.525);
\draw[tcRed,line width=1.2pt] (1.785,9.525) -- (1.915,9.395);
\draw[tcGrid,-{Stealth[length=3pt]},line width=.55pt] (1.670,8.130) -- (2.030,8.130);
\draw[tcGrid,-{Stealth[length=3pt]},line width=.55pt] (1.670,10.790) -- (2.030,10.790);
\draw[tcGrid,-{Stealth[length=3pt]},line width=.55pt] (0.520,9.280) -- (0.520,9.640);
\draw[tcGrid,-{Stealth[length=3pt]},line width=.55pt] (3.180,9.280) -- (3.180,9.640);
\node[] at (1.850,7.780) {$t=0$};
\node[] at (5.600,11.060) {$\ket{\circ}_{e_0}\otimes\ket{\eta_{00,1}}$};
\fill[tcFace] (5.000,8.860) rectangle (5.600,9.460);
\fill[tcFace] (5.000,9.460) rectangle (5.600,10.060);
\draw[tcGrid,line width=.45pt] (4.400,8.260) -- (4.400,10.660);
\draw[tcGrid,line width=.45pt] (5.000,8.260) -- (5.000,10.660);
\draw[tcGrid,line width=.45pt] (5.600,8.260) -- (5.600,10.660);
\draw[tcGrid,line width=.45pt] (6.200,8.260) -- (6.200,10.660);
\draw[tcGrid,line width=.45pt] (6.800,8.260) -- (6.800,10.660);
\draw[tcGrid,line width=.45pt] (4.400,8.260) -- (6.800,8.260);
\draw[tcGrid,line width=.45pt] (4.400,8.860) -- (6.800,8.860);
\draw[tcGrid,line width=.45pt] (4.400,9.460) -- (6.800,9.460);
\draw[tcGrid,line width=.45pt] (4.400,10.060) -- (6.800,10.060);
\draw[tcGrid,line width=.45pt] (4.400,10.660) -- (6.800,10.660);
\draw[tcPurple,line width=2.1pt] (4.400,9.460) -- (5.000,9.460);
\draw[tcPurple,line width=2.1pt] (5.000,8.860) -- (5.000,9.460);
\draw[tcPurple,line width=2.1pt] (5.000,9.460) -- (5.000,10.060);
\draw[white,line width=3.4pt] (5.000,9.460) -- (5.600,9.460);
\draw[tcHole,densely dashed,line width=.8pt] (5.000,9.460) -- (5.600,9.460);
\draw[tcHole,fill=white,line width=1.1pt] (5.300,9.460) circle (.092);
\draw[tcRed,line width=1.2pt] (4.935,9.395) -- (5.065,9.525);
\draw[tcRed,line width=1.2pt] (4.935,9.525) -- (5.065,9.395);
\draw[tcRed,line width=1.2pt] (5.535,9.395) -- (5.665,9.525);
\draw[tcRed,line width=1.2pt] (5.535,9.525) -- (5.665,9.395);
\draw[tcGrid,-{Stealth[length=3pt]},line width=.55pt] (5.420,8.130) -- (5.780,8.130);
\draw[tcGrid,-{Stealth[length=3pt]},line width=.55pt] (5.420,10.790) -- (5.780,10.790);
\draw[tcGrid,-{Stealth[length=3pt]},line width=.55pt] (4.270,9.280) -- (4.270,9.640);
\draw[tcGrid,-{Stealth[length=3pt]},line width=.55pt] (6.930,9.280) -- (6.930,9.640);
\node[text=tcPurple] at (5.600,7.780) {$t=1$};
\node[text=tcBlue] at (10.600,11.020) {$t=0$};
\node[text=tcPurple] at (13.650,11.020) {$t=1$};
\node[draw=tcBlue!40,fill=tcBlue!5,rounded corners=2pt,minimum width=2.55cm,minimum height=.49cm,text=tcBlue] at (10.600,10.450) {$\ket{\Psi_{00,0}}$};
\node[draw=tcPurple!40,fill=tcPurple!5,rounded corners=2pt,minimum width=2.55cm,minimum height=.49cm,text=tcPurple] at (13.650,10.450) {$\ket{\Psi_{00,1}}$};
\node[draw=tcBlue!40,fill=tcBlue!5,rounded corners=2pt,minimum width=2.55cm,minimum height=.49cm,text=tcBlue] at (10.600,9.770) {$\ket{\Psi_{10,0}}$};
\node[draw=tcPurple!40,fill=tcPurple!5,rounded corners=2pt,minimum width=2.55cm,minimum height=.49cm,text=tcPurple] at (13.650,9.770) {$\ket{\Psi_{10,1}}$};
\node[draw=tcBlue!40,fill=tcBlue!5,rounded corners=2pt,minimum width=2.55cm,minimum height=.49cm,text=tcBlue] at (10.600,9.090) {$\ket{\Psi_{01,0}}$};
\node[draw=tcPurple!40,fill=tcPurple!5,rounded corners=2pt,minimum width=2.55cm,minimum height=.49cm,text=tcPurple] at (13.650,9.090) {$\ket{\Psi_{01,1}}$};
\node[draw=tcBlue!40,fill=tcBlue!5,rounded corners=2pt,minimum width=2.55cm,minimum height=.49cm,text=tcBlue] at (10.600,8.410) {$\ket{\Psi_{11,0}}$};
\node[draw=tcPurple!40,fill=tcPurple!5,rounded corners=2pt,minimum width=2.55cm,minimum height=.49cm,text=tcPurple] at (13.650,8.410) {$\ket{\Psi_{11,1}}$};
\draw[tcGrid,-{Stealth[length=5pt]},line width=.8pt] (7.420,9.380) -- (8.750,9.380);
\node[font=\scriptsize] at (8.080,9.950) {$\overline X_1^\alpha\overline X_2^\beta$};
\node[font=\scriptsize,align=center,text width=1.05cm] at (8.080,8.850) {gauge average};
\node[] at (7.950,7.060) {$\displaystyle\ket{\Psi_{\alpha\beta,t}}\ \propto\ \ket{\circ}_{e_0}\otimes\sum_{g\in G_{e_0}}g\,X_{\mathrm{def}}^{\,t}\overline X_1^{\,\alpha}\overline X_2^{\,\beta}\ket{\eta_{00,0}}$};
\draw[tcRule,line width=.6pt] (0.000,6.480) -- (15.850,6.480);
\node[anchor=west,font=\small] at (0.050,6.080) {\textbf{(c) Two fixed holes}};
\node[anchor=east,font=\small,text=tcBlue] at (15.800,6.080) {$g_2=2^{a+b-2}$};
\node[] at (2.500,5.430) {\textbf{Separated}};
\fill[tcFace] (1.525,2.850) rectangle (2.175,3.500);
\fill[tcFace] (1.525,3.500) rectangle (2.175,4.150);
\fill[tcFace] (2.825,3.500) rectangle (3.475,4.150);
\fill[tcFace] (2.825,4.150) rectangle (3.475,4.800);
\draw[tcGrid,line width=.45pt] (0.875,2.850) -- (0.875,4.800);
\draw[tcGrid,line width=.45pt] (1.525,2.850) -- (1.525,4.800);
\draw[tcGrid,line width=.45pt] (2.175,2.850) -- (2.175,4.800);
\draw[tcGrid,line width=.45pt] (2.825,2.850) -- (2.825,4.800);
\draw[tcGrid,line width=.45pt] (3.475,2.850) -- (3.475,4.800);
\draw[tcGrid,line width=.45pt] (4.125,2.850) -- (4.125,4.800);
\draw[tcGrid,line width=.45pt] (0.875,2.850) -- (4.125,2.850);
\draw[tcGrid,line width=.45pt] (0.875,3.500) -- (4.125,3.500);
\draw[tcGrid,line width=.45pt] (0.875,4.150) -- (4.125,4.150);
\draw[tcGrid,line width=.45pt] (0.875,4.800) -- (4.125,4.800);
\draw[white,line width=3.4pt] (1.525,3.500) -- (2.175,3.500);
\draw[tcHole,densely dashed,line width=.8pt] (1.525,3.500) -- (2.175,3.500);
\draw[tcHole,fill=white,line width=1.1pt] (1.850,3.500) circle (.092);
\draw[white,line width=3.4pt] (2.825,4.150) -- (3.475,4.150);
\draw[tcHole,densely dashed,line width=.8pt] (2.825,4.150) -- (3.475,4.150);
\draw[tcHole,fill=white,line width=1.1pt] (3.150,4.150) circle (.092);
\draw[tcRed,line width=1.2pt] (1.460,3.435) -- (1.590,3.565);
\draw[tcRed,line width=1.2pt] (1.460,3.565) -- (1.590,3.435);
\draw[tcRed,line width=1.2pt] (2.110,3.435) -- (2.240,3.565);
\draw[tcRed,line width=1.2pt] (2.110,3.565) -- (2.240,3.435);
\draw[tcRed,line width=1.2pt] (2.760,4.085) -- (2.890,4.215);
\draw[tcRed,line width=1.2pt] (2.760,4.215) -- (2.890,4.085);
\draw[tcRed,line width=1.2pt] (3.410,4.085) -- (3.540,4.215);
\draw[tcRed,line width=1.2pt] (3.410,4.215) -- (3.540,4.085);
\node[] at (2.500,2.230) {$a=4,\quad b=4$};
\node[text=tcBlue,font=\small] at (2.500,1.650) {$\boldsymbol{g_2=2^{6}=64}$};
\node[] at (7.900,5.430) {\textbf{Collinear}};
\fill[tcFace] (6.925,2.850) rectangle (7.575,3.500);
\fill[tcFace] (6.925,3.500) rectangle (7.575,4.150);
\fill[tcFace] (7.575,2.850) rectangle (8.225,3.500);
\fill[tcFace] (7.575,3.500) rectangle (8.225,4.150);
\draw[tcGrid,line width=.45pt] (6.275,2.850) -- (6.275,4.800);
\draw[tcGrid,line width=.45pt] (6.925,2.850) -- (6.925,4.800);
\draw[tcGrid,line width=.45pt] (7.575,2.850) -- (7.575,4.800);
\draw[tcGrid,line width=.45pt] (8.225,2.850) -- (8.225,4.800);
\draw[tcGrid,line width=.45pt] (8.875,2.850) -- (8.875,4.800);
\draw[tcGrid,line width=.45pt] (9.525,2.850) -- (9.525,4.800);
\draw[tcGrid,line width=.45pt] (6.275,2.850) -- (9.525,2.850);
\draw[tcGrid,line width=.45pt] (6.275,3.500) -- (9.525,3.500);
\draw[tcGrid,line width=.45pt] (6.275,4.150) -- (9.525,4.150);
\draw[tcGrid,line width=.45pt] (6.275,4.800) -- (9.525,4.800);
\draw[white,line width=3.4pt] (6.925,3.500) -- (7.575,3.500);
\draw[tcHole,densely dashed,line width=.8pt] (6.925,3.500) -- (7.575,3.500);
\draw[tcHole,fill=white,line width=1.1pt] (7.250,3.500) circle (.092);
\draw[white,line width=3.4pt] (7.575,3.500) -- (8.225,3.500);
\draw[tcHole,densely dashed,line width=.8pt] (7.575,3.500) -- (8.225,3.500);
\draw[tcHole,fill=white,line width=1.1pt] (7.900,3.500) circle (.092);
\draw[tcRed,line width=1.2pt] (6.860,3.435) -- (6.990,3.565);
\draw[tcRed,line width=1.2pt] (6.860,3.565) -- (6.990,3.435);
\draw[tcRed,line width=1.2pt] (7.510,3.435) -- (7.640,3.565);
\draw[tcRed,line width=1.2pt] (7.510,3.565) -- (7.640,3.435);
\draw[tcRed,line width=1.2pt] (8.160,3.435) -- (8.290,3.565);
\draw[tcRed,line width=1.2pt] (8.160,3.565) -- (8.290,3.435);
\node[] at (7.900,2.230) {$a=3,\quad b=4$};
\node[text=tcBlue,font=\small] at (7.900,1.650) {$\boldsymbol{g_2=2^{5}=32}$};
\node[] at (13.300,5.430) {\textbf{Right angle}};
\fill[tcFace] (11.675,3.500) rectangle (12.325,4.150);
\fill[tcFace] (12.325,2.850) rectangle (12.975,3.500);
\fill[tcFace] (12.325,3.500) rectangle (12.975,4.150);
\draw[tcGrid,line width=.45pt] (11.675,2.850) -- (11.675,4.800);
\draw[tcGrid,line width=.45pt] (12.325,2.850) -- (12.325,4.800);
\draw[tcGrid,line width=.45pt] (12.975,2.850) -- (12.975,4.800);
\draw[tcGrid,line width=.45pt] (13.625,2.850) -- (13.625,4.800);
\draw[tcGrid,line width=.45pt] (14.275,2.850) -- (14.275,4.800);
\draw[tcGrid,line width=.45pt] (14.925,2.850) -- (14.925,4.800);
\draw[tcGrid,line width=.45pt] (11.675,2.850) -- (14.925,2.850);
\draw[tcGrid,line width=.45pt] (11.675,3.500) -- (14.925,3.500);
\draw[tcGrid,line width=.45pt] (11.675,4.150) -- (14.925,4.150);
\draw[tcGrid,line width=.45pt] (11.675,4.800) -- (14.925,4.800);
\draw[white,line width=3.4pt] (12.325,3.500) -- (12.975,3.500);
\draw[tcHole,densely dashed,line width=.8pt] (12.325,3.500) -- (12.975,3.500);
\draw[tcHole,fill=white,line width=1.1pt] (12.650,3.500) circle (.092);
\draw[white,line width=3.4pt] (12.325,3.500) -- (12.325,4.150);
\draw[tcHole,densely dashed,line width=.8pt] (12.325,3.500) -- (12.325,4.150);
\draw[tcHole,fill=white,line width=1.1pt] (12.325,3.825) circle (.092);
\draw[tcRed,line width=1.2pt] (12.260,3.435) -- (12.390,3.565);
\draw[tcRed,line width=1.2pt] (12.260,3.565) -- (12.390,3.435);
\draw[tcRed,line width=1.2pt] (12.260,4.085) -- (12.390,4.215);
\draw[tcRed,line width=1.2pt] (12.260,4.215) -- (12.390,4.085);
\draw[tcRed,line width=1.2pt] (12.910,3.435) -- (13.040,3.565);
\draw[tcRed,line width=1.2pt] (12.910,3.565) -- (13.040,3.435);
\node[] at (13.300,2.230) {$a=3,\quad b=3$};
\node[text=tcBlue,font=\small] at (13.300,1.650) {$\boldsymbol{g_2=2^{4}=16}$};
\draw[tcRule,line width=.6pt] (0.000,1.120) -- (15.850,1.120);
\draw[tcBlue,line width=2.1pt] (0.200,0.620) -- (0.640,0.620);
\node[anchor=west,font=\scriptsize] at (0.750,0.620) {active edge in $\ket{1}$};
\draw[tcHole,fill=white,line width=1.1pt] (5.0,.62) circle (.092);
\node[anchor=west,font=\scriptsize] at (5.200,0.620) {hole};
\draw[tcRed,line width=1.2pt] (7.155,0.555) -- (7.285,0.685);
\draw[tcRed,line width=1.2pt] (7.155,0.685) -- (7.285,0.555);
\node[anchor=west,font=\scriptsize] at (7.420,0.620) {removed star};
\fill[tcFace] (11.15,.49) rectangle (11.51,.75);
\node[anchor=west,font=\scriptsize] at (11.650,0.620) {removed plaquette};
\end{tikzpicture}%
\endgroup%
}
\caption{Ground states of the toric code and of fixed hole sectors after the minimal SUSY lift. Qubits occupy edges. Panels (a) and (b) show periodic fundamental domains, with opposite boundaries identified. Thick coloured edges carry $\ket{1}$ in a product configuration, and all other active edges carry $\ket{0}$. The drawings are starting configurations for the indicated gauge averages. (a) The four flat configurations differ by two noncontractible dual strings. Averaging over the full star group $G$ gives the four toric-code ground states. (b) The same hole $e_0$ removes two stars and two plaquettes. For $\alpha=\beta=0$, the two displayed configurations differ on the three active edges of $X_{\mathrm{def}}$. Applying the global strings and averaging over the surviving star group $G_{e_0}$ gives the eight states in Eq.~\eqref{eq:toric-eight-states}. (c) Local patches illustrate three two-hole patterns on a larger torus, with every unshown edge active. Counting distinct removed stars $a$ and plaquettes $b$ gives sector GSDs of $64$, $32$ and $16$. These dimensions are within fixed patterns and do not sum over hole positions. For separated holes, individual merged checks have unfixed signs, so original fully flat configurations alone do not exhaust the ground space.}
\label{fig:toric-sector-ground-states}
\end{figure*}

We first construct the ground space with one hole, then compare different two-hole geometries and finally give the count for arbitrary hole patterns. Figure~\ref{fig:toric-sector-ground-states} illustrates the original four toric-code ground states alongside the one-hole and two-hole cases. The fragmentation and active-region arguments in the main text do not require this additional spectral analysis.

\subsection{One hole}
\label{app:toric-one-hole}
Place a hole on edge $e_0$, with endpoints $v_1,v_2$ and adjoining plaquettes $p_1,p_2$. The two corresponding stars and two corresponding plaquettes vanish, as shown in Fig.~\ref{fig:toric-sector-ground-states}(b). Omitting the fixed one-dimensional hole factor, the restricted Hamiltonian is
\begin{equation}
 H_{e_0}
 =-\sum_{v\neq v_1,v_2}J_v\mathsf S_v
  -\sum_{p\neq p_1,p_2}K_p B_p.
 \label{eq:toric-one-hole-hamiltonian}
\end{equation}
There remain $N_E-1$ active qubits. All $N_V-2$ surviving stars are independent. Any relation among them would also be a relation among the original stars, whose only nontrivial relation involves every vertex, including the two removed ones. The same argument gives $N_P-2$ independent surviving plaquettes. Removing the hole qubit introduces no new relations because none of these surviving operators acts on it. A mixed relation would require the $X$ and $Z$ supports to cancel separately, so it supplies no additional dependence. Hence
\begin{equation}
 \dim\mathcal G_{e_0}
 =2^{(N_E-1)-(N_V-2)-(N_P-2)}=8,
 \label{eq:toric-one-hole-gsd}
\end{equation}
where $\mathcal G_{e_0}$ denotes the sector ground space. Removing four Hamiltonian terms releases only two independent constraints, because the original generating set had two redundancies. Together with removing one physical qubit, this leaves one additional logical qubit.

This additional degree of freedom can be exhibited locally. Choose the endpoint $v_1$ and the adjoining plaquette $p_1$, and define operators on their remaining three edges
\begin{equation}
 X_{\mathrm{def}}
 =\prod_{e\in\operatorname{star}(v_1)\setminus\{e_0\}}\sigma_e^x,
 \qquad
 Z_{\mathrm{def}}
 =\prod_{e\in\partial p_1\setminus\{e_0\}}\sigma_e^z.
 \label{eq:toric-defect-operators}
\end{equation}
Both commute with every surviving term in Eq.~\eqref{eq:toric-one-hole-hamiltonian}. The original star and plaquette shared two edges, and deleting $e_0$ leaves exactly one shared edge. Thus
\[
 X_{\mathrm{def}}^2=Z_{\mathrm{def}}^2=\id,
 \qquad
 X_{\mathrm{def}}Z_{\mathrm{def}}
 =-Z_{\mathrm{def}}X_{\mathrm{def}}.
\]
They act as Pauli operators on the extra qubit. The three-edge $Z_{\mathrm{def}}$ string connects $v_1$ to $v_2$ around the other sides of $p_1$. Its endpoints cost no energy because their star terms are absent. The dual interpretation holds for $X_{\mathrm{def}}$.

The two usual toric-code logical qubits can still be represented by noncontractible loops avoiding $e_0$. Although a missing plaquette no longer supplies its elementary loop deformation, the boundary of the combined pair remains constrained. On the active edges,
\begin{equation}
 C_Z
 =\prod_{e\in\partial(p_1\cup p_2)}\sigma_e^z
 =\prod_{p\neq p_1,p_2}B_p.
 \label{eq:toric-one-hole-merged}
\end{equation}
The union boundary contains no $e_0$, and $C_Z$ has eigenvalue $+1$ throughout $\mathcal G_{e_0}$. An analogous identity holds for the merged pair of stars. Loops can therefore be deformed around the combined defect. The extra doubling comes from the open strings ending at missing checks. As logical degrees of freedom, the ground space is $\mathbb C^4_{\mathrm{toric}}\otimes\mathbb C^2_{\mathrm{def}}$. The defect factor has local logical operators, so its degeneracy can be split by a local perturbation such as $-\epsilon Z_{\mathrm{def}}$ while preserving the hole pattern. Related damaged-star and damaged-plaquette operators describe an additional local qubit in treatments of physical qubit loss~\cite{Stace2009}.

\paragraph*{The eight states as gauge averages.}
The states can be constructed by averaging flat product configurations over the surviving vertex transformations. Use $\sigma^z\ket{0}=\ket{0}$ and $\sigma^z\ket{1}=-\ket{1}$. Choose logical $X$ strings $\overline X_1,\overline X_2$ on noncontractible dual-lattice loops avoiding $e_0$. On the original qubit lattice, take
\begin{equation}
 \ket{s_{\alpha\beta,t}}
 =\mathsf S_{v_1}^{\,t}
   \overline X_1^{\,\alpha}\overline X_2^{\,\beta}
   \ket{0}^{\otimes N_E},
 \qquad \alpha,\beta,t\in\{0,1\}.
 \label{eq:toric-flat-seeds}
\end{equation}
Every seed has $B_p=+1$ for every original plaquette. The labels $\alpha,\beta$ specify the two global holonomies, which flatness does not fix. The original star $\mathsf S_{v_1}$ preserves these holonomies and all plaquette eigenvalues, while flipping $e_0$ and the other three incident edges. Consequently,
\begin{equation}
 \ket{s_{\alpha\beta,t}}
 =\ket{t}_{e_0}\otimes\ket{\eta_{\alpha\beta,t}},
 \qquad
 \ket{\eta_{\alpha\beta,t}}
 =X_{\mathrm{def}}^{\,t}\ket{\eta_{\alpha\beta,0}}.
 \label{eq:toric-seed-factorization}
\end{equation}
This explicitly displays the dependence on $t$ on the remaining edges. For example, at $\alpha=\beta=0$ the two configurations on the three other edges of $\operatorname{star}(v_1)$ are $\ket{000}$ and $\ket{111}$, with every other active edge in $\ket{0}$. Replacing $\ket{t}_{e_0}$ by $\ket{\circ}_{e_0}$ leaves this difference intact. The use of $\mathsf S_{v_1}$ in Eq.~\eqref{eq:toric-flat-seeds} is an auxiliary construction on the original lattice, before the hole is inserted. Its lifted counterpart vanishes in the hole sector.

Let $G_{e_0}$ be the group generated by the surviving stars $\mathsf S_v$, $v\neq v_1,v_2$. These are $N_V-2$ independent spin flips, so $|G_{e_0}|=2^{N_V-2}$. The normalized states are
\begin{equation}
 \begin{aligned}
 \ket{\Psi_{\alpha\beta,t}}
 &=\frac{\ket{\circ}_{e_0}}{\sqrt{2^{N_V-2}}}
      \otimes\sum_{g\in G_{e_0}}g\ket{\eta_{\alpha\beta,t}}\\
 &=\frac{\ket{\circ}_{e_0}}{2^{(N_V-2)/2}}
      \otimes\prod_{v\neq v_1,v_2}(\id+\mathsf S_v)
                          \ket{\eta_{\alpha\beta,t}}.
 \end{aligned}
 \label{eq:toric-eight-states}
\end{equation}
The gauge orbit contains $2^{N_V-2}$ distinct orthogonal product configurations, giving the displayed normalization. Every surviving star has eigenvalue $+1$ after the average, and every surviving plaquette retains eigenvalue $+1$. Choose conjugate Wilson loops $\overline Z_1,\overline Z_2$ avoiding $e_0$, with the usual intersection pairing with $\overline X_1,\overline X_2$. Then
\[
\begin{gathered}
\overline Z_1\ket{\Psi_{\alpha\beta,t}}
   =(-1)^\alpha\ket{\Psi_{\alpha\beta,t}},\\
 \overline Z_2\ket{\Psi_{\alpha\beta,t}}
   =(-1)^\beta\ket{\Psi_{\alpha\beta,t}},
 \\
 Z_{\mathrm{def}}\ket{\Psi_{\alpha\beta,t}}
   =(-1)^t\ket{\Psi_{\alpha\beta,t}}.
\end{gathered}
\]
The last equality follows from $B_{p_1}=\sigma_{e_0}^zZ_{\mathrm{def}}$ on the original flat seeds. The three commuting labels distinguish eight orthogonal states, exhausting Eq.~\eqref{eq:toric-one-hole-gsd}. The hole state is identical in all eight. Figure~\ref{fig:toric-sector-ground-states}(b) displays the two local configurations and their eight gauge-averaged states. Before inserting the hole, $t=0,1$ belonged to one gauge orbit. Afterwards, the surviving gauge average does not identify them, and $X_{\mathrm{def}}$ exchanges the two resulting states.

\subsection{Two holes and their relative geometry}
\label{app:toric-two-holes}
Now fix holes on two distinct edges. Let $a$ count the distinct removed stars and $b$ the distinct removed plaquettes. The same independence argument applies to the surviving stabilizers, giving
\begin{equation}
 g_2
 =2^{(N_E-2)-(N_V-a)-(N_P-b)}
 =2^{a+b-2}.
 \label{eq:toric-two-hole-gsd}
\end{equation}
If the two hole edges share neither a vertex nor an adjoining plaquette, then $a=b=4$ and $g_2=64$. Consecutive collinear hole edges share one vertex but no plaquette, giving $a=3$, $b=4$ and $g_2=32$. If they meet at a right angle, they share one vertex and one plaquette, giving $a=b=3$ and $g_2=16$. Figure~\ref{fig:toric-sector-ground-states}(c) illustrates these geometries, and Table~\ref{tab:toric-two-holes} records their counts. Thus the same number of holes can release different numbers of independent constraints, depending on how their interaction supports overlap.

\begin{table*}[tp]
\centering
\begin{tabular}{lccc}
\hline
Two-hole arrangement & Removed stars $a$ & Removed plaquettes $b$ & Sector GSD\\
\hline
Separated & $4$ & $4$ & $64$\\
Consecutive and collinear & $3$ & $4$ & $32$\\
Meeting at a right angle & $3$ & $3$ & $16$\\
\hline
\end{tabular}
\caption{Ground-state degeneracies for three fixed two-hole patterns in the minimally lifted toric code. Separated means that the two hole edges share neither a vertex nor an adjoining plaquette. The count is within each pattern and does not sum over hole positions.}
\label{tab:toric-two-holes}
\end{table*}

The separated case also clarifies why the degeneracy grows faster than simply adding a second copy of the one-hole defect qubit. For two well-separated holes, define $C_Z^{(1)},C_Z^{(2)}$ as the merged plaquette boundaries around them, and $C_X^{(1)},C_X^{(2)}$ as the corresponding merged stars. Here a merged star is the product of the two original endpoint stars, with their common $\sigma_{e_i}^x$ factors cancelled before acting on the remaining edges. On the active space,
\[
 C_Z^{(1)}C_Z^{(2)}=\prod_{\text{surviving }p}B_p,
 \qquad
 C_X^{(1)}C_X^{(2)}=\prod_{\text{surviving }v}\mathsf S_v.
\]
The ground-state constraints fix both products to $+1$, but each pair can have individual signs $(+1,+1)$ or $(-1,-1)$. These are two additional binary labels. Together with the two local defect qubits and the two toric-code qubits, they account for the six binary degrees of freedom in $g_2=64$. This distinguishes the present restricted Hamiltonian from a code in which every merged star and plaquette is separately imposed as a $+1$ constraint, as in qubit-loss constructions~\cite{Stace2009}. Such merged constraints follow automatically for one hole on a torus, but need not follow individually for several holes.

\subsection{Arbitrary hole patterns and sector energies}
\label{app:toric-many-holes}
For a nonempty hole set $D$ with $m=|D|$, let $a(D)$ count vertices incident on at least one hole and $b(D)$ count plaquettes containing at least one hole. The preceding stabilizer count immediately gives
\begin{equation}
 g(D)=\dim\mathcal G_D=2^{a(D)+b(D)-m},
 \qquad m\geq1.
 \label{eq:toric-arbitrary-hole-gsd}
\end{equation}
The no-hole sector instead has the original GSD of four. For $m$ holes sharing neither vertices nor plaquettes, $a=b=2m$ and $g(D)=8^m$. The count is not monotone for arbitrary patterns. If all incident terms on an active edge have already vanished, that edge is a free qubit, and replacing it by a hole halves the sector GSD. The all-hole sector contains just one state. Within this prescription, geometry enters the exact dimension through $m,a,b$, while the spatial form of the logical operators can depend on finer details of the pattern.

For uniform positive $J_v=J$ and $K_p=K$, the sector minimum lies above the fully active minimum by
\begin{equation}
 E_{\min}(D)-E_{\min}(\varnothing)
 =J\,a(D)+K\,b(D).
 \label{eq:toric-sector-energy}
\end{equation}
For spatially varying positive couplings, the right-hand side is the sum of the couplings of all removed terms, counted once each. Thus the local SUSY zero mode does not make these sector minima global ground states. The full minimally lifted Hamiltonian still has its four global ground states in the fully active sector. Its other invariant sectors can have large and geometry-dependent lowest-energy degeneracies, including local contributions with no topological protection. A systematic study of many-hole patterns, their logical operators, topological properties and dynamics deserves a separate investigation. The counts above demonstrate how the site-wise lifting rule converts the geometry of inactive edges into exact information about the sector ground spaces.

\section{Representations of the seed matrix algebra}
\label{app:matrix-representations}

\begin{proposition}
Every nonzero linear, multiplicative, adjoint-preserving map $\phi\colon\End(\mathbb C^n)\to\End(\widetilde{\mathcal H})$, with $\dim\widetilde{\mathcal H}=d<\infty$, has the form
\begin{equation}
\begin{gathered}
\widetilde{\mathcal H}\simeq(\mathbb C^n\otimes\mathbb C^r)\oplus\mathbb C^s,
 \\
 \phi(O)\simeq(O\otimes\id_r)\oplus0_s,
 \\ d=nr+s,
\end{gathered}
 \label{eq:app-rep-map}
\end{equation}
where $r\geq1$ and $s\geq0$. The equivalence uses the same unitary change of basis for every $O$. The map need not preserve the identity.
\end{proposition}
\begin{proof}
Write $\widehat E_{ab}=\phi(\ket a\bra b)$ and $P=\sum_a\widehat E_{aa}$. The matrix-unit relations $\widehat E_{ab}\widehat E_{cd}=\delta_{bc}\widehat E_{ad}$ and $\widehat E_{ab}^\dagger=\widehat E_{ba}$ make the diagonal images orthogonal projectors. The off-diagonal images identify their ranges $W_a=\operatorname{Ran}\widehat E_{aa}$ unitarily, since $\widehat E_{ab}^\dagger\widehat E_{ab}=\widehat E_{bb}$ and $\widehat E_{ab}\widehat E_{ab}^\dagger=\widehat E_{aa}$. Thus every $W_a$ has the same dimension $r$. From an orthonormal basis $\ket{0,\mu}$ of $W_0$, define $\ket{a,\mu}=\widehat E_{a0}\ket{0,\mu}$. These vectors form an orthonormal basis of $\operatorname{Ran}P$ and obey $\widehat E_{ab}\ket{c,\mu}=\delta_{bc}\ket{a,\mu}$, giving $r$ copies of the defining representation. Every matrix unit annihilates $\ker P$, because $\widehat E_{ab}=\widehat E_{ab}P$. Taking $s=\dim\ker P$ proves the stated form. Conversely, that form preserves multiplication and adjoints.
\end{proof}

This statement concerns representations of the full seed matrix algebra. It does not classify all representations satisfying only the abstract SUSY or paraSUSY relations. The site-wise theorem chooses $s\geq1$ to introduce an inactive subspace, although the classification permits $s=0$. For bond-wise lifting, the embedding into the physical tensor-product space remains a separate choice, as discussed in Sec.~\ref{subsec:bond-classification}.

\phantomsection


\begin{thebibliography}{99}

\bibitem{Deutsch1991}
J.~M.~Deutsch,
``Quantum statistical mechanics in a closed system,''
\href{https://doi.org/10.1103/PhysRevA.43.2046}{Phys. Rev. A \textbf{43}, 2046--2049 (1991)}.

\bibitem{Srednicki1994}
M.~Srednicki,
``Chaos and quantum thermalization,''
\href{https://doi.org/10.1103/PhysRevE.50.888}{Phys. Rev. E \textbf{50}, 888--901 (1994)}.

\bibitem{Sala2020}
P.~Sala, T.~Rakovszky, R.~Verresen, M.~Knap and F.~Pollmann,
``Ergodicity breaking arising from Hilbert space fragmentation in dipole-conserving Hamiltonians,''
\href{https://doi.org/10.1103/PhysRevX.10.011047}{Phys. Rev. X \textbf{10}, 011047 (2020)}.

\bibitem{Khemani2020}
V.~Khemani, M.~Hermele and R.~Nandkishore,
``Localization from Hilbert space shattering: From theory to physical realizations,''
\href{https://doi.org/10.1103/PhysRevB.101.174204}{Phys. Rev. B \textbf{101}, 174204 (2020)}.

\bibitem{Moudgalya2021}
S.~Moudgalya, A.~Prem, R.~Nandkishore, N.~Regnault and B.~A.~Bernevig,
``Thermalization and its absence within Krylov subspaces of a constrained Hamiltonian,''
in \emph{Memorial Volume for Shoucheng Zhang}
(World Scientific, 2021),
\href{https://doi.org/10.1142/9789811231711_0009}{pp.~147--209}.

\bibitem{MoudgalyaReview2022}
S.~Moudgalya, B.~A.~Bernevig and N.~Regnault,
``Quantum many-body scars and Hilbert space fragmentation: A review of exact results,''
\href{https://doi.org/10.1088/1361-6633/ac73a0}{Rep. Prog. Phys. \textbf{85}, 086501 (2022)}.

\bibitem{Menon1997}
G.~I.~Menon, M.~Barma and D.~Dhar,
``Conservation laws and integrability of a one-dimensional model of diffusing dimers,''
\href{https://doi.org/10.1007/BF02183622}{J. Stat. Phys. \textbf{86}, 1237--1263 (1997)}.

\bibitem{Maassarani1998}
Z.~Maassarani,
``The $XXC$ models,''
\href{https://doi.org/10.1016/S0375-9601(98)00322-3}{Phys. Lett. A \textbf{244}, 160--164 (1998)}.

\bibitem{Yang2008}
Z.~Yang, L.~Yang, J.~Dai and T.~Xiang,
``Rigorous solution of the spin-1 quantum Ising model with single-ion anisotropy,''
\href{https://doi.org/10.1103/PhysRevLett.100.067203}{Phys. Rev. Lett. \textbf{100}, 067203 (2008)}.

\bibitem{Caha2018}
L.~Caha and D.~Nagaj,
``The pair-flip model: A very entangled translationally invariant spin chain,''
\href{https://arxiv.org/abs/1805.07168}{arXiv:1805.07168 [quant-ph] (2018)}.

\bibitem{Pai2019}
S.~Pai, M.~Pretko and R.~M.~Nandkishore,
``Localization in fractonic random circuits,''
\href{https://doi.org/10.1103/PhysRevX.9.021003}{Phys. Rev. X \textbf{9}, 021003 (2019)}.
Erratum, \href{https://doi.org/10.1103/PhysRevX.9.049901}{Phys. Rev. X \textbf{9}, 049901 (2019)}.

\bibitem{Rakovszky2020}
T.~Rakovszky, P.~Sala, R.~Verresen, M.~Knap and F.~Pollmann,
``Statistical localization: From strong fragmentation to strong edge modes,''
\href{https://doi.org/10.1103/PhysRevB.101.125126}{Phys. Rev. B \textbf{101}, 125126 (2020)}.

\bibitem{DeTomasi2019}
G.~De Tomasi, D.~Hetterich, P.~Sala and F.~Pollmann,
``Dynamics of strongly interacting systems: From Fock-space fragmentation to many-body localization,''
\href{https://doi.org/10.1103/PhysRevB.100.214313}{Phys. Rev. B \textbf{100}, 214313 (2019)}.

\bibitem{Yang2020}
Z.-C.~Yang, F.~Liu, A.~V.~Gorshkov and T.~Iadecola,
``Hilbert-space fragmentation from strict confinement,''
\href{https://doi.org/10.1103/PhysRevLett.124.207602}{Phys. Rev. Lett. \textbf{124}, 207602 (2020)}.

\bibitem{Zadnik2021}
L.~Zadnik and M.~Fagotti,
``The folded spin-$1/2$ $XXZ$ model: I. Diagonalisation, jamming, and ground state properties,''
\href{https://doi.org/10.21468/SciPostPhysCore.4.2.010}{SciPost Phys. Core \textbf{4}, 010 (2021)}.

\bibitem{Pozsgay2021}
B.~Pozsgay, T.~Gombor, A.~Hutsalyuk, Y.~Jiang, L.~Pristy\'{a}k and E.~Vernier,
``Integrable spin chain with Hilbert space fragmentation and solvable real-time dynamics,''
\href{https://doi.org/10.1103/PhysRevE.104.044106}{Phys. Rev. E \textbf{104}, 044106 (2021)}.

\bibitem{McClarty2020}
P.~A.~McClarty, M.~Haque, A.~Sen and J.~Richter,
``Disorder-free localization and many-body quantum scars from magnetic frustration,''
\href{https://doi.org/10.1103/PhysRevB.102.224303}{Phys. Rev. B \textbf{102}, 224303 (2020)}.

\bibitem{Hahn2021}
D.~Hahn, P.~A.~McClarty and D.~J.~Luitz,
``Information dynamics in a model with Hilbert space fragmentation,''
\href{https://doi.org/10.21468/SciPostPhys.11.4.074}{SciPost Phys. \textbf{11}, 074 (2021)}.

\bibitem{Lee2021}
K.~Lee, A.~Pal and H.~J.~Changlani,
``Frustration-induced emergent Hilbert space fragmentation,''
\href{https://doi.org/10.1103/PhysRevB.103.235133}{Phys. Rev. B \textbf{103}, 235133 (2021)}.

\bibitem{Khudorozhkov2022}
A.~Khudorozhkov, A.~Tiwari, C.~Chamon and T.~Neupert,
``Hilbert space fragmentation in a 2D quantum spin system with subsystem symmetries,''
\href{https://doi.org/10.21468/SciPostPhys.13.4.098}{SciPost Phys. \textbf{13}, 098 (2022)}.

\bibitem{Hart2022}
O.~Hart and R.~Nandkishore,
``Hilbert space shattering and dynamical freezing in the quantum Ising model,''
\href{https://doi.org/10.1103/PhysRevB.106.214426}{Phys. Rev. B \textbf{106}, 214426 (2022)}.

\bibitem{Will2024}
M.~Will, R.~Moessner and F.~Pollmann,
``Realization of Hilbert space fragmentation and fracton dynamics in two dimensions,''
\href{https://doi.org/10.1103/PhysRevLett.133.196301}{Phys. Rev. Lett. \textbf{133}, 196301 (2024)}.

\bibitem{MoudgalyaMotrunich2022}
S.~Moudgalya and O.~I.~Motrunich,
``Hilbert space fragmentation and commutant algebras,''
\href{https://doi.org/10.1103/PhysRevX.12.011050}{Phys. Rev. X \textbf{12}, 011050 (2022)}.

\bibitem{Balasubramanian2024}
S.~Balasubramanian, S.~Gopalakrishnan, A.~Khudorozhkov and E.~Lake,
``Glassy word problems: Ultraslow relaxation, Hilbert space jamming, and computational complexity,''
\href{https://doi.org/10.1103/PhysRevX.14.021034}{Phys. Rev. X \textbf{14}, 021034 (2024)}.

\bibitem{Danieli2022}
C.~Danieli, A.~Andreanov and S.~Flach,
``Many-body localization transition from flat-band fine tuning,''
\href{https://doi.org/10.1103/PhysRevB.105.L041113}{Phys. Rev. B \textbf{105}, L041113 (2022)}.

\bibitem{Nicolau2023}
E.~Nicolau, A.~M.~Marques, R.~G.~Dias and V.~Ahufinger,
``Flat band induced local Hilbert space fragmentation,''
\href{https://doi.org/10.1103/PhysRevB.108.205104}{Phys. Rev. B \textbf{108}, 205104 (2023)}.

\bibitem{Flach2014}
S.~Flach, D.~Leykam, J.~D.~Bodyfelt, P.~Matthies and A.~S.~Desyatnikov,
``Detangling flat bands into Fano lattices,''
\href{https://doi.org/10.1209/0295-5075/105/30001}{EPL \textbf{105}, 30001 (2014)}.

\bibitem{Maimaiti2017}
W.~Maimaiti, A.~Andreanov, H.~C.~Park, O.~Gendelman and S.~Flach,
``Compact localized states and flat-band generators in one dimension,''
\href{https://doi.org/10.1103/PhysRevB.95.115135}{Phys. Rev. B \textbf{95}, 115135 (2017)}.

\bibitem{Kwan2025}
Y.~H.~Kwan, P.~H.~Wilhelm, S.~Biswas and S.~A.~Parameswaran,
``Minimal Hubbard models of maximal Hilbert space fragmentation,''
\href{https://doi.org/10.1103/PhysRevLett.134.010411}{Phys. Rev. Lett. \textbf{134}, 010411 (2025)}.

\bibitem{Stephen2024}
D.~T.~Stephen, O.~Hart and R.~M.~Nandkishore,
``Ergodicity breaking provably robust to arbitrary perturbations,''
\href{https://doi.org/10.1103/PhysRevLett.132.040401}{Phys. Rev. Lett. \textbf{132}, 040401 (2024)}.

\bibitem{Khudorozhkov2025}
A.~Khudorozhkov, C.~Stahl, O.~Hart and R.~Nandkishore,
``Robust Hilbert space fragmentation in group-valued loop models,''
\href{https://doi.org/10.1103/PhysRevB.111.024310}{Phys. Rev. B \textbf{111}, 024310 (2025)}.

\bibitem{Scherg2021}
S.~Scherg, T.~Kohlert, P.~Sala, F.~Pollmann, B.~Hebbe Madhusudhana, I.~Bloch and M.~Aidelsburger,
``Observing non-ergodicity due to kinetic constraints in tilted Fermi--Hubbard chains,''
\href{https://doi.org/10.1038/s41467-021-24726-0}{Nat. Commun. \textbf{12}, 4490 (2021)}.

\bibitem{Adler2024}
D.~Adler, D.~Wei, M.~Will, K.~Srakaew, S.~Agrawal, P.~Weckesser,
R.~Moessner, F.~Pollmann, I.~Bloch and J.~Zeiher,
``Observation of Hilbert space fragmentation and fractonic excitations in 2D,''
\href{https://doi.org/10.1038/s41586-024-08188-0}{Nature \textbf{636}, 80--85 (2024)}.

\bibitem{Wang2025}
Y.-Y.~Wang, Y.-H.~Shi, Z.-H.~Sun, C.-T.~Chen, Z.-A.~Wang, K.~Zhao,
H.-T.~Liu, W.-G.~Ma, Z.~Wang, H.~Li, J.-C.~Zhang, Y.~Liu,
C.-L.~Deng, T.-M.~Li, Y.~He, Z.-H.~Liu, Z.-Y.~Peng, X.~Song,
G.~Xue, H.~Yu, K.~Huang, Z.~Xiang, D.~Zheng, K.~Xu and H.~Fan,
``Exploring Hilbert-space fragmentation on a superconducting processor,''
\href{https://doi.org/10.1103/PRXQuantum.6.010325}{PRX Quantum \textbf{6}, 010325 (2025)}.

\bibitem{Zhao2025}
L.~Zhao, P.~R.~Datla, W.~Tian, M.~M.~Aliyu and H.~Loh,
``Observation of quantum thermalization restricted to Hilbert space fragments and $\mathbb{Z}_{2k}$ scars,''
\href{https://doi.org/10.1103/PhysRevX.15.011035}{Phys. Rev. X \textbf{15}, 011035 (2025)}.

\bibitem{Datla2026}
P.~R.~Datla, L.~Zhao, W.~W.~Ho, N.~Klco and H.~Loh,
``Statistical localization of $U(1)$ lattice gauge theory in a Rydberg simulator,''
\href{https://doi.org/10.1038/s41567-026-03183-w}{Nat. Phys. \textbf{22}, 355--361 (2026)}.

\bibitem{Takano2009}
K.~Takano, H.~Suzuki and K.~Hida,
``Exact spin-cluster ground states in a mixed diamond chain,''
\href{https://doi.org/10.1103/PhysRevB.80.104410}{Phys. Rev. B \textbf{80}, 104410 (2009)}.

\bibitem{Hida2009}
K.~Hida, K.~Takano and H.~Suzuki,
``Finite temperature properties of mixed diamond chain with spins 1 and 1/2,''
\href{https://doi.org/10.1143/JPSJ.78.084716}{J. Phys. Soc. Jpn. \textbf{78}, 084716 (2009)}.

\bibitem{Katsura2025}
H.~Katsura, C.~Matsui, C.~Paletta and B.~Pozsgay,
``Weak ergodicity breaking with isolated integrable sectors,''
\href{https://doi.org/10.1103/PhysRevResearch.7.023099}{Phys. Rev. Research \textbf{7}, 023099 (2025)}.

\bibitem{Witten1981}
E.~Witten,
``Dynamical breaking of supersymmetry,''
\href{https://doi.org/10.1016/0550-3213(81)90006-7}{Nucl. Phys. B \textbf{188}, 513--554 (1981)}.

\bibitem{Witten1982}
E.~Witten,
``Constraints on supersymmetry breaking,''
\href{https://doi.org/10.1016/0550-3213(82)90071-2}{Nucl. Phys. B \textbf{202}, 253--316 (1982)}.

\bibitem{Cooper1995}
F.~Cooper, A.~Khare and U.~Sukhatme,
``Supersymmetry and quantum mechanics,''
\href{https://doi.org/10.1016/0370-1573(94)00080-M}{Phys. Rep. \textbf{251}, 267--385 (1995)}.

\bibitem{Cooper2001}
F.~Cooper, A.~Khare and U.~Sukhatme,
\href{https://doi.org/10.1142/4687}{\emph{Supersymmetry in Quantum Mechanics}}
(World Scientific, Singapore, 2001).

\bibitem{Rubakov1988}
V.~A.~Rubakov and V.~P.~Spiridonov,
``Parasupersymmetric quantum mechanics,''
\href{https://doi.org/10.1142/S0217732388001616}{Mod. Phys. Lett. A \textbf{3}, 1337--1347 (1988)}.

\bibitem{Khare1992}
A.~Khare,
``Parasupersymmetric quantum mechanics of arbitrary order,''
\href{https://doi.org/10.1088/0305-4470/25/12/008}{J. Phys. A: Math. Gen. \textbf{25}, L749--L754 (1992)}.

\bibitem{Nicolai1976}
H.~Nicolai,
``Supersymmetry and spin systems,''
\href{https://doi.org/10.1088/0305-4470/9/9/010}{J. Phys. A: Math. Gen. \textbf{9}, 1497--1506 (1976)}.

\bibitem{EsslerKorepin1992}
F.~H.~L.~Essler and V.~E.~Korepin,
``Higher conservation laws and algebraic Bethe Ans\"atze for the supersymmetric $t$--$J$ model,''
\href{https://doi.org/10.1103/PhysRevB.46.9147}{Phys. Rev. B \textbf{46}, 9147--9162 (1992)}.

\bibitem{FendleySchoutensDeBoer2003}
P.~Fendley, K.~Schoutens and J.~de Boer,
``Lattice models with $\mathcal N=2$ supersymmetry,''
\href{https://doi.org/10.1103/PhysRevLett.90.120402}{Phys. Rev. Lett. \textbf{90}, 120402 (2003)}.

\bibitem{FendleyNienhuisSchoutens2003}
P.~Fendley, B.~Nienhuis and K.~Schoutens,
``Lattice fermion models with supersymmetry,''
\href{https://doi.org/10.1088/0305-4470/36/50/004}{J. Phys. A: Math. Gen. \textbf{36}, 12399--12424 (2003)}.

\bibitem{FendleySchoutens2005}
P.~Fendley and K.~Schoutens,
``Exact results for strongly correlated fermions in $2+1$ dimensions,''
\href{https://doi.org/10.1103/PhysRevLett.95.046403}{Phys. Rev. Lett. \textbf{95}, 046403 (2005)}.

\bibitem{SannomiyaKatsura2019}
N.~Sannomiya and H.~Katsura,
``Supersymmetry breaking and Nambu--Goldstone fermions in interacting Majorana chains,''
\href{https://doi.org/10.1103/PhysRevD.99.045002}{Phys. Rev. D \textbf{99}, 045002 (2019)}.

\bibitem{YangFendley2004}
X.~Yang and P.~Fendley,
``Non-local spacetime supersymmetry on the lattice,''
\href{https://doi.org/10.1088/0305-4470/37/38/003}{J. Phys. A: Math. Gen. \textbf{37}, 8937--8948 (2004)}.

\bibitem{HagendorfFendley2012}
C.~Hagendorf and P.~Fendley,
``The eight-vertex model and lattice supersymmetry,''
\href{https://doi.org/10.1007/s10955-012-0430-0}{J. Stat. Phys. \textbf{146}, 1122--1155 (2012)}.

\bibitem{Hagendorf2013}
C.~Hagendorf,
``Spin chains with dynamical lattice supersymmetry,''
\href{https://doi.org/10.1007/s10955-013-0709-9}{J. Stat. Phys. \textbf{150}, 609--657 (2013)}.

\bibitem{Padmanabhan2017}
P.~Padmanabhan, S.-J.~Rey, D.~Teixeira and D.~Trancanelli,
``Supersymmetric many-body systems from partial symmetries: Integrability, localization and scrambling,''
\href{https://doi.org/10.1007/JHEP05(2017)136}{J. High Energy Phys. \textbf{05} (2017), 136}.

\bibitem{Maity2026}
S.~Maity, P.~Padmanabhan, J.~Hietarinta and V.~Korepin,
``Almost local integrable models from supersymmetry algebras,''
\href{https://doi.org/10.1088/1751-8121/ae6a4f}{J. Phys. A: Math. Theor. \textbf{59}, 205202 (2026)}.

\bibitem{Maity2025NonHermitian}
S.~Maity, P.~Padmanabhan and V.~Korepin,
``Non-hermitian integrable systems from constant non-invertible solutions of the Yang--Baxter equation,''
\href{https://doi.org/10.1007/JHEP05(2025)206}{J. High Energy Phys. \textbf{05} (2025), 206}.

\bibitem{Beckers1990}
J.~Beckers and N.~Debergh,
``Parastatistics and supersymmetry in quantum mechanics,''
\href{https://doi.org/10.1016/0550-3213(90)90469-T}{Nucl. Phys. B \textbf{340}, 767--776 (1990)}.

\bibitem{Durand1991}
S.~Durand, M.~Mayrand, V.~Spiridonov and L.~Vinet,
``Higher order parasupersymmetric quantum mechanics,''
\href{https://doi.org/10.1142/S0217732391003651}{Mod. Phys. Lett. A \textbf{6}, 3163--3170 (1991)}.

\bibitem{KhareMishraRajasekaran1993}
A.~Khare, A.~K.~Mishra and G.~Rajasekaran,
``Parasupersymmetric quantum mechanics and its infinite order limit,''
\href{https://doi.org/10.1142/S021773239300012X}{Mod. Phys. Lett. A \textbf{8}, 107--113 (1993)}.

\bibitem{QuesneVansteenkiste1998}
C.~Quesne and N.~Vansteenkiste,
``$C_\lambda$-extended harmonic oscillator and (para)supersymmetric quantum mechanics,''
\href{https://doi.org/10.1016/S0375-9601(98)00046-2}{Phys. Lett. A \textbf{240}, 21--28 (1998)}.

\bibitem{Tanaka2007}
T.~Tanaka,
``Parasupersymmetry and $\mathcal N$-fold supersymmetry in quantum many-body systems I. General formalism and second order,''
\href{https://doi.org/10.1016/j.aop.2006.11.009}{Ann. Phys. \textbf{322}, 2350--2373 (2007)}.

\bibitem{Faddeev1996}
L.~D.~Faddeev,
``How algebraic Bethe ansatz works for integrable model,''
\href{https://arxiv.org/abs/hep-th/9605187}{arXiv:hep-th/9605187 (1996)}.

\bibitem{KorepinBogoliubovIzergin1993}
V.~E.~Korepin, N.~M.~Bogoliubov and A.~G.~Izergin,
\href{https://doi.org/10.1017/CBO9780511628832}{\emph{Quantum Inverse Scattering Method and Correlation Functions}}
(Cambridge University Press, Cambridge, 1993).

\bibitem{Baxter1982}
R.~J.~Baxter,
\href{https://physics.anu.edu.au/research/ftp/mpg/baxter_book.php}{\emph{Exactly Solved Models in Statistical Mechanics}}
(Academic Press, London, 1982).

\bibitem{Sutherland1975}
B.~Sutherland,
``Model for a multicomponent quantum system,''
\href{https://doi.org/10.1103/PhysRevB.12.3795}{Phys. Rev. B \textbf{12}, 3795--3805 (1975)}.

\bibitem{Pfeuty1970}
P.~Pfeuty,
``The one-dimensional Ising model with a transverse field,''
\href{https://doi.org/10.1016/0003-4916(70)90270-8}{Ann. Phys. \textbf{57}, 79--90 (1970)}.

\bibitem{McCoyWu1973}
B.~M.~McCoy and T.~T.~Wu,
\href{https://doi.org/10.4159/harvard.9780674180758}
{\textit{The Two-Dimensional Ising Model}}
(Harvard University Press, Cambridge, Massachusetts, 1973).

\bibitem{Chiba2024}
Y.~Chiba,
``Proof of absence of local conserved quantities in the mixed-field Ising chain,''
\href{https://doi.org/10.1103/PhysRevB.109.035123}{Phys. Rev. B \textbf{109}, 035123 (2024)}.

\bibitem{Kitaev2003}
A.~Yu.~Kitaev,
``Fault-tolerant quantum computation by anyons,''
\href{https://doi.org/10.1016/S0003-4916(02)00018-0}{Ann. Phys. \textbf{303}, 2--30 (2003)}.

\bibitem{Uimin1970}
G.~V.~Uimin,
``One-dimensional problem for $S=1$ with modified
antiferromagnetic Hamiltonian,''
JETP Lett. \textbf{12}, 225--228 (1970).

\bibitem{Lai1974}
C.~K.~Lai,
``Lattice gas with nearest-neighbor interaction in one dimension
with arbitrary statistics,''
\href{https://doi.org/10.1063/1.1666522}
{J. Math. Phys. \textbf{15}, 1675--1676 (1974)}.

\bibitem{Sklyanin1988}
E.~K.~Sklyanin,
``Boundary conditions for integrable quantum systems,''
\href{https://doi.org/10.1088/0305-4470/21/10/015}{J. Phys. A: Math. Gen. \textbf{21}, 2375--2389 (1988)}.

\bibitem{Shannon1948}
C.~E.~Shannon,
``A mathematical theory of communication,''
\href{https://doi.org/10.1002/j.1538-7305.1948.tb01338.x}{Bell Syst. Tech. J. \textbf{27}, 379--423 (1948)}.

\bibitem{YangYang1969}
C.~N.~Yang and C.~P.~Yang,
``Thermodynamics of a one-dimensional system of bosons
with repulsive delta-function interaction,''
\href{https://doi.org/10.1063/1.1664947}{J. Math. Phys. \textbf{10}, 1115--1122 (1969)}.

\bibitem{McCoy2010}
B.~M.~McCoy,
\href{https://doi.org/10.1093/acprof:oso/9780199556632.001.0001}
{\textit{Advanced Statistical Mechanics}}
(Oxford University Press, Oxford, 2010).

\bibitem{McCoyPerkShrock1983}
B.~M.~McCoy, J.~H.~H.~Perk and R.~E.~Shrock,
``Time-dependent correlation functions of the transverse Ising
chain at the critical magnetic field,''
\href{https://doi.org/10.1016/0550-3213(83)90132-3}
{Nucl. Phys. B \textbf{220}, 35--47 (1983)}.

\bibitem{PerkCapelQuispelNijhoff1984}
J.~H.~H.~Perk, H.~W.~Capel, G.~R.~W.~Quispel and F.~W.~Nijhoff,
``Finite-temperature correlations for the Ising chain
in a transverse field,''
\href{https://doi.org/10.1016/0378-4371(84)90102-X}
{Physica A \textbf{123}, 1--49 (1984)}.

\bibitem{DuttaBhattacharjee2001}
A.~Dutta and J.~K.~Bhattacharjee,
``Phase transitions in the quantum Ising and rotor models
with a long-range interaction,''
\href{https://doi.org/10.1103/PhysRevB.64.184106}
{Phys. Rev. B \textbf{64}, 184106 (2001)}.

\bibitem{Koffel2012}
T.~Koffel, M.~Lewenstein and L.~Tagliacozzo,
``Entanglement entropy for the long-range Ising chain
in a transverse field,''
\href{https://doi.org/10.1103/PhysRevLett.109.267203}
{Phys. Rev. Lett. \textbf{109}, 267203 (2012)}.

\bibitem{FeySchmidt2016}
S.~Fey and K.~P.~Schmidt,
``Critical behavior of quantum magnets with long-range
interactions in the thermodynamic limit,''
\href{https://doi.org/10.1103/PhysRevB.94.075156}
{Phys. Rev. B \textbf{94}, 075156 (2016)}.

\bibitem{ClassenHowes2024}
J.~Classen-Howes,
``Quantum and classical Hilbert space fragmentation,''
\href{https://ora.ox.ac.uk/objects/uuid:75484692-9176-43ce-8adb-3553ee162a58}{D.Phil. thesis, University of Oxford (2024)}, Sec.~2.3.1.

\bibitem{Hietarinta1992}
J.~Hietarinta,
``All solutions to the constant quantum Yang--Baxter equation in two dimensions,''
\href{https://doi.org/10.1016/0375-9601(92)90044-M}{Phys. Lett. A \textbf{165}, 245--251 (1992)}.

\bibitem{Hietarinta1993}
J.~Hietarinta,
``Solving the two-dimensional constant quantum Yang--Baxter equation,''
\href{https://doi.org/10.1063/1.530185}{J. Math. Phys. \textbf{34}, 1725--1756 (1993)}.

\bibitem{Fendley2004}
P.~Fendley, K.~Sengupta and S.~Sachdev,
``Competing density-wave orders in a one-dimensional hard-boson model,''
\href{https://doi.org/10.1103/PhysRevB.69.075106}{Phys. Rev. B \textbf{69}, 075106 (2004)}.

\bibitem{Ljubotina2023}
M.~Ljubotina, J.-Y.~Desaules, M.~Serbyn and Z.~Papi\'{c},
``Superdiffusive energy transport in kinetically constrained models,''
\href{https://doi.org/10.1103/PhysRevX.13.011033}{Phys. Rev. X \textbf{13}, 011033 (2023)}.

\bibitem{Logaric2025}
L.~Logari\'{c}, J.~Goold and S.~Dooley,
``Hilbert subspace ergodicity,''
\href{https://doi.org/10.1103/PhysRevB.111.144310}{Phys. Rev. B \textbf{111}, 144310 (2025)}.

\bibitem{SalbergerKorepin2017}
O.~Salberger and V.~Korepin,
``Entangled spin chain,''
\href{https://doi.org/10.1142/S0129055X17500313}{Rev. Math. Phys. \textbf{29}, 1750031 (2017)}.

\bibitem{Stace2009}
T.~M.~Stace, S.~D.~Barrett and A.~C.~Doherty,
``Thresholds for topological codes in the presence of loss,''
\href{https://doi.org/10.1103/PhysRevLett.102.200501}{Phys. Rev. Lett. \textbf{102}, 200501 (2009)}.

\end{thebibliography}
\end{document}